\documentclass[binding = 0.6cm]{sapthesis}
\usepackage{microtype}
\usepackage[utf8]{inputenc}
\usepackage{hyperref}
\hypersetup{pdftitle={PhD_thesis_Silvetti},pdfauthor={Federico Silvetti}}
\usepackage[toc, page]{appendix}
\usepackage{cite}
\usepackage{slashed}
\usepackage{amssymb}
\usepackage{subcaption}

\allowdisplaybreaks
\newcommand{\Nc}{N_{\mathrm{c}}}

\newcommand{\sigh}{{\hat \sigma}}
\newcommand{\Acal}{{\cal A}}
\newcommand{\Mcal}{{\cal M}}

\newcommand{\order}[1]{{\cal O}\(#1\)}
\newcommand{\xbar}{\bar{x}}
\newcommand{\ybar}{\bar{y}}
\newcommand{\as}{\alpha_s}

\newcommand{\iu}{\mathrm{i}}
\newcommand{\dd}{\mathrm{d}}

\newcommand{\calC}{\mathcal{C}}
\newcommand{\calL}{\mathcal{L}}
\newcommand{\calF}{\mathcal{F}}
\newcommand{\calO}{\mathcal{O}}

\newcommand{\barQ}{\bar{Q}}
\newcommand{\tilA}{\tilde{A}}
\newcommand{\tilF}{\tilde{F}}
\newcommand{\tilN}{\tilde{N}}
\newcommand{\tilC}{\tilde{C}}
\newcommand{\tilK}{\tilde{K}}

\newcommand{\CE}{C_{\text{eff}}}

\newcommand{\pt}[1]{p_{{\rm t}#1}}
\newcommand{\kb}{\mathbf{k}}
\newcommand{\pb}{\mathbf{p}}
\newcommand{\dCH}{d_{\text{OFF}}}
\newcommand{\spd}[2]{\(#1\cdot#2\)}
\newcommand{\hell}{\href{https://www.roma1.infn.it/~bonvini/hell/}{\texttt{HELL}}}

\newcommand{\intv}{\int \frac{\dd^Dv}{(2\pi)^D}}
\newcommand{\eu}{\mathrm{e}}
\newcommand{\PlusTau}{\(\frac1{\frac{1}{1+\xi}-\tau}\)_+}

\newcommand{\kvec}{\textbf{k}}
\newcommand{\pvec}{\textbf{p}}
\newcommand{\qvec}{\textbf{q}}
\newcommand{\qt}{q_{\rm t}}
\newcommand{\kt}{k_{\rm t}}

\newcommand{\qthat}{\hat q_{\rm t}}
\newcommand{\pthat}{\hat p_{\rm t}}

\newcommand{\HW}{{HW}}

\newcommand{\NNLO}{\text{NNLO}~}
\newcommand{\NNLL}{\text{NNLL}~}

\newcommand{\NNNLO}{\text{N3LO}~}
\newcommand{\NNNLL}{\text{N3LL}~}
\newcommand{\NNNLLp}{\text{N3LL'}~}
\newcommand{\NNNNLL}{\text{N4LL}~}
\newcommand{\RadISH}{\texttt{RadISH}}
\newcommand{\MCFM}{\texttt{MCFM}~}
\newcommand{\rC}{\mathrm{C}}
\newcommand{\dZ}{\dd \mathcal{Z}}
\newcommand{\muR}{\mu_R}
\newcommand{\muF}{\mu_F}
\newcommand{\LambdaQ}{\Lambda_{\rm QCD}}

\def\({\!\left(}
\def\){\right)}
\def\[{\!\left[}
  \def\]{\right]}
\newcommand{\lb}{\bigg\lbrace}
\newcommand{\rb}{\bigg\rbrace}
\def\nn{\nonumber \\}
\def\permille{\ensuremath{{}^\text{o}\mkern-5mu/\mkern-3mu_\text{oo}}}
\newcommand{\xb}{\bar{x}}
\newcommand{\qb}{\bar{q}}
\newcommand{\ub}{\bar{u}}
\newcommand{\db}{\bar{d}}
\newcommand{\sbar}{\bar{s}}
\newcommand{\Ab}{\bar{A}}
\newcommand{\Cb}{\bar{C}}
\newcommand{\Kb}{\bar{K}}
\newcommand{\bb}{\bar{b}}
\newcommand{\btil}{\tilde{b}}
\newcommand{\Atil}{\tilde{A}}
\newcommand{\rmf}{\mathrm{f}}

\newcommand{\xfitter}{\texttt{xfitter}}
\newcommand{\HERAPDF}{\texttt{HERAPDF}$2.0$}
\newcommand{\QSPDF}{\texttt{QSPDF}}

\newcommand{\uda}{{\uparrow\!\downarrow}}
\newcommand{\dua}{{\downarrow\!\uparrow}}
\newcommand{\uparr}{\uparrow}
\newcommand{\dwarr}{\downarrow}

\newcommand{\MSbar}{\overline{\text{MS}}}

\newcommand{\hypF}{{}_2F_1}
\newcommand{\gE}{\gamma_{\text{E}}}
\newcommand{\polyg}{\psi^{(0)}}
\newcommand{\polygp}{\psi^{(1)}}
\newcommand{\dilog}{\mathrm{Li}_2}
\newcommand{\NN}{\nonumber \\ & \qquad}
\newcommand{\calP}{\mathcal{P}}
\DeclareMathOperator\arcoth{arcoth}
\newcommand{\dvec}[2]{\left(
    \begin{array}{c}
      #1 \\ #2
    \end{array}
\right)}
\newcommand{\thetapair}{\psi}
\newcommand{\Dvec}{\mathbf{\Delta}}

\title{Resummation phenomenology and PDF determination for precision QCD at the LHC}
\author{Federico Silvetti}
\IDnumber{1591304}
\course{Dottorato di Ricerca in Fisica}
\cycle{XXXV}
\courseorganizer{Scuola Dottorale in Scienze Astronomiche, Chimiche, Fisiche, Matematiche e della Terra "Vito Volterra"}
\AcademicYear{2022/2023}
\copyyear{2023}
\advisor{Dr. Marco Bonvini}
\submitdate{September 2023}
\reviewerlabel{Referees}
\reviewer{Prof. Simone Alioli, Università degli Studi Milano-Bicocca}
\reviewer{Prof. Giancarlo Ferrera, Università degli Studi di Milano}
\examdate{18 September 2023}
\examiner{Prof. Roberto Bonciani, Sapienza Università di Roma}
\examiner{Prof. Giancarlo Ferrera, Università degli studi di Milano}
\examiner{Prof. Alessandro Papa, Unversità della Calabria}
\cycle{XXXV}
\versiondate{March 2024}
\copyrightstatement{ This work is licensed under Attribution-NonCommercial 4.0 International }
\authoremail{federico.silvetti@uniroma1.it}

\begin{document}
	\frontmatter
	\maketitle
	\dedication{%Scemo chi legge.
          the hard path of thought\\
          your former self destroyed\\
          the dreaming way is eased\\
          down to the crushing center\\
          and spared the dance of forever\\
        {\rm Marathon: Infinity, Electric Sheep One}}

	\begin{abstract}
          With the ongoing Run 3 of the LHC and its upcoming High-Luminosity upgrade, there is a growing need to study observables that can be both experimentally measured and theoretically predicted with high precision.
          To match the demand for increased precision on the theory side, improvements of fixed-order perturbative predictions, resummation of logarithmic enhancements and accurate determination of proton structure are required.
          
          This thesis provides an exploration of the latter two topics. 
          We discuss high-energy logarithms and their resummation techniques, introducing an extension of the \hell~formalism (High Energy Large Logarithms) for multi-differential distributions in transverse momentum, rapidity and invariant mass. As a phenomenological study, we apply this framework to heavy-quark pair production at the LHC, studying the kinematics of both a single quark and the final-state pair. 
          Part of the thesis is dedicated to studying an extension of the $\kt{}$-factorisation framework, which underlies high-energy resummation, to capture next-to-leading logarithmic corrections. To put this hypothesis to the test, we delve into the computation of a NLO off-shell coefficient function. We use Higgs-induced Deep Inelastic Scattering in the limit $m_t\rightarrow\infty$ as a benchmark process and report a partial result.

          Beside high-energy logarithms, we consider the determination of transverse-momentum distributions from a high-invariant mass system with additional QCD radiation and exclusive cuts on the final state. Specifically, we focus on $HW^+$ associate production with a veto on the leading jet and analyse the Higgs transverse momentum spectrum at NNLO, using $\qt$-subtraction. We complement the fixed order study  with NNLL resummation of jet-veto logarithms and linear power correction in $\pt{HW}$ using the $\RadISH$ (RADiation off Initial State Hadrons) formalism.

          Finally, in the last project pertaining to this thesis we consider the problem of Parton Distribution Function determination. We propose a minimal parametrisation guided by physical arguments and investigate its performance in fitting the HERA dataset with NLO QCD theory predictions.
          \end{abstract}

	\tableofcontents

	\mainmatter
                
	\chapter*{Introduction} \addcontentsline{toc}{chapter}{Introduction}
	A decade after the discovery of the Higgs Boson, the research programme of the Large Hadron Collider (LHC) has shifted into its precision era. While measurements from the previous LHC runs have confirmed validity of the Standard Model as a fundamental theory of Nature, some phenomena outside the direct domain of collider physics, such as the nature of dark matter, matter-antimatter asymmetry, the origin and pattern of neutrino masses and the lack of CP violation in strong interaction are still largely not understood.
Explaining these phenomena demands the introduction of new physics Beyond the Standard Model (BSM). Yet, the search for clear signals remains inconclusive.
The ongoing Run 3 of the LHC, along with its planned High-Luminosity upgrade, is expected to yield a wealth of new experimental data.
More than ever before, it is essential to define and study observables that can be both experimentally measured and theoretically predicted with percent uncertainty.

On the theory side, achieving finer control over Quantum Chromodynamics (QCD) effects, which dominate at hadron colliders like the LHC, requires several inputs. Fixed-order calculations have seen significant advancements, with the computation of higher-order corrections such as next-to-next-to-leading order (NNLO) being commonplace and promising developments for third-order corrections (N3LO).
Beside fixed order theory, resummed computations are necessary to control the dependence of observables in kinematical region where logarithmic enhancements are important and all-order contributions must be taken into account.

In the first part of this thesis we consider small-$x$ resummation. This technique addresses high-energy logarithms of the form $\alpha_s^n \frac{1}{x}\log^k\frac{1}{x}$, with $x$ being the ratio between a hard scattering scale and collider energy. These become relevant when the latter is large. We present a framework to perform this kind of resummation at leading logarithmic accuracy on distributions differential in invariant mass, transverse momentum and rapidity of an arbitrary process at the LHC and showcase a phenomenological study of its impact for heavy-quark pair production. On the same topic, we present an attempt of extending the underlying resummation strategy to next-to-leading accuracy, using Higgs-induced Deep Inelastic Scattering in the large top-quark mass limit as a working case. Alongside the improvement of collider observables, the interest in this type of resummation is linked to the problem of Parton Distribution Function (PDF) determination, where small-$x$ effects are known to play an important role to describe low-invariant mass and high-energy data.

Generally speaking, resummation becomes mandatory also when one considers increasingly differential observables. Indeed, as the number of energy scales grows larger, so will the number of the kinematic regions which may develop logarithmic enhancements. 
One such case is the transverse-momentum distribution of a system with high invariant mass $Q$, produced with extra QCD radiation and when applying exclusive cuts on the final state.
This kind of system is known to require  resummation as the convergence of the perturbative expansion is hindered by large logarithmic terms of the form ${\log\(\frac{Q}{p_{t}}\)}$, where $p_{t}$ is some transverse momentum scale and $Q$ is the hard scale of the process.
In this thesis we present a study of the Higgs $\pt{}$ spectrum in $W^+H$ associated production at NNLO with a veto on the leading-jet transverse momentum. The fixed-order prediction is obtained via $\qt{}$-subtraction and we complement the fixed order with NNLL resummation of the jet veto, as well as linear power corrections in $\pt{HW}$, with the \RadISH\ formalism.

On top of perturbative computations, robust and accurate determinations of the proton structure in the form of PDFs are necessary to obtain reliable theory predictions.
Since PDFs encode the non-perturbative information in the hadron scattering dataset they are extracted from, their functional expression is not known from first principle. State-of-the-art PDF sets avoid introducing biases by parametrising PDFs with general polynomials or neural networks.  
Instead, within the scope of this thesis, we investigate the opportunity of building a PDF parametrisation biased with physical arguments, using a simple model of the proton as a quantum-statistical ensemble. In order to validate the viability of this alternative, we test this construction by fitting it against the HERA dataset, using NLO theory and comparing the best fit with the performance of \HERAPDF\ under the same conditions.

\begin{figure}
  \centering
  \includegraphics[width= 0.8\linewidth]{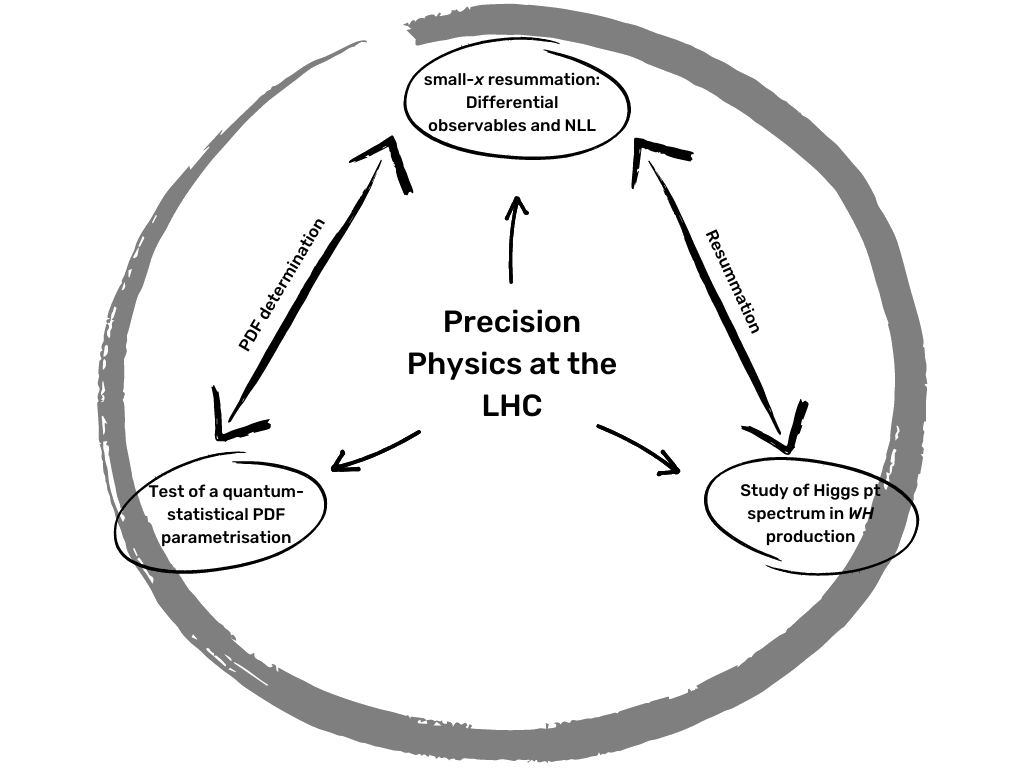}
  \captionsetup{labelformat=empty}
  \caption{\textbf{Figure 0.} Synthetic diagram of the thesis contents. The three main topics of the research at hand are contained in the black circles. The two double-headed arrows highlight overlapping applications.}
  %\label{fig:thesis_diagram}
\end{figure}
This thesis is organised as shown in figure 0. %~\ref{fig:thesis_diagram}.
Chapter~\ref{ch1} serves as an introductory chapter, providing an overview of perturbative QCD features and the running coupling. It discusses the logarithmic enhancements that arise in certain processes and the necessity of resummation techniques to accurately account for them. The chapter also explores topics such as the initial-state infrared singularity, factorisation of splitting functions, and presents the collinear-factorisation master formula for cross-section calculations.

In chapter~\ref{ch2}, the focus shifts towards high-energy logarithms and their resummation techniques. The chapter introduces the HELL (High Energy Large Logarithms) formalism and examines its application in resumming differential cross-sections. To illustrate the practical implications, a phenomenological analysis of Heavy-Quark pair production is presented.

Chapter~\ref{ch2a} provides a more detailed overview of the $\kt{}$-factorisation approach for high-energy resummation at the leading logarithmic (LL) level. The chapter proposes an argument for extending the resummation theorem to next-to-leading logarithmic (NLL) accuracy. Furthermore, it explores the application of $\kt{}$-factorisation in the context of Higgs-induced Deep Inelastic Scattering as the test-bed for the improvement of the resummation.

Moving to chapter~\ref{ch3}, a comprehensive review of the resummation of transverse observables is presented within the RadISH framework. The chapter examines the theoretical foundations of the framework and showcases its application in studying $HW^+$ associate production.

In the final chapter~\ref{ch4}, the focus shifts towards PDF determination. The chapter provides a review of current state-of-the-art techniques and proposes a simplified PDF parametrisation rooted in a quantum-statistical model of the proton. To validate the proposed parametrisation, a test is conducted using the HERA dataset and NLO QCD theory.

The tail of the thesis contains appendix material. Specifically, appendix~\ref{app:NLO-corr} gathers common formulas and expression. We collect in appendix~\ref{app:HQ} the  detailed expressions for the differential coefficient functions discussed in chapter~\ref{ch2}. Finally, appendix~\ref{app:HDIS-NLO} is devoted to a detailed analysis of the computation behind the contents of chapter~\ref{ch2a}.

In the concluding chapter, we recap the results and prospects of each project, as well as draw some link between them to build upon in future works.

	\chapter{Perturbation theory and QCD}
        \label{ch1}
	QCD, or Quantum Chromo-Dynamics, is the modern theory that describes the fundamental degrees of freedom of the strong nuclear force, quarks and gluons, and how they manifest into observable hadrons.
In the scope of using hadron colliders as laboratories for studying precision physics, it is mandatory to have a precise theoretical understanding of QCD dynamics at all energy scales, ranging from low energy hadronisation (around the mass of a proton) to the energy of hard-scattering processes, probing the TeV scale and beyond.

This is enabled by one essential features of QCD: asymptotic freedom. That is, the progressive shrinking of the effective QCD coupling at high energy.
Thus, at a suitable energy scale, the individual elements forming a composite object, called partons, can be treated as free particles and physical observables can be computed as power series in the strong coupling.

Since hadron colliders \textit{collide hadrons} instead of their underlying fundamental constituents directly, the theoretical description of a high-energy scattering involves both short and long-distance behaviour. Unfortunately, low-energy large-distance QCD phenomena escape the perturbative  framework as the strong coupling grows larger.
Collinear factorisation solves this issue by allowing a systematic separation the short-distance, perturbative process-dependent physics from the long-distance, non-perturbative physics, which is encoded in universal objects called Parton Distribution Functions (PDFs) and obtained through non-perturbative techniques or extracted from experimental data.

In this chapter we will elucidate a brief overview of QCD and introduce some notation we shall use in this thesis, mostly parroting Refs.~\cite{Bonvini:2012sh,Rottoli:2018nma}. The many interesting properties and intricacies of QCD are covered by a very extensive literature, for example Refs.~\cite{Ellis:1991qj,Peskin:1995ev,Muta:1998vi,Collins:2011zzd,Schwartz:2014sze,Campbell:2017hsr}.
\section{QCD and strong interaction}
At its core QCD is a renormalisable Yang-Mills gauge theory~\cite{Yang:1954ek} based on the gauge group $SU(3)_{\textrm {colour}}$.
The particle content is organised in matter fields, called quarks, which come in six different flavours, and eight massless gauge bosons, the gluons, to mediate the interactions.
\subsection{The QCD Lagrangian}
At the classical level QCD dynamics is encoded in the Lagrangian 
\begin{subequations} \label{1:eq:QCDlagr}
\begin{align}
        \mathcal L_{\textrm{YM}} & = -\frac{1}{2} {\text{Tr}}\[ F^{\mu\nu} F_{\mu\nu}\] + \sum_{j=1}^{n_f} \bar q_{j} (\iu \slashed D -m_j ) q_j, \, \\
        D_\mu & = \partial_\mu + i g_s A_\mu^a t^a \, ,\\
        F_{\mu \nu} & = (\partial_\mu A^a_\nu - \partial_\nu A^a_\mu -g_s f^{abc} A_\mu^b A_\nu^c ) t^a,
\end{align}
\end{subequations}      
where $q_j$ are the fermionic fields in the fundamental representation of $SU(3)_{\textrm {colour}}$, with masses $m_j$ and $n_f$ different flavours, $A_\mu^a$ are the gluon fields in the adjoint representation of $SU(3)_{\textrm {colour}}$ and $g_s$ is the gauge coupling. 
The fundamental representation of the $SU(3)_{\textrm {colour}}$ Lie algebra includes a set of generators $t^a$ which satisfy
\begin{align}
	&[t^a,t^b]=\iu f^{abc}t^c, & {\text{Tr}} [t^a, t^b] = T_R \delta^{ab},
\end{align}
where $f^{abc}$ are the real structure constants of the algebra, $a,b,c=1,\ldots 8$, and it is customary to set the Dynkin index $T_R=1/2$. 
On the other hand, the Casimir invariants $C_F$ and $C_A$ are defined by
\begin{align}
	& \sum_a t^a_{ik} t^a_{kj} = C_F  \delta_{ij}, & \sum_{ab} f^{abc} f^{abd} = C_A \delta^{cd} , \\
        & C_F = \frac{\Nc^2-1}{2 \Nc} = \frac{4}{3}, & C_A = \Nc = 3 \, .
\end{align}
The classical Lagrangian in Eqs.~\eqref{1:eq:QCDlagr} is invariant under local transformation of $SU(3)_{\textrm {colour}}$. However the switch from the classical field theory and its quantum counterpart requires the introduction of a gauge-fixing term $\mathcal L_{\textrm{gauge-fixing}}$.
The most common and versatile choice is using a covariant gauge defined by
\begin{align}
	&\mathcal L_{\textrm{gauge-fixing}}= -\frac{1}{2\xi}(\partial^\mu A_\mu)^2,
	&\mathcal L_{\textrm{ghost}}= \partial^\mu {\bar \eta}^a (\partial_\mu \delta^{ab} + g f^{abc} A^c_\mu )\eta^b,
\end{align}
where $\xi$ is an arbitrary gauge parameter and $\mathcal L_{\textrm{ghost}}$ is an artefact of the gauge-fixing procedure. The newly introduced ghost fields $\eta$ are complex scalar fields which obey Fermi-Dirac statistics and are necessary to avoid the propagation of nonphysical gluon polarisations.

Of course there is an entire functional degree of freedom in setting up the gauge fixing. For example, a common alternative is the family of axial gauges $n^\mu A_\mu^a =0$, where $n$ is an arbitrary four-vector and the gauge-fixing and the ghost terms have the form (see e.g.~\cite{Leibbrandt:1987qv,Bassetto:1991ue})
\begin{equation}
	\mathcal L_{\textrm{gauge-fixing}}= -\frac{1}{2\xi}(n^\mu A_\mu)^2, \qquad
	\mathcal L_{\textrm{ghosts}}= \bar \eta^a n^\mu  (\partial_\mu \delta^{ab} + g f^{abc} A^c_\mu ) \eta^b.	
\end{equation}
In the common light-cone gauge, one sets $n^2=0$, $\xi =0$.
This class of gauges has the important property of explicitly allowing only two polarisation states for the gluons and completely decoupling ghosts so that they can be ignored.

Naturally  the gauge-fixing and the ghost terms explicitly break gauge-invariance. However, the full, gauge-fixed action fulfils the Becchi-Rouet-Stora-Tyutin (BRST) symmetry~\cite{Becchi:1975nq,Tyutin:1975qk}, which guarantees the renormalisability of the theory at quantum level~\cite{Weinberg:1996kr}.

\subsection{Perturbation theory and running coupling}
The interactions in the QCD Lagrangian Eq.~\eqref{1:eq:QCDlagr} are governed by a single parameter, the gauge coupling $g_s$, customarily recast as the strong coupling parameter
\begin{align}
 \as = \frac{g_s^2}{4 \pi}\, ,
\end{align}
in analogy with QED.

Physical observables (i.e. squared amplitudes and their corresponding cross-sections) can be extracted from the QCD Lagrangian Feynman rules and form perturbative expansions in terms of $\as$. However, when loop diagrams are calculated in this way,  infinities arise and the theory must be regularised and renormalised to obtain meaningful predictions.
Several techniques can be used, for example dimensional regularisation (dimreg) or Principal Value (PV).
The former consists in the observation that many ultra-violet (UV) divergent integrals in four dimensions are instead finite if the theory is embedded instead in $D=4-2\epsilon$ dimensions (assuming $\epsilon > 0$)\footnote{Infra-red (IR) singularities can be controlled in the same way with $\epsilon<0$. Indeed, analytical continuation often allows to use dimreg to tame both these regions with one parameter.}, and the divergences are isolated in poles in the parameter $\epsilon$,
\begin{equation}
\int_0^1\dd x \frac{1}{x} \rightarrow \int_0^1 \dd x x^{\epsilon -1} = \frac{1}{\epsilon}\,.
\end{equation}
Principal value instead instructs to turn a pole in a integral into a finite quantity by shifting a real singularity into two imaginary ones. Explicitly
\begin{equation}
\int_0^1\dd x \frac{1}{x} \rightarrow \frac{1}{2}\int_0^1 \dd x \[ \frac{1}{x+\iu \delta}+\frac{1}{x-\iu \delta} \] = \int_0^1\dd x \frac{x}{x^2 + \delta^2} \simeq \log(\delta)\, .
\end{equation}
 Either way, the crucial property of a renormalisable field theory is that the dependence on the cutoff parameter can be reabsorbed into a redefinition of the gauge coupling, particle masses, and wave function normalisation.

In principle, QCD would be scale invariant as long as quarks are considered massless, however this does not actually hold  at the quantum level and the renormalised Lagrangian parameters acquires a logarithmic dependence on a renormalisation scale $\mu$, ruled by Renormalisation Group Equations (RGEs)
\begin{subequations}
        \begin{align}
                & \mu^2 \frac{\dd \as( \mu^2)}{\dd \mu^2} = \beta \(\as ( \mu^2)\), \\
	        & \beta\(\as ( \mu^2)\) = - \as^2 ( \mu^2)  \(\beta_0+ \beta_1 \as ( \mu^2) + \beta_2 \as^2 ( \mu^2)+ \ldots\),
        \end{align}
\end{subequations}
where $ \beta \(\as( \mu^2) \)$ is known as the QCD $\beta$-function and its coefficients are computable in perturbation theory. At the current state of the art, the $\beta$-function is known up to five loops~\cite{Herzog:2017ohr,Luthe:2017ttg}. 

The RGE for the running coupling can be solved by iteration. 
At the lowest (leading logarithmic) order the solution is
\begin{equation}\label{1:eq:as1loop}
	\as(\mu^2)  = \frac{\as(\mu_0^2) }{1 + \beta_0 \as(\mu_0^2)\log \frac{\mu^2}{\mu_0^2}},
\end{equation}
where $\mu_0$ is an initial scale at which the coupling is known or measured.

The $\beta$ function controls the slope of the running of $\alpha$ across values of the scale $\mu^2$. Indeed, the positive sign of its first term\footnote{At least as long as there are no more than 16 active quark flavours.} $\beta_0 = \frac{11 C_A-2n_f}{12 \pi}$, ensures that the coupling is vanishing at large scales. This property is called \emph{asymptotic freedom} and it is an exclusive property of local non-Abelian gauge theories among renormalisable field theories in four-dimensional spacetime~\cite{Coleman:1973sx}.
Conversely, the coupling is divergent in the $ \mu^2 \rightarrow 0$ limit and  perturbative methods can no longer be trusted. The boundary between these two regimes is customarily identified with the Landau pole, which is defined by the equation
\begin{align}
        \frac{1}{\as(\LambdaQ^2)}  = 0.
\end{align}
At this scale, which is typically of the order of a few hundred MeV, QCD becomes strongly coupled and the impossibility of detecting colour charges, or \textit{confinement}, becomes apparent. 

\subsection{Logarithmic enhancements and resummation} \label{1:ssec:logs-and-res}
As we have just seen, the structure of perturbation theory in QCD generates recursive patterns of singularities that need to be systematically subtracted to guaranteed mathematical consistency. As a consequence of this subtraction, a regular pattern of powers of $\as(\mu_0^2) \beta_0 \log\(\frac{\mu^2}{\mu_0^2}\)$ is left. In principle, if the ratio between the two scales is large, the logarithms could compensate the smallness of $\as(\mu_0^2)$. This is avoided, in practice, by the RGE equation~\eqref{1:eq:as1loop}, which consists in a geometric series of all these (leading) logarithmic contributions.

We can make this argument more general by considering a physical observable, (i.e. a cross section $\sigma$). This will depend on all the relevant energy scales and we compute it as a power series of the coupling $\as$.
\footnote{Here we assume QCD to be the interaction driving the process, however at the Electro-Weak scale all coupling constants of the Standard Model are at least $\calO\(10^{-1}\)$, or smaller, and, in fact, computing mixed QCD-EW corrections is an open field of investigation.}
Naturally, the coefficients of the power expansion will be functions of all the arguments of $\sigma$ except for $\as$. In our example we denote this as
\begin{align}
	\sigma \(Q^2, S, \hdots ,  \as\)  =\sigma_0 \(Q^2, S, \hdots , \as\) \[1+\sum_{n=1}^\infty c_n(Q^2, S , \hdots ) \as^n \]\, , \label{1:eq:pertexp}
\end{align}
where $\sigma_0$ is the Born-level cross section and $c_n$ are the higher-order corrections from the $n$-th power of $\as$. The dots denote all the other additional variables which the observable may depend upon.
	
Since the $c_n$ are not dimensionful quantities, they may contain only terms formed by dimensionless combinations of their arguments, which we will generally denote as $x$.
Logarithms $\log(x)$ are a primary concern. Indeed, the structure of radiative corrections will generate these terms recursively, with each $c_n$ containing up to one extra power of the logarithm for each extra order in $\as$. This pattern is called a single-logarithmic enhancement\footnote{This is not the only recurrent structure that can emerge in radiative corrections as we will see in more detail in chapters~\ref{ch2a} and~\ref{ch3}. For another example, threshold effects in QCD \cite{Catani:1989ne, Idilbi:2006dg} are characterised by a double-logarithmic enhancement, with up to two extra powers of the logarithm for each extra order in $\as$.} and schematically we can write it down as 
\begin{samepage}
\begin{align}
	\sigma \(Q^2, S , \cdots,  \as\) = \sigma_0 \(Q^2, S , \cdots,  \as\)\, \bigg( & \, 1 + \nonumber \\
		& \, \as\[   a_1 \log^{~}(x) +  b_1\] \nonumber \\
		& \, \as^2\[ a_2 \log^2(x) +  b_2 \log(x) +  c_2 \] \nonumber \\
		& \, \:\vdots \qquad\qquad \ddots  \qquad\qquad \ddots \nonumber \\ 
		& \, \as^n \[ a_n \log^n (x) +  b_{n} \log(x) +  \cdots  \] \bigg) \, . \label{1:eq:logblock}
\end{align}
\end{samepage}\\
Roughly speaking, when $\alpha \log(x) \simeq \order{1}$ all terms along the first column of \eqref{1:eq:logblock} will be of the same order. In such kinematic conditions, truncating the series at any $n$ will generate errors as large as the truncated series itself, instead of the desired $\order{\alpha^{n+1}}$.
Restoring the validity of the perturbative expansion can be achieved grouping together large terms by the power of the logarithms. In our example we can formally write  
\begin{align}
\sigma \(Q^2, S , \ldots,  \as \) & = \sigma_0 \(Q^2, S , \ldots,  \as \) \nn
& \times \[ \sum_{k=0}^{\infty}\( \as a_k \log(x) \)^k +\as \sum_{k=0}^{\infty}\( \as b_k \log(x) \)^k + \ldots (\text{non log})  \] \,  , \label{1:eq:logexpansion}
\end{align}
and noting that we have reconstructed a series, with each successive terms suppressed by an extra power of $\as$ and all logarithmic pieces confined in the coefficients of each order. Starting from $n=0$, we call each inner series leading logarithms (LL), next-to-leading logarithms (NLL) and so on. Then, in order to restore the validity of perturbation theory, it is necessary to identify a suitable strategy to capture all terms inside the inner series up to the desired precision.
\begin{align}
\sigma \(Q^2, S , \ldots,  \as \) \sim  \sigma_0 \(Q^2, S , \ldots,  \as\) \[ g_0 (\text{LL}) + \alpha g_1 (\text{NLL}) + \ldots + (\text{non log terms}) \]  \,  .
\end{align}
This is as far as this hand-wave argument can get us. Indeed, explicitly building the resummed expression $g_k$ requires specifying the actual origin of the enhanced logarithms inside the scattering amplitude.
\section{Initial state singularities and DGLAP equations}
As just introduced, studying high-energy processes involving hadrons requires introducing a technique to bridge the gap between the interaction at hadron and parton level: the \emph{parton model}\footnote{For simplicity, we will assume hadron to mean proton, but what we will say can be in principle applied to any other hadron.}.

The original (or naive) parton model was proposed by Feynman by assuming an ultra-relativistic proton with momentum $p$ much larger than its mass. In this \emph{infinite momentum frame}, we can neglect the effects of mass both from the proton itself and from its constituents, each carrying a fraction $z_i$ of the total momentum $p$ of the proton,
\begin{equation}\label{1:eq:pm_assumption}
 p_i = z_i p,
\qquad 0\leq z_i \leq 1\, .
\end{equation}
These momentum fractions are not fixed by the model and obey instead a PDF $f_i(z_i)$.

Then, the  interaction between a proton and an elementary particle is described by an incoherent sum of the interaction between each parton and the elementary particle, weighted with the PDFs, 
\begin{equation}\label{1:eq:parton_model}
\sigma(p) = \sum_i \int_0^1 \dd z\, f_i(z)\, \hat\sigma_i(zp),
\end{equation}
where $\hat\sigma_i$  is the cross section for the partonic process.

Intuitively, PDFs describe the probability to pick up a parton $i$ with momentum fraction between $z$ and $z+\dd z$.
Some additional integral constraints apply in order to enforce the correct charge for the proton.
First the difference between quarks and antiquarks PDFs, integrated in $z$, counts the number of constituents quarks of the proton:
\begin{equation}\label{1:eq:valence_sr}
\int_0^1 dz\, \[ f_u(z) - f_{\bar u}(z) \] = 2,
\qquad
\int_0^1 dz\, \[ f_d(z) - f_{\bar d}(z) \] = 1,
\end{equation}
and all the other partons give zero. This is to enforce the correct flavour and charge combination for the proton.
By the same token, the sum of the momenta of all partons must equal the proton momentum:
\begin{equation}\label{1:eq:momentum_sr}
\sum_i \int_0^1 dz\, z\, f_i(z) = 1.
\end{equation}
While the partonic cross-section $\hat\sigma_i(zp)$ is computable from first principles, the PDFs are instead intrinsically non-perturbative objects, and are usually extracted from experimental measurements. The sum rules provide important constraints on the extraction process.

The way the parton model is formulated here is very simple. However, when we take into account the perturbative corrections of QCD, several issues arise. The resulting solution is known as the ``improved parton model'' and the parton distribution functions start depending on an energy scale.
\subsection{Radiative corrections}\label{1:ssec:Radiative-corrections}

\begin{figure}
	\centering
	\begin{subfigure}{0.48 \linewidth}
        \centering
        \includegraphics[width= 0.7\linewidth]{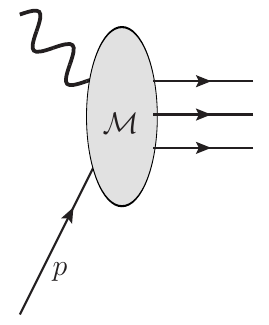}
        \caption{}
       	\label{1:fig:RadL}
	\end{subfigure}
	\begin{subfigure}{0.48\linewidth}
	\centering
        \includegraphics[width=0.7\linewidth]{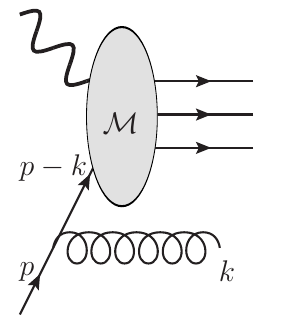}
        \caption{}       
       	\label{1:fig:RadR}
	\end{subfigure}
	\caption{Diagrams of a general hard scattering with an initial state quark. On the left is depicted the tree-level contribution, while on the right the correction from the emission of an on-shell gluon.}
\end{figure}
We will follow Ref.~\cite{Nason:2002int} to build an example of how initial state radiative corrections in QCD work.
We consider a general quark-initiated process and its first-order corrections.
The tree level amplitude is synthetically factorised as
\begin{subequations}
        \begin{align}
                \Acal_0 & = \mathcal{M} \(p\) u \(p\) \, , \\
                \sigh_0 (p) &= \frac{1}{ 2 S} \int \dd \Phi \sum \mathcal{M}   (p) u (p)  \overline{u} (p)  \mathcal{M}^\dagger (p) = \frac{1}{2 S} \int \dd \Phi  \mathrm{Tr} \[ \mathcal{M}(p) \slashed{p} \mathcal{M}^\dagger (p) \],
        \end{align}
\end{subequations}
with $\mathcal{M}$ representing the Born amplitude of the process, $\dd Phi$ its phase space and $\sigh_0$  the unpolarised cross section  (Fig.~\ref{1:fig:RadL} ). Finally, with $S$ stands for total centre-of-mass energy of the system. 
The radiative corrections\footnote{Another contribution, representing the conversion of a gluon into a quark by the emission of an antiquark, should also be accounted for in principle. However we disregard it as we are only interested in showing the simplest example of initial-state collinear singularity.} include the emission of a real gluon (Fig.~\ref{1:fig:RadR}) and a virtual correction to the amplitude. Leaving the details of the computation to appendix~\ref{app:NLO-corr}, we find
\begin{align}
\sigh_1 & = \frac{\alpha_s C_F}{2 \pi } \int \[ \sigh_0 (z p) - \sigh_0 (p) \] \frac{1+z^2}{1-z}  \frac{\dd \kt{}^2 }{ \kt{}^2} \dd z \nn
        & = \frac{\as C_F}{2 \pi } \int \(\frac{1+z^2}{1-z}\)_+ \sigh_0 (zp)   \log(\frac{Q^2}{\LambdaQ^2}) \dd{z}  \, .
\end{align}
The two terms in the square brackets correspond to the contributions from the emission of a gluon and the virtual correction respectively. Both of them would be singular in the soft limit, $z\rightarrow 1$, but their combination is finite.
On the other hand, the collinear limit $\kt{}^2 \rightarrow 0$ is still divergent and requires the introduction of a low energy cut-off like $\LambdaQ$. 
Finally the presence of the low scale in the cross section can be removed by introducing the factorisation scale $\muF^2 \simeq Q^2$, order by order in perturbation theory
\begin{align}
\sigh_{q} (p) & = \sigh_0 + \sigh_1  \nn
             & = \int_{0}^{1} \[\delta (1-z_1) + \frac{\as C_F}{2 \pi} \(  \frac{1+z_1^2}{1-z_1} \)_+ \log \(\frac{Q^2}{\LambdaQ^2}\)  \] \sigh_{0} (z_1 p)  \dd z_1   \nn
             & = \int_{0}^{1} \int_{0}^{1} \[ \delta \(1-z_1\) + \frac{\as C_F}{2 \pi} \(  \frac{1+z_1^2}{1-z_1} \)_+ \log \( \frac{Q^2}{\muF^2} \) \] \nn
             & \: \times \[\delta (1-z_2) + \frac{\as C_F}{2 \pi} \(  \frac{1+z_2^2}{1-z_2} \)_+ \log \( \frac{\muF^2}{\LambdaQ^2} \) \]  \sigh_{0} (z_1 z_2 p)  \dd z_1  \dd z_2 + \order{\as^2} \, .
\end{align}
Crucially, the term including $\log \( \frac{Q^2}{\muF^2} \)$ can now be used to include the radiative corrections in the cross section, while the low-energy cut-off scale $\LambdaQ$ and its collinear singularity  can be subtracted away by redefining the PDF. Then, Eq. \eqref{1:eq:parton_model} becomes
\begin{subequations} \label{1:eq:collinearfactorizationord1}
\begin{align}
\sigma \(x, Q^2\) & = \int_{x}^{1} \frac{ \dd z }{z} \sigh_q \(\frac{x}{z}, Q^2, \muF^2\) f_q\(z, \muF^2\) \, , \label{1:eq:cfo1a} \\
\sigh_q \(\frac{x}{z}, Q^2, \muF^2\) & = \int_{0}^{1} \dd z_1 \[ \delta (1-z_1) + \frac{\as C_F}{2 \pi} \(  \frac{1+z_1^2}{1-z_1} \)_+ \log \( \frac{Q^2}{\muF^2} \) \] \sigh_0 \( \frac{x}{z_1 z}, Q^2\) \, , \\
f_q(z, \muF^2) & = \int_{0}^{1} \frac{\dd z_2 }{z_2} \[ \delta (1-z_2) + \frac{\as C_F}{2 \pi} \(  \frac{1+z_2^2}{1-z_2} \)_+ \log \( \frac{\muF^2}{\LambdaQ^2}\) \] f \(\frac{z}{z_2}\)  \, ,
\end{align}
\end{subequations}
where $x = \frac{Q^2}{S}$, is the lowest momentum fraction probed in the scattering.\footnote{
With a small abuse of notation with respect to Eq. \eqref{1:eq:parton_model}, we made replaced the dependence on the incoming vector $p$ with $x$ in $\sigma$ and in its parton level counterpart. In practice this is equivalent to the replacement $\sigh_0 (zp) \rightarrow \int \dd w \delta (x -  z w ) \sigh_0 (w)$.} 
We can see that in \eqref{1:eq:collinearfactorizationord1} the dependence on the low energy scale $\LambdaQ$ is absorbed in the PDF at the price of introducing a factorisation scale $\muF$ in the PDF itself and in the partonic cross section. We have shown explicitly this result at the first order perturbation theory, but the result holds to all orders \cite{Ellis:1978ty, Ellis:1978sf, Altarelli:1977zs}.

\subsection{DGLAP equations}
In the last section we have seen that the inclusion of radiative corrections in proton-initiated processes involves the subtraction of some singularities from the parton cross-section into the PDFs. 
As a result, both objects develop a dependence on a new scale $\muF$, and an integro-differential equation governing the scale dependence of $f \(z, \muF^2\)$ on $\muF^2$.
%This is done by applying $\frac{\dd }{\dd \log \(\muF^2\)}$ to \eqref{1:eq:cfo1a} and taking advantage of the fact that the total hadronic cross section does not depend on the factorization scale. 
The result of this operation in the general case forms the Dokshitzer-Gribov-Lipatov-Altarelli-Parisi (DGLAP) equation~\cite{Gribov:1972ri,Altarelli:1977zs,Dokshitzer:1977sg}
\begin{equation} \label{1:eq:DGLAP}
	\muF^2 \frac{\dd f_{i}\( x, \muF^2\)}{\dd \muF^2} = \sum_j \int_{x}^{1} \frac{\dd z }{z} P_{ij}\( z, \as\( \muF^2\) \) f_j \( \frac{x}{z}, \muF^2 \) \, ,
\end{equation}
where the splitting functions $P_{ij}$ represent the singular part, in the collinear limit, of the transformation of parton $j$ into parton $i$ with longitudinal momentum fraction $z$ that may happen between the proton state and the hard scattering.

Beyond this distinction, we can show that the evolution of PDF via DGLAP equations is itself a resummation of a series of logarithms. For example, consider a single flavour\footnote{The evolution equation must account for different flavours, but this aspect is not relevant right now and will be addressed in appendix~\ref{app:2:DGLAP}.} equation 
\begin{equation}
	\muF^2 \frac{\dd}{\dd \muF^2} f \(x, \muF^2\)  =  \int_{x}^{1} \frac{\dd z }{z} P \(\frac{x}{z}, \as \(\muF^2\) \) f \(z, \muF^2\) \, ,
\end{equation}
and take a Mellin transformation (see appendix~\ref{app2:Mellintransf} for definitions)
\begin{equation}
       	\muF^2 \frac{\dd }{ \dd \muF^2} f \(N, \muF^2\) =  \gamma \(N, \as (\muF^2) \) f \(N, \muF^2\) \, , \label{1:eq:NSDGLAP}
\end{equation}
where $f (N)$ and $\gamma (N)$ are, respectively, the Mellin transform of the PDF and of the splitting function (the latter is commonly called DGLAP anomalous dimension). Equation \eqref{1:eq:NSDGLAP} can be integrated easily if we limit our computation to the lowest order.
Then, $\gamma \(N, \as\) = \as \gamma_0 (N)$ and $\dd{\as ( \muF^2 )} = - \frac{\beta_0 \as^2 \(\muF^2\)}{\muF^2} \dd \muF^2$. Using these approximations results in
\begin{subequations}
        \begin{align}
		f \(N, \as (\muF^2)\) & = f \(N, \as (\mu_{0}^2)\) \( \frac{\as (\muF^2)}{\as (\mu_{0}^2)}\)^{- \frac{\gamma_0 (N)}{\beta_0}} \, , \nn
                                      & =  f \(N, \as (\mu_{0}^2)\) \[1 + \as (\mu_{0}^2) \beta_0 \log \( \frac{\muF^2}{\mu_{0}^2} \) \]^{ \frac{\gamma_0 (N)}{ \beta_0}} \, . \label{1:eq:NSEVOLORD0}
        \end{align}
\end{subequations}
%where $\beta_0 = \frac{11 C_A - 2 n_f}{12 \pi}$.
It is clear that \eqref{1:eq:NSEVOLORD0} resums to all orders in the coupling powers of $ \as \log \( \frac{\muF^2}{\mu_{0}^2}\)$.
This represents another explicit example of resummation for a single logarithmic enhancement.
One last crucial feature of the PDFs and DGLAP equations is that they do not depend explicitly on any specific process and are \emph{universal}. Once the PDF are given, and the DGLAP splitting functions are known to some order, they can be used to compute any hadronic observable. We defer to chapter~\ref{ch4} a discussion of the state of the art on the determination of DGLAP splittings and their role in PDF analysis.

\subsection{Collinear factorisation at proton proton colliders}
\begin{figure}
	\centering
	\includegraphics[width=\linewidth]{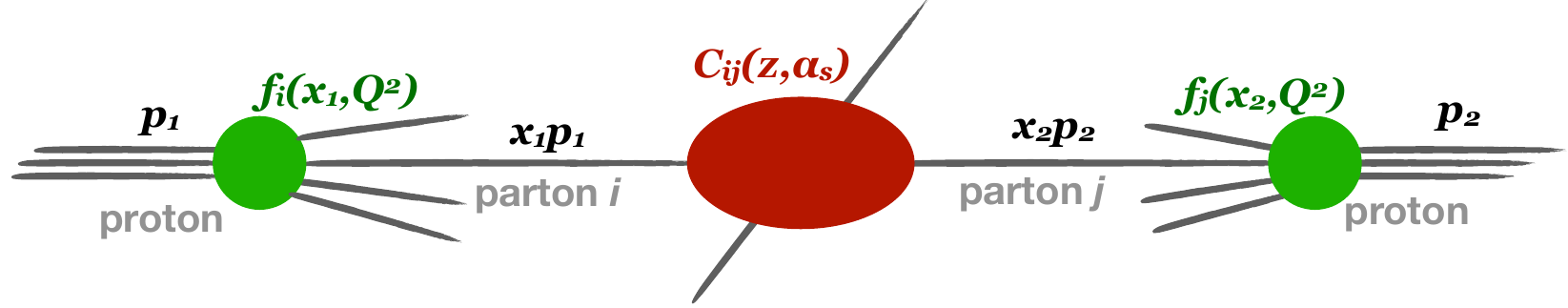}
	\caption{Simplified diagram of a proton collision in collinear factorisation.}
	\label{1:fig:pp}
\end{figure}
In the previous sections we have introduced some of the features a scattering process involving one initial-state proton, which we will now collect together into the general master formula for scattering at the LHC.
Consider the process shown in figure \ref{1:fig:pp}
\begin{equation}
	\mathrm{p}\(p_1\) + \mathrm{p}\(p_2\) \rightarrow \mathrm{F} + \textnormal{remnants} \, , 
\end{equation}
in which two protons collide to produce some final state F. The process is inelastic and inclusive over the remnants of the protons.
Therefore, we can use the same collinear factorisation structure from the previous example and extend it to deal with two hadrons this time. The cross section will then be written in the form
\begin{equation}
        \sigma \(x, Q^2\) = \sum_{ij} \int \int  \dd{x_1} \dd{x_2} \sigh_{ij} \( \frac{x}{x_1 x_2} , \as(\muF^2), \frac{Q^2}{\muF^2}\) f_{i}\(x_1, \muF^2\) f_{j}\(x_2, \muF^2\) + \order{\frac{\LambdaQ^2}{Q^2}}\, , \label{1:eq:2protonsigma}  
\end{equation}    
with $Q^2$ being a hard scale for the process, usually the invariant mass of the final state, $x=\frac{Q^2}{S}$, and $S=(p_1+p_2)^2$ the collider centre-of-mass energy squared. 
The dependence on the factorisation scale $\muF^2$ is introduced, just like in the case of a single proton, to shift the initial state collinear singularities from the coefficient function to the PDFs.
For a general process at hadron-hadron colliders, there is no formal proof of the validity of collinear factorisation to all order of $\as$. Instead, such proof exists for lepton-hadron collisions~\cite{Catani:1994sq, Curci:1980uw} or some inclusive observables \cite{Collins:1989gx,Gross:2022hyw}.

Beyond the total cross-section, we now establish the notation of its differential counterpart in collinear factorisation.
Given a generic final-state momentum $q$ (either the momentum of a single particle or the total momentum of different particles) in the collider centre-of-mass frame, we denote  its invariant mass squared $Q^2\equiv q^2$, rapidity $Y=\frac12\log\(\frac{q^0+q^3}{q^0-q^3}\)$ and transverse component $\qt^2=-\[(q^1)^2+(q^2)^2\]$. Then, the cross-section, differential in these variables, reads
\begin{align}\label{2:eq:collfactdiff}
\frac{\dd\sigma}{\dd Q^2\dd Y\dd \qt^2} \(x, Q^2,Y,\qt^2\) &= x \sum_{ij} \int_x^1 \frac{\dd{z}}z\int \dd y\, \frac{\dd C_{ij}}{\dd Q^2\dd y\dd \qt^2} \(z, Q^2,y,\qt^2,\as, \frac{Q^2}{\muF^2}\) \nn
& \times L_{ij}\(\frac{x}{z}, Y-y, \muF^2\), \\
  L_{ij}\(\xbar, \ybar, \muF^2\) & = f_{i}\(\sqrt{\xbar}\eu^{\ybar}, \muF^2\) f_{j}\(\sqrt{\xbar}\eu^{-\ybar}, \muF^2\) \vartheta\(\eu^{-2\left|\ybar\right|}-\xbar\) \,, \label{eq:lumi}     
\end{align}
again with $x=Q^2/S$ ($S$ is the collider energy squared) and the sum extends over all possible partons $i,j$ in each proton. We also rescaled the partonic cross section as
\begin{equation}
\hat\sigma_{ij}\( \frac{x}{x_1 x_2}, \as \( \muF^2 \) \) =x  \frac{C_{ij}}{Q^2} \( \frac{x}{x_1 x_2}, \as \( \muF^2 \), \frac{Q^2}{\muF^2} \)\, , \label{1:eq:coeffunexpansion}
\end{equation}
to introduce the coefficient function $C$. Indeed, in Eq.~\eqref{2:eq:collfactdiff}, the function is the parton-level coefficient function, which depends on $z=Q^2/ s=\frac{x}{x_1 x_2}$ (the parton-level analogue of $x$) where $s$ is the partonic centre-of-mass energy, and on $y$ which is the rapidity of $q$ with respect to the partonic centre-of-mass frame, and is related to the proton-level rapidity $Y$ by a longitudinal boost.
Finally we can observe that the triple-differential cross section from Eq.~\eqref{2:eq:collfactdiff} has the form of a Mellin-Fourier convolution and it can be diagonalised by taking a suitable transform with respect to $x$ and $Y$,
\begin{subequations}\label{2:eq:MellFourColl}
\begin{align}
& \int_0^1\dd x \, x^{N-1}\int_{-\infty}^\infty \dd Y\,e^{\iu bY}\frac{\dd\sigma}{\dd Q^2\dd Y\dd \qt^2}
= \sum_{ij}\frac{\dd C_{ij}}{\dd Q^2\dd y\dd\qt^2}\(N,Q^2,b,\qt^2,\as, \frac{Q^2}{\muF^2}\)\, L_{ij}\(N,b,\muF^2\), \\
& L_{ij}\(N,b,\muF^2\) = \int_0^1\dd x \, x^N \int_{-\infty}^\infty \dd y \,e^{\iu b y} L_{ij}\(x, y,\muF^2\) = f_i\(N+i\frac b2,\muF^2\)\, f_j\(N-i\frac b2,\muF^2\).
\end{align}     
\end{subequations}
We further observe that the dependence on the transverse momentum does not affect the structure of the cross section formula, and thus impacts only the kinematics.

	\chapter{Heavy-quark pair production and differential small-$x$ resummation}
	\label{ch2}
	In this chapter, we return to the concept of single-logarithmic enhancement introduced in section~\ref{1:ssec:logs-and-res} and focus on the specific class of high-energy logarithms.
We start with Sec.~\ref{2:sec:HElog}, giving a brief overview on  how small-$x$ logarithms arise and how they can be resummed in the DGLAP splitting and coefficient functions. Then, in Sec.~\ref{2:sec:differential} the focus shifts on the topic of resummation in differential observables. We discuss in detail how $\kt{}$-factorisation must be modified in this case and present a numerical implementation for this class of distributions. In Sec.~\ref{2:sec:HQ}, we apply the aforementioned technique to heavy-quark pair production at the LHC as a demonstration that the methodology works and that it can be used for phenomenology~\cite{Silvetti:2022hyc,Silvetti:2022yed}.

\section{High-energy logarithms}\label{2:sec:HElog}
We call high-energy logarithms  the terms in the form $\as^n\frac1x\log^k\frac1x$, $k<n$, where $x$ is the positive dimensionless scaling variable of Eq.~\eqref{1:eq:2protonsigma}.
These perturbative corrections arise beyond the leading order in both the partonic cross sections and the DGLAP splitting functions~\eqref{1:eq:DGLAP} governing PDF evolution (in $\MSbar$-like schemes).
For a given hard scale $Q^2$, larger collider energy $S$ allows to probe decreasing values of $x$. So, for any given process, there will be a kinematical regime where these logarithms compensate the smallness of $\as$ and perturbation theory breaks down.

\begin{figure}[t]
  \centering
  \includegraphics[width=0.7\textwidth]{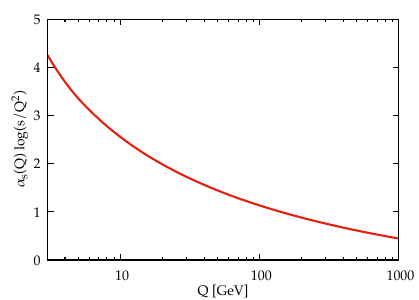}
  \caption{Size of the logarithmic enhanced contributions as a function of the final state scale $Q$ for LHC at $\sqrt{S}=13$~TeV.}
  \label{2:fig:logx}
\end{figure}

At the LHC, at current energy $\sqrt{S}=13$~TeV, the smallest accessible values of $x$ for the production of an Electroweak (EW) final state with $Q\sim80$~GeV is $x_{\rm min} = Q^2/S \sim 3 \times10^{-5}$. For final states characterised by smaller energy scales, the smallest value of $x$ gets even smaller.
A rough way to assess the necessity of resumming these logarithms is studying the value of $\as(Q^2)\log\(\frac{1}{x_{\rm min}}\)=\as(Q^2)\log\( \frac{S}{Q^2} \)$.
This is shown in figure~\ref{2:fig:logx} as a function of $Q$ for fixed $\sqrt{S}=13$~TeV.
From the plot we see that the region where $\as(Q^2)\log\frac1{x_{\rm min}}\gtrsim1$ corresponds to $Q\lesssim 100$~GeV.

Of course, this does not mean that resummation of small-$x$ logarithms is mandatory there: it may be that the region where the logarithms are large gives only a small contribution to the full cross section, Eq.~\eqref{1:eq:2protonsigma}, so that the effect of resummation is negligible.
However, it clearly shows that the small-$x$ terms get larger and larger at small $Q$, so at some point the resummation will certainly have an effect.

\subsection{Resummation of DGLAP splitting functions} 
Starting from leading logarithmic (LL) accuracy, small-$x$ logarithms emerge from kinematics.
Specifically, they appear when multiple gluon emissions occur from a gluon initial-state line, each one carrying away a significant amount of energy.
Therefore, small-$x$ logarithms always involve gluons and only the singlet sector of DGLAP evolution is affected.

\begin{figure}[t]
  \centering
  \includegraphics[width=0.7\textwidth]{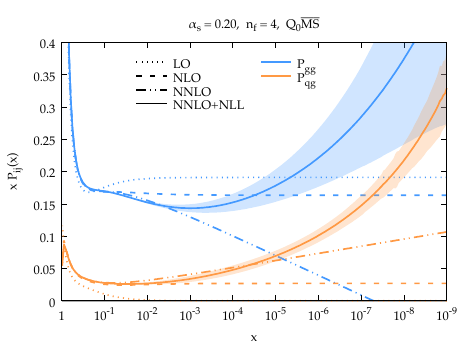}
  \caption{$P_{gg}$ and $P_{qg}$ splitting functions at fixed order and with resummation~\cite{Bonvini:2018xvt}. The error band represents an estimate of the missing sub-leading logarithmic contributions.}
  \label{fig:Pgg}
\end{figure}

In particular, the splitting functions $P_{gg}(x,\as)$ and $P_{gq}(x,\as)$ already contain a logarithmic enhancement at LL, while $P_{qg}(x,\as)$ and $P_{qq}(x,\as)$ are sub-leading, starting only at NLL.
The structure of the small-$x$ singular contribution can be expressed explicitly by
\begin{equation}\label{eq:PijSX}
x\,P_{ij}(x,\as) \overset{x\to0}= \sum_{n=\delta_{iq}}^\infty\as^{n+1}\sum_{k=0}^\infty p_{nk} \(\as\log\frac1x\)^k, \qquad i,j=g,q,
\end{equation}
where $n=0$ is the LL (which is absent if $i=q$), then $n=1$ is the NLL and so on.

The  expansion in Eq.~\ref{eq:PijSX} is of $x$ times the splitting function, so in the actual splitting functions the high-energy terms will be enhanced by a $1/x$ factor.
Explicitly, the LO splitting functions of Eq.~\eqref{eq:Pij0} behave as
\begin{equation}\label{2:eq:PijFactor}
P_{gg}(x,\as) \overset{x\to0}= \frac{\as C_A}{\pi x} + \calO(\as^2),
\qquad
P_{gq}(x,\as) \overset{x\to0}= \frac{\as C_F}{\pi x} + \calO(\as^2),
\end{equation}
while at this order the $P_{qg}$ and $P_{qq}$  are not enhanced.
At this order, the difference between $P_{gg}$ and $P_{gq}$  is just a colour factor. Likewise, at NLL also $P_{qg}$ and $P_{qq}$ are related by a simple colour factor. This property holds at LL to all orders, but does not survive at NLL. 

A Mellin transform maps the high-energy logarithms of the splitting function into rational polynomials for the anomalous dimension, explicitly
\begin{equation}
        \int_{0}^{1} \dd{x} x^{N-1} \frac{1}{x} \[\as \log(\frac{1}{x})\]^k  = \frac{\as^k}{\(N-1\)^k} \label{ch2:eq:mellinedlogs} \, .
\end{equation}

Let us examine the behaviours of $P_{gg}(x,\as)$ and $P_{qg}(x,\as)$ to illustrate the scenarios mentioned above. Figure~\ref{fig:Pgg} displays these functions across values of $x$ at fixed LO, NLO and NNLO for a chosen value of $\as=0.2$, corresponding to a scale of $Q\sim 5$~GeV. 
For both cases, it is evident that at NNLO, a logarithmic growth arises, rendering the three-loop correction susceptible to perturbative instability. 
It is worth noting that since $P_{gg}$ is a LL quantity, the logarithmic term should already be present at NLO, with an additional squared term at NNLO. However, these terms are absent due to a fortuitous cancellation. Starting from N3LO, the LL pattern is reinstated.
A notable deviation from the NLO result occurs when $x$ is less than or approximately equal to $10^{-3}$ for both cases.

The same plot also shows the resummed result. For $P_{gg}$, it is close to the NNLO one down to $x\sim10^{-3}$, and it deviates from it both quantitatively and qualitatively for smaller values of $x$. In the case of $P_{qg}$, the resummed result is close to the NLO result (and consequently smaller than the NNLO result) for values of $x$ greater than or approximately equal to $10^{-3}$. Subsequently, it gradually increases and approaches the NNLO result, until a significant deviation occurs for $x$ less than or around $10^{-5}$. Hence, comprehending the necessity and impact of incorporating resummation is not a straightforward matter based solely on the magnitude of $\as\log\frac1x$. However, it becomes evident that for sufficiently small values of $x$, there is always a region where the fixed-order result becomes unstable and the inclusion of resummation becomes imperative.

We now recall how resummation is achieved for the splitting functions.
The key observation is the existence of the BFKL equation~\cite{Lipatov:1976zz,Fadin:1975cb,Kuraev:1976ge,Kuraev:1977fs,Balitsky:1978ic,Fadin:1998py},
which is an evolution equation similar to DGLAP, but describing the evolution of PDFs in the variable $x$.
Similarly to the way in which DGLAP, which is an evolution in energy scale, resums logarithms of the energy scale,
the solution to the BFKL equation resums the logarithms of $x$.
In particular, requiring that PDFs satisfy both evolution equations leads to a consistency relation between the
splitting functions and their BFKL counterpart, the so-called BFKL kernel. This consistency relation, called \emph{duality},
allows to resum the small-$x$ logarithms in the splitting functions given the knowledge of the BFKL kernel at fixed order.

However, life is not that simple. The BFKL evolution kernel is itself perturbatively unstable, and using it to directly resum the small-$x$ logarithms results in a perturbatively unstable resummed result (that is to say, NLL is a large correction to the LL).
Stabilising the BFKL kernel and thus the resummation requires using duality in reverse to resum part of the unstable contributions in BFKL using DGLAP at fixed order, and then exploiting a symmetry property of the BFKL kernel to resum the other part. Finally, it has also been realised that the resummation of a class of contributions originating by the running of the strong coupling, despite being formally sub-leading, is necessary in order to remove another source of instability.
The whole complicated machinery has been developed by several groups in more than fifteen years of activity~\cite{Salam:1998tj,Ciafaloni:1999yw,Ciafaloni:2003rd,Ciafaloni:2007gf,Rojo:2009us,Thorne:1999sg,Thorne:1999rb,Thorne:2001nr,White:2006yh}.\footnote{Recently there has also been an interesting activity in achieving the same results in the context of effective field theories~\cite{Rothstein:2016bsq}.}
Recently, after approximately ten years from the latest publication on small-$x$ resummation of splitting functions, the results of the Altarelli-Ball-Forte (ABF)~\cite{Ball:1995vc,Ball:1997vf,Altarelli:2001ji,Altarelli:2003hk,Altarelli:2005ni,Altarelli:2008aj} group have been further developed and improved, and made publicly available through the numerical code HELL (High-Energy Large Logarithms)~\cite{Bonvini:2016wki,Bonvini:2017ogt,Bonvini:2018xvt}.

\subsection{Resummation of cross sections}
Alongside the splitting functions, partonic cross sections are affected by high-energy logarithms.
The structure is similar to that of splitting functions, Eq.~\eqref{eq:PijSX}.
However, since the small-$x$ contributions are generated at LL in gluon-initiated hard scattering, an extra power of $\as$ always arises without a logarithmic term.
This implies that coefficient functions always start from NLL and are zero at LL with respect to the fixed-order power counting.
For this reason, a \emph{relative} counting scheme in used for small-$x$ logarithms by  labelling the first non-vanishing logarithmic order as the LL$x$.

\begin{figure}[t]
  \centering
  \includegraphics[width=0.7\textwidth]{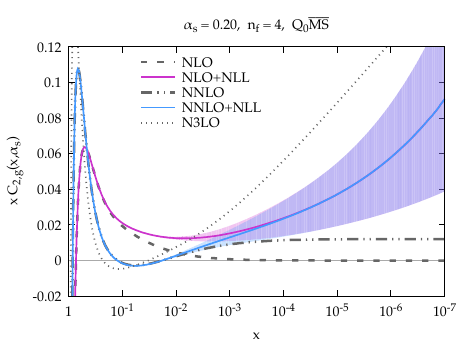}
  \caption{The coefficient function $C_{2g}$ contributing to the DIS structure function $F_2$ at fixed order and with resummation~\cite{Bonvini:2017ogt}. The error band on the resummed results represents a rough estimate of the uncertainty from unknown sub-leading logarithmic contributions.}
  \label{fig:C2g}
\end{figure}

%These considerations also show that the first power of the small-$x$ logarithm appears at N$^3$LO, namely at $\calO \(\as^3\)$\footnote{This is the same order in which the logarithms appear in the splitting functions, however there the LO corresponds to $\calO \(\as\)$, so the $\calO \(\as^3\)$ is the NNLO for splitting functions.}.
This logarithm gives rise to the perturbative instability, and determines the need for resummation.
As an example, we show in figure~\ref{fig:C2g} a DIS coefficient function, specifically $C_{2,g}$, namely the gluon contribution to the structure function $F_2$.\footnote{It is well known that the DIS cross section can be decomposed in structure functions, each corresponding to a specific Lorentz structure of the hadronic tensor, and thus interacting differently with the leptonic part of the cross section.}
Also in this case we see that, for the chosen value of $\as=0.2$, the logarithm starts to diverge at $x\lesssim10^{-3}$.
The plot also shows the resummed result. In particular, the one matched to NNLO (blue curve) starts deviating from the fixed-order
below $x\sim10^{-3}$, growing steadily but slower than the N$^3$LO logarithmic growth.

%%%%%%%%%%%%%%%%%%%%%%%%%%%%%%%%%%%%%%%%%%%%%%%%%%%%%%%%%%%%%%%%%%%%%%%%%%%%%
%%%%%%%%%%%%%%%%%%%%%%%%%%%%%%%%%%%%%%%%%%%%%%%%%%%%%%%%%%%%%%%%%%%%%%%%%%%%%

The strategy to resum the small-$x$ logarithms in the coefficient functions is rather different from that of DGLAP evolution and revolves around expressing the hadron level cross-section~\eqref{1:eq:2protonsigma} using the $\kt{}$-factorisation formalism~\cite{Catani:1990xk,Catani:1990eg,Collins:1991ty,Catani:1993ww,Catani:1993rn,Catani:1994sq} to separate the small-$x$ contributions between the parton level coefficient function and the PDF.
While this resummation formalism is central to the contents of both this chapter and the next one, we postpone a detailed review to chapter~\ref{ch2a} and, for now, limit ourselves to elucidate its main features.

The basic observation is that the leading small-$x$ logarithms arise, in a physical gauge, from $\kt{}$ integration over gluon exchanges in the $t$ channel.
Therefore, in the small-$x$ limit, the generic amplitude squared can be decomposed into contributions that are two-gluon irreducible (2GI) in the $t$ channel and thus do not contain any logarithmic enhancement.
\begin{subequations}
   \begin{align}
     \sigma \(x, Q^2\) &= \frac{x}{Q^2} \int_x^1 \frac{\dd{z}}z \dd y \, \calC \(z,y, \xi_1, \xi_2, \as(Q^2), \frac{Q^2}{\muF^2}\) \calL \(\frac xz, y, \xi_1, \xi_2\), \label{2:eq:ktfac}\\
     \calL_{ij} \(z, y, \xi_1,\xi_2\) & = {\cal F}_g\(\sqrt{z}\eu^{y},\xi_1\)\, {\cal F}_g\(\sqrt{z}\eu^{-y},\xi_2\)\, \vartheta\(\eu^{-2|y|}-z\)
   \end{align}
\end{subequations}      
In this way the cross section of the process factorises~\cite{Catani:1990xk,Catani:1990eg,Catani:1994sq} into a process-dependent off-shell coefficient function ($\calC$) and process-independent ``unintegrated'' luminosity ($\calL$). The latter in turn is given by a combination of off-shell ``unintegrated'' PDFs ($\calF$).
Crucially, $\calC$ is defined to be 2GI by construction, so small-$x$ can not arise in there at LL. On the other hand,  $\calF$ contains the sum over any number of 2GI gluon emission, starting from a collinear parton and ending in an off-shell gluon. Since every extra emission entails an integral over a $t$-channel propagator, an extra high-energy leading logarithm will be produced as well. Therefore, summing over all number of extra emissions will capture the entire LL contribution.
By making explicit the dependence of unintegrated PDF on collinear PDFs and comparing the result with the standard collinear factorisation, one finally obtains an expression for the LL resummation of small-$x$ logarithms in the collinear partonic coefficient functions.

The last step of this procedure was traditionally performed in Mellin moment space, which allows to obtain rather simple analytical expressions.
Despite the elegance of this method, it was soon realised that subleading effects from the running of the strong coupling are important and required to obtain perturbatively stable results~\cite{Ball:2007ra,Altarelli:2008aj}.
However, such modification in Mellin space is complicated and unsuitable for efficient numerical evaluation.
An alternative but equivalent formulation of the resummation was proposed more recently~\cite{Bonvini:2016wki}, solving this technical limitation by working directly in $\kt{}$-space.
This novel approach has an efficient numerical implementation in the public code \hell,
and was used for many phenomenological applications~\cite{Bonvini:2017ogt,Bonvini:2018iwt}, including the first PDF fits with consistently resummed theory predictions at small-$x$~\cite{Ball:2017otu,Abdolmaleki:2018jln,Bonvini:2019wxf}.

\section{Resummation in differential observables in $pp$ collisions} \label{2:sec:differential}

Until now, the applications of the \hell\ method have focused on inclusive observables, such as DIS structure functions~\cite{Bonvini:2016wki,Bonvini:2017ogt} and the total Higgs production cross section~\cite{Bonvini:2018iwt}. However, there is a need to extend the resummation to differential observables, which are particularly relevant for LHC phenomenology. The resummation of such observables has been previously investigated using the Mellin-space formalism. This includes the resummation of rapidity distributions~\cite{Caola:2010kv}, transverse momentum distributions~\cite{Forte:2015gve}, and double differential distributions in both rapidity and transverse momentum~\cite{Muselli:2017fuf}. It is the purpose of this section to reformulate these results in the new \hell\ language, thereby supplementing them with the running coupling contributions and thus providing a ready-to-use numerical implementation.

We focus on processes at hadron-hadron colliders that are gluon-gluon initiated at lowest order, i.e. Higgs production, jet production, or heavy-quark pair production.
The latter especially will be considered as a practical application in section~\ref{2:sec:HQ}.
Focusing on this group of processes makes the resummation simpler, because there are no collinear singularities to subtract at LL.
This is not the case when the process is initiated by (massless) quarks, because small-$x$ logarithms at LL appear from chains of emissions ending with a gluon. This would force the off-shell coefficient function to contain at least one gluon to (massless) quark splitting and induce a collinear singularity\footnote{Drell-Yan and DIS are examples of this mechanism, where the singularity must be subtracted at the resummed level (for DIS, Ref.~\cite{Bonvini:2017ogt}, shows a way to tackle this issue).}.

\subsection{Extension of $\kt$ factorisation to differential observables}
Refs.~\cite{Caola:2010kv,Forte:2015gve,Muselli:2017fuf} have established a resummation formula for a differential observable at LL accuracy with fixed coupling, using the ladder-expansion approach. This differs somewhat from the approach taken in the original works~\cite{Catani:1990xk,Catani:1990eg,Catani:1994sq}, where the resummation was obtained by proving $\kt{}$-factorisation and comparing it with the standard collinear one. Despite the different language used in the two approaches, they are based on the same underlying factorisation property and lead to the same result.

The steps of the derivation in Refs.~\cite{Caola:2010kv,Forte:2015gve,Muselli:2017fuf} can be followed to construct a formula using the ingredients of $\kt{}$-factorisation.
Rather than repeating the derivation, we will instead formulate the result in terms of $\kt{}$ factorisation, showing that it corresponds to the results of Refs.~\cite{Caola:2010kv,Forte:2015gve,Muselli:2017fuf} at LL and fixed coupling.

Similarly to the inclusive case~\eqref{2:eq:ktfac}, the differential cross section in $\kt$ factorisation is a straightforward extension of the collinear factorisation Eq.~\eqref{2:eq:collfactdiff},
where the partons are replaced by off-shell gluons and integration over this off-shellness is added.
The result reads
\begin{align}
\frac{\dd\sigma}{\dd Q^2\dd Y\dd \qt^2} & = x\int_x^1\frac{\dd z}z\int \dd\eta \int_0^\infty \dd\xi_1 \int_0^\infty \dd\xi_2\, \frac{\dd{\cal C}}{\dd Q^2\dd\eta\dd\qt^2}(z,\xi_1,\xi_2,Q^2,\eta,\qt^2,\as)\nn
&\times \calL \(\frac x z,Y-\eta,\xi_1,\xi_2\), \label{2:eq:ktfact}
\end{align}
where $\xi_{1,2}=\vec{k}^2_{1,2}/Q^2$ are the off-shellnesses of the gluons normalised to the hard scale $Q^2$, and $\vec{k}_{1,2}$ are the transverse components of the off-shell gluon momenta (for more details on the kinematics, see App.~\ref{app:XS}).
The (differential) off-shell coefficient function ($\dd{\cal C}$) represents the process-dependent hard scattering and it is initiated by off-shell gluons.  
%More precisely, it corresponds to the last 2GI part (in the $t$ channel) of the amplitude squared of the process, saturating the off-shell gluon indices with a suitable projector~\cite{Catani:1990xk,Catani:1990eg,Catani:1994sq}.
Everything else is collected into the two unintegrated gluon PDFs ${\cal F}_g$, that include the standard collinear PDFs and the chain of emissions from the initial parton to the last gluon (the ladder in the language of Refs.~\cite{Caola:2010kv,Forte:2015gve,Muselli:2017fuf}).
The integration variables are the analogue of $z$ and $y$ of Eq.~\eqref{2:eq:collfactdiff}, but referred to the centre-of-mass frame of the off-shell partons.
The parton-level centre-of-mass frame in $\kt$-factorisation is equivalent to the collinear one up to setting the off-shellness equal to zero, and performing a longitudinal boost.

Eq.~\eqref{2:eq:ktfact} is equivalent to the result of Ref.~\cite{Muselli:2017fuf}.\footnote{Notice that Ref.~\cite{Muselli:2017fuf} considers the Higgs production process so it shows only the double differential distribution in rapidity and transverse momentum and the invariant mass is clearly fixed to the Higgs mass. However, the derivation there is general for invariant mass distributions.}
\begin{align}\label{2:eq:MellFourkt}
\int_0^1\dd x\, x^{N-1}\int_{-\infty}^\infty &\dd Y\,\eu^{\iu bY}\frac{\dd\sigma}{\dd Q^2\dd Y\dd \qt^2}\nonumber\\
&= \int_0^\infty \dd\xi_1 \int_0^\infty \dd\xi_2\,
\frac{\dd{\cal C}}{\dd Q^2\dd \eta\dd\qt^2}(N,\xi_1,\xi_2,Q^2,b,\qt^2,\as)\, \calL\(N,b,\xi_1,\xi_2\),
\end{align}
with
\begin{align}
\frac{\dd{\cal C}}{\dd Q^2\dd \eta\dd\qt^2}(N,\xi_1,\xi_2,Q^2,b,\qt^2,\as)
  &= \int_0^1\dd z\,z^N \int_{-\infty}^\infty \dd \eta\,\eu^{\iu b \eta}\frac{\dd{\cal C}}{\dd Q^2\dd \eta\dd\qt^2} (z,\xi_1,\xi_2,Q^2,\eta,\qt^2,\as) \nonumber\\
\calL\(N,b,\xi_1,\xi_2\) &= \int_0^1\dd z\,z^N \int_{-\infty}^\infty \dd \eta \, \eu^{\iu b \eta} \calL\(z, \eta,\xi_1,\xi_2\) \nonumber\\
&= {\cal F}_g\(N+\iu\frac b2,\xi_1\)\, {\cal F}_g\(N-\iu \frac b2,\xi_2\),
\end{align}
where we have used the definition Eq.~\eqref{2:eq:lumioffnonY}, changed variable from $z,\eta$ to $x_{1,2}=\sqrt{z}\eu^{\pm\eta}$
(the longitudinal proton's momentum fractions carried by each off-shell gluon) and used the $\vartheta$ function to obtain the product of two Mellin transforms
\begin{equation}
{\cal F}_g(N,\xi) = \int_0^1\dd x_{1,2}\,x_{1,2}^N {\cal F}_g(x_{1,2},\xi).
\end{equation}
Following Ref.~\cite{Catani:1990xk,Catani:1990eg,Catani:1994sq}, at LL and fixed coupling the unintegrated PDF is
\begin{equation}\label{2:eq:FgLLFC}
{\cal F}_g\(N,\xi\) = R(N,\as) \gamma(N,\as)\(\frac{Q^2}{\muF^2}\)^{\gamma(N,\as)}\xi^{\gamma(N,\as)-1} f_g(N,\muF^2),
\end{equation}
where $\gamma(N,\as)$ is the resummed (gluon) anomalous dimension at LL and $R(N,\as)$ is a scheme-dependent factor.
For simplicity, we will ignore the quark contributions at this time and refer to appendix~\ref{sec:channels}. 
Plugging Eq.~\eqref{2:eq:FgLLFC} into Eq.~\eqref{2:eq:MellFourkt}, we immediately recover the result of Ref.~\cite{Muselli:2017fuf}.
Integrating over $\qt^2$ we also reproduce the result of Ref.~\cite{Caola:2010kv}.

To reproduce the result of Ref.~\cite{Forte:2015gve}, which is not differential in rapidity,
it is simpler to integrate Eq.~\eqref{2:eq:ktfact} over $Y$ and then take simply a Mellin transform
before using Eq.~\eqref{2:eq:FgLLFC}.
The first step leads to
\begin{equation}\label{2:eq:ktfactnonY}
\frac{\dd\sigma}{\dd Q^2\dd \qt^2} = \tau\int_\tau^1\frac{\dd z}z \int_0^\infty \dd\xi_1 \int_0^\infty \dd\xi_2\, \frac{\dd{\cal C}}{\dd Q^2\dd\qt^2}(z,\xi_1,\xi_2,Q^2,\qt^2,\as)\, \calL\(\frac x z,\xi_1,\xi_2\),
\end{equation}
with
\begin{align}\label{2:eq:lumioffnonY}
\calL\(z,\xi_1,\xi_2\)
  &= \int \dd \eta \, {\cal F}_g(\sqrt{z}\eu^{\eta},\xi_1)\, {\cal F}_g(\sqrt{z}\eu^{-\eta},\xi_2)\, \theta(\eu^{-2|\eta|}-z) \nonumber\\
  &= \int_{z}^1 \dd x_2\, {\cal F}_g\(\frac{z}{x_2},\xi_1\)\, {\cal F}_g(x_2,\xi_2).
\end{align}
This simplifies the expression to a Mellin convolution, which can be transformed into
\begin{equation}\label{2:eq:Mellkt}
\int_0^1\dd\tau\, \tau^{N-1}\frac{\dd\sigma}{\dd Q^2\dd \qt^2}
= \int_0^\infty \dd\xi_1 \int_0^\infty \dd\xi_2\,
\frac{\dd{\cal C}}{\dd Q^2\dd\qt^2}(N,\xi_1,\xi_2,Q^2,\qt^2,\as)\, {\cal F}_g\(N,\xi_1\)\, {\cal F}_g\(N,\xi_2\),
\end{equation}
with
\begin{align}
\frac{\dd{\cal C}}{\dd Q^2\dd\qt^2}(N,\xi_1,\xi_2,Q^2,\qt^2,\as)
  &= \int_0^1\dd z\,z^N \frac{\dd{\cal C}}{\dd Q^2\dd\qt^2} (z,\xi_1,\xi_2,Q^2,\qt^2,\as).
\end{align}
Plugging now Eq.~\eqref{2:eq:FgLLFC} into Eq.~\eqref{2:eq:Mellkt}, we finally obtain the result of Ref.~\cite{Forte:2015gve}.

The unintegrated PDF depends on $\xi$ through $\xi^{\gamma-1}$, which means that the integrals over $\xi_{1,2}$ can be expressed as Mellin transforms. The results mentioned earlier can be expressed (up to factors) as the $\gamma$-th Mellin moments with respect to $\xi_{1,2}$ of the off-shell coefficient functions of the partons, which are commonly referred to as impact factors. Beyond fixed coupling, the effect of running coupling can be included according to the ABF method from Refs.~\cite{Ball:1995vc,Ball:1997vf,Altarelli:2001ji,Altarelli:2003hk,Altarelli:2005ni,Ball:2007ra,Altarelli:2008aj}. However, adding running coupling effects to the impact factors is not ideal for a numerical implementation. Very briefly, the reason is that the ABF strategy requires computing $\gamma$ by a recursive expression of its derivatives with respect to $\as$. In turn, this transforms the impact factor into an asymptotic series to be numerically returned to direct space with an inverse Mellin transform over $N$. The combination of these steps proved to be a computational bottleneck both for the lack of closed expression and for the complexity of the expressions involved~\cite{Bonvini:2016wki}.

In the next section, we return to from Eq.~\eqref{2:eq:ktfact} to rebuild the resummed expression at the differential level using the \hell\ language. This approach makes it straightforward to include running-coupling effects and enables a stable and numerical implementation.

\subsection{Small-$x$ resummation of differential distributions in the \texttt{HELL} language}\label{2:ssec:diff-hell}
The approach to small-$x$ resummation described in Refs.~\cite{Bonvini:2016wki,Bonvini:2017ogt,Bonvini:2018iwt} and used in the \hell\ code has a significant advantage in terms of numerical implementation. This advantage stems from two main reasons.
First, including running-coupling effects in the resummation can be done straightforwardly and accurately, unlike in the impact-factor approach described in Refs.~\cite{Ball:2007ra,Altarelli:2008aj}, where it leads to a divergent series that must be treated approximately. Secondly, the result can be expressed directly in momentum space in terms of the off-shell coefficient function, unlike in the impact-factor formulation, where a double Mellin transform in both $z$ and $\xi$ is required for each initial-state off-shell gluon. When these Mellin transforms can be computed analytically, the \hell\ formulation has a minor disadvantage of requiring numerical integration over $\xi$. However, when the Mellin transform in $\xi$ cannot be computed analytically, the impact-factor formulation becomes problematic, while the \hell\ approach does not encounter any such problems.

The key step of the \hell\ approach is to represent the unintegrated PDF in terms of the collinear gluon and quark-singlet PDFs while incorporating running coupling effects. The general form of this expression, at least at LL accuracy, is described in Refs.~\cite{Bonvini:2016wki,Bonvini:2017ogt,Bonvini:2018iwt}.
\begin{equation}\label{2:eq:Foff}
{\cal F}_g(N,\xi) = U'\(N,Q^2\xi,\muF^2\) f_g(N,\muF^2) + \frac{C_F}{C_A} \[U'\(N,Q^2\xi,\muF^2\) - \delta(\xi)\] f_q(N,\muF^2),
\end{equation}
where
\begin{equation}
U'\(N,Q^2\xi,\muF^2\) \equiv \frac{\dd}{\dd\xi}U\(N,Q^2\xi,\muF^2\)
\end{equation}
and $U(N,\vec{k}^2,\muF^2)$ is the evolution function of the collinear gluon
\footnote{
The quark part of \eqref{2:eq:Foff} uses the same evolutor of the gluon part.   This is motivated by the fact that the leading term in $P_{gg}$ and $P_{gq}$ is identical up to a factor $\frac{C_F}{C_A}$.  Moreover,  the $\delta\(\xi\)$ in the quark part stands for the no-emission event in which the parton remains collinear. It must be subtracted to convert the quark into a gluon  and so this contribution must start at order $\as$.}
from the scale $\muF^2$ to the scale $\vec{k}^2$, times the scheme-dependent function $R(N,\as)$.
The evolutor $U(N,\kvec^2,\muF^2)$ can be identified with the solution of the DGLAP equation up to using the small-$x$ LL anomalous dimension,
which involves only gluons at LL in place of the fixed-order one.
Then, including running coupling effects in the DGLAP evolution equation provides the necessary handle to capture them in the resummation as well~\cite{Bonvini:2016wki,Bonvini:2017ogt,Bonvini:2018iwt}.
Conversely, evaluating the evolution function at fixed coupling, we get back Eq.~\eqref{2:eq:FgLLFC}.

%In practice, to simplify the numerical implementation and avoid potential numerical issues, the evolution function is approximated in a way that reproduces exactly the results of Refs.~\cite{Ball:2007ra,Altarelli:2008aj}, namely it is valid at LL and at ``leading running coupling'' (i.e.\ leading $\beta_0$ terms are retained).
%Within this approximation it takes the form~\cite{Bonvini:2017ogt,Bonvini:2018iwt}
To make the numerical implementation of the evolution function easier and avoid potential numerical problems, it is approximated to exactly reproduce the results of Refs.~\cite{Ball:2007ra,Altarelli:2008aj}. This approximation (given by Refs.~\cite{Bonvini:2017ogt,Bonvini:2018iwt}) is only valid at LL and "leading running coupling" (which means only the leading $\beta_0$ terms are included) and reads
\begin{subequations}
\begin{align}
& U\(N,Q^2\xi,\muF^2\) \simeq R(N,\as)\, D_\text{higher-twist}\(\frac{Q^2}{\muF^2}\xi\)\,  U_{\rm ABF}\(N,\frac{Q^2}{\muF^2}\xi\), \label{2:eq:UABFht} \\
& D_\text{higher-twist}(\xi) =
\begin{cases}
1 & \xi\geq1\\
1-\(-\as\beta_0\log\xi\)^{1+\frac1{\as\beta_0}}\qquad & \xi_0<\xi<1 \\
0 & \xi\leq\xi_0, \qquad \xi_0=\exp\frac{-1}{\as\beta_0},
\end{cases}\\
& U_{\rm ABF}(N,\xi) = \Big(1+r(N,\as)\log\xi\Big)^\frac{\gamma(N,\as)}{r(N,\as)},\quad\quad r(N,\as) = \as^2\beta_0\frac{\dd}{\dd\as}\log\[\gamma(N,\as)\], \label{2:eq:UABF} 
\end{align}
\end{subequations}
the damping function $D_\text{higher-twist}(\xi)$ is designed to make the evolution function vanishing at the Landau pole $\xi_0$ as it would do at LL with full running coupling~\cite{Bonvini:2017ogt} while leaving  the perturbative expansion  unaffected.
Conversely, $U_{\rm ABF}$ is the approximated evolution function and the anomalous dimension $\gamma$ appearing would be LL anomalous dimension.
However, it is more convenient to include subleading contributions here to make the result consistent with the resummation in DGLAP evolution.
This modification produces subleading effects in the resummation and  is not central for the result at end, so we defer to Ref.~\cite{Bonvini:2018iwt} for further detail.

The scheme factor $R(N,\as)$ can be addressed by using $Q_0\MSbar$ scheme where by definition is set to unity.
Since this scheme differs from the usual $\MSbar$ only from relative order $\as^3$ (at LL), it can be safely used in conjunction with $\MSbar$ fixed-order computations up to NNLO~\cite{Catani:1993ww,Catani:1994sq,Ciafaloni:2005cg,Marzani:2007gk}.

Plugging Eq.~\eqref{2:eq:Foff} into Eq.~\eqref{2:eq:MellFourkt} and considering only the gluon contribution for simplicity, we get
\begin{align}
\int_0^1\dd x\, x^{N-1}\int_{-\infty}^\infty \dd Y\,\eu^{\iu bY}\frac{\dd\sigma}{\dd Q^2\dd Y\dd \qt^2}
&= \int_0^\infty \dd\xi_1 \int_0^\infty \dd\xi_2\, \frac{\dd{\cal C}}{\dd Q^2\dd \eta\dd\qt^2}(N,\xi_1,\xi_2,Q^2,b,\qt^2,\as) \nonumber\\
  &\times U'\(N+\iu\frac b2,Q^2\xi_1,\muF^2\)\, f_g\(N+\iu\frac b2,\muF^2\) \nonumber\\
  &\times U'\(N-\iu\frac b2,Q^2\xi_2,\muF^2\)\, f_g\(N-\iu\frac b2,\muF^2\). \label{2:eq:MellFourkt2}
\end{align}
This expression matches with the gluon-gluon channel of the collinear factorisation expression Eq.~\eqref{2:eq:MellFourColl}.
\begin{align}
  \frac{\dd C_{gg}}{\dd Q^2\dd y\dd\qt^2}\(N,Q^2,b,\qt^2,\as, \frac{Q^2}{\muF^2}\)
&= \int_0^\infty \dd\xi_1 \int_0^\infty \dd\xi_2\, \frac{\dd{\cal C}}{\dd Q^2\dd \eta\dd\qt^2}(N,\xi_1,\xi_2,Q^2,b,\qt^2,\as) \nonumber\\
  &\times U'\(N+\iu\frac b2,Q^2\xi_1,\muF^2\) U'\(N-\iu\frac b2,Q^2\xi_2,\muF^2\). \label{2:eq:resCggNb}
\end{align}
So far this is not dissimilar to the approach of older works; in particular, if one replaces $U'$
with the LL fixed-coupling expression from Eq.~\eqref{2:eq:FgLLFC}, one recognises the
definition of the impact factor.
Here instead, we keep a more generic expression for $U'$ and further manipulate the result.
Indeed, we notice that the $N,b$ dependence of the right-hand side of Eq.~\eqref{2:eq:resCggNb}
has the same form of the right-hand side of Eq.~\eqref{2:eq:MellFourColl} or Eq.~\eqref{2:eq:MellFourkt}.
We thus recognise Eq.~\eqref{2:eq:resCggNb} as the Mellin-Fourier transform of
\begin{align}\label{2:eq:resCggzy}
  \frac{\dd C_{gg}}{\dd Q^2\dd y\dd\qt^2}\(x,Q^2,y,\qt^2,\as, \frac{Q^2}{\muF^2}\)
  &= \int_0^\infty \dd\xi_1 \int_0^\infty \dd\xi_2 \int_x^1\frac{\dd z}z \int_{-\frac12\log\frac zx}^{\frac12\log\frac zx} \dd\eta\nonumber\\
  &\times \frac{\dd{\cal C}}{\dd Q^2\dd \eta\dd\qt^2}(z,\xi_1,\xi_2,Q^2,y-\eta,\qt^2,\as) \nonumber\\
  &\times U'\(\sqrt{\frac xz}\eu^{\eta},Q^2\xi_1,\muF^2\) U'\(\sqrt{\frac xz}\eu^{-\eta},Q^2\xi_2,\muF^2\),
\end{align}
which is expressed as a 4-dimensional integral (to be performed numerically in general) over simple quantities, namely the differential off-shell coefficient function and the evolution factors in physical momentum space.

This result is advantageous for numerical evaluation. Indeed, the two additional integrations over $z$ and $\eta$ are much simpler to compute than the inverse Mellin-Fourier transform over $N$ and $b$ of Eq.~\eqref{2:eq:resCggNb}.
This is especially true in \hell, because the anomalous dimension in the definition of $U'$ is available only as a table with precomputed values of $N$ along a fixed inversion contour. This would not be compatible with the Mellin inversion due to the $\pm ib$ imaginary shift.
Instead, the evolution factor is computed once and for all in \hell\ directly in momentum space, and it can be used in an expression like Eq.~\eqref{2:eq:resCggzy} since it is process-independent.
On top of this, this has the advantage, compared to the impact-factor formulation, of incorporating the running coupling contributions through Eq.~\eqref{2:eq:UABFht}.

We would like to highlight one final difference between our current approach to resummation in the \hell\ code and previous formulations.
In earlier works, we would set the $N$ dependence of the off-shell coefficient function to zero before computing the inverse Mellin transform, as the resulting analytical expressions were simpler.
However, this approach would ignore physical kinematic constraints and could lead to inaccurate results, especially when dealing with differential distributions.
In contrast, we choose to retain the subleading $N$ dependence, which preserves these constraints and avoids any numerical issues.
Though it may not always be possible to compute the Mellin transform of the off-shell coefficient function analytically, we believe it is better to keep the full $N$ dependence to ensure accuracy and consistency.
This is similar to the approach taken in Ref.~\cite{Bonvini:2017ogt} for DIS, where setting $N$ to zero caused a loss of quark mass effects on kinematic constraints and required their manual restoration.

%\subsection{Effect of resummation in PDF fits}-> MOVED TO chapter4

\section{Heavy-quark production at the LHC} \label{2:sec:HQ}

After explaining the general method for calculating small-$x$ resummation of differential distributions using \hell, we now shift the focus to the production of heavy-quark pairs in collisions between protons. This process is significant because it is measured at the LHC, particularly at LHCb, in the forward region where one parton is at small $x$, and therefore, it can offer crucial constraints on PDFs, particularly the gluon, in a region of $x$ that remains unexplored. Furthermore, this process has fixed-order results available up to NNLO~\cite{Nason:1987xz,Frixione:2007nw,Catani:2020kkl,Mazzitelli:2023znt}, making it an appropriate candidate for precise studies.

The process can be schematised as
\begin{equation} \label{hfhadroproduction}
\mathrm{p}\(p_1\) + \mathrm{p}\(p_2\) \rightarrow Q \(p_3\) + \bar Q \(p_4\) + X ,
\end{equation}
the two incoming protons have light-cone momenta $p_{1,2}$ with $\(p_1+p_2\)^2=S$, the outgoing heavy quarks  momenta are  $p_{3,4}$ with their mass being $p^2_{3,4}=m^2$.
As usual, $X$ represents any additional radiation together with the remnants of the protons.
For simplicity, we do not account for the hadronisation and decay chain of the final state and leave the heavy quarks open.\footnote{The mass of the heavy quarks acts as a regulator for the purposes of infrared safety.}
Since these effects factorise at LL with respect to the hard scattering process, they should not modify the impact of resummation.

Heavy-quark pair production has been studied in the literature of high-energy resummation at the level of the total cross section~\cite{Catani:1990eg,Ball:2001pq}
and for some differential observables~\cite{Baranov:2002cf,Kniehl:2006sk,Bolognino:2019yls,Celiberto:2022rfj,Celiberto:2022dyf}.

In our approach, we need to compute the coefficient function of the partonic sub-process where two off-shell gluons produce the final state.
At lowest order, as appropriate for LL resummation, the process is
\begin{equation}
g^*(k_1) + g^*(k_2) \to Q \(p_3\) + \bar Q \( p_4\),
\end{equation}
where the off-shell gluon momenta are parametrised as
%\footnote{Here we are using a slightly inconsistent notation.  Indeed, we assume that the bold vectors $\kvec_{1,2}$ are 2-dimensional Euclidean vectors in the transverse plane.   However, when they are summed to 4-dimensional Minkowski vectors, we mean them to be the 4-vector with the same spatial components.   The confusion may only arise when they appear in a scalar product, because the two interpretations would differ by a sign.   In these cases, we always consider them as 2-vectors.}
\begin{subequations}
\begin{align}
k_1 &= x_1 p_1 + \vec{k}_1, \\
k_2 &= x_2 p_2 + \vec{k}_2.
\end{align}
\end{subequations}
In this way, the off-shellness of the gluons is given by a transverse component while the longitudinal momentum fractions $x_{1,2}$ correspond to the first argument of the unintegrated PDFs. These
are related to the longitudinal boost of the partonic reference frame $Y-\eta=\frac12\log\frac{x_1}{x_2}$ which appears in the coefficient function, additional information is given in appendix~\ref{app:XS}.

To calculate the off-shell coefficient function, we need to choose a vector $q$ with respect to which we want to differentiate. There are two alternatives: either $q$ can be one of the two heavy quark momenta $p_{3,4}$ , or it can be the sum of the two momenta, representing the combined heavy-quark pair.
We will now present the results for each configuration.

\subsection{Results differential in the single heavy-quark}\label{2:ssec:singlekin}

In this section we consider the final state to be one of the heavy quarks, and identify the momentum $q=p_3$.
The details of the computation of the partonic off-shell coefficient function are given in App.~\ref{app:singlekin}.

\begin{figure}[t]
  \centering
  \includegraphics[width=0.49\textwidth,page=9 ]{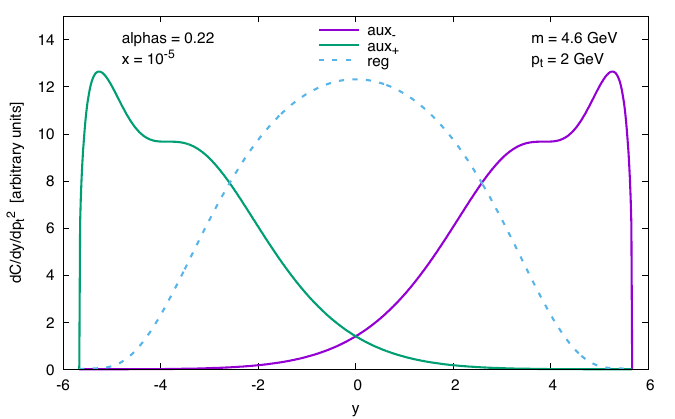}
  \includegraphics[width=0.49\textwidth,page=10]{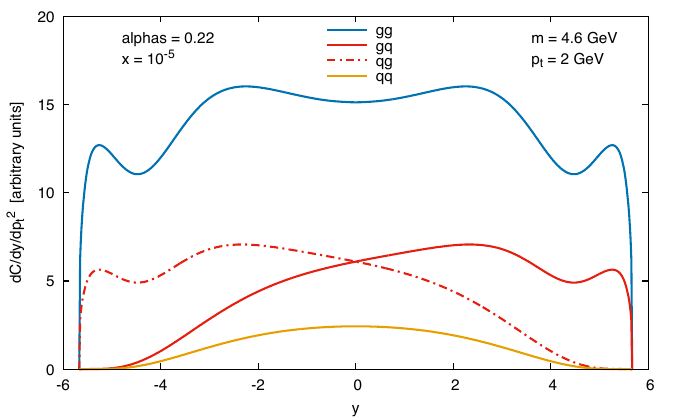}
  \caption{The auxiliary Eq.~\eqref{eq:resCaux} and regular Eq.~\eqref{eq:resCreg} functions as a function of partonic rapidity $y$
    for single quark production of mass $m=4.6$~GeV at $p_t=2$~GeV and $x=10^{-5}$ (left plot).
    The resummed coefficient functions at parton level for each partonic channel
    constructed according to Eq.~\eqref{eq:resC_reg_aux} for the same kinematics (right plot).}
  \label{2:fig:SQDoubleDiffPartonic}
\end{figure}

We start by presenting the resummed result at parton level, computed according to Eqs.~\eqref{eq:resC_reg_aux}.
We consider the case of  bottom pair production, with $m_b=4.6$~GeV, starting from  the double differential case in partonic rapidity $y$ and transverse momentum $p_t$ of the bottom quark.
\footnote{As we consider the bottom quarks to be on-shell, the invariant mass distribution is a delta function and the triple differential distribution is of no interest.}
In Fig.~\ref{2:fig:SQDoubleDiffPartonic}, we plot this distribution at fixed $p_t=2$~GeV.
\footnote{The choice of this value is justified by the interest in constraining the gluon PDF at small-$x$ using heavy-flavour forward production~\cite{PROSA:2015yid,Gauld:2015yia,Gauld:2016kpd}. Indeed, measurements from the LHCb experiment on D meson production~\cite{LHCb:2013xam,LHCb:2015swx,LHCb:2016ikn} are available in 4 $\pt{}$ bins between $\(0,\, 8\)$ GeV. Moreover, the most extreme configuration of high rapidity and low $\pt{}$ was shown to be where Monte Carlo generators and prediction based on models of the $\kt$-factorisation disagreed the most~\cite{Feng:2022inv}.}  

The regular Eq.~\eqref{eq:resCreg} and auxiliary Eq.~\eqref{eq:resCaux} contributions are showed in the left panel.
These can be used to build the various quark and gluon channels according to Eqs.~\eqref{eq:resC_reg_aux}, which we plot in the right plot.
The shapes of these functions are quite peculiar, mostly due to the peak of the auxiliary contribution at large rapidity.
On the other hand, since these are parton-level, results they should not be expected to share all the characteristics of the hadron-level distribution.
Indeed, they may present some new features missing in the fixed order due to the all-order nature of the resummation.

\begin{figure}[t]
  \centering
  \includegraphics[width=0.49\textwidth,page=3]{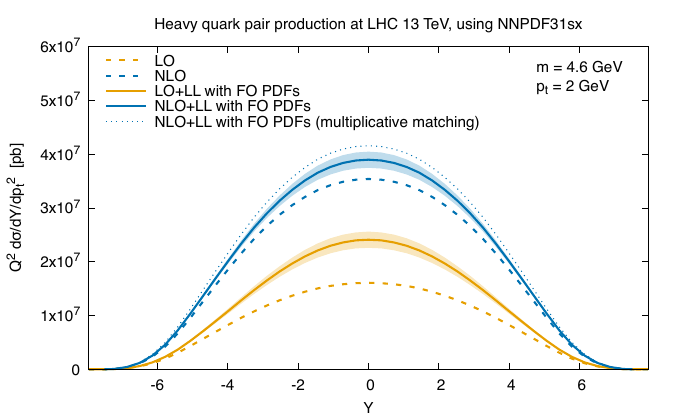}
  \includegraphics[width=0.49\textwidth,page=4]{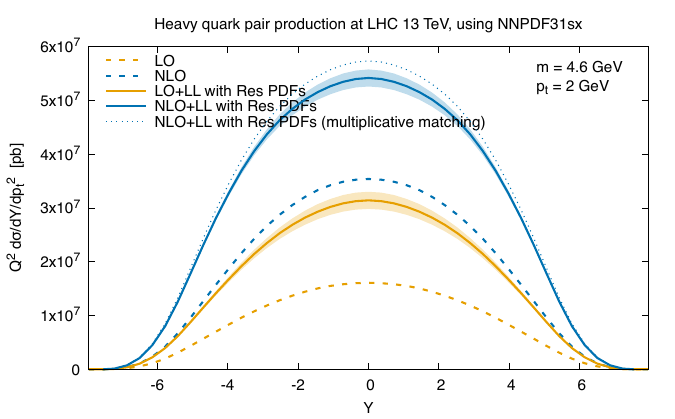}\\
  \includegraphics[width=0.49\textwidth,page=7]{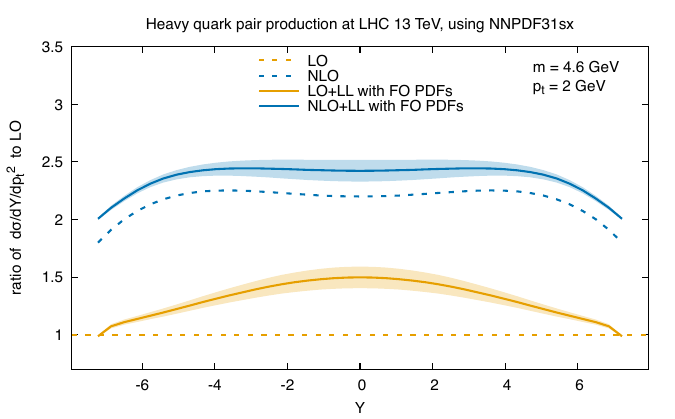}
  \includegraphics[width=0.49\textwidth,page=8]{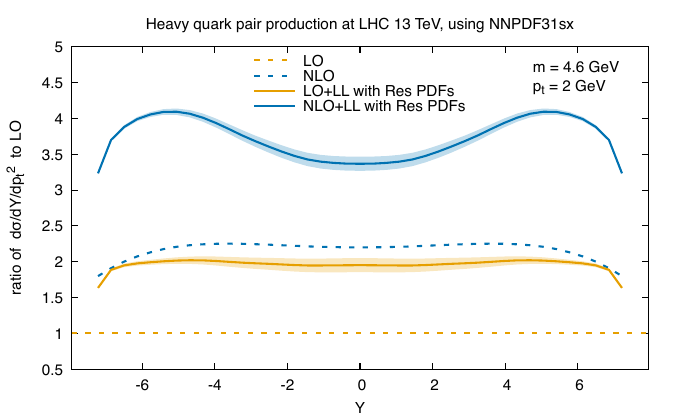}
  \caption{The double differential distribution in rapidity and transverse momentum of the bottom quark,
    plotted as a function of the rapidity for $p_t=2$~GeV, for bottom pair production at LHC $13$~TeV.
    The left plots are obtained using NNPDF31sx at fixed order, while in the right plot the resummed result
    is computed with the resummed PDFs from the same family.
    The uncertainty band represents an estimate of NLL corrections.}
  \label{2:fig:SQDoubleDiff}
\end{figure}

Fig.~\ref{2:fig:SQDoubleDiff} shows the differential distributions after convolution with the PDFs at LHC $13$~TeV.
This is a way to assess the effect of the resummed contributions on physical cross sections.
The NNPDF31sx PDF set is utilised, which was obtained as part of a study on the integration of small-$x$ resummation in PDF fits. This PDF set is advantageous because it provides PDFs that were consistently obtained with and without small-$x$ resummation. To highlight the impact of resummation on the perturbative coefficient, the same fixed-order PDFs are used to compute both the fixed-order and resummed results. The resummed results obtained using resummed PDFs are also provided to determine the extent to which resummation in PDFs affects the cross-section.
%However, the paper's main goal is to present the formalism and not to conduct a comprehensive phenomenological study, which is deferred to future work.

The figures in Fig.~\ref{2:fig:SQDoubleDiff} display the double differential distribution in rapidity $Y$ and transverse momentum $p_t$ at different orders, along with their ratios to LO, for a fixed $p_t=2$~GeV.
The left plots use fixed-order PDFs for both fixed-order and resummed results, while the LO and NLO cross sections are shown in orange and blue dashed lines, respectively.
The latter, obtained from POWHEG-box~\cite{Nason:2004rx,Frixione:2007vw,Alioli:2010xd}, is about twice as large as the LO result, which is partly due to the large value of $\as$ at this low scale.
\footnote{We use $\muR=\muF=m_b$ to simplify the matching with the resummed contribution. This may come at the expense of the stability of the perturbative expansion.}
We plot the LO+LL (orange) and NLO+LL (blue) as solid curves and observe that resummation accounts for a positive correction at LO (about 50\% at central rapidity).
Conversely, this effect is still positive, albeit weaker, at NLO.
This suggests that the inclusion of small-$x$ resummation improves the perturbative series convergence.
%Overall, the NLO+LL result is approximately a 140\% correction over the LO across the whole rapidity range except towards the endpoints, where it goes down a bit following the analogous behaviour of the NLO.
%Most of this effect is driven by the NLO correction alone, but at large rapidities the contribution of the
%resummation is larger, reaching approximately 50\%.
%
In the plots on the right the resummed results are computed with resummed PDFs.
In this case, the impact of resummation becomes much larger. This change is driven by the larger size the resummed gluon PDF at small $x$~\cite{Ball:2017otu,Abdolmaleki:2018jln,Bonvini:2019wxf}.
The NLO+LL curve shows a high K-factor at large rapidity, where the contribution from the gluon at small $x$ is significant, indicating that the observable is sensitive to PDFs at small $x$ and is therefore an essential process to provide additional constraints to PDF fits, consistent with the previous studies reported in Refs.~\cite{PROSA:2015yid,Gauld:2015yia,Gauld:2016kpd}.

\begin{figure}[t]
  \centering
  \includegraphics[width=0.49\textwidth,page=1]{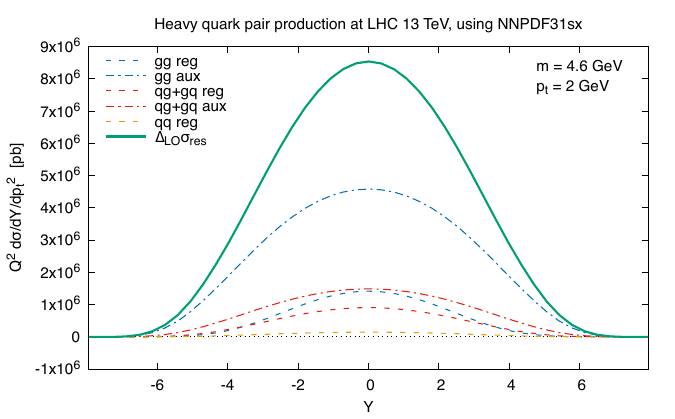}
  \includegraphics[width=0.49\textwidth,page=2]{images/plot_QQbarSQ}
  \caption{Breakdown of the individual contributions to the resummed result from the $gg$, $gq+qg$ and $qq$ channels
    separating the regular and auxiliary parts. The left plot focuses on the resummed contribution to be matched to the LO,
    while the right plot focuses on the resummed contribution to be matched to NLO.
    The results in these plots are obtained using NNPDF31sx with resummation.}
  \label{2:fig:SQDoubleDiff2}
\end{figure}

The plots in Fig.~\ref{2:fig:SQDoubleDiff2} offer a breakdown of the various contributions that form the resummed result.
A $gg$ channel and  $q\bar q$ channel form the LO result, with the former being way larger than the latter.
The decomposition of the resummed result requires distinguishing regular and auxiliary contributions on top of parton flavour channel, as given in Eqs.~\eqref{eq:resC_reg_aux}.
The breakdown of the individual resummed contributions for the LO+LL matching is shown in the left panel of Fig.~\ref{2:fig:SQDoubleDiff2}.
The dominant contributions come from the auxiliary part, both in the $gg$ channel and in the $qg+gq$ channel.
Instead, the regular contributions are smaller and localised in a region of central rapidity.
Also, there is a visible flavour hierarchy in the contributions with the $gg$ channel dominating over the $qg+gq$
\footnote{The $qg+gq$ channel is symmetric because we plot them together, but the individual $qg$ and $gq$ contributions are obviously asymmetric (see Fig.~\ref{2:fig:SQDoubleDiffPartonic}).}
, and the $qq$ being the smallest.
Fig.~\ref{2:fig:SQDoubleDiff2} presents the breakdown of the resummed contribution to the NLO+LL result, where the the regular contribution is unaffected as it starts from $\calO(\as^2)$.
The right plot shows that the auxiliary contributions dominate in the forward region and become comparable with the regular ones at mid rapidity due to the subtraction at $\calO(\as)$.

%The right plot of Fig.~\ref{2:fig:SQDoubleDiff2} shows the analogous breakdown for the resummed contribution to be added to the NLO to construct the NLO+LL result.
%The difference here is only in the auxiliary contributions,as the regular contribution starts at relative $\Ord(\as^2)$ and is thus unaffected when subtracting the expansion at $\Ord(\as)$.
%Because of this subtraction, the auxiliary contributions become comparable with the regular ones at mid rapidities,
%but they still dominate in the forward region, as expected.

To assess the stability of our resummed results, we must consider the uncertainties associated with them. As our resummed results are limited to LL accuracy, the primary source of uncertainty we examine is that arising from the leftover subleading logarithmic contributions.
In previous \texttt{HELL} works~\cite{Bonvini:2016wki,Bonvini:2017ogt,Bonvini:2018iwt,Bonvini:2018xvt} this source was studied by varying subleading ingredients in the construction of the resummed anomalous dimension in Eq.~\eqref{2:eq:UABFht},
and by replacing $r(N,\as)$ in Eq.~\eqref{2:eq:UABF} with $\as\beta_0$, this affecting the form of the evolution function itself.\footnote{Briefly, one contribution come from a degree of freedom in the way of implementing the resummation of subleading running coupling in the  anomalous dimension~\cite{Bonvini:2017ogt}. Another, and by far larger, contribution comes from the use of the LL$^\prime$ anomalous dimension  introduced in Ref.~\cite{Bonvini:2016wki} in place of the full NLL one. (see also Ref.~\cite{Bonvini:2018iwt}).}
These deformations of the subleading logarithms in the resummed result are then added in quadrature to form a representative uncertainty which we show as an error band in Fig.~\ref{2:fig:SQDoubleDiff}.
Conceptually, estimating the uncertainties in this way underestimates the correct size of NLL contributions, but it  suffices to show that the difference between LO+LL and NLO+LL can not be generated only by missing subleading terms, as it is much larger than the bands. This forces the conclusion that, at least for this combination of scale and  $\pt{}$, subleading-power contributions play a large role in the small-$x$ region.
To further probe this difference, we compare an additive matching, used as a default choice in past works, to a multiplicative one
\footnote{Without going in detail, additive matching involves subtracting the doubly counted contributions from the expansion of the resummed result up to the order at which the fixed order is computed, while multiplicative matching involves multiplying the fixed order by the resummed result divided by its expansion.}.
In the plot, the difference between the two curves is shown as a dotted line. This difference, which is related to the ratio between the exact NLO and its small-$x$ approximation, includes the effect of subleading-power contributions and falls outside the uncertainty band from subleading logarithms.

\begin{figure}[t]
  \centering
  \includegraphics[width=0.49\textwidth,page=6]{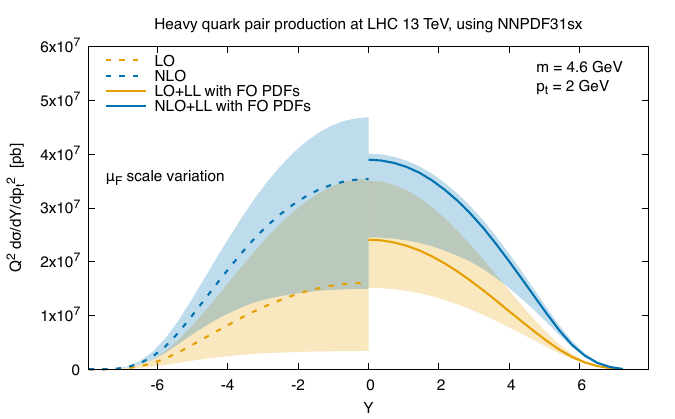}
  \includegraphics[width=0.49\textwidth,page=5]{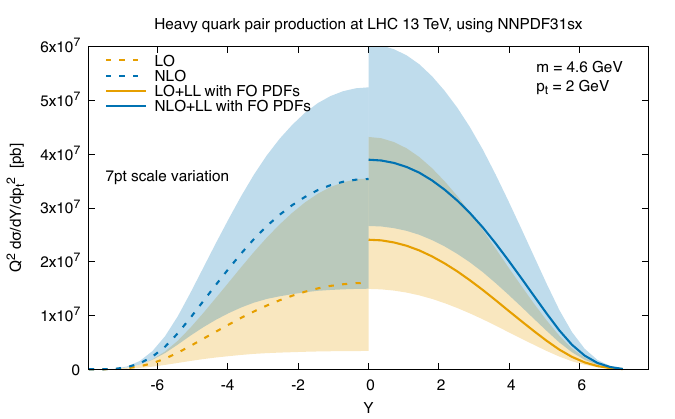}
  \caption{Scale uncertainty for the double differential distribution in rapidity and transverse momentum of the bottom quark,
    plotted as a function of the rapidity for $\pt=2$~GeV, for bottom pair production at LHC $13$~TeV.
    The left plot shows factorisation scale uncertainty only, while the right plot shows the standard 7-point uncertainty envelope.}
  \label{2:fig:SQDoubleDiffScaleUnc}
\end{figure}

Fig.~\ref{2:fig:SQDoubleDiffScaleUnc} shows the scale uncertainty band of our results, taking advantage of the symmetry of the rapidity distribution across negative and positive values to depict fixed-order and resummed results  respectively.
The left plot shows the impact of a factorisation scale variation of a factor of 2 up and down, while the right plot shows the envelope of the 7-point variation of $\muF$ and $\muR$.
The inclusion of small-$x$ resummation leads to a reduction of the uncertainty due to $\muF$ variation. However, the uncertainty of the resummed result becomes comparable to the fixed-order result when $\muR$ variations are also considered.
This is not unexpected as the process happens at a low energy scale, thus driving sizeable variations of $\as$ with changes of $\muR$ (this same mechanism makes the NLO uncertainty larger than the LO one). Since the LL resummation does not include $\muR$ logarithms from the coefficient functions, the running of $\as$ is not compensated completely in the matching.
Arguably, a full NLL resummation would be required to obtain reduced 7-points uncertainty.
          
\begin{figure}[t]
  \centering
  \includegraphics[width=0.69\textwidth,page=3]{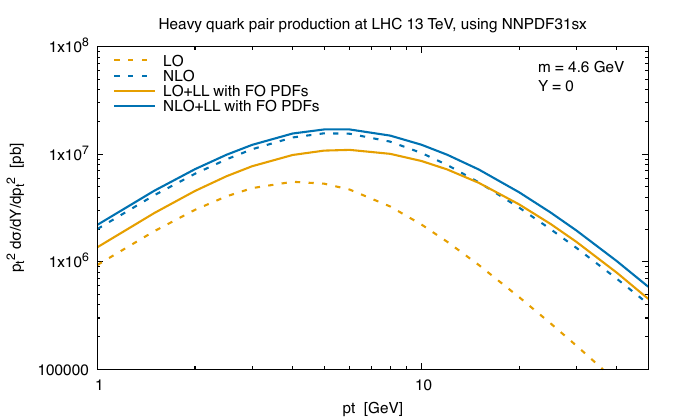}
  \caption{The double differential distribution in rapidity and transverse momentum of the bottom quark,
    plotted as a function of $p_t$ for central rapidity $Y=0$, for bottom pair production at LHC $13$~TeV.}
  \label{2:fig:SQDoubleDiffPt}
\end{figure}

\begin{figure}[t]
  \centering
  \includegraphics[width=0.49\textwidth,page=3]{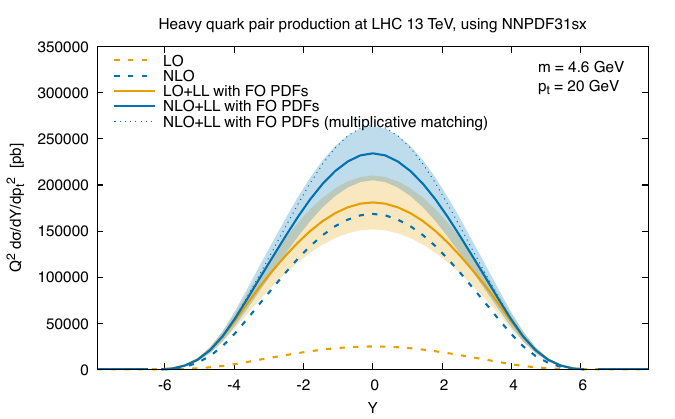}
  \includegraphics[width=0.49\textwidth,page=4]{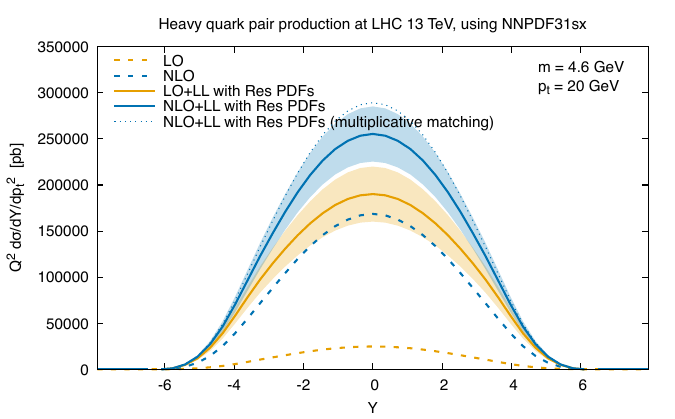}
  \caption{The double differential distribution in rapidity and transverse momentum of the bottom quark,
    plotted as a function of the rapidity for $p_t=20$~GeV, for bottom pair production at LHC $13$~TeV.
    The left plots are obtained using NNPDF31sx at fixed order, while in the right plot the resummed result
    is computed with the resummed PDFs from the same family.
    The uncertainty band represents an estimate of NLL corrections.}
  \label{2:fig:SQDoubleDiff_highpt}
\end{figure}

Finally, in  Fig.~\ref{2:fig:SQDoubleDiffPt} we show the same double differential distribution as a function of $\pt{}$ at fixed $Y=0$ and using fixed-order PDFs.
Interestingly, in the larger $\pt{}$ region, the size of the NLO correction grows as well as the impact of resummation over the LO.
Conversely, matching resummation to NLO gives a smaller correction, implying that small-$x$ terms form a dominant part of the NLO at large $\pt{}$.
Given the lack of direct dependence on the transverse momentum in the resummation, this effect is simply a result of the underlying kinematical constraints.
Specifically, it is reasonable to think that the smaller phase space available at large transverse momentum means that contributions from the low-$x$ region become more prominent, even at central rapidity. On the other hand, at large rapidity, this phenomenon is expected to occur at all values of $\pt{}$.
Also, we can show the compensation effect at high-$\pt{}$ across all values of $Y$ in Figs.~\ref{2:fig:SQDoubleDiff_highpt}, where the double differential distribution is plotted at $\pt{} =20$~GeV. Compared to the $\pt{}=2$~GeV plot from Figs.~\ref{2:fig:SQDoubleDiff}, we can see that this time the NLO curve and the LO+LL are significantly closer to each other, with the NLO+LL one slightly above. This is consistent with the picture given by Fig.~\ref{2:fig:SQDoubleDiffPt} for the high-$\pt{}$ region.

Moreover, results for differential heavy-quark pair production at NNLO recently became available~\cite{Catani:2020kkl,Mazzitelli:2023znt}.\footnote{To our present knowledge, the is still no open-source implementation of these results.} In principle, this development allows to study in greater depth to which extent fixed-order perturbation theory manages to capture the same effect of resummation. However, it must be stressed that small-$x$ resummation for coefficient functions in the $k_t$-factorisation formalism is known only to LL accuracy and any attempt at studying a NNLO+LL matching would suffer from an incomplete subtraction of the first subleading logarithmic terms, which can be sizeable. Currently, the NLL resummation of high-energy logarithms in coefficient functions is not completely understood as we will see in chapter~\ref{ch2a}. Thus, we need to postpone the study of this aspect of heavy-quark phenomenology to future works. 

\subsection{Results differential in the heavy-quark pair} \label{2:ssec:pairkin}

In this section, we focus on the final state being a pair of heavy quarks, where we assume that $q=p_3+p_4$.
This choice loosely corresponds to the production of a heavy quark bound state ($J/\psi$, $\Upsilon$ or heavier resonances).
The computation of the partonic off-shell coefficient function is given in detail in appendix~\ref{app:pairkin}.
A major consideration is that this kinematic setting is a 2 to 1 process at the lowest order, so the differential coefficient function contains delta functions (see equation~\eqref{eqP:triplediffxs2}).
Thus, some of the integrals defining the resummed collinear coefficient functions, as described in section~\ref{sec:channels}, can be computed analytically~\cite{Silvetti:2022hyc}.

The simplified structure of the coefficient function, however, complicates the plotting for the triple differential distribution.
The regular coefficient function Eq.~\eqref{eq:resCreg} is a proper function, while the auxiliary coefficient Eq.~\eqref{eq:resCaux} is a distribution and cannot be plotted together.
A workaround is showing the hadron-level cross section, which we plot at LHC energy scale $13$~TeV, with bottom mass $m_b=4.6$~GeV and using the same NNPDF31sx~\cite{Ball:2017otu} PDF.

\begin{figure}[t]
  \centering
  \includegraphics[width=0.49\textwidth,page=3]{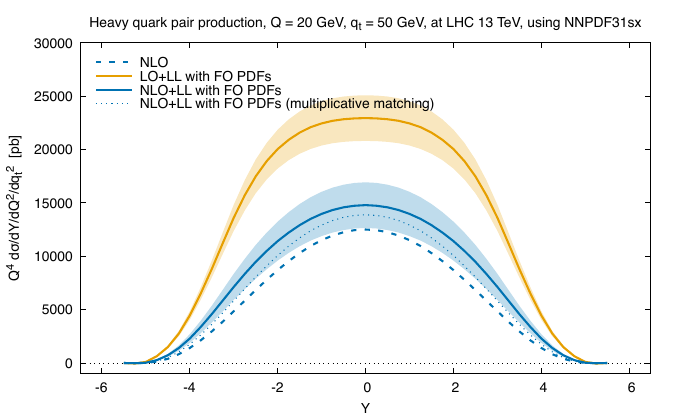}
  \includegraphics[width=0.49\textwidth,page=4]{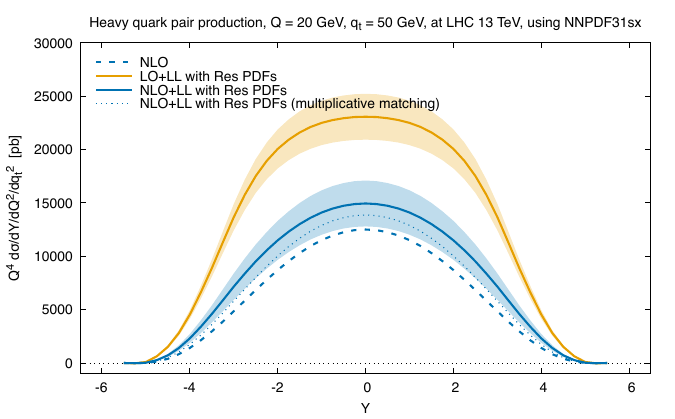}
  \caption{The triple differential distribution in invariant mass, rapidity and transverse momentum of the bottom pair,
    plotted as a function of the rapidity for $Q=20$~GeV and $p_t=50$~GeV, for bottom pair production at LHC $13$~TeV.
    The left plots are obtained using NNPDF31sx at fixed order, while in the right plot the resummed result
    is computed with the resummed PDFs from the same family.
  The uncertainty band represents an estimate for the NLL corrections.}
  \label{2:fig:TripleDiff}
\end{figure}

In Fig.~\ref{2:fig:TripleDiff} we show the triple differential distribution, plotted as a function of the rapidity $Y$ of the pair at fixed invariant mass $Q=20$~GeV and fixed transverse momentum $q_t=50$~GeV.
The LO curve cannot be plotted  as it is proportional to $\delta(\qt^2)$, likewise no ratio plot can be obtained.
The NLO (blue dashed curve) is smaller than the LL curve (solid orange), which is effectively identical to a LO+LL matching.
Instead, the resummed NLO+LL curve (solid blue) gives a small positive correction to the NLO result, pointing toward the still larger LL prediction.
The solid blue curve of the resummed NLO+LL provides a slight positive adjustment to the NLO result, which implies that resummation is likely to slightly improve perturbative expansion convergence.
As with Fig.~\ref{2:fig:SQDoubleDiff}, the left panel shows the resummed result with the same fixed-order PDFs used for NLO, while the right panel shows the resummed contribution with resummed PDFs.
The difference between the two options is not significant, which could be due to the larger value of $x$ and the higher invariant mass, indicating that this observable is not very effective in constraining the PDFs at small $x$. Details are provided in the figure caption.

\begin{figure}[t]
  \centering
  \includegraphics[width=0.49\textwidth,page=1]{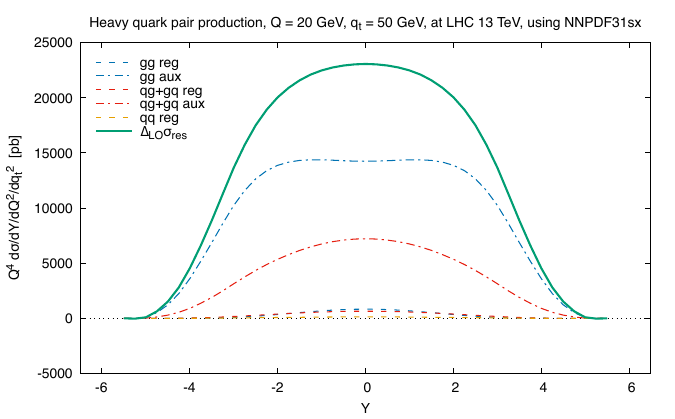}
  \includegraphics[width=0.49\textwidth,page=2]{images/plot_QQbarTripleDiff_qt50}
  \caption{Breakdown of the individual contributions to the resummed triple differential distribution
    in invariant mass, rapidity and transverse momentum of the bottom pair
    from the $gg$, $gq+qg$ and $qq$ channels separating the regular and auxiliary parts.
    The left plot focuses on the resummed contribution to be matched to the LO,
    while the right plot focuses on the resummed contribution to be matched to NLO.
    The results in these plots are obtained using NNPDF31sx with resummation at LHC 13 TeV,
    as a function of the rapidity, for invariant mass $Q=20$~GeV and for transverse momentum
    $q_t=50$~GeV.}
  \label{2:fig:TripleDiff2}
\end{figure}

Fig.~\ref{2:fig:TripleDiff2} contains the breakdown of the individual contributions to the cross section
The matching to LO (left plots) is ordered like the single-quark distributions in Fig.~\ref{2:fig:SQDoubleDiff2}.
The resummed contribution is positive, which aligns with the effective identity LL and  LO+LL distribution.
When performing the subtraction of the $\calO(\as)$ expansion to match the resummation to NLO (right plots), we observe a smaller contribution from resummation.
It is worth noting that the auxiliary contributions now reach a similar magnitude as the regular contributions at moderate rapidity, but they still dominate when the latter is larger.
\begin{figure}[t]
  \centering
  \includegraphics[width=0.49\textwidth,page=6]{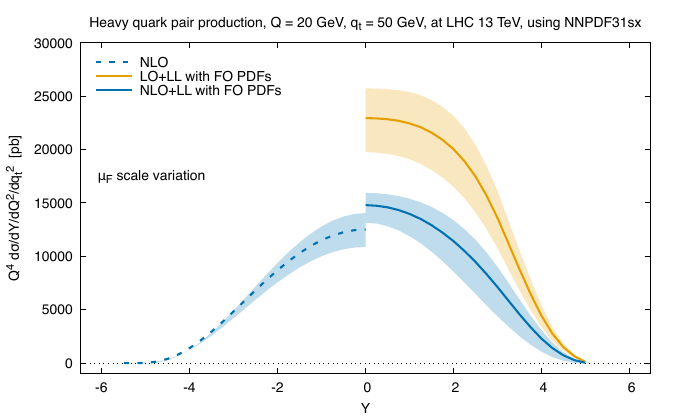}
  \includegraphics[width=0.49\textwidth,page=5]{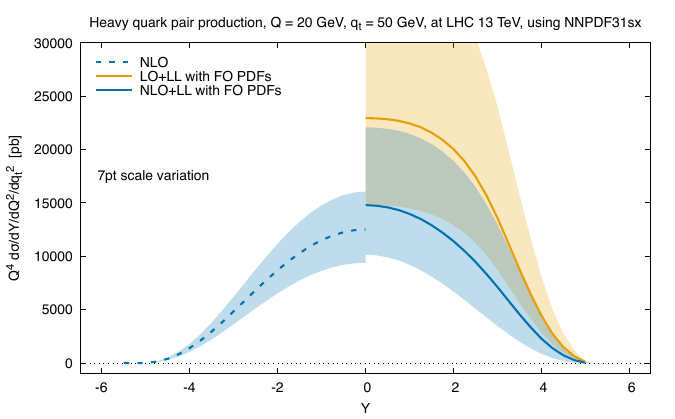}
  \caption{Scale uncertainty for the triple differential distribution in rapidity and transverse momentum of the bottom quark pair,
    plotted as a function of the rapidity for $\pt=50$~GeV, for bottom pair production at LHC $13$~TeV.
    The left plot shows factorisation scale uncertainty only, while the right plot shows the standard 7-point uncertainty envelope.}
  \label{2:fig:PairTripleDiffScaleUnc}
\end{figure}

In Fig.~\ref{2:fig:TripleDiff} we perform an analysis of uncertainty from subleading logarithms using the same technique from the previous section.
The resulting band is slightly larger than the single-quark case, but it's still insufficient to cover the gap between LO+LL and NLO+LL.
This implies that non-small-$x$ effects play a large role.
Similarly, a comparison between multiplicative and additive matching at NLO+LL gives a degree of variation comparable with the uncertainty band from subleading logarithms.
Fig.~\ref{2:fig:PairTripleDiffScaleUnc} contains the study of scale variations, again, similarly to the single quark case.
The $\mu_F$ variations on the left plot and a full 7-point one on the right plot.
It is worth noting that in both plots, a noticeable reduction is observed when transitioning from LO+LL to NLO+LL, indicating a stabilisation of the perturbative expansion with the inclusion of resummation.

In conclusion, we summarise the new result we discussed in this chapter.
First, we have obtained small-$x$ resummed formulae for distributions of partonic coefficient functions, valid for any gluon-gluon initiated process at LO. We applied this to heavy-quark pair production at the LHC and produced a number of numerical results as demonstration of the framework. We found evidence that small-$x$ resummation has sizeable effect, as we expect from the low-$x$ values the process can reach at the LHC. 
A more complete study of small-$x$ phenomenology in heavy quark production can be pursued in future work as well as the application of the formalism to other processes, especially differential Drell-Yan.

        \chapter[HDIS: small-$x$ resummation beyond LL]{Higgs-Induced Deep Inelastic scattering: small-$x$ resummation beyond leading logarithms}
	\label{ch2a}
	In this chapter we return to the structure of $\kt{}$-factorisation to discuss an attempt to push the logarithmic accuracy of small-$x$ resummation in the coefficient function to relative next-to-leading logarithms.

As discussed in Sec.~\ref{2:sec:HElog}, the effects of small-$x$ resummation are known to NLL$x$ only for the QCD anomalous dimensions, while the counterpart coefficient function resummation is consistently known only to relative LL$x$ accuracy.
The improvement of logarithmic accuracy has already been achieved in other resummation schemes. For example, in the the hybrid factorisation framework, the effects of resummation in Mueller--Navelet jets~\cite{Mueller:1986ey} was subject to extensive study, as well as exclusive electroproduction of light vector mesons~\cite{Ivanov:2005gn,Ivanov:2006gt}.
A partial NLL treatment was achieved for  multi-jet detection~\cite{Caporale:2016soq,Caporale:2016zkc,Celiberto:2017ius}, and on final states where a $J/\psi$~\cite{Boussarie:2017oae}, a Drell--Yan pair~\cite{Golec-Biernat:2018kem}, or a heavy-flavoured jet~\cite{Bolognino:2021mrc} is inclusively emitted in association with a light-flavoured jet, provided that the two objects are well separated in rapidity. Similar analyses on inclusive heavy-quark pair photo- and hadroproductions were respectively proposed in Refs.~\cite{Celiberto:2017nyx,Bolognino:2019yls}.
More recently, Ref.~\cite{Celiberto:2022fgx} featured a full computation of the next-to-leading order correction to the impact factor (vertex) for the production of a forward Higgs boson, obtained in the infinite top-mass limit. This was a missing ingredient for a NLL$x$ description of the inclusive hadroproduction of a forward Higgs in the small-$x$ regime in the language of the BFKL equation.

This motivates the interest in searching a similar extension for the resummation scheme summarised in section~\ref{2:sec:HElog} and reformulated in the \hell\ formalism. Thus, the chapter is organised as follows. The first section~\ref{2:1:1} is devoted a step-by-step derivation of the $\kt{}$-factorisation at LL$x$ using  heavy-quark photo-production as an explicit example. In section~\ref{2:sec:HDIS}, we switch our focus to Higgs-induced Deep Inelastic Scattering (HDIS), which will act as the test case for the NLL$x$ expansion. Section~\ref{2:sec:strategy} elucidates the structure of the computation and collects the partial results obtained. More details are collected in appendix~\ref{app:HDIS-NLO}.

\section{Structure of the $k_t$ factorisation}\label{2:1:1}
We start by going over the formulation of $\kt{}$-factorisation along the lines of Refs.~\cite{Catani:1990xk,Catani:1990eg,Catani:1994sq} and consider the process
\begin{subequations} \label{2:eq:HQphoto}
   \begin{align}
        & \mathrm{p}\(p_1\) +\gamma\(p_2\) \rightarrow  Q\(p_3\) + \barQ \(p_4\) + X \, . \\
        & g/q\(k\) +\gamma\(p_2\) \rightarrow  Q\(p_3\) + \barQ \(p_4\) + X \, ,
   \end{align}
\end{subequations}
with $p_{1,2}=\frac{\sqrt{S}}{2}\(1,0,0,\pm1\)$ and $S$ the collider energy. If we take the high-energy limit $S=\spd{p_1}{p_2}\gg Q^2=( 2m )^2$, or alternatively $x=\frac{Q^2}{S} \ll 1$, the presence of logarithms $\log\(\frac{1}{x}\)$  spoils the perturbativity of $\as(Q^2)\ll 1$ in both the anomalous dimensions and the hard scattering of Eq.~\eqref{2:eq:HQphoto}.
To manage the logs in the latter, we need to rewrite the cross section as
\begin{equation}\label{2:eq:ktfactgammag}
\sigma \( x, Q^2\)= \int \dd^2\kt \int_x^1 \frac{\dd z}{z} \hat\sigma_\text{off} \(\frac{x}{z},\frac{\kt^2}{Q^2}\) \calF \(z,\kt; Q^2\) \, ,
\end{equation}
and give a prescription to compute the parton cross-section $\hat\sigma_\text{off}$ with an incoming gluon of momentum $k=z p_1 + \kt$. To do so, we need to leverage the regime
\begin{equation}
 S \gg \(p_3 + p_4\)^2 \equiv s\, , \:  S \gg \(k - p_1\)^2 \equiv \kt^2 \; . 
\end{equation}

\subsection{Scaling of LO off-shell coefficients and the high-energy projector}\label{2:ssec:ktfact1}
First we show that a single gluon polarisation contributes in the high-energy limit. To do so we rewrite the cross section of Eq.~\eqref{2:eq:HQphoto} as 
\begin{equation}\label{2:eq:tensorfact}
  4 m^2 \sigma = \frac{2m^2}{S}\int\frac{\dd^4k}{(2\pi)^4} A_{\mu\nu}\( k,p_2 \) d^{\mu\mu'}\(k,n\)d^{\nu\nu'}\(k,n\)G_{\mu'\nu'}\(p_1,k\)\,,
\end{equation}
emphasising the integral over the gluon connecting the lowest-order $\gamma + g^* \rightarrow Q + \barQ$ squared amplitude ($A^{\mu\nu}$) to the rest of the $g/q \rightarrow g^* + \text{ISR}$ amplitude $G^{\mu\nu}$.
 $d^{\mu\nu}\(k,n\)$ is numerator from the intermediate propagator in the light-cone gauge (LCG) with axis $n=p_2$. The corresponding denominators are hidden inside $G$
\footnote{In this specific example, current conservation in $A$ implies that we could dispense with writing $d$ explicitly as all gauge terms drop. This may turn out not to be true beyond the LO factorisation, so we keep them explicitly instead.}.
We then note that the lowest-order Abelian absorptive part $A$ is independent of the gauge vector and can be decomposed as
\begin{equation}\label{2:eq:HEapprox}
  A^{\mu\nu}\(k,p_2\) = D_1\(-g^{\mu\nu} + \frac{k^\mu k^\nu}{k^2}\) -\frac{D_2}{k^2}\(\frac{k^2}{\spd{k}{p_2}}p_2^\mu-k^\mu\)\(\frac{k^2}{\spd{k}{p_2}}p_2^\nu-k^\nu\)\, ,
\end{equation}
with $D_{1,2}$ being scalar functions of the Minkowski invariants $\frac{x}{z}$ and $\frac{k^2}{z S}$.
An explicit computation yields that they are indeed small and $\calO\(\frac{x}{z}\)$ in the region identified by $x\rightarrow 0$ at fixed $\frac{k^2}{z S}$.
Then, in the same limit, $z \rightarrow 0$ follows from $\calO\(\frac{x}{z}\) \sim 1$ and the tensor structure associated with $D_2$ provides a net $\(z\)^{-2}$ power enhancement over $D_1$. So, Eq.~\ref{2:eq:HEapprox} can be approximated as
\begin{equation}\label{2:eq:HEscaling}
         A^{\mu\nu}\(k,p_2\) \simeq -\frac{D_2 \kt^2}{(zS)^2}p_2^\mu p_2^\nu\,.
\end{equation}
Using this result we can rewrite Eq.~\eqref{2:eq:tensorfact} as
  \begin{align}
    4 m^2 \sigma & = \frac{2m^2}{S}\int\frac{\dd^4k}{(2\pi)^4} A_{\mu\nu}\( k,p_2 \) d^{\mu\mu'}\(k,n\)d^{\nu\nu'}\(k,n\)G_{\mu'\nu'}\(p_1,k\) \nonumber \\
     & \simeq  -\frac{2m^2}{S}\int \frac{\dd z \dd k^2}{2z}\int\frac{\dd^2\kt}{(2\pi)^4} \frac{D_2 \kt^2}{(zS)^2}p_2^\mu p_2^\nu d_{\mu\mu'}\(k,n\)d_{\nu\nu'}\(k,n\)G^{\mu'\nu'}\(p_1,k\) \nonumber \\
     & \simeq  \int \frac{\dd z}{z}\int \dd^2\kt \(\frac{x}{z} D_2 \) \[ \int \frac{\dd k^2}{(2\pi)^4} \frac{-\kt^2}{z S^2} p_2^\mu p_2^\nu G_{\mu \nu}\(p_1,k\)\]\label{2:eq:calF}\, ,
  \end{align}
which matches the form of Eq.~\eqref{2:eq:ktfactgammag}. In practice, the easiest way to extract the $A_2$ coefficient is to couple eikonal vertices to the polarisation tensors in Eq.~\eqref{2:eq:tensorfact} such that
\begin{align}
       D_2 & = A_{\mu\nu}\( k,p_2 \) d_{\mu\mu'}\(k,n\)d_{\nu\nu'}\(k,n\)\(-\frac{z^2}{\kt^2}p_1^{\mu'}  p_1^{\nu'}\) \nonumber \\
       & = A_{\mu\nu}\( k,p_2 \) \(-\frac{\kt^\mu \kt^\nu}{\kt^2}\) \nonumber \\
       & = \frac{z}{x} \hat\sigma_{\text{off}} \(\frac{x}{z},\frac{\kt^2}{Q^2}\)\equiv \frac{x}{Q^2} \calC\(\frac{x}{z},\xi\) \, .
\end{align}
In this way, we find the prescription to compute the off-shell coefficient function $\calC$ by defining a replacement for the momentum and polarisation tensor of the incoming gluon. Explicitly
\begin{subequations}
\begin{align}
& p_1 \xrightarrow{\text{off-shell}} k = z p_1 + \kt \, ,\\ 
& d^{\mu\nu}\(p_1,n\) \xrightarrow{\text{off-shell}} \dCH^{\mu\nu}\(\kt\)=-\frac{\kt^\mu \kt^\nu}{\kt^2} \, .   \label{2:eq:dCH} 
\end{align}
\end{subequations}
This derivation effectively shows that the high energy limit of $\sigma$ is probed by the exchange of a soft gluon, $z\rightarrow 0$.
Moreover, the on-shell process $\gamma g \rightarrow Q \barQ $ at Born level can be recovered smoothly by setting $\kt \rightarrow 0$, ensuring that $\hat\sigma_{\text{off}}$ is gauge invariant at this level.

\subsection{The soft emission chain and unintegrated gluon distribution}
Having just defined the off-shell cross section at LO, we now move to the off-shell gluon distribution $\calF$ in Eq.~\eqref{2:eq:ktfactgammag}, which can be identified with the square bracket in Eq.~\eqref{2:eq:calF}. A formal proof of this relation is complicated and beyond the scope of this discussion, but more detail can be found in Ref.~\cite{Catani:1994sq}. Instead, we use a simpler argument under the assumptions of fixed coupling, modelled after the discussion in Ref.~\cite{Caola:2010kv}

We begin by returning to Eq.~\eqref{2:eq:calF} and comparing it to its counterpart in collinear factorisation. If we consider only  a single extra emission from the collinear initial state, we can identify the collinear coefficient function $C_1\equiv \frac{x}{Q^2}\sigh$ as
\begin{equation}\label{app:1gluon_xspace}
C_{1,\text{bare}}\(x,\frac{\muF^2}{Q^2}, \as; \epsilon \) = \int_x^1 \frac{\dd z}{z} \int \frac{\dd \xi}{\xi^{1+\epsilon}} \calC\( \frac{x}{ z }, \xi,\as; \epsilon\) K^1\( z, \frac{\muF^2}{Q^2 \xi}, \as; \epsilon \) ,
\end{equation}
where $K^1$ stands for the amplitude for a single gluon emission contracted with the expression in square brackets in Eq.~\eqref{2:eq:calF} and the polarisation tensor of the initial collinear gluon. Given the presence of the IR singularity, we use dimensional regularisation with $D=4-2\epsilon$ and $\epsilon<0$. Then, the additional factor $\xi^{-\epsilon}$ is introduced by dimensional regularisation and is factorised for convenience.
After taking a Mellin transform over $x$ the convolution reduces to the product
\begin{equation}\label{1g_d}
C_{1,\text{bare}}\( N, \frac{\muF^2}{Q^2}, \as; \epsilon\) = \int_0^{\infty} \frac{d\xi}{\xi^{1+\epsilon}} \calC\( N, \xi, \as; \epsilon \) \[K^1\( N,\(\frac{\muF^2}{Q^2\xi}\)^{\epsilon} \as; \epsilon\) \].
\end{equation}
We pin down the $\epsilon$ singularity using the expansion
\begin{equation}\label{epsilon_expansion}
  \frac{1}{\xi^{1+\epsilon}} = -\frac{\delta(\xi)}\epsilon  + \sum_{k=0}^{{\infty}} \left[\frac{\log^k \xi}{\xi} \right]_+\frac{(-\epsilon)^k}{k!}\,,
\end{equation}
and retrieve from Eq.~\eqref{1g_d} 
\begin{equation}\label{coll_limit}
C_{1,\text{bare}}\( N, \frac{\muF^2}{Q^2}, \as; \epsilon\)=-\frac{1}{\epsilon} \calC\(N,0,\as; \epsilon\) \times \left[ K^1\( N, \as \)\] + \calO\(\epsilon^0\),
\end{equation}
At this point we  must identify this singular contribution as the leading $N$ pole of the gluon anomalous dimension $K^1(N,\as) = \frac{\as C_A}{ \pi \(N-1\)}= \as \gamma_0(N)$.

If now we consider the same structure repeated $n$ times for the case of chained emissions, we get
\begin{align}\label{ngluon_kt}
& C_{n,\textnormal{bare}}\( N, \frac{\muF^2}{Q^2}, \as; \epsilon\) =  \[\as\(\frac{\muF^2}{Q^2}\)^{\epsilon} \gamma_0(N)\] \int_0^{\infty} \frac{\dd \xi_n}{\xi_n^{1+\epsilon}} \calC\( N, \xi_n, \as; \epsilon\) \times \nonumber\\
& \times \int_0^{\xi_n} \[\as\(\frac{\muF^2}{Q^2}\)^{\epsilon} \gamma_0(N)\] \frac{\dd\xi_{n-1}}{\xi_{n-1}^{1+\epsilon}}\times \ldots \times\int_0^{\xi_2}\[\as\(\frac{\muF^2}{Q^2}\)^{\epsilon} \gamma_0(N)\]\frac{\dd\xi_{1}}{\xi_{1}^{1+\epsilon}}.
\end{align}
The collinear singularities can be removed by subtracting the collinear pole before each integration with the same tools of normal collinear factorisation, in the language of Ref.~\cite{Curci:1980uw}
\begin{align}\label{CFPsub}
\int_0^{\xi_2}\[ \as \(\frac{\muF^2}{Q^2}\)^{\epsilon} \gamma_0(N)\] \frac{\dd\xi_{1}}{\xi_{1}^{1+\epsilon}}\rightarrow & \(1-\calP_{\MSbar}\)\int_0^{\xi_2}\[\as\(\frac{\muF^2}{Q^2 }\)^{\epsilon} \gamma_0(N)\] \frac{\dd\xi_{1}}{\xi_{1}^{1+\epsilon}}  \nonumber\\
& = \( \as \gamma_0(N) \) \( -\frac{1}{\epsilon}\frac{(4\pi)^{\epsilon}}{\Gamma(1-\epsilon)} \(\frac{\muF^2}{Q^2\xi_2}\)^{\epsilon} + \frac{S_{\epsilon}}{\epsilon} \),
\end{align}
this subtraction is, of course, scheme-dependent. For example in the $\MSbar$ case we have $$\calP_{\MSbar} f\(\epsilon\) \equiv \sum_{k>0} \lim_{\epsilon \rightarrow 0} \[\epsilon^k f\(\epsilon\)\]\frac{S_{\epsilon}^k}{\epsilon^k} \, , $$ with $S_{\epsilon} = \(\frac{\eu^{-\gE{}}}{4 \pi}\)^{\epsilon}$.
This operations can be carried out recursively to factorise the collinear singularities out of the coefficient function and into the PDFs to all orders of $\as$.
We show this schematically by writing the bare forward amplitude ($\bar M$) in $d=4-2\epsilon$ dimensions  as the chain  convolution of a hard part $H$ and the iteration of a kernel $K$. The latter represents a single gluon emission 
\begin{equation}\label{kernel_exp}
\bar M = H \otimes_{x,k_T} (1 + K + K\otimes_{x,k_T} K + ... ) \equiv 
M (1 + K + K^2 + K^3 + ... ) = H \frac{1}{1-K},
\end{equation}
and  $\otimes_{x,k_T}$ stands both for convolution in $x$ space and $k_T$ integration. 
\begin{subequations}
\begin{align}
&  1-K = 1 - \calP K - (1-\calP)K = \[1-\calP K (1-(1-\calP )K)^{-1}\]\[1-(1-\calP )K\],   \label{geometric_K} \\
  &  \frac{1}{1-K} = \[\frac{1}{1-(1-\calP)K}\]\[\frac{1}{1-\calP K (1-(1-\calP)K)^{-1}}\]. \label{sub_K}
\end{align}
\end{subequations}
Then finally
\begin{align}
   \bar M & = \[H \frac{1}{1-(1-\calP)K}\]\otimes\[\frac{1}{1-\calP K \(1-\(1-\calP\)K\)^{-1}} \bar \Gamma\] \nn
   & \rightarrow M\otimes\Gamma \, ,
\end{align}
with the forward amplitude now being free of collinear singularities, which are shifted inside the transition function $\Gamma$ which, once more, is absorbed into the the PDF definition to introduce scale dependence.

Beyond this, we return to Eq.~\eqref{CFPsub} and observe that the insertion of $n-1$ iterative subtractions looks like
\begin{align}
& C_{n}\( N, \frac{\muF^2}{Q^2}, \as; \epsilon\) = \[\as\(\frac{\muF^2}{Q^2}\)^{\epsilon} \gamma_0(N)\] \int_0^{\infty} \frac{\dd \xi_n}{\xi_n^{1+\epsilon}} \calC\( N, \xi_n, \as; \epsilon\) \times \(1-\calP \)\nonumber\\
& \times \int_0^{\xi_n} \[\as\(\frac{\muF^2}{Q^2}\)^{\epsilon} \gamma_0(N)\] \frac{\dd\xi_{n-1}}{\xi_{n-1}^{1+\epsilon}}\times... \times\(1-\calP\)\int_0^{\xi_2}\[\as\(\frac{\muF^2}{Q^2}\)^{\epsilon} \gamma_0(N)\]\frac{\dd\xi_{1}}{\xi_{1}^{1+\epsilon}}\, .
\end{align}
Note that we write $\calP$ for $\calP_{\MSbar}$ and omit all scheme-dependent finite terms even before the dimensional regularisation is relaxed for brevity, as they do not affect the end result we seek to show in this argument.
Now we can safely perform the first $n-1$ integrals and then sum over the number of emissions $n$
\begin{align}\label{ngluon_msb}
  &C_{n}\( N, \frac{\muF^2}{Q^2}, \as; \epsilon\) = \[\as\(\frac{\muF^2}{Q^2}\)^{\epsilon} \gamma_0(N)\]\times \nonumber\\
  &\quad\times \int_0^{\infty} \frac{\dd \xi_n}{\xi_n^{1+\epsilon}} \calC\( N, \xi_n, \as; \epsilon\) \frac{1}{(n-1)!}\frac{1}{\epsilon^{n-1}} \[\as \gamma_0(N) \(1-\( \frac{\muF^2}{Q^2 \xi_n} \)^{\epsilon}\)\]^{n-1},\\
  &C\(N,\frac{\muF^2}{Q^2}, \as; \epsilon\)  = \sum_{n=0}^{\infty} C_{n}\(N,\frac{\muF^2}{Q^2}, \as; \epsilon\) = \nonumber \\
  &\quad = \[\as\(\frac{\muF^2}{Q^2}\)^{\epsilon} \gamma_0(N)\] \int_0^{\infty}\frac{\dd \xi}{\xi^{1+\epsilon}} \calC\(N, \xi, \as{}; \epsilon\) \exp \[ \frac{\as \gamma_0(N)}{\epsilon} \(1-\( \frac{\muF^2}{Q^2 \xi_n} \)^{\epsilon}\) \] \nonumber .
  \end{align}
  This leaves us with a resummed expression for $C$ in Mellin space
\begin{equation}\label{2:eq:CcollMellin}
  C\(N,\frac{\muF^2}{Q^2}, \as\) =  \int_0^{\infty} \dd \xi \,   \calC\(N, \xi, \as{}\)  \[ \as \gamma_0(N) \xi^{\as \gamma_0(N)-1}\]\,, 
  \end{equation}
when $\epsilon \rightarrow 0$ and $\muF^2 = Q^2$.
\begin{equation}
  C\(N, \as\)  =  \int_0^{\infty} \dd \xi \,  \calC\(N, \xi, \as\)  \frac{\dd}{\dd \xi} U\(N,\xi\) \quad \text{with}\quad U\(N,\xi\) = \xi ^{\as \gamma_0(N)} \, .
\end{equation}
This last equation expression has a similar structure to Eq.~\ref{2:eq:resCggNb}, but with only one off-shell gluon rather than two and it evaluates the total coefficient function rather than its triple differential distribution. 

Now, if we multiply by the Mellin transform of the PDF, we achieve a fixed-coupling LL definition of $\calF$
\begin{align}
 C\(N, Q^2, \as\) f_g\(N, Q^2\) & = \int_0^{\infty} \dd \xi \calC\(N, Q^2, \kt{}^2, \as\) \[ U'\(N, \xi, Q^2 \) f_{g}\( N, Q^2\) \]\, , \nn
 & = \int_0^{\infty} \dd \kt{}^2 \calC\(N, Q^2, \xi, \as\) \calF\(N,\xi,Q^2\)\, , \label{2:eq:Foffbuild}
 \end{align}
and, after returning to direct space
\begin{equation}
\int_x^1\frac{\dd z}{z}  C\(\frac{x}{z}, Q^2, \as\) f\(z, Q^2\) = \int_x^1\frac{\dd z}{z} \int_0^{\infty} \dd \xi \calC_{g}\(\frac{x}{z}, Q^2, \xi, \as\) \calF\(z,\xi,Q^2\) \, ,
\end{equation}
we obtain the factorised expression like Eq.~\eqref{2:eq:ktfactgammag}.
Two observations are in order. First, we defined the evolutor $U$ in this example considering only LL$x$ and fixed coupling contribution, but the same structure holds with running coupling and NLL$x$ by using the \hell\ evolutor discussed in section~\ref{2:ssec:diff-hell} at Eqs.~\eqref{2:eq:UABFht}. Second, in this argument we considered only gluon in the initial state and specifically in the definition of $\calF$ in Eq.~\eqref{2:eq:Foffbuild}. However, as argued in Sec.~\ref{2:sec:HElog}, the contributions from a initial-state collinear quark are related to the gluon ones only by a colour factor. So, we can account for both by using Eq.~\eqref{2:eq:Foff} to define $\calF$.

For completeness we also point out (again for fixed coupling) the Mellin space analogue of the resummation formula. We can recognise that Eq.~\eqref{2:eq:CcollMellin} has again the structure of a Mellin transform over $\xi$, this defines the so-called impact factor
 \begin{equation}\label{2:eq:ImpFac}
 h\(N,M\) = M \int_0^{\infty}\dd \xi\, \xi^{M-1} \frac{\dd}{\dd\xi}\calC\(N,Q^2,\xi,\as\) \, ,
 \end{equation}
which captures the entire LL$x$ tower of contributions when evaluated in $M=\as\gamma_0(N)$.
and specifically one evaluated at $\as \gamma_0(N)$.
While this technique is very elegant, to obtain physical predictions for phenomenology the Mellin transform over $N$ must be undone. As argued in section~\ref{2:sec:differential}, this is often impossible to do analytically and cumbersome to perform numerically.

\section{Higgs-induced Deep Inelastic Scattering} \label{2:sec:HDIS}
\begin{figure}
\centering
\begin{subfigure}{0.4\linewidth}
\centering
\includegraphics[scale=1.5,page = 1]{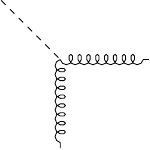}
\caption{Feynman diagram for the $ggH$ vertex in the effective theory.}
\label{2:fig:LOdia}
\end{subfigure}
\begin{subfigure}{0.4\linewidth}
\centering
\includegraphics[page = 6]{images/HDIS-real-diagrams.pdf}
\caption{Feynman diagram for the $\as gggH$ vertex from the effective theory.}
\label{2:fig:NLOcdia}
\end{subfigure}
\end{figure}

While the operation elucidated in the previous subsection can be carried out in principle with any process, there are some advantages in considering effective gluon-scalar scattering as our test-case.
First, the addition of this coupling allows to restrict ourselves to the pure gluon sector of QCD by setting the number of massless quark flavours to zero, simplifying the overall computation.
Second, the Born level kinematics are those of a $2\rightarrow 1$ scattering and their NLO counterpart contain only one-loop calculations.
Thus, we consider the effective Lagrangian
\begin{equation}
\mathcal{L}_{\text{EFT}} =  \mathrm{Tr}[F^{\mu \nu} F_{\mu \nu}]\frac{\as \sqrt{\sqrt{2} G_F} }{12 \pi}\( 1 + \frac{\as}{4 \pi} \frac{19}{3} C_A \)  \frac{H}{1+\delta'} + \cdots \, , \label{2:eq:eff-lagrangian}
\end{equation}
where $H$ stands for the Higgs boson field, $\delta' = \frac{\as C_A}{4\pi} \frac{8}{3}  $ and $F^{\mu \nu}$ is the usual gluon field strength tensor.
We extract the extra Feynman rules from the effective interactions, depicted in Figs.~\ref{2:fig:LOdia} and~\ref{2:fig:NLOcdia}.
The rules associated with the $g\(k\)H\(q\)\rightarrow g\(k+q\)$ vertex read
\begin{equation}
M^{\mu \nu}_{ab}\(k, k+q\) =  \iu  \CE \( k^\nu \(k+q\)^\mu  - \spd{k}{k+q} g^{\mu \nu}\) \delta_{ab}\,  , 
\end{equation}
with  $a, b $ being the colour gluons, $k_1$ and $q$ are assumed to enter the vertex while $k+q$  is exiting. The effective coupling is defined as $\CE = \frac{\as \sqrt{G_F \sqrt{2}} }{3 \pi}$.
Similarly, $g\(k\)H\(q\)\rightarrow g\(p_3\)g\(p_4\)$ has an effective vertex
\begin{equation}\label{2:eq:EFF3vertex}
E^{\mu \nu \rho}_{abc}\(k,-p_4,-p_3,\) =  \CE g_sf_{abc} \[ g^{\mu\nu}\(k+p_4\)^\rho + g^{\nu\rho}\(p_3-p_4\)^\mu + g^{\rho\nu}\(-p_3-k\)^\nu \] \, , 
\end{equation}
where this time $k$ is assumed incoming and $p_{3,4}$ are outbound and $f_{abc}$ is the usual QCD structure constant. These two, together with the normal gluon QCD vertices, allows us to build all the amplitudes we need at NLO. 

\section{Born-level computation}
Before jumping into the strategy for the NLL$x$ computation, it is worth to retrace quickly the same steps from the definition of the off-shell coefficient function (Sec.~\ref{2:1:1}) to highlight some additional properties of this specific case.
We start by writing the Born-level tensor from Eq.~\eqref{2:eq:HEapprox}.
\begin{equation}\label{2:eq:HDISA0assembly}
 A_0^{\mu \nu}\(k,q\) =\frac{2\pi}{2C_AC_F}\delta_{ab}\int \dd^4 v \delta\(v^2\)\delta\(k+q-v\) M^{\mu\rho}_{ac}\(k,v\)d_{\rho \sigma}\(v,n\)\delta_{cd}M^{*\nu\sigma}_{db}\(k,v\) \, .
\end{equation}
Just like in the previous case, $$k=z p_1 +\kt{}\,,$$ and $n=p_2$. On the other hand, the incoming Higgs is off-shell and carries momentum $$q=p_2-x p_1\, .$$  The off-shellness $q^2$ also sets the process hard scale as $Q^2=-q^2$ and as usual $x=\frac{Q^2}{S}$.
In Eq.~\eqref{2:eq:HDISA0assembly} we have written the squared amplitude of the diagram in Fig.~\ref{2:fig:LOdia}, integrated over the outbound gluon momentum $v$. Moreover, we wrote the Light-cone gauge (LCG) polarisation tensor
\begin{equation}
d^{\mu\nu}\(v,n\) = -g^{\mu\nu} + \frac{v^\mu n^\nu + v^\nu n^\mu}{\spd{v}{n}}\, ,
\end{equation}
 but it is inconsequential for this tree-level amplitude as it is gauge invariant and only the $-g^{\mu\nu}$ term survives. If we carry out the integration we get
\begin{align}\label{2:eq:HDISA0}
    & A_0^{\mu \nu}\(k,q\)  = \CE^2 \(k^2 + \spd{k}{q}\)^2 \delta\[\(k+q\)^2\]\bigg\lbrace -g^{\mu \nu} + \frac{Q^2}{\(k^2 + \spd{k}{q}\)^2} k^\mu k^\nu \nonumber \\
              &\qquad - \frac{k^2}{\(k^2 + \spd{k}{q}\)^2} q^\mu q^\nu  + \frac{\spd{k}{q}}{\(k^2 + \spd{k}{q}\)^2} \(k^\mu q^\nu + k^\nu q^\mu\) \bigg\rbrace \nonumber \\
              &\quad = \delta\[z-x\(1+\xi\)\] \CE^2Q^2x \left\lbrace -g^{\mu\nu}Q^2 +k^\mu k^\nu +\xi q^\mu q^\nu + x\(1+\xi\) \(k^\mu q^\nu + k^\nu q^\mu\) \right\rbrace\, ,
 \end{align}
 where we introduce once more the dimensionless scale $\xi= \frac{-\kt{}^2}{Q^2}$.
 
 Moving on, $A_0$ can be recast in the same form of Eq.~\eqref{2:eq:HEapprox} and expanded for small-$x$
 \begin{align}
 & A_0^{\mu\nu}\(k,q\) = \frac{\pi \CE^2 Q^4}{2}\delta\(z-x(1+\xi)\) \bigg\lbrace -\frac{ (z-2 \xi  x )^2}{ x ^2} \[-g^{\mu\nu}+\frac{k^\mu k^\nu}{k^2}\] \nonumber \\
   &\quad -\frac{z^2}{
   x^2 k^2}\[\frac{k^2}{\spd{k}{q}}q^\mu - k^\mu \]\[\frac{k^2}{\spd{k}{q}}q^\nu - k^\nu \]\bigg\rbrace \, .
 \end{align}
 We can now see explicitly that it obeys the scaling properties behind Eq.~\eqref{2:eq:HEscaling}.
This offers a simple proof of the universality of $\kt{}$-factorisation.

If we stick to LL$x$ resummation we can go a step further and combine the previous expression with the one in Eq.~\eqref{2:eq:dCH} and normalise over the initial state flux and spin, yielding the complete LO off-shell coefficient function
 \begin{equation}\label{2:eq:HDISLL}
 \calC_0 \(\frac{x}{z},\xi\)= \frac{1}{2} A_{0,\mu\nu}\(-\frac{\kt^\mu \kt^{\nu} }{\kt^2}\) = \frac{\pi  \CE^2 Q^2 z^2 }{4 x^2} \delta \left(-\frac{z}{x }+\xi +1\right) \, .
 \end{equation}
 We can use this expression to highlight some properties of the process at hand. Using the impact factor approach to resummation of Refs.~\cite{Marzani:2008az,Marzani:2008qri}, we can extract the logarithms structure of this process by computing the impact factor like in Eq.~\eqref{2:eq:ImpFac},
\begin{align}
h\(N,M\) & = M \int_0^{\infty} \dd \xi \xi^{M-1}\int_0^1 \dd \tau \tau^{N-1} \calC_0\(\tau,\xi\) \nonumber \\
& =  \frac{\pi  \CE^2 Q^2}{4} M \int_0^{\infty} \dd\xi \xi^{M-1}\(\frac{1}{1+\xi}\)^{N-1} \nonumber \\
& = \frac{\pi \CE^2 Q^2}{4} \frac{\Gamma\(M+1\) \Gamma\(N-1-M \)}{\Gamma\(N-1\) } \, ,\label{2:eq:HDISLLimp}
\end{align}	
with the convergence constraint $0 < \text{Re}(M) <\text{Re}(N)-1$.
Then, the LL$x$ terms can be reconstructed by expanding around $M = 0$, as in Mellin space the high-energy logarithms $\frac{\log^k(x)}{x}$ are mapped to poles $\(\frac{1}{N-1}\)^k$.
\begin{equation}\label{2:eq:hexp}
	h\(N,M\) \simeq \frac{\pi \CE^2 Q^2}{4}\bigg[ 1  + \frac{M}{N-1} + \(\frac{M}{N-1}\)^2 + \(\frac{M}{N-1}\)^3 + \ldots	\bigg]\, .
\end{equation}
Now, setting $ M = \gamma \(\as, N\) $ with the NLL expansion for the anomalous dimension from appendix A of \cite{Bonvini:2018xvt} given by\footnote{Customarily results in Mellin space are computed with a shift $N \rightarrow N+1$ so that the small-$x$ poles in $N$-space are located at $N=0$ rather than $N=1$. We do not use this simplification in order to keep the definition of the Mellin transformation consistent across all chapters.}
\begin{align}
\gamma \(\as, N\) & = \as \[ \frac{C_A}{\pi (N-1)} - \frac{11 C_A + 2 n_f\( 1-2\frac{C_F}{C_A} \) }{12 \pi }\] \nonumber \\
& + \as^2 \frac{n_f \(26 C_F - 23 C_A\)}{36 \pi^2 (N-1)} + \calO\(N^0, \as^3\) \, ,
\end{align}
we obtain  the contribution in the resummed impact factor
\begin{align}
h\(N, \gamma \(\as,N\)\) & = \frac{\pi \CE^2 Q^2 }{4}\bigg\lbrace 1 + \frac{\as}{4 \pi} \[\frac{4C_A}{(N-1)^2} - \frac{11C_A}{3(N-1)} \] + \nonumber \\
& + \(\frac{\as}{4\pi}\)^2\[\(\frac{4 C_A}{(N-1)^2}\)^2 - \frac{88 C_A^2}{3 (N-1)^3}  \] +\calO\(\(\frac{\as}{N-1}\)^2,\as^3\) \bigg\rbrace\, .\label{2:eq:hFAIL}
\end{align}
We can compare this result with the high-energy expansion of the impact factor obtained by Mellin transformation of the NNLO coefficient function from \cite{Soar:2009yh},
\begin{align}
h_{\text{FO}}\(\as,N\) & = \frac{\pi \CE^2 Q^2 }{4}\bigg\lbrace 1 +\frac{\as}{4\pi}\[\frac{4C_A}{(N-1)^2} - \frac{11 C_A}{3 (N-1)} \] \nonumber \\
& + \(\frac{\as}{4\pi}\)^2\[\(\frac{4C_A}{(N-1)^2}\)^2 - \frac{220 C_A^2}{3 (N-1)^3} \] + \calO\(\as^3\) \bigg\rbrace  \,. \label{2:eq:hFO}
\end{align}
The first observation is the presence of a double-logarithmic enhancement.
A term with two extra powers of the logarithms appearing at each new order in $\as$ as well as expected single logarithm, as discussed in chapter~\ref{ch2}. This difference is generated by the simplified $2\rightarrow1$ kinematics of HDIS and the non-vanishing of Eq.~\eqref{2:eq:HDISLL} in the high-energy limit~\cite{Hautmann:2002tu, Caola:2010kv}.
Since the extra power of $\frac{1}{N-1}$, is systematically generated at all-orders as shown in Eq.~\eqref{2:eq:hexp}, one could hope that the NLL$x$ information from the anomalous dimension alone could intercept both LL$x$ and NLL$x$ terms.
Unfortunately Eqs.~\eqref{2:eq:hFAIL} and~\eqref{2:eq:hFO} show that using the NLL-resummed $\gamma$ alone is insufficient to predict the NLL contributions beyond at NLO.

\section{Cornering the NLL correction in the coefficient function} \label{2:sec:strategy}
Let us  consider the resummed expression for the coefficient function in the \hell\ language
\begin{equation}
C\(x\) = \int_0^{\infty}\dd \xi \int_x^1\frac{\dd z}{z} \calC\(\frac{x}{z},\xi,\as\)U'\(z,\xi,\as\)\,,
\end{equation}
 and observe that, the evolutor $U$ is effectively known to NLL$x$ thanks to the knowledge of the NLL$x$ anomalous dimension (as discussed in chapter~\ref{ch2}). So, the only missing NLL$x$ information is in the NLO correction inside $\calC$.   

Unfortunately, we cannot simply take the prescription of Eq.~\eqref{2:eq:dCH} and apply it to the NLO scattering amplitude for HDIS. As argued in section~\ref{2:ssec:ktfact1}, the definition of the off-shell projector relies on the gauge invariance of the Born level amplitude and the scaling properties in the high-energy limit.
Instead, we return to the same decomposition of Eq.~\eqref{2:eq:tensorfact}, this time considering both the Born-level $A_0$ and a next-to-leading order contribution $A_1^{\mu\nu}$.
\begin{equation}\label{2:eq:tensorfact2}
 Q^2 \sigma = \frac{Q^2}{2S}\int\frac{\dd^4k}{(2\pi)^4}\[ A_{0,\mu\nu}\( k,q \) + \frac{\as}{4\pi} A_{1,\mu\nu}\( k,q,n \)\] d^{\mu\mu'}\(k,n\)d^{\nu\nu'}\(k,n\)G_{\mu'\nu'}\(p_1,k\)\,.
\end{equation}
This new term is defined as the two gluon irreducible (2GI) part of the squared NLO amplitude for $H+g^* \rightarrow g$, integrated over loops and the suitable phase space, and summed(averaged) over all other indices of the final(initial) state except for the Lorentz index of the incoming off-shell gluon.
Just like $A_0$, $A_1^{\mu\nu}$ is meant to be 2GI in the sense that it cannot be factorised again by cutting across a pair of gluon propagators. While $A_0$ is completely gauge invariant and thus can not depend on $n$ at all, the same does not hold for $A_1^{\mu\nu}$. Indeed, the latter can include scalar combinations of the gauge axis. On the other hand, since $A_1^{\mu\nu}$ is saturated on both its indices by the LCG polarisation tensors (which are orthogonal to $n$ by construction), all its terms proportional to $n^\mu$ or $n^\nu$ will vanish. It follows then that $A_1^{\mu\nu}$ can be represented by a combination of all outer products of $k$ and $q$.

Before moving on to the computation of $A_1^{\mu\nu}$, we rewrite $A_0^{\mu\nu}$ using the tensor decomposition
\begin{align}\label{2:eq:mytenred}
 T_1^{\mu\nu} = g^{\mu\nu}, \: T_2^{\mu\nu} = p_1^\mu p_1^\nu,\: T_4^{\mu\nu}=\kt^\mu \kt^\nu,\: T_7^{\mu\nu} = p_1^\mu \kt^\nu + p_1^\nu \kt^\mu \, ,
\end{align}
and obtain
\begin{align}
A_{0}^{\mu\nu}\(k,q,n\) & = \frac{\pi \CE^2 Q^2 }{4}\delta\(\frac{1}{1+\xi}-\tau\)\bigg[-\frac{(\xi -1)^2}{(\xi +1)^2} g^{\mu\nu} + \frac{4 \xi  x ^2}{(\xi +1)^2} \frac{p_{1}^{\mu}p_{1}^{\nu}}{Q^2} \nn
                      & + \frac{2 \sqrt{\xi } x }{\xi +1} \frac{\(p_{1}^\mu \kt^\nu + p_{1}^\nu \kt^\mu\)}{\sqrt{\xi}Q^2} + \frac{4 \xi }{(\xi   +1)^2} \frac{\kt^\mu \kt^\nu}{\xi Q^2}\bigg] \,,
\end{align}
where we introduced the rescaled variable $\tau=\frac{x}{z}$.
\subsection{Strategy of the NLO off-shell computation}\label{2:ssec:NLO-strat}
The actual computation of $A_1^{\mu \nu}$ is rather cumbersome and involves a lot of details better left to appendix~\ref{app:HDIS-NLO}. Here, we summarise the main steps.
We start by recasting the squared amplitude of each diagram (tree-level or loop) entering $A_1^{\mu\nu}$ as a combination of tensors~\eqref{2:eq:mytenred}. To do so, we use the following projectors and  in this way eliminate any vector dependence on final state momenta from the tensor
\begin{align}
& P_1^{\mu\nu} = \frac{1}{D-3}\[g^{\mu\nu} - \frac{\kt^\mu \kt^\nu}{\kt^2}- \frac{p_1^\mu n^\nu+p_1^\nu n^\mu}{\spd{p_1}{n}}\],\: n^{\mu\nu} = \frac{n^\mu n^\nu}{\spd{p_1}{n}^2} \nn
& P_4^{\mu\nu} = \frac{1}{(D-3) \kt^2}\[(D-2)\frac{kt^\mu\kt^\nu}{\kt^2} - g^{\mu\nu} + \frac{p_1^\mu n^\nu+ p_1^\nu n^\mu}{\spd{p_1}{n}}\], \: P_7^{\mu\nu} = \frac{\kt^\mu n^\nu  + \kt^\nu n^\mu}{2 \kt^2 \spd{p_1}{n}}\, .
\end{align}
 Then, the scalar function given by every combination of tensor $T_k$ and diagram is integrated over the two-body phase space (one-body phase space and loop momenta) respectively for real emission (loop) diagrams. These integrals require a pattern of different regulators to account for the different types of singularities present in the overall computation. First, the introduction of the initial state off-shellness $\xi=\frac{\vec{k_t}^2}{Q^2}$ will regulate the collinear singularity. At the same time dimensional regularisation is still required to track the UV divergence in the loop corrections and construct a suitable counterterm to renormalise them. Finally, working in the light-cone gauge, according to the prescriptions of Ref.~\cite{Catani:1994sq}, will induce some extra spurious singularities. These are non-physical and are known to vanish in gauge-invariant objects~\cite{Curci:1980uw}. However, since the off-shell 2GI subset of the NLO correction is not guaranteed to be gauge-invariant on its own, we introduce a Principal Value (PV) prescription to explicitly keep track of these terms.
\begin{equation}
d_{\mu\nu}\(k,n\) = -g_{\mu\nu} + \frac{k_\mu n_\nu+k_\nu n_\mu}{\spd{k}{n}} \xrightarrow[PV]{}-g^{\mu\nu}+ \frac{z_{\text{gauge}} \[ \(k_\mu n_\nu+k_\nu n_\mu\)\]}{\spd{p_1}{n}\(z_{\text{gauge}}^2+\delta^2\)}\, ,
\end{equation}
so that for $\delta \neq 0$ the singularity for $z_{\text{gauge}} = \frac{\spd{k}{n}}{\spd{p_1}{n}} = 0$ is regularised.
\subsection{Incomplete result and issues with singularity cancellation}
In the end we get the expression
\begin{equation}\label{2:eq:A1-decomp}
A_{1}^{\mu\nu}\(k,q,n\) = C_A \frac{\pi \CE^2 Q^2 }{4}\[D_0 g^{\mu\nu} + D_1 \frac{p_{1}^{\mu}p_{1}^{\nu}}{Q^2} + D_2 \frac{\(p_{1}^\mu \kt^\nu + p_{1}^\nu \kt^\mu\)}{\sqrt{\xi}Q^2} + D_3 \frac{\kt^\mu \kt^\nu}{\xi Q^2}\] \, ,
\end{equation}
with every $D_k$ being a scalar function of the remaining invariants $\(\tau,\xi,x\)$ and the aforementioned regulators. 

Now we show the singular contributions in $\epsilon$ and $\delta$ for each coefficient.
\begin{subequations}\label{2:eq:Wrong-Coef}
\begin{align}
& D_0 = \frac{\xi-1}{ (\xi+1)^2} \bigg[ 4 I_0(\delta)\frac{(\xi-1)}{\epsilon} -4 I_0(\delta) \frac{1}{\xi}\left(\left(\xi ^2+2 \xi +2\right) | \xi -1| +2 (\xi -1) \xi \right) \log
   (\xi +1) \nn
   & \quad + \frac{1}{3\epsilon}\(3 \left(\xi ^2+2 \xi +2\right) \log \left(\frac{\xi }{\xi +1}\right)-11 (\xi-1)\) \bigg]\delta\(\frac{1}{1+\xi}-\tau\) \nn
& \quad -\frac{2 I_0\(\delta\)}{(\xi   +1) \tau^3} \bigg[ -\frac{(\xi  \tau +\tau -1) \sqrt{1-4 \xi  \tau ^2}}{\xi \left(\xi  \tau ^2+\tau -1\right)} \bigg(2 \xi ^3 \tau ^6-\xi ^2 \tau ^5-2 \xi ^3 \tau ^4 \nn
& \quad +\left(-3 \xi ^2-2 \xi +1\right) \tau^3+\left(2 \xi ^2+\xi -4\right) \tau ^2+(\xi +5) \tau -2\bigg) -\frac{\sqrt{1-4 \xi  \tau ^2}}{\xi  (\tau -1) \left(\xi  \tau ^2+\tau -1\right)}  \nn
& \quad \times \bigg(\xi ^2 \left(\xi ^2-1\right) \tau ^7 +\left(-\xi ^4+3 \xi ^3+5 \xi^2+\xi \right) \tau ^6-\xi  \left(2 \xi ^2+9 \xi +5\right) \tau^5 \nn
& \quad +\left(3 \xi ^3+3 \xi ^2+7 \xi +3\right) \tau ^4+\left(4 \xi ^2-5 \xi -12\right) \tau ^3-2 \left(\xi ^2-2 \xi -9\right) \tau ^2-2 (\xi +6) \tau +3\bigg) \nn
& \quad +\frac{2 \tau^2 \left((\xi -1) \tau ^2+2 \tau -1\right)^2}{(\tau -1)^2} \bigg] \PlusTau , \\
& D_1 = \frac{2 \xi  x^2}{(\xi +1)^2} \bigg[- \frac{8 I_0(\delta)}{\epsilon} -\frac{4 I_0(\delta)}{\xi}((\xi +1) | \xi -1| -4 \xi ) \log (\xi +1) \nn
& \quad + \frac{1}{3 \epsilon} \(22 -3 (\xi+1) \log\(\frac{\xi}{\xi+1}\) \)\bigg]\delta\(\frac{1}{1+\xi}-\tau\) \nn
& \quad  -\frac{2 x^2I_0\(\delta\)}{(\xi +1) \tau ^3} \bigg[\tau  (2 \tau -1) (2 \tau +1) (\xi  \tau +\tau -1) \sqrt{1-4 \xi  \tau ^2}-8\xi  \tau ^4 -\frac{2 (\tau -1) \sqrt{1-4 \xi  \tau ^2}}{\xi  \left(\xi  \tau ^2+\tau   -1\right)} \nn
& \quad \times \bigg(\xi  \left(\xi ^2-\xi-2\right) \tau ^4+\left(\xi ^2+3 \xi -2\right) \tau ^3-\left(\xi ^2+\xi   -5\right) \tau ^2-4 \tau +1\bigg)\bigg]\PlusTau \, ,\\
& D_2 = \frac{x\sqrt{\xi}}{(1+\xi)} \bigg[-\frac{8 I_0\(\delta\)}{\epsilon} - \frac{2 I_0\(\delta\)}{\xi} ((3 \xi +5) | \xi -1| -8 \xi ) \log (\xi +1) \nn
& \quad + \frac{44-3 \sqrt{2} (3 \xi +5) \log \left(\frac{\xi }{\xi +1}\right)}{6
   \epsilon } \bigg] \delta \(\frac{1}{1+\xi} - \tau\) \nn
&\quad - \frac{x I_0\(\delta\)}{\xi^{3/2}(1+\xi)}\bigg[ \frac{8 \xi ^2}{\tau -1} \left((\xi +1) \tau ^2-2 \tau +1\right) + \frac{\sqrt{1-4 \xi  \tau ^2}}{\tau ^4 \left(\xi  \tau ^2+\tau -1\right)}\bigg( \xi  \left(\xi ^3-14 \xi ^2-17 \xi -2\right) \tau ^6 \nn
& \quad +\left(3 \xi ^3+28 \xi ^2+7 \xi -2\right) \tau ^5+\left(-9 \xi ^2-6 \xi +9\right) \tau ^4-\left(\xi ^2+\xi +16\right) \tau ^3 \nn
& \quad +\left(-\xi ^2+2 \xi +14\right) \tau ^2-6 \tau +1 \bigg) \bigg]\PlusTau \, ,  \\
& D_3 = \frac{1}{\(1+\xi\)^2} \bigg[ -\frac{16 I_0\(\delta\)\xi}{\epsilon} -4 \text{I0} \left(\left(2 \xi ^2+5 \xi +5\right) | \xi -1| -8 \xi \right) \log (\xi +1) \nn
& \quad \frac{\xi  \left(44-3 \left(2 \xi ^2+5 \xi +5\right) \log \left(\frac{\xi}{\xi +1}\right)\right)}{3 \epsilon }  \bigg] \delta\(\frac{1}{1+\xi}-\tau\)\nn
& \quad  - \frac{2 I_0\(\delta\)}{\xi \(1+\xi\)\tau^3} \bigg[-8\xi^2 \tau^4 + \frac{\sqrt{1-4 \xi  \tau ^2}}{\xi  \tau ^2+\tau -1} \bigg( -2 \xi ^3 (\xi +1) \tau ^7-\xi ^2 \left(\xi ^2-3 \xi -2\right) \tau ^6 \nn
& \quad +\xi \left(2 \xi ^3+6 \xi ^2+9 \xi +6\right) \tau ^5-\left(-5 \xi ^3+3 \xi^2+14 \xi +2\right) \tau ^4-2 \left(\xi ^3+3 \xi ^2-5 \xi -2\right) \tau^3 \nn
& \quad -2 \xi  (\xi +1) \tau ^2-4 \tau +2\bigg) \bigg] \PlusTau \, ,
\end{align}
\end{subequations}
with the additional definition $I_0\(\delta\)= \int_0^1\dd z\frac{z}{z^2+\delta^2}$.

Some issues emerge already from this partial result. First, if we average over the direction of the transverse momentum $\kt{}$ and then expand around $\xi = 0$, we find
\begin{align}
\bigg\langle A_{1}^{\mu\nu}\(k,q,n\) \bigg\rangle & \simeq C_A \frac{\pi C^2 Q^2}{4} g^{\mu\nu}\bigg\lbrace \frac{1}{3\epsilon}\[12 I_0\(\delta\) -11 - 6 \log(\xi)\]\delta\(1-\tau\) \nn
& \quad  + \frac{6 \left(2 \tau ^3-4 \tau^2-3 \tau +1\right) I_0\(\delta \)}{3 \tau ^2 (1-\tau )_+}+8 I_0\(\delta\) \bigg\rbrace \nn
& \quad  + \calO \(\epsilon^0, \log(\xi)\) \, ,\label{2:eq:2GIcollinearLim}
\end{align}
which does not match the same quantity computed from scratch in collinear factorisation by setting $\kt{} = 0$
\begin{equation}
A_{1}^{\mu\nu}\(k,q,n\)\bigg|_{\kt{}=0} = C_A \frac{\pi C^2 Q^2}{4} g^{\mu\nu}\[\frac{4}{\epsilon^2}\delta\(1-\tau\) +\frac{23}{3\epsilon}\delta\(1-\tau\) +4 I_0\(\delta\) \frac{1-\tau}{\tau} \] \, . \label{2:eq:2GIcollinear}
\end{equation}
The most concerning issue with this mismatch lies in the impossibility of cancelling the mixed singularities $\frac{\log(\xi)}{\epsilon}$ and $\frac{I_0(\delta)}{\epsilon}$.

In the collinear case of Eq.~\eqref{2:eq:2GIcollinear} the double $\epsilon$ pole is cancelled by an equal and opposite contribution in the 2GR diagrams. However, in our off-shell scheme the missing 2GR contributions can not generate any $\epsilon$ poles. This emerges naturally from Eq.~\eqref{2:eq:last-int}, where the dimensional regularisation phase-space Jacobian is proportional to $s^{-\epsilon}$ and implies that only amplitudes proportional to $\frac{1}{s}= \frac{1}{\(k+q\)^2}$ can produce collinear poles. No such contribution is included in the 2GR amplitude.
On the other hand, the mixed $\frac{I_0\(\delta\)}{\epsilon}$ term in Eq.~\eqref{2:eq:2GIcollinearLim} does not have a counterpart in Eq.~\eqref{2:eq:2GIcollinear}. For the same reason as before, this divergence cannot be cancelled by 2GR terms.

\subsection{Outlook}
At this time we are unable to confirm if this unphysical results are a consistent feature of how the computation is set-up. One possibility is an incompatibility between the simultaneous presence of the off-shellness (de facto a hard-cutoff regulator) and the Principal Value prescription. The latter is already known to induce failures in the power-counting classification of singularities when deployed in loop integrals (see Ref.~\cite{Bassetto:1991ue} for an in-depth review). Another possible point of failure of the computation presented lies in the direct integration approach used to obtain Eq.~\eqref{2:eq:Wrong-Coef}. Indeed, the verbosity and sheer size of the output expressions may obfuscate the physical structure of the result and intuitive ways to fix the clash between the regulators. One possible solution to this issue is to re-frame the NLO off-shell coefficient along the lines of Ref.~\cite{Blanco:2022iai,vanHameren:2022mtk}, where general NLO schemes for the $\kt{}$-factorisation are devised.

Unfortunately, it is not possible to directly leverage the results in the impact factor formalism from Ref.~\cite{Celiberto:2022fgx} to extract the tensor in Eq.~\eqref{2:eq:A1-decomp}.
Indeed, computing the forward Higgs impact factor along the lines of Refs.~\cite{Fadin:2001dc,Celiberto:2022fgx} requires assigning a dummy polarisation, $\epsilon^\mu_{\text{GT}}\(k\) = -\frac{p_1^\mu}{S}$, to the off-shell gluon. This operation runs counter to the decomposition at the heart of our computation. Moreover, as opposed to our attempt at computing the fully inclusive NLO coefficient function, the known result for the Higgs impact factor is differential in the kinematical variable of the final state boson, obscuring the details of how the regulators map the IR singularities and their cancellation in momentum space.\footnote{Of course, Ref.~\cite{Celiberto:2022fgx} still offers a convincing proof that the impact factor is  free from singularities by studying its representation in Mellin space. }
On the other hand, it is worth noticing that the impact factor calculation is carried out in Feynman gauge and using dimensional regularisation for both IR and UV poles. In our case, using this setup of gauge and regulators would help simplifying the intermediate steps of the computation and remove the necessity of introducing the PV-prescription altogether. The price of this simplification however lies in the 2GI definition of $A_1^{\mu\nu}$, which makes it potentially non gauge-invariant and susceptible to undesired contamination if computed with different gauge fixing~\cite{Catani:1994sq}.

Hopefully, these insights and workarounds will offer a way to solve the issues presented thus far. Then, once the tensor decomposition of Eq.~\eqref{2:eq:A1-decomp} is computed consistently, the next steps will be to examine the coefficient and isolate the ones containing next-to-leading high-energy logarithms.
If the same scaling of Eq.~\eqref{2:eq:HEscaling} still holds, then it can be conjectured that including the NLO 2GI off-shell contribution in the off-shell coefficient function is the only missing output to achieve a complete NLL resummation of high-energy logarithms in the \hell\ language~\eqref{2:eq:ktfact}.
%Otherwise, it will be necessary to devise a new projector to generalize the one defined by Eq.~\eqref{2:eq:dCH}.

	\chapter{The Higgs spectrum in $HW$ associate production at NNLO+NNLL }
	\label{ch3}
	In perturbative QCD, cancellation theorems ensure that soft-gluon singularities are eliminated for Infra-Red and Collinear-safe observable~\cite{Banfi:2004yd}. Indeed, observables are not sensitive to arbitrary soft radiation in the final state because particle detectors have a finite energy resolution. As a result, the undetected real gluon emission cancels out the singularities that appear in virtual contributions. However, even though singularities are cancelled out explicitly, real and virtual contributions can still become imbalanced in situations where the real radiation is severely limited by kinematics, resulting in large logarithms. These are commonly observed at the exclusive boundary of the phase space or when strict cuts are used to enhance sensitivity in experimental searches. Then, to maintain the predictive power of perturbation theory, these logarithmically enhanced terms must be resummed to all orders.

One such case is the transverse-momentum distribution of a system with high-invariant mass $Q$, produced with extra QCD radiation and exclusive cuts on the final state.
In this configuration large logarithmic terms of the form ${\log\(\frac{Q}{p_{t}}\)}$, where $p_{t}$ is a transverse momentum scale introduced by the cuts and $Q$ is the hard scale of the process, may spoil perturbation theory and require resummation to restore predictivity.

This chapter is organised as follows.
Section~\ref{3:sec:soft-resum} covers some introductory examples about the resummation of transverse observables. In particular, from section~\ref{3:ssec:RadISH} to~\ref{3:ssec:qtsubtraction} introduce the \RadISH\ and $\qt$-subtraction formalism. 
Indeed, in section~\ref{3:sec:WHsection} the focus switches to the study of $HW^+$ associate production, which is a novel application of the aforementioned techniques.
We begin by validating our setup by computing the total $HW^+$ cross-section and $\pt{H}$ spectrum at NNLO in $\qt$-subtraction against fixed-order codes.
Then, we produce an analysis of the $\pt{H}$ spectrum in presence of a veto on the leading jet transverse momentum and study the impact of resumming the corresponding logarithms at NNLL. Moreover, we complement the resummation with an evaluation of the role of linear power corrections in $\pt{HW}$ in the same process.

\section{Sudakov resummation}\label{3:sec:soft-resum}
We start by showcasing an  example borrowed from Ref.\cite{Rottoli:2018nma,Catani:1997xc}.
Consider a generic infrared and collinear safe observable $v \in \(0,1\)$ and assume without loss of generality that it is dimensionless. 
We want to model the impact of the emission of extra soft-gluon radiation on a distribution or cross section $\sigma^{(n)}$.
The real and virtual gluon emission ``probability''  can be written as
\begin{subequations} \label{3:eq:softradreal}
\begin{align}
	\frac{\dd w_{\text{real}} (z)}{\dd z} & = 2C \frac{\as}{\pi} \frac{1}{1-z} \log \frac{1}{1-z} \vartheta(1-z-\eta)\, , \label{3:eq:softradreal-real}\\
	\frac{\dd w_{\text{virt}} (z)}{\dd z} & = - 2C \frac{\as}{\pi} \delta(1-z) \int_0^{1-\eta} \frac{\dd \zeta }{1-\zeta} \log \frac{1}{1-\zeta}\, , \label{3:eq:softradreal-virt}
\end{align}
\end{subequations}
where $C$ is a process-specific (colour) coefficient, $1-z$ the fraction of energy carried by unobserved soft particles in the final state. Eq.~\eqref{3:eq:softradreal-real} is valid in the soft and collinear limit, and we can recognise the double logarithmic divergence, which is regularised by  an (artificial) lower cut-off $\eta$.
However, in case of infrared and collinear safe observables, the total emission probability is finite and one can take the limit $\eta \rightarrow 0$:
\begin{equation}\label{3:eq:softradtot}
	\frac{\dd w (z)}{\dd z}  = \lim_{\eta\rightarrow 0}\left[\frac{\dd w_{\text{real}} (z)}{\dd z} + \frac{\dd w_{\text{virt}} (z)}{\dd z} \right]=   2C \frac{\as}{\pi}\left(\frac{1}{1-z}\log \frac{1}{1-z }\right)_+\, ,
\end{equation}
Where we used the plus-prescription as defined in appendix~\ref{app1:ssec:Ppresc}.
In particular, if the emission probability is integrated from $z=0$ up to $z=1$ we have that
\begin{equation}
	\int_0^1 \dd z \frac{\dd w (z)}{\dd z} = 0 \, . 
\end{equation} 
However, kinematic constraints may unbalance virtual and real contributions and induce singularities in the coefficients of the perturbative expansion.
For example, if an observable $v$ enhances soft radiation like $\delta v =\mathcal{O}(1-z)$, $\sigma^{(n+1)}$  can be written as
\begin{equation}
	\sigma^{(n+1)} (v) = \int_0^v \dd y\,  \sigma^{(n)} (v-y) \left( \frac{\dd w (z)}{\dd z} \right)_{z=1-y} + \ldots \xrightarrow[v \rightarrow 0]{}  -C \frac{\as}{\pi} \sigma^{(n)} (v) \log^2 v + \ldots \ .
\end{equation}  
As the real emission is hampered in the region $v \rightarrow 0$, the partonic cross section develops a double-logarithmic divergent contribution, as opposed to soft and hard-collinear ones that develop at most a single-logarithmic divergence.
This effect is known as Sudakov suppression~\cite{Sudakov:1954sw} and becomes large in the $v \rightarrow 0$ regime. As a result, despite the coupling being in the perturbative regime $\as \ll 1$, the cross section may not have a well behaved series expansion.

This simple example shows that, despite being free of singularities, fixed-order predictions of IRC-safe observables can still become unreliable if soft gluon effects in real and virtual terms are kinematically unbalanced. Our example explicitly considered a singularity at the exclusive boundary of phase-space, but similar mechanisms can also appear inside the physical region of phase space and give rise to so-called Sudakov shoulders~\cite{Catani:1997xc}. To resolve this issue, all-order calculations of Sudakov logarithms are necessary to get reliable results.

Primary examples in hadron collisions are the production of a system of high invariant mass close to threshold (where $1-v \sim \frac{M}{\sqrt{s}} $) and the cumulative cross section
\begin{equation}\label{3:eq:cumulative}
  \Sigma(v) \equiv \int_0^v dv' \frac{\dd \sigma(v')}{\dd v'}
\end{equation}
for the distribution of the transverse momentum $p_t$ in colour-singlet production, where $v\sim \frac{p_t}{M}$. 
The first studies for the all-order resummation of Sudakov logarithms for transverse-momentum distributions date back to over forty year ago~\cite{Dokshitzer:1978hw,Bassetto:1984ik}. Various techniques to achieve the resummation of the logarithmically-enhanced terms have since emerged, either by exploiting the properties of the QCD matrix elements and of gluon radiation (angular ordering and coherence)~\cite{Catani:1990rr} or using Soft-Collinear Effective Theory~\cite{Becher:2014oda} (SCET).
%In most cases, resummation is obtained in a suitable conjugate space, where the phase-space constraints can be tamed, with the disadvantage of either having process-dependent computations or different prescription for different observable.

\subsection{A showcase of transverse-momentum resummation}\label{sec:tmomres}
Now we consider a specific observable affected by large Sudakov logarithms: the transverse-momentum distribution of systems with a high invariant mass $M \gg p_t$, where the transverse momentum $p_t$ vanishes at the Born level. In such cases, like Higgs boson or Drell-Yan pair production, the LO transverse-momentum distribution is a delta function at $\pt{}=0$. If the heavy system is produced with a transverse momentum much smaller than $M$ the emission of real radiation is strongly suppressed and cannot balance the virtual contributions. Then, double logarithms of $p_t/M$ appear at all orders and the convergence of the series is spoiled for this configuration.
Ref.~\cite{Dokshitzer:1978hw} proposed an exponentiation of the most leading logarithmic contributions at small $p_t$ with the Dokshitzer-Dyakonov-Troyan formula. This results in a cross section which is exponentially suppressed in the limit $p_t \rightarrow 0 $. Briefly, this approach requires considering the leading soft and collinear contributions from an ensemble of $n$ gluons whose transverse momenta $k_{t,i}$ are {\it strongly ordered}:
\begin{equation}
	k^2_{t,n } \ll \cdots \ll  k^2_{t,2 } \ll  k^2_{t,1 } \lesssim p_t^2 \ll s.
\end{equation}
As a consequence, the cross section becomes naturally suppressed if $p_t \ll M$ as there is no phase space left for soft gluon production.

However, there are other configurations leading to a system with small transverse momentum, as  the only requirement for having a system with $p_t \sim 0$ is that the vector combination $\sum_{i=1}^n \vec k_{t,i}$ is small.
Indeed, around and below the peak of the distribution, the DDT-resummed spectrum vanishes as $\frac{\dd \sigma}{\dd p_t}\sim p_t$ rather than exponentially due to kinematic cancellations becoming predominant, as shown in Ref.~\cite{Parisi:1979se}.
In the same  paper, Parisi and Petronzio suggested to perform the resummation in the impact-parameter ($b$) space where the two competing effects leading to a vanishing $p_t$ are correctly handled through a Fourier transform,
\begin{equation}\label{3:eq:fourier}
	\vartheta_{\text{PS}} (\vec p_t, \vec k_{t,1},\ldots \vec k_{t,n} )=\delta^{(2)} \left(\vec p_t - \sum_{i=1}^n \vec k_{t,i} \right) = \int \dd^2 b \frac{1}{4 \pi^2} e^{\iu \vec b \cdot \vec p_t  } \prod_{i=1}^n e^{-\iu \vec b \cdot \vec k_{t,i} },
\end{equation}
which explicitly shows how, in Fourier space, the constraints factorise and transverse-momentum conservation is respected.

Using the $b$-space formulation, Collins, Soper and Sterman (CSS) established a formalism to resum the transverse momentum in Drell-Yan pair production in Ref.~\cite{Collins:1984kg}.
In this formalism, the partonic cross section is written as a convolution of parton-in-parton distributions ${\mathcal P}_{i/j}(x,{\vec k})$, at momentum fraction $x$ and transverse momentum ${\vec k}$, with an additional eikonal function $U$ which describes coherent soft-gluon emission~\cite{Collins:1981mv,Collins:1981tt,Collins:1982wa,Collins:1983ju},
\begin{align}
\frac{\dd \sigma_{ab\to F}}{\dd M^2 \dd{ p}^2_t} \simeq &  \sum_{c}\ \hat \sigma_{c\bar c \to F}^{({0)}}(M^2) \; H_{c \bar c}(M)\; \int \dd x_a \dd^2{\vec k}_a\,{\mathcal P}_{c/a}(x_a,{\vec k}_a,M) \int \dd x_b d^2{\vec k}_b\, {\mathcal P}_{\bar c/b}(x_b,{\vec k}_b,M) \nn
& \times \int  \dd^2{\vec q}\ U_{c\bar c}({\vec q})\ \delta(M^2-x_ax_bs)\ \delta^{(2)} \left(\vec p_t + {\vec q}- {\vec k}_a-{\vec k}_b\right)  , \label{eq:CSSfact}
\end{align}
where $\hat \sigma^{(0)}_{c \bar c \to F}$ is the LO cross section for the process $c \bar c \rightarrow F$, where $F$ can be a electroweak boson (Drell-Yan pair production, $c,\bar c = q, \bar q$) or a Higgs boson ($c,\bar c = g, g$). The factor $H_{c\bar c} (M)=1 + {\cal O}(\alpha_s(M))$ absorbs hard-gluon corrections and is computable in perturbation theory. Under Fourier transform, the cross section factorises and RGEs are developed for the separate pieces and solved.
In $b$-space, the logarithms of $p_t/M$ become logarithms of $bM$ and exponentiate. The final result is usually written as (here we use the notation of Ref.~\cite{Catani:2000vq}) 
\begin{align}\label{3:eq:CSSCatani}
	\frac{\dd \sigma_{p p \to F}}{\dd M^2 \dd \pt{} } =& \sum_{a,b} \int_0^1 \dd x_1 \dd x_2\, f_{a/h_1} (x_1, \muF^2)  f_{b/h_2} (x_2, \muF^2) \int_0^\infty \dd b\, b p_t J_0(p_t b) \nn
	  \times & \sum_{c} \int_0^1 \dd z_1 \dd z_2\, C_{c a} (\as(b^2_0/b^2),z_1) C_{\bar c b} (\as(b^2_0/b^2),z_2)  \delta(M^2-z_1 z_2 x_1 x_2 s) \nn
	  \times &\hat \sigma^{(0)}_{c \bar c \to F} (M) H_{\text{CSS}} (M) \exp\( - R_{\text{CSS},c}(b) \),
\end{align}
where the Bessel function $J_0$ descends from the integration over the azimuth angles in Eq.~\eqref{3:eq:fourier} and $b_0 = 2e^{-\gamma_E}$.
The Sudakov form factor $R_{\text{CSS}}(b)$ is defined as
\begin{align}
R_{\text{CSS},c}(b) &= \int_{b_0^2/b^2}^{M^2}\frac{\dd k_t^2}{k_t^2}
  {R}_{{\text{CSS}},c}'\left(k_{t}\right)=\int_{b_0^2/b^2}^{M^2}\frac{\dd
                    k^2_{t}}{k^2_t}
                    \left(A_{{\text{CSS}},c}(\as(k^2_t))\log\frac{M^2}{k_t^2} +
                    B_{{\text{CSS}},c}(\as(k^2_t))\right).
\end{align}
The anomalous dimensions $A_{{\text{CSS}},c}$ and $B_{{\text{CSS}},{c}}$, the coefficient functions $C_{ab} (\as, z)$, and the process-dependent hard function $H_{\text{CSS}}$ admit an expansion in the strong coupling and are analytically known with enough precision to perform a N$3$LL resummation~\cite{Catani:2010pd,Catani:2011kr,Catani:2012qa,Gehrmann:2014yya,Li:2016ctv,Vladimirov:2016dll}. The very first results at N$4$LL are also starting to appear in literature~\cite{Camarda:2023dqn}.
Using the $b$-space formulation, the $p_t$ spectrum for colour-singlet production has been resummed up to NNLL accuracy both for Higgs~\cite{Bozzi:2005wk} and for Drell-Yan pair production~\cite{Bozzi:2010xn,Banfi:2012du}. Similar results were also obtained in the SCET approach~\cite{Becher:2012yn,Neill:2015roa,Becher:2010tm,Chen:2018pzu}.
%Combined threshold and transverse-momentum resummation have been considered in Refs.~\cite{Laenen:2000ij,Kulesza:2002rh,deFlorian:2005fzc} and more recently in Refs.~\cite{Lustermans:2016nvk,Marzani:2016smx,Muselli:2017bad}.\FS{Update citations}

To obtain theoretical predictions, one has to integrate over $b$ in Eq.~\eqref{3:eq:CSSCatani}. However, this procedure hits the Landau pole in the coupling $\as\(\frac{b_0^2}{b^2}\)$ at large values of $b$.
To sidestep this issue and compute the observable for arbitrary values of $p_t$, a prescription must be introduced to regularise the integral. One option is using the $b_*$-prescription~\cite{Collins:1984kg} to replace the impact parameter $b$ with
\begin{equation}
	b_* = \frac{b}{\sqrt{1+(b/b_{\text{lim}})^2}}, 	\qquad b_* < b_{\text{lim}},
\end{equation} 
so that the parameter $b_{\text{lim}} \sim 1/\LambdaQ$ enforces a separation between the perturbative and non-perturbative regimes of QCD.
An additional factor $\sim \eu^{-gb^2}$ is introduced to model the non-perturbative region, with the extra parameter $g$ usually tuned to data.\footnote{
More specifically, two drawbacks are associated with working in impact parameter space.
The first is the difficulty of matching the resummed and fixed–order predictions, as the resummation is performed in $b$–space does not allow to control exactly which logarithmic terms (in $p_t$–space) are taken into account. 
Secondly, the detailed form of the non–perturbative input is not completely settled on theoretical theoretical ground~\cite{Ellis:1997sc}
Moreover, the necessity of inferring the non-perturbative parametres as well as the $b_{\text{lim}}$ by fitting creates difficulties in disentangling the quantitative difference between predictions of the resummation and the information input in the fitting. We leave the reader to Refs.~\cite{Catani:2000jh,Qiu:2000hf,Kulesza:2002rh}, for a more thorough discussion of these issues.}.

Other prescriptions avoid this issue. For example: one can set a suitable soft scale before the Landau pole to preserve the definiteness of the $b$-integral~\cite{Kulesza:2002rh}, resort to Borel summation~\cite{Bonvini:2008ei} or introduce scales depending on the hadronic variables ~\cite{Becher:2011xn,Becher:2012yn}.

Beside the $b$-space formalism, the problem of resummation of transverse-momentum distributions in $p_t$ space was studied across the past decade~\cite{Kulesza:1999sg,Kulesza:1999gm,Kulesza:2001jc,Ebert:2016gcn,Kang:2017cjk,Monni:2016ktx,Bizon:2017rah,Monni:2019yyr,Re:2021con}.
In particular, the approach of Refs.~\cite{Monni:2016ktx,Bizon:2017rah,Monni:2019yyr,Re:2021con} will be the focus of the next section.

\subsection{Resummation with \RadISH} \label{3:ssec:RadISH}
Our starting point is the inclusive hadronic production of a colour-singlet 
\begin{equation}\label{3:eq:hadron-noc}
\mathrm{p} \mathrm{p}\rightarrow \mathrm{F}+X,
\end{equation}
where the collision of the two protons produces a collection of colour neutral particles (i.e. $H$, $W^\pm$, leptons) F. At parton level and leading order, this is driven either by  $q{\bar q}$ annihilation, as in the case of the Drell--Yan process, or gluon fusion, as in the case of Higgs boson production.
We focus on the cumulative cross section for such process, which we denote as
\begin{equation} \label{3:eq:cum-cs}
\Sigma(v) \, \equiv \, \int_0^\infty \dd V\; \frac{\dd \sigma(V)}{\dd V} \vartheta \[v - V\(p_1,p_2; k_{1},\ldots,k_{n}\)\]\, ,
\end{equation}
where $V\(p_1,p_2; k_{1},\ldots,k_{n}\)$ is a generic observable in the range $\(0,v\)$, while $p_{1,2}$ define the beam axis of the initial state at Born level and $k_{n}$ are the transverse momenta of additional radiation of the initial state legs. While the \RadISH\ formalism has been shown to work for generic \textit{recursive Infrared and Collinear} (rIRC) safe observables~\cite{Monni:2016ktx,Bizon:2017rah}, in this example we will follow the same argument of Refs.~\cite{Bizon:2017rah,Re:2021con} to $\NNNLL$ accuracy. Thus, we consider just an inclusive and transverse rIRC observable
\begin{equation}
V\(p_1,p_2; k_{1},\ldots,k_{n}\) \xrightarrow{\text{inclusive}} V\(p_1,p_2; k_{1}+\ldots +k_{n}\) \xrightarrow{\text{transverse}} d_\ell g_\ell(\phi) \(\frac{\kt}{M}\)^a \, ,
\end{equation}
where $\kt$ is the total transverse momentum of the radiated partons, while $d_\ell$ is a normalisation constant and $g_\ell(\phi)$ encodes the dependence on the angle $\phi$ between $\kt$ and an arbitrary reference vector in the transverse plane $r$, and $M$ is the hard reference scale for the colour singlet (i.e. the mass of the final state). Finally, $a>0$ is a consequence of collinear safety. The prime example of an observable with all this properties is the total transverse momentum of the colour singlet, in which case $d_\ell = g_\ell(\phi) = a = 1$.

In the soft limit, the cumulative cross section in \eqref{3:eq:cum-cs} can be recast as
\begin{equation} \label{3:eq:Sigma-2}
\Sigma(v)  = \int \dd \Phi_B \, {\cal V}(\Phi_B) \sum_{n=0}^{\infty} \int\prod_{i=1}^n [\dd k_i] |M( p_1,p_2; k_1,\dots,k_n)|^2 \vartheta\[v-V( p_1,p_2;k_1+\ldots+k_n)\] \, ,
\end{equation}
where $M$ is the renormalised matrix element for $n$ real emissions (the case with $n = 0$ reduces to the Born contribution $|M_B(p_1,p_2)|^2$), $[\dd k_i]$ denotes the phase space for the $i$-th emission with momentum $k_i$, and the $\vartheta$ function represents the measurement function for the observable under study.
By $\Phi_B$ we denote the Born phase space, while ${\cal V}(\Phi_B)$ is the all-order hard-virtual form factor.
Crucially, in the soft limit, it is possible to establish a logarithmic power counting in the matrix element by decomposing it in blocks of correlated emissions ${\cal M}(k_1,\ldots,k_n)$.
Indeed,
\begin{align}
|M(p_1,p_2;k_1)|^2 =&  |M_B(p_1,p_2)|^2 |{\cal M}(k_1)|^2,\nn 
|M(p_1,p_2;k_1,k_2)|^2 =&  |M_B(p_1,p_2)|^2 \[ |{\cal M}(k_1,k_2)|^2  +\frac1{2!}|{\cal M}(k_1)|^2|{\cal M}(k_2)|^2\],\nn
|M(p_1,p_2;k_1,k_2,k_3)|^2 =& |M_B(p_1,p_2)|^2 \bigg[ |{\cal M}(k_1,k_2,k_3)|^2 + \frac1{3!}|{\cal M}(k_1)|^2|{\cal M}(k_2)|^2|{\cal M}(k_3)|^2\nn
 +|{ \cal M}(k_1,k_2)|^2|{\cal M}(k_3)|^2 & -|{\cal M}(k_1,k_3)|^2|{\cal M}(k_2)|^2-|{\cal M}(k_2,k_3)|^2|{\cal M}(k_1)|^2 \bigg],
\end{align}
and so on. Each ${\cal M}(k_1,\ldots,k_n)$ represents the matrix element for the emission of $n$-particles that can not be factorised in terms of
lower-multiplicity squared amplitudes.
Then, we can use a perturbative expansion
\begin{equation} \label{3:eq:nPC-def}
|{\cal M}(k_1, \ldots,k_n)|^2~\equiv~ \sum_{j=0}^{\infty}\left(\frac{\alpha_s(\mu)}{2\pi}\right)^{n+j}n\mbox{PC}^{(j)}(k_1, \dots,k_n),
\end{equation}
where $\mu$ is a common renormalisation scale, and $\alpha_s$ is the strong coupling constant in the $\overline{\rm MS }$ scheme. The notation $n\mbox{PC}$ in Eq.~\eqref{3:eq:nPC-def} stands for ``$n$-particle correlated'' block and allows to conveniently truncate the logarithmic power counting as each $n\mbox{PC}^{(j)}\(k_1,\ldots, k_n\)$ with $n$ emissions and $j$ loops will contributes only to N$^{n+j-1}$LL.
On the other hand, Eq.~\eqref{3:eq:Sigma-2} still contains IRC singularity from the virtual form factor ${\cal V}(\Phi_B)$. Given the overall IRC safety of the observable, these will cancel against the rest of the QCD radiation. To showcase this mechanism, we introduce a regularisation scale $q_0$ to divide the real emissions in two regions of transverse momentum
\begin{itemize}
\item[$\kt < q_0$] defines the \textit{unresolved} region, containing kinematical configurations which give small (indeed, vanishing with small $q_0$) contributions to the observable $V$. These will be factorised out of the phase space integral and exponentiated to cancel the IRC singularities from ${\cal V}(\Phi_B)$.
\item[$\kt > q_0$] gives the \textit{resolved} region, which instead gives sizeable contributions to the measurement function $\vartheta\[v-V\(p_1,p_2;k_1,\ldots,k_n\)\]$. However, once more, IRC-safety of the observable guarantees that these configurations will have a power like dependence on the resolution scale $q_0$, which can be relaxed to zero.
\end{itemize}
For observables which solely depend on the total transverse momentum of QCD radiation, it is most convenient to set  $q_0 = \epsilon k_{t1}$, where $0<\epsilon \ll 1$, while $k_{t1}$ is the total transverse momentum of the hardest resolved block.
Moreover, the same strategy can be applied for the resummation of different observables, enabling the development of a joint resummation framework showcased in Ref.~\cite{Monni:2019yyr}.

Now, we can return to the all-order cumulative cross section and take a Mellin transform\footnote{In the previous chapters we denoted the Mellin transform only by renaming the argument of the transformed function $f(N) = \int_0^1\dd z z^{N-1} f(z)$. Instead, in this chapter we adopt the notation of the \RadISH\ works~\cite{Bizon:2017rah}, $f_N = \int_0^1\dd z z^{N-1}f(z)$. }, as defined in appendix~\ref{app2:Mellintransf}, this reduces convolutions with parton densities to algebraic products.
\begin{align} \label{3:eq:DGLAP}
{\bf  f}_{N}^{\text{T}}(\mu) & = \(f_{N,g}(\mu),f_{N,q}(\mu),f_{N,\bar{q}}(\mu)\)\, , \nn
{\bf  f}_{N}(\mu) & = {\cal P} \exp\left[-\int_{\mu}^{\mu_0}\frac{\dd k_t}{k_t} \, \frac{\as(k_t)}{\pi} \, {\bf\Gamma}_{N}(\as(k_t)) \right] \, {\bf  f}_{N}(\mu_0) \, ,\nn
\Big[{\bf \Gamma}_{N}(\as)\Big]_{ab} & = \int_0^1 \dd z \, z^{N-1} \, {P}_{a b}(z,\as) = \sum_{n=0}^\infty\left(\frac{\as}{2\pi}\right)^n\Big[{\bf \Gamma}^{(n)}_{N}(\as)\Big]_{ab} \, , \nn
{P}_{ij}(z,\as) & = \sum_{n=0}^\infty\left(\frac{\as}{2\pi}\right)^n {P}^{(n)}_{ij}(z)\, ,
\end{align}
where ${\cal P}$ is the path-ordering symbol, $P_{ab}$ the collinear splitting functions from appendix~\ref{app:2:DGLAP}.
In principle the ${\bf \Gamma}$ matrix is not diagonal and the path ordering is mandatory to obtain the correct result. However, for the purpose of demonstrating the resummation formalism the details  of the flavour structure can be disregarded by assuming flavour-conserving real-emission kernels. This allows to drop the path ordering. This simplification will be relaxed later to obtain the full result.

Following Ref.~\cite{Bizon:2017rah}, the cumulative cross section differential in the Born variables can be rewritten by factorising the Born matrix element away from the real emissions
\begin{align} \label{3:eq:hadxs}
& \frac{\dd\Sigma(v)}{\dd\Phi_B}  = \int_{{\cal C}_1} \frac{\dd N_1}{2\pi i} \int_{{\cal C}_2}\frac{\dd N_2}{2\pi i} \, \,
 x_1^{-N_1} \, x_2^{-N_2} \sum_{c_1, c_2} \frac{\dd|{ M}_B|_{c_1c_2}^2}{\dd\Phi_B} \, \, {\bf f}^{T}_{N_1}(\mu_0) \, \hat{\bf \Sigma}^{c_1,c_2}_{N_1,N_2}(v) \, {\bf f}_{N_2}(\mu_0)\, , \nn
& \frac{\dd|{\cal M}_B|_{c_1c_2}^2}{\dd\Phi_B}  \equiv \int {\dd}\Phi'_B \,|{\cal M}_B|_{c_1c_2}^2 \,\delta(x_1-x_1') \, \delta(x_2-x_2') \, \delta(\Omega_B-\Omega_B')\, ,
\end{align}
where the sum runs over all allowed Born flavour combinations, $\Omega_B$ denotes a set of internal phase-space variables of the colour-singlet system, and the integration contours  ${\cal C}_1$ and ${\cal C}_2$ in the double inverse Mellin transform  lie along the imaginary axis to the right of all singularities of the integrand. Instead,  $\hat{\bf \Sigma}$ is a matrix containing the action of DGLAP evolution from scale $\mu_0$ to $\mu$ of the parton distribution and the partonic cross section evolution from the flavour-conserving radiation.
For inclusive observables, its all-order expression under the above assumption on
${\bf\Gamma}_N$ is\footnote{The last two lines of Eq.~\eqref{3:eq:partxs-mellin} reduce to $ \vartheta \(v-V(p_1,p_2;k_1) \)$ for $n=0$. } 
\begin{align} \label{3:eq:partxs-mellin}
\hat{\bf \Sigma}^{c_1,c_2}_{N_1,N_2}(v)& = \bigg[{\bf C}^{c_1; T}_{N_1}(\as(\mu_0)) \, H(\mu_R) \, {\bf C}^{c_2}_{N_2}(\as(\mu_0)) \bigg] \,
\int_0^{M}\frac{\dd k_{t1}}{k_{t1}} \int_0^{2\pi} \frac{\dd\phi_1}{2\pi} \, \eu^{-{\bf R}(\epsilon k_{t1})} \nn
&\times \, \exp\left[-\sum_{\ell=1}^{2} \left(\int_{\epsilon k_{t1}}^{\mu_0}\frac{\dd k_t}{k_t} \, \frac{\as(k_t)}{\pi}{\bf\Gamma}_{N_\ell}(\as(k_t)) + \int_{\epsilon k_{t1}}^{\mu_0}\frac{\dd k_t}{k_t} \, {\bf \Gamma}_{N_\ell}^{(\rC)}(\as(k_{t}))\right)\right] \nn
&\times \, \sum_{\ell_1=1}^2 \left({\bf R}_{\ell_1}'\left(k_{t1}\right) + \frac{\as(k_{t1})}{\pi}{\bf\Gamma}_{N_{\ell_1}}(\as(k_{t1})) + {\bf \Gamma}_{N_{\ell_1}}^{(\rC)}(\as(k_{t1}))\right) \nn
&
\times \, \sum_{n=0}^{\infty}\frac{1}{n!} \prod_{i=2}^{n+1}\int_{\epsilon k_{t1}}^{k_{t1}}\frac{\dd k_{ti}}{k_{ti}}\int_0^{2\pi} \frac{\dd\phi_i}{2\pi} \, \, \vartheta\left(v-V(p_1,p_2;k_1,\dots, k_{n+1})\right)\nn
& \times \, \sum_{\ell_i=1}^2 \left({\bf R}_{\ell_i}'\left(k_{ti}\right) \frac{\as(k_{ti})}{\pi}{\bf\Gamma}_{N_{\ell_i}}(\as(k_{ti})) + {\bf \Gamma}_{N_{\ell_i}}^{(\rC)}(\as(k_{ti}))\right)\, ,
\end{align}
where $H(\mu_R)$ admits itself  a perturbative expansion  
\begin{equation}\label{3:eq:hard-virtual}
H(\mu_R)   =   1 + \sum_{n=1}^{\infty} \left(\frac{\as(\mu_R)}{2\pi}\right)^n \,H^{(n)}(\mu_R) \, ,
\end{equation}
and represents the finite contribution from the virtual form factor at the renormalisation scale $\mu_R$. Conversely, ${\bf C}^{c_\ell}$ is a
$(2n_f+1)\times(2n_f+1)$ diagonal matrix defined as
\begin{equation}
[{\bf C}^{c_\ell}]_{a b} = C_{c_\ell a} \delta_{a b} \, ,
\end{equation}
using the  collinear coefficient functions $C_{ij}$. It satisfies a flavour-conserving renormalisation-group evolution equation stemming from the running of its coupling
\begin{align} \label{3:eq:RGEC}
{\bf C}^{c_\ell}(\as(\mu)) & = \exp\left[-\int_\mu^{\mu_0}\frac{\dd k_t}{k_t} \, {\bf \Gamma}^{(\rm C)}(\as(k_{t}))\right] \, {\bf C}^{c_\ell}(\alpha_s(\mu_0)) \nn
& = \delta(1-z) \, {\bf 1} + \sum_{n=1}^\infty \left(\frac{\as(\mu)}{2\pi}\right)^n {\bf C}^{(n)}(z) \, , \nn
{\bf \Gamma}^{(\rC)}(\as(k_{t}))  & = 2 \, \beta(\as(k_{t})) \, \frac{\dd \log{\bf C}^{c_\ell}(\as(k_{t}))}{\dd \as(k_{t})} \nn 
& = \, \sum_{n=1}^\infty \left(\frac{\as(k_{t})}{2\pi}\right)^{n+1} {\bf \Gamma}^{({\rC}, n)}(\as(k_{t})) \, .
\end{align}
Finally, the ${\bf R}_{\ell}'$ function represents the contribution from radiation emitted from leg $\ell$ conserving the momentum fraction of the incoming partons and the emitter's flavour
$c_\ell$, explicitly written as
\begin{equation}
[{\bf R}_{\ell}']_{ab} = R_\ell' \, \delta_{ab} \, .
\end{equation}
Its definition descends from the Sudakov radiator ${\bf R}$
\begin{align} \label{3:eq:rad-mat}
[{\bf R}]_{ab} & = R \, \delta_{ab} \, ,\nn
R(k_{t1}) &  = \sum_{\ell=1}^2 \, R_\ell(k_{t1}) = \int_{k_{t1}}^{M}\frac{\dd k_{t}}{k_{t}}  \sum_{\ell=1}^2  {R}_\ell'(k_t) \nn
& = \int_{k_{t1}}^M \frac{\dd k_t}{k_t} \sum_{\ell=1}^2 \left[ A_{\ell}(\as(k_t))\log\frac{M^2}{k_t^2} + B_{\ell}(\as(k_t)) \right] \, , \nn
{R}'(k_{t1}) & = \sum_{\ell=1}^2 \, {R}_\ell'(k_{t1}) \, ,\qquad \quad {R}_\ell'(k_{t1}) = \frac{\dd R_\ell(k_{t1})}{\dd L} \, , \qquad \quad L \, = \, \log\frac{M}{k_{t1}} \, ,
\end{align}
where, the anomalous dimensions $A_\ell$ and $B_\ell$ define the inclusive probability $|{\cal M}(k)|_{\rm inc}^2$ \cite{Bizon:2017rah} for a correlated block of arbitrary
multiplicity to have total transverse momentum $\kt{}$. They admit a perturbative expansion as
\begin{align}
A_\ell(\as) & = \sum_{n=1}^\infty \left(\frac{\as}{2\pi}\right)^n \, A_{\ell}^{(n)} \, , \nn
B_\ell(\as) & = \sum_{n=1}^\infty \left(\frac{\as}{2\pi}\right)^n \, B_{\ell}^{(n)} \, .
\end{align}
For colour-singlet production, the coefficients entering the Sudakov radiator
%satisfy $A^{(n)}_{1}=A^{(n)}_{2}=A^{(n)}$, and $B^{(n)}_{1}=B^{(n)}_{2}=B^{(n)}$ and
are available in the literature up to N$^3$LL in Refs.~\cite{deFlorian:2001zd,Davies:1984hs,Becher:2010tm,Banfi:2012jm,Li:2016ctv,Vladimirov:2016dll,Bizon:2017rah,Henn:2019swt,Huber:2019fxe,vonManteuffel:2020vjv}.
At this point we can summarise the meaning of Eq.~\eqref{3:eq:partxs-mellin} by emphasising the role of its parts. On one hand, the last three lines of the equation encode the contribution from resolved radiation with transverse momentum harder than $\epsilon k_{t1}$ with the inclusion of DGLAP evolution (with diagonal flavour kernels). On the other hand, the hard-virtual corrections and collinear splitting functions are in the first square bracket. Both these objects are evaluated at an arbitrary starting scale $\mu_0$ and then evolved to $\epsilon k_{t1}$. Finally, the all-order virtual form factor, combined with the exponentiated unresolved radiation, combines in the Sudakov exponential $\eu^{-{\bf R}(\epsilon k_{t1})}$.

As argued in Ref.~\cite{Catani:2010pd},  one needs to include the contribution from the collinear coefficient functions ${\bf G}$. These are required to describe the azimuthal correlations with the initial-state gluons.
In practice, this means including in  Eq.~\eqref{3:eq:partxs-mellin} an additional term  where one makes the replacements
\begin{equation}
\left[{\bf C}^{c_1; T}_{N_1}(\alpha_s(\mu_0)) H(\mu_R) {\bf C}^{c_2}_{N_2}(\alpha_s(\mu_0)) \right]\to\left[{\bf G}^{c_1; T}_{N_1}(\alpha_s(\mu_0)) H(\mu_R) {\bf G}^{c_2}_{N_2}(\alpha_s(\mu_0)) \right],
\end{equation}
and
\begin{equation} \label{3:eq:RGE-G}
{\bf \Gamma}_{N_\ell}^{(C)}(\alpha_s(k_{t}))\to{\bf \Gamma}_{N_\ell}^{(G)}(\alpha_s(k_{t})),
\end{equation} 
where ${\bf \Gamma}_{N_\ell}^{(G)}$ is defined analogously to Eq.~\eqref{3:eq:RGEC}, and the flavour structure of ${\bf G}$ is analogous to the one of the ${\bf C}$ matrix.
For the purpose of this discussion, this extra contribution is not relevant and can be disregarded. Further details can be found in Ref.~\cite{Re:2021con,Bizon:2017rah}.

Crucially, we can now simplify Eq.~\eqref{3:eq:partxs-mellin} by making the IRC cancellation explicit. The first step is considering the transverse momentum for each block of correlated emissions and  rescaling it to the one of the hardest one $k_{tn} = \zeta_n k_{t1}$. Then, each quantity can be expanded around $k_{t1}$, yielding
\begin{align} \label{3:eq:expanded_res}
R'(k_{ti}) & = \sum_{j=0}^2 R^{(j+1)}(k_{t1}) \, \frac1{j!} \, \log^j\frac1{\zeta_i} \, + \, \dots \, , \nn
\frac{\as(k_{ti})}{\pi} \, {\bf\Gamma}_{N_\ell}(\as(k_{ti})) & = \sum_{j=0}^1 \, \frac{\dd^j}{\dd L^{\,j}} \, \frac{\as(k_{t1})}{\pi} \, {\bf\Gamma}_{N_\ell}(\as(k_{t1})) \, \frac1{j!} \, \log^j\frac1{\zeta_i}
\, + \, \dots \, , \nn
{\bf\Gamma}_{N_\ell}^{(\rC)}(\as(k_{ti})) & = \sum_{j=0}^0 \, \frac{\dd^j}{\dd L^{\,j}} \, {\bf\Gamma}_{N_\ell}^{(\rC)}(\as(k_{t1})) \, \frac1{j!} \, \log^j\frac1{\zeta_i} \, + \, \dots \, , \nn
\end{align}
where $R^{(j)}(k_{t1}) = \dd^j R(k_{t1})/\dd L^{\,j},\, L=\log(M/k_{t1})$, and the ellipses denote neglected N3LL terms.
Resolved contributions are suppressed by one extra logarithmic order with respect to the corresponding unresolved ones, which  can be expanded around $k_{t1}$ in Eq.~\eqref{3:eq:partxs-mellin}, giving rise to logarithms $\log\(\frac{1}{\epsilon}\)$ to cancel against the IRC poles to all orders.
\begin{align} \label{3:eq:expanded_unres}
R(\epsilon k_{t1}) & = \sum_{j=0}^3 R^{(j)}(k_{t1}) \, \frac1{j!} \, \log^j\frac1\epsilon \, + \, \dots \, , \nn
\int_{\epsilon k_{t1}}^{\mu_0} \frac{\dd k_t}{k_t} \, \frac{\as(k_t)}{\pi} \,  {\bf\Gamma}_{N_\ell}(\as(k_t)) & = \sum_{j=0}^2 \, \frac{\dd^j}{\dd L^{\,j}} \, \int_{k_{t1}}^{\mu_0} \frac{\dd k_t}{k_t} \, \frac{\as(k_t)}{\pi} \, {\bf\Gamma}_{N_\ell}(\as(k_t)) \, \frac1{j!} \, \log^j\frac1\epsilon \, + \, \dots \, , \nn
\int_{\epsilon k_{t1}}^{\mu_0} \frac{\dd k_t}{k_t} \, {\bf\Gamma}_{N_\ell}^{(\rC)}(\as(k_t)) & = \sum_{j=0}^1 \, \frac{\dd^j}{\dd L^{\,j}} \, \int_{k_{t1}}^{\mu_0} \frac{\dd k_t}{k_t} \, {\bf\Gamma}_{N_\ell}^{(\rC)}(\as(k_t)) \, \frac1{j!} \, \log^j\frac1\epsilon \, + \, \dots \, . \nn
\end{align}
The loop expansion of the anomalous dimensions obeys an analogous perturbative counting. A further significant simplification arises in the subleading terms. Indeed, at N$^k$LL logarithmic accuracy, only blocks up to $k-1$ resolved emissions need to feature a $\log1/{\zeta_i}$ correction in $R'$, as the simultaneous correction of $k$ factors of $R'$ affects N$^{k+1}$LL. Unresolved contributions are expanded correspondingly, in order to cancel the $\epsilon$ divergences of the modified resolved blocks to the same logarithmic order.

After these steps the master formula Eq.~\eqref{3:eq:partxs-mellin} can be rewritten, to \NNNLL accuracy, as
\begin{align} \label{3:eq:mellin_expanded_final}
\hat{\bf \Sigma}^{c_1,c_2}_{N_1,N_2}(v) & = \bigg[ {\bf C}^{c_1; T}_{N_1}(\as(\mu_0)) \, H(\mu_R) \, {\bf C}^{c_2}_{N_2}(\as(\mu_0))  \bigg] \, \int_0^{M}\frac{\dd k_{t1}}{k_{t1}} \int_0^{2\pi} \frac{\dd\phi_1}{2\pi} \, \nn
& \times \, \eu^{-{\bf R}(k_{t1}) - {\bf R}'(k_{t1})\log\frac1\epsilon - \frac1{2!}{\bf R}''(k_{t1})\log^2\frac1\epsilon }\nn %- \frac1{3!}{\bf R}'''(k_{t1})\log^3\frac1\epsilon} \nn
& \times \, \exp \Bigg[ -\sum_{\ell=1}^{2} \bigg( \int_{k_{t1}}^{\mu_0} \frac{\dd k_t}{k_t} \, \frac{\as(k_t)}{\pi} \, {\bf \Gamma}_{N_\ell}(\as(k_t)) \nn
& \hspace{5mm} + \frac{\dd}{\dd L} \int_{k_{t1}}^{\mu_0} \frac{\dd k_t}{k_t} \, \frac{\as(k_t)}{\pi} \, {\bf \Gamma}_{N_\ell}(\as(k_t)) \, \log\frac1\epsilon \nn %& \hspace{10mm}+ \, \frac1{2!} \frac{\dd^2}{\dd L^2} \int_{k_{t1}}^{\mu_0} \frac{\dd k_t}{k_t} \, \frac{\as(k_t)}{\pi} \, {\bf \Gamma}_{N_\ell}(\as(k_t)) \, \log^2\frac1\epsilon \nn
& \hspace{5mm} + \int_{k_{t1}}^{\mu_0} \frac{\dd k_t}{k_t} \, {\bf \Gamma}_{N_\ell}^{(\rC)}(\as(k_{t})) + \frac{\dd}{\dd L} \int_{k_{t1}}^{\mu_0} \frac{\dd k_t}{k_t} \, {\bf \Gamma}_{N_\ell}^{(\rC)}(\as(k_{t})) \, \log\frac1\epsilon \bigg) \Bigg] \nn
& \times \, \sum_{\ell_1=1}^2 \left( {\bf R}_{\ell_1}'(k_{t1}) + \frac{\as(k_{t1})}{\pi} \, {\bf\Gamma}_{N_{\ell_1}}(\as(k_{t1})) + {\bf \Gamma}_{N_{\ell_1}}^{(\rC)}(\as(k_{t1})) \right) \nn
& \times \, \sum_{n=0}^{\infty}\frac1{n!} \prod_{i=2}^{n+1} \int_{\epsilon}^1 \frac{\dd \zeta_i}{\zeta_i} \int_0^{2\pi} \frac{\dd\phi_i}{2\pi} \sum_{\ell_i=1}^2 \Bigg[ {\bf R}_{\ell_i}'(k_{t1}) + {\bf R}_{\ell_i}''(k_{t1}) \, \log\frac1{\zeta_i} \nn %& + \frac1{2!}{\bf R}_{\ell_i}'''(k_{t1}) \, \log^2\frac1{\zeta_i} \nn
& \hspace{5mm} + \, \frac{\as(k_{t1})}{\pi} {\bf \Gamma}_{N_{\ell_i}}(\as(k_{t1})) + \frac{\dd}{\dd L} \left(\frac{\as(k_{t1})}{\pi} {\bf \Gamma}_{N_{\ell_i}}(\as(k_{t1}))\right) \log\frac1{\zeta_i} + {\bf \Gamma}_{N_{\ell_i}}^{(\rC)}(\as(k_{t1})) \Bigg] \nn
& \times \, \vartheta\left(v-V(p_1,p_2;k_1,\dots, k_{n+1})\right) \, .
\end{align}

After performing the expansion, we can return to direct space by undoing the Mellin transformations. In particular, at N3LL, only up to two hard-collinear resolved emissions are needed.
Moreover,  we can relax the assumption of flavour-conserving real radiation by including flavour-changing kernels in the DGLAP-evolution contributions in momentum space directly. In practice, we perform the following replacements
\begin{align} \label{3:eq:conversions_to_direct_space}
& \frac{\dd |{\cal M}_B|_{c_1c_2}^2}{\dd \Phi_B} \, \, {\bf f}^{T}_{N_1}(k_{t1}) \Bigg[ \sum_{\ell=1}^2 \frac{\as(k_{t1})}{\pi} {\bf\Gamma}_{N_{\ell}}(\as(k_{t1})) \Bigg] {\bf f}_{N_2}(k_{t1}) \nn
&\qquad  \longrightarrow ~~ \frac{\as(k_{t1})}{\pi} \, {P}(z,\as(k_{t1})) \otimes {\cal L}_{\rm NLL}(k_{t1})   = -\partial_L \, {\cal L}_{\rm NLL}(k_{t1})\, , \nn
& \frac{\dd |{\cal M}_B|_{c_1c_2}^2}{\dd \Phi_B} \, \, {\bf f}^{T}_{N_1}(k_{t1}) \, {\bf C}^{c_1;T}_{N_1}(\as(k_{t1})) \, H(\mu_R) \nn
& \qquad \qquad \times \, \Bigg[ \sum_{\ell=1}^2 \left( \frac{\as(k_{t1})}{\pi} {\bf \Gamma}_{N_{\ell}}(\as(k_{t1})) + \,{\bf \Gamma}_{N_{\ell}}^{(\rC)}(\as(k_{t1})) \right) \Bigg] {\bf C}^{c_2}_{N_2}(\as(k_{t1})) \,{\bf f}_{N_2}(k_{t1}) \nn
& \qquad \longrightarrow ~~ - \, \partial_L \, {\cal L}_{\rm NNLL}(k_{t1}) \, , \nn %in the paper NNLL subscript is missing, is it a typo?
& \frac{\dd |{\cal M}_B|_{c_1c_2}^2}{d\Phi_B} \, \, {\bf f}^{T}_{N_1}(k_{t1}) \Bigg[ \sum_{\ell=1}^2 \frac{\dd }{\dd  L}  \left( \frac{\as(k_{t1})}{\pi} {\bf\Gamma}_{N_{\ell}}(\as(k_{t1})) \right) \Bigg] {\bf f}_{N_2}(k_{t1}) \nn
& \qquad \longrightarrow ~~ 2 \, \frac{\beta_0}{\pi} \, \as^2(k_{t1}) \, {P}^{(0)} \otimes {\cal L}_{\rm NLL}(k_{t1})\, ,\nn
&\frac{\dd |{\cal M}_B|_{c_1c_2}^2}{\dd \Phi_B} \, \, {\bf f}^{T}_{N_1}(k_{t1})\Bigg[\sum_{\ell_i=1}^2\frac{\as(k_{t1})}{\pi} {\bf\Gamma}_{N_{\ell_i}}(\as(k_{t1})) \Bigg]\Bigg[ \sum_{\ell_j=1}^2\frac{\as(k_{t1})}{\pi}{\bf\Gamma}_{N_{\ell_j}}(\as(k_{t1})) \Bigg] {\bf f}_{N_2}(k_{t1}) \nn
& \qquad \longrightarrow ~~ \frac{\as^2(k_{t1})}{\pi^2} \, {P}(z,\as(k_{t1})) \otimes {P}(z,\as(k_{t1})) \otimes {\cal L}_{\rm NLL}(k_{t1}) \nn
& \qquad \simeq \frac{\as^2(k_{t1})}{\pi^2} \, {P}^{(0)} \otimes {P}^{(0)} \otimes {\cal L}_{\rm NLL}(k_{t1}) \, ,
\end{align}
where we defined $\partial_L \, \equiv \, \dd /\dd  L$, $L=\log(M/k_{t1})$,
$\beta_0$ is the lowest-order contribution to the QCD beta function, ${\cal L}$ is the parton luminosity at the various logarithmic orders (see the appendices of \cite{Re:2021con} for its explicit expression), and its convolution with ${P}^{(0)}$ is defined as
\begin{align} \label{3:eq:Pluminosity-NLL}
{P}^{(0)} \otimes {\cal L}_{\rm NLL}(k_{t1})  & \equiv \sum_{c,c'} \frac{\dd |{\cal M}_{B}|_{cc'}^2}{d\Phi_B} \,  \bigg[ \big({P}^{(0)} \otimes f \big)_c(k_{t1},x_1) \, f_{c'}(k_{t1},x_2) \\
& \hspace{20mm} + \, f_c(k_{t1},x_1) \, \big({P}^{(0)} \otimes f\big)_{c'}(k_{t1},x_2) \bigg] \, , \nn
{P}^{(0)} \otimes {P}^{(0)} \otimes {\cal L}_{\rm NLL}(k_{t1}) & \equiv  \sum_{c,c'} \frac{\dd |{\cal M}_{B}|_{cc'}^2}{d\Phi_B} \, \bigg[ \big({P}^{(0)} \otimes {P}^{(0)} \otimes f \big)_c (k_{t1},x_1) \, f_{c'}(k_{t1},x_2) \nn
&\hspace{20mm} + \, f_c(k_{t1},x_1) \, \big({P}^{(0)} \otimes {P}^{(0)} \otimes f \big)_{c'}(k_{t1},x_2) \nn
&\hspace{20mm} + \, 2 \, \big({P}^{(0)} \otimes f \big)_c (k_{t1},x_1) \, \big({P}^{(0)} \otimes f \big)_{c'} (k_{t1},x_2)\bigg]\, . \nonumber
\end{align}
The $\log\(\frac{1}{\epsilon}\)$ can be rewritten as radiative integrals like in Ref.~\cite{Banfi:2014sua}
\begin{equation}
\log^k\frac{1}{\epsilon} = k \int_\epsilon^1 \frac{\dd \zeta}{\zeta} \, \log^{k-1}\frac1\zeta \, , \qquad \quad k\geq 1\, .
\end{equation}
Then, we define the average of a function $G(p_1,p_2;\{k_i\})$ over the measure $\dd  {\cal Z}$
\begin{equation} \label{3:eq:dZ}
\int \dZ \, G(p_1,p_2;\{k_i\}) = \epsilon^{R'(k_{t1})} \, \sum_{n=0}^{\infty}\frac1{n!} \, \prod_{i=2}^{n+1} \int_\epsilon^1 \frac{\dd \zeta_i}{\zeta_i} \int_0^{2\pi} \frac{\dd \phi_i}{2\pi} \, R'(k_{t1}) \, G(p_1,p_2; k_1,\dots,k_{n+1}) \, , 
\end{equation}
which no longer depends on the $\epsilon$ regulator to any order.
We finally arrive at
\begin{subequations} \label{3:eq:master-kt-space}
\begin{align} 
\frac{\dd\Sigma^{\rm N^3LL}(v)}{\dd\Phi_B} & = \int\frac{\dd k_{t1}}{k_{t1}} \frac{\dd\phi_1}{2\pi} \, \partial_{L} \left(- \, \eu^{-R(k_{t1})} {\cal L}_{\rm N^3LL}(k_{t1}) \right) \int \dZ \, \vartheta \left(v-V(p_1,p_2;k_1,\dots, k_{n+1})\right) \label{3:eq:master-kt-space-NLL}\\
& + \, \int \frac{\dd k_{t1}}{k_{t1}} \frac{\dd\phi_1}{2\pi} \, \eu^{-R(k_{t1})} \int \dZ \int_0^1 \frac{\dd\zeta_s}{\zeta_s} \frac{\dd\phi_s}{2\pi} \nn
& \times \, \Bigg\{ \bigg( R' (k_{t1}) {\cal L}_{\rm NNLL}(k_{t1}) - \partial_L {\cal L}_{\rm NNLL}(k_{t1}) \bigg) \bigg( R''(k_{t1}) \log\frac1{\zeta_s} + \frac{1}{2} R'''(k_{t1}) \log^2\frac1{\zeta_s}\bigg) \nn
& \quad - \, R'(k_{t1}) \left( \partial_L {\cal L}_{\rm NNLL}(k_{t1}) - 2 \, \frac{\beta_0}{\pi} \, \as^2(k_{t1}) {P}^{(0)} \otimes {\cal L}_{\rm NLL}(k_{t1}) \log\frac1{\zeta_s} \right) \nn
& \quad + \, \frac{\as^2(k_{t1})}{\pi^2} {P}^{(0)} \otimes {P}^{(0)} \otimes {\cal L}_{\rm NLL}(k_{t1}) \Bigg\} \nn
& \times \, \bigg[ \vartheta\left(v-V(p_1,p_2;k_1,\dots, k_{n+1},k_s)\right) - \vartheta\left(v-V(p_1,p_2;k_1,\dots, k_{n+1})\right) \bigg] \label{3:eq:master-kt-space-NNLL}\\
& + \, \frac12\int\frac{\dd k_{t1}}{k_{t1}} \frac{\dd\phi_1}{2\pi} \eu^{-R(k_{t1})} \int \dZ \int_0^1 \frac{\dd \zeta_{s1}}{\zeta_{s1}} \frac{\dd\phi_{s1}}{2\pi} \int_0^1 \frac{\dd \zeta_{s2}}{\zeta_{s2}} \frac{\dd\phi_{s2}}{2\pi} \, R'(k_{t1}) \nn
& \times \, \Bigg\{ {\cal L}_{\rm NLL}(k_{t1}) \big(R''(k_{t1})\big)^2 \log\frac{1}{\zeta_{s1}} \log\frac{1}{\zeta_{s2}} \nn
& - \partial_L {\cal L}_{\rm NLL}(k_{t1}) R''(k_{t1}) \bigg( \log\frac{1}{\zeta_{s1}} + \log\frac{1}{\zeta_{s2}} \bigg)  + \,  \frac{\as^2(k_{t1})}{\pi^2} {P}^{(0)} \otimes {P}^{(0)} \otimes {\cal L}_{\rm NLL}(k_{t1}) \Bigg\} \nn
& \times \, \bigg[ \vartheta\left(v-V(p_1,p_2;k_1,\dots,k_{n+1},k_{s1},k_{s2})\right) -  \vartheta\left(v-V(p_1,p_2;k_1,\dots,k_{n+1},k_{s1})\right) \nn
& \quad  - \,  \vartheta\left(v-V(p_1,p_2;k_1,\dots,k_{n+1},k_{s2})\right) + \vartheta\left(v-V(p_1,p_2;k_1,\dots,k_{n+1})\right)  \bigg] \, , \label{3:eq:master-kt-space-N3LL}
\end{align}
\end{subequations}
where $\as(k_{t1})$ is shorthand for $\as(k_{t1}) \, = \, \as/(1-2\as\beta_0 L)$, and $\as = \as(\mu_R)$ unless stated otherwise.

Eq.~\eqref{3:eq:master-kt-space} encodes the resummation of all logarithms $\log(1/v)$  up to N$^3$LL, therefore neglecting subleading-logarithmic terms of order $\alpha_s^n \log^{2n-6}(1/v)$.
The expression is separated in three terms, the first one~\eqref{3:eq:master-kt-space-NLL} starts at LL and contains the full NLL corrections. The second term ~\eqref{3:eq:master-kt-space-NNLL}, in the second to fourth lines, is necessary to achieve NNLL accuracy, while the third term ~\eqref{3:eq:master-kt-space-N3LL} (fifth to ninth lines) is purely N$^3$LL.
Of course, Eq.~\eqref{3:eq:master-kt-space} still contains subleading-logarithmic terms (i.e.~starting at N$^4$LL in $\log(1/v)$). One could, even if not strictly required, perform further expansions on each of the terms of Eq.~\eqref{3:eq:master-kt-space} in order to discard some of undesired terms.
For instance, for a N$^3$LL resummation, the full N$^3$LL radiator is necessary in the first term of Eq.~\eqref{3:eq:master-kt-space} and can be instead be replaced with its NNLL truncation in the second term ~\eqref{3:eq:master-kt-space-NNLL} and NLL one in the third one ~\eqref{3:eq:master-kt-space-NLL}.
Analogously, for a NNLL resummation, the NLL radiator suffices in the second term of Eq.~\eqref{3:eq:master-kt-space-NNLL} and it's possible to split $R'(k_{t1})$ into
\begin{equation}
R'(k_{t1}) = \hat{R}'(k_{t1}) + \delta \hat{R}'(k_{t1}) \, ,
\end{equation}
where $\hat{R}'(k_{t1})$ is a strictly NLL term and $\delta \hat{R}'(k_{t1})$ a NNLL.
Then, after expanding Eq.~\eqref{3:eq:master-kt-space} about the former and retaining only contributions linear in $\delta \hat{R}'(k_{t1})$ one obtains
\begin{align}
\frac{\dd\Sigma^{\rm NNLL}(v)}{\dd\Phi_B} & =
\int \frac{\dd k_{t1}}{k_{t1}} \frac{\dd \phi_1}{2\pi} \int \dZ  \bigg\{\partial_L \left[-\eu^{-R(k_{t1})}{\cal L_{\rm NNLL}} \right] \vartheta\( v-V(p_1,p_2;k_1,\dots, k_{n+1} \) \nn
&+ \eu^{-R(k_{t1})}\hat{R}'(k_{t1})\!\int_{\epsilon}^{1}\frac{ \dd \zeta_{s}}{\zeta_s}   \left[  \left( \delta\hat{R}'(k_{t1})+\hat{R}''(k_{t1})    \log\frac{1}{\zeta_s}\right) {\cal    L}_{\rm NLL}- \partial_{L}{\cal    L}_{\rm NLL}    \right] \nn
&\times \[ \vartheta\left(v-V(p_1,p_2;k_1,\dots, k_{n+1},k_s)\right) - \vartheta\left(v-V(p_1,p_2;k_1,\dots, k_{n+1})\right)  \]\bigg\}\, .\label{3:eq:Sigma-NNLL}
\end{align}
This reproduces the result from Ref.~\cite{Monni:2016ktx} where the \RadISH\ approach was first formulated at NNLL for the Higgs-boson transverse-momentum distribution.

The discussion of this section reproduced the main steps in the construction of the \RadISH\ resummation method, landing to Eqs.\eqref{3:eq:master-kt-space} and \eqref{3:eq:Sigma-NNLL}, which are implemented into the code of the same name for efficient numerical evaluation of resummed differential observables for a number of different processes at hadron colliders.

Before moving on to discussing $HW$ associated production as a phenomenological application, a few extra observations are necessary.
First, in our review we considered the logarithm of the form $L = \log(M/k_{t1})$. While this definition is useful to simplify the setup of the resummation, in practice it is more versatile to introduce a separate resummation scale $Q$, of the same size of $M$, as a way to probe the size of neglected logarithmic corrections and resum logarithms $\log(Q/k_{t1})$ instead. This is formally achieved by writing $L = \log(Q/k_{t1}) + \log(M/Q)$, with the implicit hierarchy $\log(Q/k_{t1}) \gg \log(M/Q)$, valid in the IRC limit, and then expanding $L$ around $\log(Q/k_{t1})$ at the relevant logarithmic accuracy.
Second, when the resummed results are matched to a fixed-order prediction, it is desirable to make the resummed contribution vanish in the hard region $k_{t1}\gg Q$ of the $v$ spectrum, where the fixed order alone is reliable. One way to achieve this is modifying the resummed logarithms $\log(Q/k_{t1})$ with a power suppression, negligible at small $k_{t1}$. For example
\begin{align} \label{3:eq:modified-log}
& \log\frac{Q}{k_{t1}}  ~\longrightarrow~ \tilde{L} \, = \, \frac{1}{p} \log\left[\left(\frac{Q}{k_{t1}}\right)^{p} + 1\right]\, , \qquad \int \frac{\dd k_{t1}}{k_{t1}}  ~\longrightarrow~ \int_0^\infty \frac{\dd k_{t1}}{k_{t1}} \,{\cal J}(k_{t1})\, , \nn
& \qquad \qquad {\cal J}(k_{t1}) = \frac{\left(Q/k_{t1}\right)^p}{1+\left(Q/k_{t1}\right)^p} \, = \, 1 - \left(\frac{k_{t1}}{Q}\right)^p + \dots \, ,
\end{align}
where $p$ is a positive real parameter tuned to make the resummed differential spectrum vanishing faster than
the fixed-order one at large $v$. The above prescription also induces a Jacobian $\mathcal{J}(k_{t1})$,
which ensures the absence of subleading-power corrections with fractional $\as$ powers in the final distribution, still keeping
the $k_{t1}\to 0$ region unmodified. Moreover, the final resummed result will now have an explicit $p$ dependence through power-suppressed terms, which cancels up to the accuracy of the fixed-order component in the matching.
From this point onward we will assume this procedure of logarithmic modification, which formally corresponds to considering the logarithmic region $k_{t1} < Q$ and working in the $p\to\infty$ limit of Eqs.~\eqref{3:eq:modified-log}. On top of this we redefine $L \, \equiv \, \log(Q/k_{t1})$, in order to avoid unnecessarily convoluted expressions.

Finally, in this presentation of the \RadISH\ formalism we considered explicitly terms up to N3LL, however it can be improved to \NNNLLp accuracy~\cite{Re:2021con}, which amounts to including also the complete set of constant contributions of relative order ${\cal O}(\as^3)$ with respect to the Born. These terms formally belong to the logarithmic tower $\as^nL^{n-3}$, namely they are a subset of the \NNNNLL correction, however they are of particular relevance since their inclusion completes the perturbative expansion of the resummed cumulative cross section $\Sigma(v)$ up to all terms of order $\as^n\log^{2n-6}(1/v)$.

The definition of `primed' logarithmic accuracy provides a strategy to include these constant terms in $\Sigma(v)$. Indeed, there are two sides required to achieve this upgrade.
First, while the resummation formula presented in Eq.~\eqref{3:eq:partxs-mellin} is formally valid to all logarithmic orders, its accuracy is limited by the highest known perturbative order for the anomalous dimensions and coefficient functions. On top of this, the expression \eqref{3:eq:mellin_expanded_final} in Mellin space and \eqref{3:eq:master-kt-space} in momentum space relies on the expansions in Eq.~\eqref{3:eq:expanded_unres} and \eqref{3:eq:expanded_res} for computational streamlining. Thus, achieving full \NNNLLp requires extending the subset of such approximations to include the ones that affect third-order constant contributions.
Second, in the structure of Eq.~\eqref{3:eq:mellin_expanded_final} the weight of the hardest resolved radiation $k_{t1}$ provides at least one power of $\as$. Thus, it  can be shown that including one further logarithmic derivative in the exponent of the second line, as well as in the resolved ensemble, only affects ${\cal O}(\as^4)$ terms. Then, the structure of Eq.~\eqref{3:eq:mellin_expanded_final} is sufficient as is to achieve \NNNLLp accuracy. The only required modification is to evaluate its contributions to appropriate perturbative order and to rewrite the conversions of Eqs.~\eqref{3:eq:conversions_to_direct_space} to momentum space accordingly. The details of these operations are detailed in Ref.~\cite{Re:2021con} and we will not delve further in this discussion.

\subsection{Recoil effects}\label{3:ssec:recoil-effect}

Following \cite{Catani:2015vma,Re:2021con}, we now review another relevant feature of the resummed spectrum of transverse observables. As a general example, let's consider the production of a colour singlet. Naturally, at LO in perturbation theory the  distribution of observables of the singlet decay products is constrained by the Born level production, which carries a vanishing transverse momentum.
On the other hand, when considering the same distribution at the resummed level, higher-order contributions due to  soft and collinear multi-parton radiation will  dynamically produce a finite value of the transverse momentum $\qt$ for the singlet, and this $\qt$-recoil  has to be distributed between the decay products, thereby affecting the shapes of related observables.

%Since this $\qt$-recoil effect is a non-singular contribution to the $\qt$ distribution at small values of $\qt$ and it is not directly and unambiguously computable through transverse-momentum resummation.

Neglecting these ${\cal O}(\qt /M)$ corrections is acceptable for the resummed calculation of the $\qt$ distribution alone. However, if we remain exclusive in the transverse momentum of some of the decay products or apply some fiducial cuts, then the momentum of the colour singlet must be fully specified by the one of its decay chain and $\qt$ is not vanishing.
Thus, the resummed calculation requires the specification of a prescription to distribute this $\qt$ between different configurations of the Born-level decay chain.
Empirically, this means that the resummation will produce a smearing of the LO distribution ($\delta(\qt)$).

For a generic observable $v$, this recoil procedure amounts to considering the differential spectrum with respect to $v$, and to boost its underlying Born kinematics from a rest frame of the singlet to the laboratory frame: there the singlet has transverse momentum equal to $\qt(v)$ (for example, $\qt = M v$ if $v = \pt{} / M$). Then, one can apply the fiducial selection cuts on the boosted Born kinematics instead of the rest frame ones. This is sufficient\cite{Ebert:2020dfc,Becher:2020ugp,Catani:2015vma} to capture all linear power corrections in presence of fiducial cuts, together with their resummation with the same accuracy as the leading-power terms, for observables which are azimuthally symmetric at leading power, such as $\pt{} / M$.

In the \RadISH\ language, the implementation of these recoil effects is achieved by evaluating the transverse momentum $\qt (v)$ of the colour singlet and its azimuthal angle $\phi$ for each $m$-parton contribution to Eq.~\eqref{3:eq:master-kt-space}, with $v$ being defined by the $\vartheta (v - V( p_1,p_2; k_1, \dots, k_m ))$ measurement function. Then, we apply the aforementioned boost and cut.
Indeed, to enforce the latter on the boosted Born system, we modify each measurement function in Eq.~\eqref{3:eq:master-kt-space} as
\begin{equation}
\vartheta (v - V(p_1,p_2; k_1, \dots, k_m ) ) \,\, \longrightarrow \,\, \vartheta (v - V( p_1,p_2; k_1, \dots, k_m ) ) \,\vartheta_{\rm cuts}(\Phi_{B} , \{ k_1, \dots, k_m \})\, ,
\end{equation}
where the dependence on $k_1, \dots, k_m$ in $\Theta_{\rm cuts}$ encodes the effect of the boost (i.e.~$\vartheta_{\rm cuts}$ equals 1 or 0 if the boosted Born configuration passes or not the cuts).

Instead, in absence of recoil effects, the momenta $k_1, \dots, k_m$ do not enter the computation of the cuts. Then, the constraint $\vartheta_{\rm cuts} (\Phi_{B}, \{ k_1, \dots, k_m \})$
reduces to $\vartheta_{\rm cuts} (\Phi_{B})$ and factorises out of the resummation formula entirely.

One more consideration is necessary if the resummed result must be matched with fixed-order predictions. Indeed, recoil effects will also impact the perturbative expansion of the resummation, as detailed in Ref.~\cite{Bizon:2017rah}. Since the recoil procedure entails boosts on the differential spectrum, the correct steps are to first compute the expansion at a given value $v$, and then apply fiducial cuts on the boosted kinematics, consistently with what is done in the resummation component.

\subsection{Essentials of $\qt$-subtraction at NNLO}\label{3:ssec:qtsubtraction}
Before moving on to our application for $HW$ production, we need to introduce the use of subtraction methods as a way to compute higher-order QCD corrections. Although the technology behind this kind of computation has been developed and successfully applied up to \NNNLO across for a variety of processes at LHC~\cite{Catani:2007vq,FebresCordero:2022psq,Chen:2022cgv}, we will summarise the main idea following Ref.~\cite{Catani:2007vq} for a NNLO example.

We return to the process in Eq.~\eqref{3:eq:hadron-noc} and consider a colour singlet with total invariant mass $Q^2$.
Higher order QCD corrections at NNLO will divide in three groups:
\begin{itemize}
\item {\em double real} emissions, where two extra partons recoil against the final state;
\item {\em real-virtual} corrections, where one parton recoils against the one-loop production of the singlet;
\item {\em two-loop virtual} corrections to the Born level process.
\end{itemize}
Each of these subprocess is separately divergent due to IR singularities, but (for any choice of IRC-safe observable~\cite{Banfi:2004yd}) these will disappear in their combination.
Thus, we need to rearrange the singular contributions to achieve their explicit cancellation. To this end we first observe that, naturally, the transverse momentum $\qt$ of colourless final state  is exactly zero at leading order. Then, among the NNLO contributions, the region $\qt  \neq 0$, are actually given by the NLO contributions to the final state ${\rm singlet}+{1{\rm jet}}$,
\begin{equation} \label{3:eq:Fplusjets}
\dd \sigma_{\rm NNLO}\(\qt \neq 0 \)=\dd \sigma^{{\rm singlet}+{\rm jets}}_{\rm NLO}\, .
\end{equation}
This means that, when $\qt\neq 0$, the IR divergences in our NNLO calculation are actually those of one lower order in perturbation theory and  can be handled on their own by using any method to perform NLO calculations.

In addition to those, the limit $\qt \to 0$ will be associated with one more singularity, this time properly of NNLO and an additional subtraction is required. 
Crucially, the singular behaviour of $\dd \sigma^{{\rm singlet}+{\rm jets}}_{\rm NLO}$ in the small $\qt$ regime is controlled by the same class of logarithms we discussed in the previous section. So, we can leverage our knowledge on the resummed cross-section to achieve the desired cancellation of IR singularities. Explicitly, Eq.~(\ref{3:eq:Fplusjets}) can be expanded to include the $\qt = 0$ limit by writing
\begin{equation} \label{3:eq:mainNNLO}
\dd{\sigma}_{\rm NNLO}={H}_{\rm NNLO}\otimes \dd{\sigma}_{\rm LO}+ \dd{\sigma}^{{\rm singlet}+1 { \rm jet}}_{\rm NLO}- \dd{\sigma}_{\rm LO}\otimes{\cal C}\(\frac{\qt^2}{Q^2}\)\, \dd^2 \qt \, ,
\end{equation}
where the symbol $\otimes$ stands for convolutions over momentum fractions and sum over flavour indices of the partons.
The hard virtual coefficient ${H}_{\rm NNLO}$ does not depend on $\qt$ and is obtained from the  NNLO truncation of the perturbative function from Eq.~\eqref{3:eq:hard-virtual}. Finally, the last term represents a counterterm to remove the small $\qt$ logarithmic behaviour of $\dd{\sigma}^{{\rm singlet}+1 { \rm jet}}_{\rm NLO}$
\begin{equation}\label{3:eq:sigmalimit}
{\cal C}\(\frac{\qt^2}{Q^2}\) \xrightarrow[\qt \to 0]{} \sum_{n=1}^\infty \left(\frac{\as}{2\pi}\right)^n\sum_{k=1}^{2n} {\cal C}^{(n;k)} \;\frac{Q^2}{\qt^2}\log^{k-1} \(\frac{Q^2}{\qt^2}\)  \;\; .
\end{equation}
The explicit form of the counterterm is constrained only by its small-$\qt$ limit so that it has the form given in Eq.~\eqref{3:eq:sigmalimit}.
This is so that the weight ${\cal C}\(\frac{\qt^2}{Q^2}\)$ with leading order kinematics matches a corresponding ``event'' in  $\dd{\sigma}^{{\rm singlet}+1{\rm jet}}$ with $\qt \to 0$.
%Moreover, the physical information of the virtual correction to the LO subprocess contained in the expansion of ${H}_{\rm NNLO}$ will pick up additional contributions from the specific form chosen for the counterterm.For the purpose of our application, this complication does not need to be considered here as the fixed order expansion of Eq.~\ref{3:eq:Sigma-NNLL} computed in \RadISH\ already accounts for both terms.
%The simplicity of the LO subprocess is such that final-state partons actually appear only in %the second term in Eq.~(\ref{main}).the term $d{\sigma}^{F+{\rm jets}}$ on theright-hand side of Eq.~(\ref{main}).Therefore,%For this reason, arbitrary IR-safe cuts on the jets at (N)NLO can effectively be accounted for through a (N)LO computation. Owing to this feature,our NNLO extension of the subtraction formalism is observable-independent.

\section{Study of $HW^+$ associate production} \label{3:sec:WHsection}
Studies of Higgs boson mass and couplings are a pivotal element in the high-precision physics programme at the Large Hadron Collider (LHC).
These quantities provide a handle into the mechanism of the Electro-Weak symmetry breaking (EWSB) and potential Beyond Standard Model (BSM) physics.  From a theoretical perspective, the processes in which a Higgs boson is produced in association with vector bosons, i.e. $\mathrm{pp} \rightarrow \HW$, pp $\rightarrow {ZH}$, are of special interest as they are potentially sensitive to the physical effects beyond the Standard Model. Indeed, on the experimental side, associate Higgs plus EW boson production has been at the core of several searches to study the Standard Model Higgs and puts constraints on potential BSM effects from both the ATLAS~\cite{ATLAS:2018kot,ATLAS:2020fcp,ATLAS:2020jwz,ATLAS:2022ers,ATLAS:2022tnm} and CMS~\cite{CMS:2016tad,CMS:2021nnc,Collaboration:2022mlq} collaborations.
Instead, on the theory side, QCD corrections to HW production were evaluated at in Refs.~\cite{Han:1991ia,Baer:1992vx,Ohnemus:1992bd}, the NNLO QCD order can be found in Ref.~\cite{Brein:2003wg,Brein:2012ne,Harlander:2018yio}, finally, the inclusive-level N3LO cross-section is available from Ref.~\cite{Baglio:2022wzu}. The electroweak correction are available to NLO from Ref.~\cite{Ciccolini:2003jy,Denner:2011id} and are rather sizeable.
At NNLO in QCD, many studies on exclusive cross sections and fully differential observables are available in Refs.~\cite{Ferrera:2011bk,Ferrera:2014lca,Ferrera:2013yga,Campbell:2016jau,Caola:2017xuq,Ferrera:2017zex}, while a NNLO QCD analysis with a resolved jet was performed in Refs.~\cite{Majer:2020kdg,Gauld:2021ule}. 

Thus, in this section we present a study of the $\pt{H}$ spectrum in pp$ \rightarrow HW^+$ production at NNLO order in QCD, with an additional veto on the leading jet transverse momentum $\pt{jv}$, emulating some of the analysis cuts of Ref.~\cite{ATLAS:2020fcp}. As we have seen in the preceding sections, fixed-order perturbation theory is, however, insufficient to accurately describe the observable considered here, due to the presence of the jet veto.

The overall objective of our calculation is assembling the Higgs $p_t$ spectrum at the fiducial level $pp \rightarrow W^+ H \rightarrow \(l^+ \nu_l\)  \(\gamma\gamma\)$ at \NNLO  according to the $\qt$-subtraction formalism~\cite{Catani:2007vq}, including  NNLL-accurate resummation of logarithms of leading jet transverse momentum $\pt{j}$. The main tool for this study is resummation code \RadISH~\cite{Monni:2016ktx,Bizon:2017rah,Monni:2019yyr,Kallweit:2020gva,Re:2021con}, built on the formalism summarised in section~\ref{3:ssec:RadISH}, which will be used both to produce the $\qt$-subtraction counterterm and the $\pt{j}$ resummation.

\subsection{$\qt$-cumulant and inclusive cross-section}
We start our validation by computing the total cross section for
\begin{equation}
 \mathrm{pp} \rightarrow   \mathrm{H} \mathrm{W}^+ \rightarrow \(\gamma \gamma\)  \(l^+\nu_l\) .
\end{equation}
Applying  the technique of section~\ref{3:ssec:qtsubtraction}.

We return to $q_t$-subtraction and use it to build an approximation of the total cross-section as
\begin{equation}\label{3:eq:qtsubtotal}
\bar{\sigma}_{\rm N(NLO)}^{HW}\(\qt\) =
\int_0^\infty \dd \pt{HW} \frac{\dd \sigma^{HW}_{\rm N(NLL)}}{\dd \pt{HW}} + \int_{\qt}^\infty\dd \pt{HW}\left\lbrace \frac{\dd \sigma^{HW+1{\rm jet}}_{N(LO)}}{\dd \pt{HW}} - \[\frac{\dd \sigma^{HW}_{\rm N(NLL)}}{\dd \pt{HW}} \]_{\rm N(NLO)}\right\rbrace
%\Sigma^{HW+1{\rm jet}}_{\rm (N)LO}\( \qt \) + \[ \Sigma^{HW}_{\rm (N)NLL}\( \qt \)  \]_{\rm (N)NLO} \, ,
\end{equation}
where we wrote the cumulant of Eq.~\eqref{3:eq:cum-cs} explicitly as integral over $\pt{HW}$. We introduce $\qt = 1$ GeV as a small safety cut to  regularise the real (real-virtual + double real) emission $\frac{\dd \sigma^{HW+1{\rm jet}}_{N(LO)}}{\dd \pt{HW}}$, which is generated using \MCFM \cite{Boughezal:2016wmq,Campbell:2019dru}. The same software will provide the fixed-order input for all parts of this study.
On the other hand, $\[\frac{\dd \sigma^{HW}_{\rm N(NLL)}}{\dd \pt{HW}} \]_{\rm N(NLO)}$ stands for the fixed-order expansion of the $\pt{HW}$-resummed \RadISH~ cross-section and corresponds to the third contribution from Eq.~\ref{3:eq:mainNNLO}.
These two terms would be divergent for $\qt \rightarrow 0$, but their combinations is finite.
Finally, the member $\int_0^\infty \dd \pt{HW} \frac{\dd \sigma^{HW}_{\rm N(NLL)}}{\dd \pt{HW}}$ coincides with the first one on the right-hand side of Eq.~\ref{3:eq:mainNNLO} once it is integrated over all values of $\pt{HW}$.
Then, if the subtraction is performed with sufficient accuracy, one can interpolate the small-$\qt $ to the inclusive cross-section up to errors $\calO\(\frac{\qt}{Q}\)$, with $Q=\frac{m_{HW}}{2}$.\footnote{We stress that, for this combination to work as intended, it is crucial to use the prime counting to capture the correct normalisation of the cross section, as detailed in Sec.~\ref{3:ssec:RadISH} and Refs.~\cite{Bizon:2017rah,Re:2021con}. We assume the same choice of logarithmic counting for the rest of this discussion.}

%\begin{figure}
%  \centering
%  \includegraphics[page = 1, width = 0.9\linewidth]{images/plot_slicingNNLO_fxscale.pdf}
%  \caption{Ingredients for the determination of the inclusive $HW^+$ cross-section by $\qt$ subtraction in Eq.~\eqref{3:eq:qtsubtotal}. The points in purple and green correspond to $ \[ \Sigma^{HW}_{\rm% (N)NLL}\( \qt \)  \]_{\rm (N)NLO}$ at NLO and NNLO respectively. Similarly, green and yellow are $\Sigma^{HW+1{\rm jet}}_{\rm (N)LO}\( \qt \)$ }
%  \label{3:fig:slicing-breakdown}
%\end{figure}

\begin{figure}
  \centering
  \includegraphics[page = 4, width = 0.9\linewidth]{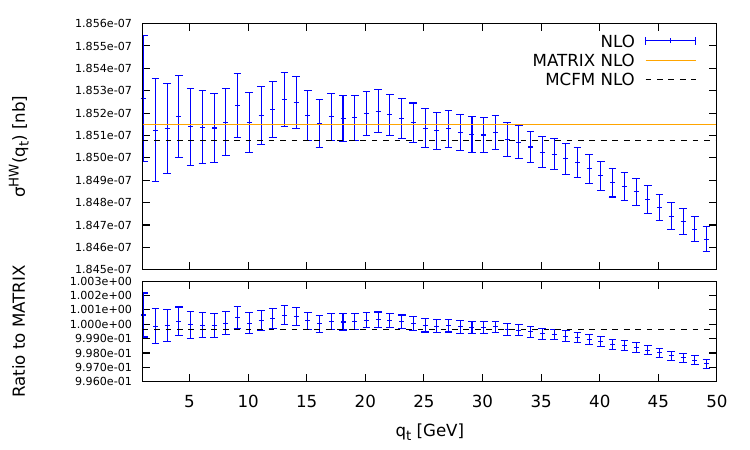}
  \caption{Determination of the inclusive $HW$ cross-section by $\qt$-subtraction at NLO, compared against the same prediction from \MCFM and \texttt{MATRIX}. The error bars in the plot estimate the residual numerical uncertainty. We can appreciate that the subtraction is remarkably stable up to $\qt\simeq25$ GeV. From the ratio plot we observe that at small $\qt$ the subtraction is accurate up to \permille. }
  \label{3:fig:slicing-NLO}
\end{figure}
\begin{figure}
  \centering
  \includegraphics[page = 3, width = 0.9\linewidth]{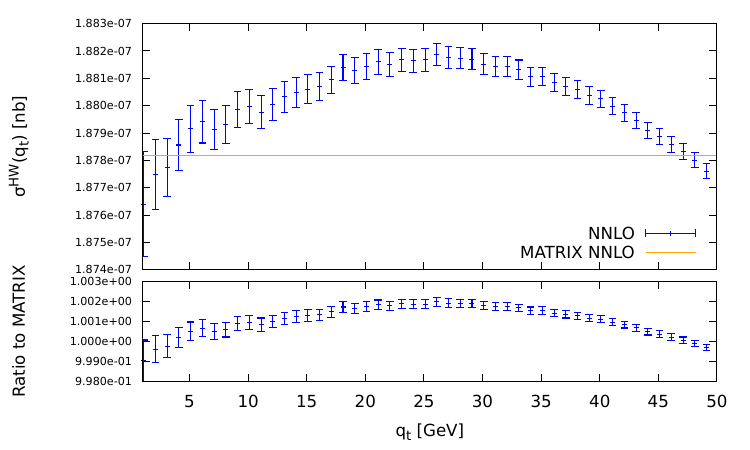}
  \caption{Determination of the inclusive $HW$ cumulant by $\qt$-subtraction at NNLO. The yellow line represents the  prediction the NNLO total cross-section from \texttt{MATRIX}. The error bars in the plot estimate the residual numerical uncertainty. The subtraction is accurate enough only in the first few bin $q_t \in \(1,5\)$ GeV.}
  \label{3:fig:slicing-NNLO}
\end{figure}

We use fixed scales $\muF=\muR=205.38$ GeV, PDFs from the set NNPDF$3.1$\_nnlo\_as\_$0118$,  and a single selection cut of the lepton from the $W^+$ decay, $\pt{} > 25$ GeV.
Figures~\ref{3:fig:slicing-NLO} and \ref{3:fig:slicing-NNLO} report our computation of $\bar{\sigma}^{HW}_{\rm N(NLO)}\(\qt\)$ at NLO and NNLO respectively. 
In the first plot, we observe that the subtraction is remarkably stable up to $25$ GeV while at NNLO only in the first few bins up to $5$ GeV.
So, we use this ranges to make a linear interpolation of $\bar{\sigma}_{\rm N(NLO)}^{HW}\(\qt \)\xrightarrow[\qt \rightarrow 0 ]{} \sigma_{\rm N(NLO)}^{HW}$. 

In order to evaluate the correctness of our subtraction, we compare our it to the total cross section obtained from the \texttt{MATRIX} fixed order code~\cite{Grazzini:2017mhc} and, at NLO only, \MCFM. The predictions from these two programs are shown in Figs.~\ref{3:fig:slicing-NLO} and \ref{3:fig:slicing-NNLO} in yellow and dashed lines respectively.
We summarise our estimate in table~\ref{3:tab:sigma-table}, finding that our subtraction setup reproduces the correct cross-section consistently within numerical accuracy.

\begin{table}[]
\centering
\caption{Summary of the calibration of the inclusive cross-section for $HW^+$ in fb.}
\label{3:tab:sigma-table}
\begin{tabular}{|c|c|c|c|}\hline
  &  $\qt$-subtraction           &  \texttt{MATRIX}     &   \texttt{MCFM} \\ \hline\hline
LO   & N.A.      &  $1.569516218\cdot 10^{-1}$  &   $1.575(7)\cdot10^{-1}$   \\      %$1.5704(2)\cdot 10^{-1}$
NLO  & $1.852(3)\cdot10^{-1}$    &  $1.85147732\cdot 10^{-1}$   &   $1.8508(11)\cdot10^{-1}$   \\   
NNLO & $1.878(2)\cdot 10^{-1}$   &  $1.878182207\cdot 10^{-1}$  &   N.A.   \\ \hline 
\end{tabular}
\end{table}

\begin{figure}
\centering
\includegraphics[page = 1, width = 0.9\linewidth]{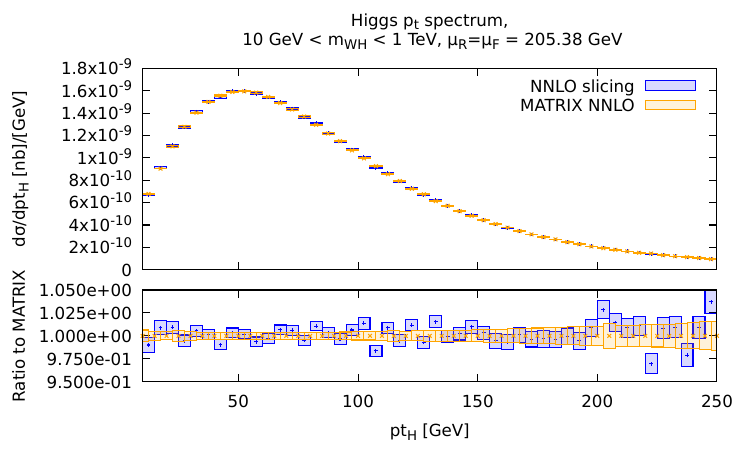}
\caption{Comparison between the NNLO $\pt{H}$ spectrum obtained by $\qt$-subtraction and \texttt{MATRIX}. We work at fixed scales $\muF=\muR=205.38$ GeV and with large cuts on the mass of the $HW$ system, $10~\text{GeV} < m_{HW} < 1~\text{TeV}$, as well as a selection cut on the lepton from the $W$ decay, $\pt{}>25$ GeV. The error bands represent the numerical uncertainty. We observe that the two distributions are superimposed up to deviations of about  1\%.}
\label{3:fig:matching-NNLO}
\end{figure}

\subsection{Study of the Higgs $p_t$ spectrum resummation in associated $HW^+$ production}

Next, we move to computing the NNLO Higgs spectrum using $\qt$-subtraction.
We keep the slicing parameter $\qt = 1~\text{GeV}$, the same scale setting and final state cut of the previous ones, but also add a constraint on the mass of the $HW$ system, $10$ GeV $< m_{W H} < 1$ TeV.
We rewrite the $\qt$-subtraction as
\begin{equation}  \label{3:eq:master}
 \frac{ \dd\sigma_{\HW}^{\rm N(NLO)}}{\dd \pt{H}} \equiv  \frac{\dd\sigma_{\HW}^{\rm N(NLL)}}{\dd \pt{H}}  + \frac{\dd \sigma_{\HW \text{+jet} }^{\rm N(LO) }}{\dd \pt{H}}\( \pt{HW}> \qt \) -\[\frac{\dd\sigma_{\HW}^{\rm N(NLL)}}{\dd \pt{H}}\(\pt{HW} > \qt \)\]_{\rm N(NLO)} ,
\end{equation}
where $\frac{\dd\sigma_{\HW}^{\rm N(NLL)}}{\dd \pt{H}}$ is the $\pt{HW}$-resummed Higgs spectrum integrated over $\pt{HW} \in \(0,\infty\)$ and $\[\frac{\dd\sigma_{\HW}^{\rm N(NLL)}}{\dd \pt{H}}\]_{\rm N(NLO)}$ its fixed order expansion integrated in $\pt{HW}\in\(\qt,\infty\)$. They correspond exactly to the first and third term on the r.h.s of Eq.~\eqref{3:eq:mainNNLO}. Likewise $\frac{\dd \sigma_{\HW \text{+jet} }}{\dd \pt{H}}\( \pt{HW}> \qt \) = \int_{\qt}^\infty \dd \pt{HW} \frac{\dd \sigma_{\HW \text{+jet} }}{\dd \pt{H} \dd \pt{HW}}$.
To confirm the correctness of this setup, we show the resulting distribution together with the corresponding prediction from \texttt{MATRIX} in figure~\ref{3:fig:matching-NNLO}. The two histograms appear to be compatible within numerical accuracy.
\footnote{Despite denoting Eq.~\eqref{3:eq:master} as a fixed order quantity, this true only when the recoil procedure of Sec.~\ref{3:ssec:recoil-effect} for power corrections is not considered. Indeed, when the latter is applied,  $\frac{ \dd\sigma_{\HW}^{\rm N(NLO)}}{\dd \pt{H}}$ will also include linear corrections $\frac{\qt}{m_{HW}}$ to all order. In order to avoid confusion with the $\pt{j}$ resummation, we will still refer to the $\qt$-subtraction spectrum as N(NLO) instead of N(NLO)+N(NLL)$\qt$.} 

\begin{figure}
  \centering
  \includegraphics[page = 1, width = 0.9\linewidth]{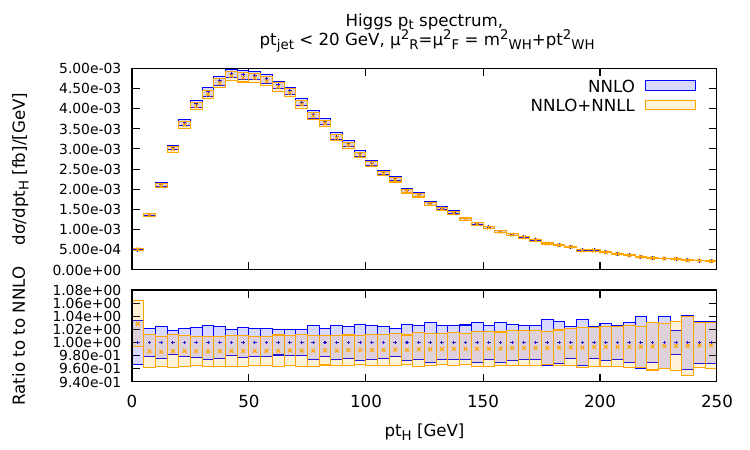}
  \caption{Comparison between the NNLO determination of the $\pt{H}$ spectrum in $HW$ associated production with a veto on the leading jet $\pt{JV}= 20$ GeV (blue) and its NNLO+NNLL matched counterpart with the resummation of $\pt{j}$. The difference amounts to a small correction of around 2\% across the peak of the distribution. The error bars represent scale variations of $\muF^2$, $\muR^2$ and $Q^2$ by a factor 2. }
  \label{3:fig:ptH-NNLO}
\end{figure}

\begin{figure}
\centering
\includegraphics[page = 6, width = 0.9\linewidth]{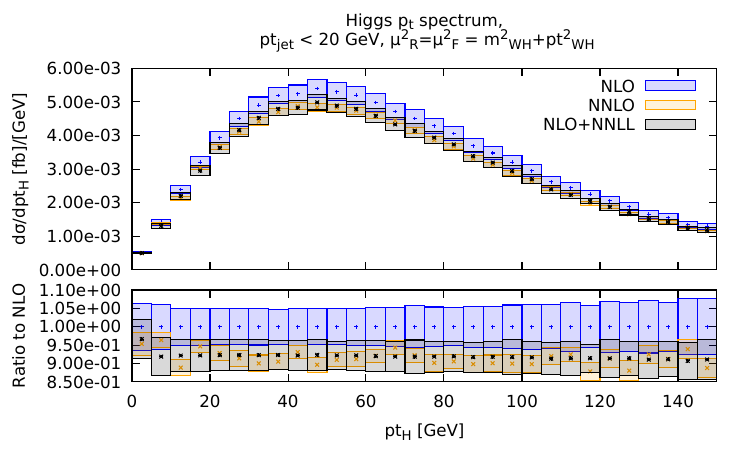}
\caption{Comparison of the determination of the $\pt{H}$ spectrum in $HW$ associated production with $\qt$-subtraction with a veto on the leading jet $\pt{JV}= 20$~GeV at NLO (blue) and NNLO (yellow).
The matching of the NNLL resummation with the NLO fixed order is shown in black. The small effect of the NNLL resummation is in line with the size of the radiative corrections. The error bars represent scale variations of $\muF^2$, $\muR^2$ by a factor 2. }
\label{3:fig:NLO-to-NNLO}
\end{figure}

\begin{figure}
  \centering
  \includegraphics[page = 4, width = 0.9\linewidth]{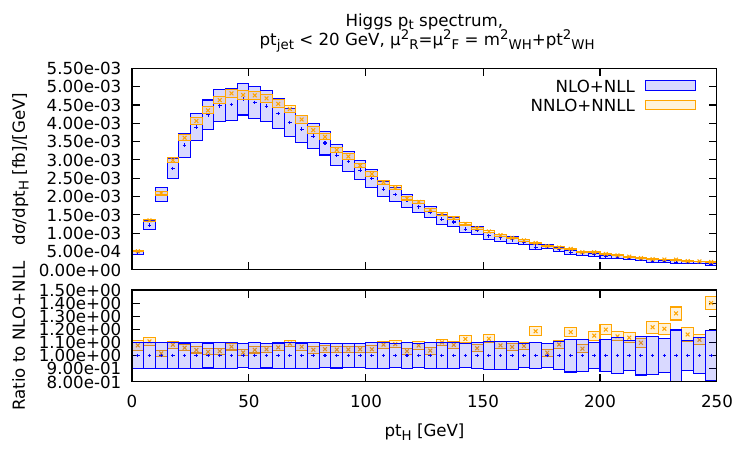}
  \caption{Comparison between the NNLO determination of the $\pt{H}$ spectrum in $HW$ associated production with a veto on the leading jet $\pt{JV}= 20$ GeV at NLO+NLL (blue) and NNLO+NNLL (orange). The two distributions show acceptable perturbative convergence up to  $\pt{H} \sim 125 $ GeV. The error bars represent scale variations of $\muF^2$, $\muR^2$ and $Q^2$ by a factor 2. }
  \label{3:fig:ptH-NNLO-3}
\end{figure}

\begin{figure}
\centering
\includegraphics[page = 7, width = 0.9\linewidth]{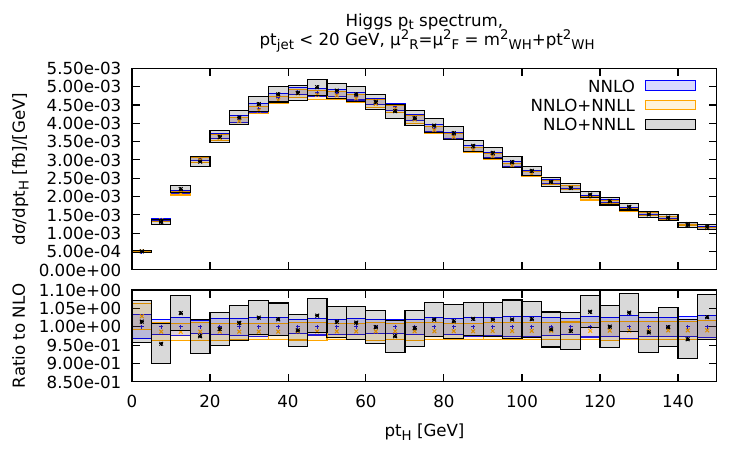}
\caption{Comparison of histograms from Fig.~\ref{3:fig:ptH-NNLO} (blue and yellow) with the NLO+NNLL matching from Fig.~\ref{3:fig:NLO-to-NNLO} (black). The error bars represent scale variations of $\muF^2$, $\muR^2$ and $Q^2$ by a factor 2. }
\label{3:fig:ptH-NNLO-2}
\end{figure}

\begin{figure}
\centering
\includegraphics[page = 13, width = 0.9\linewidth]{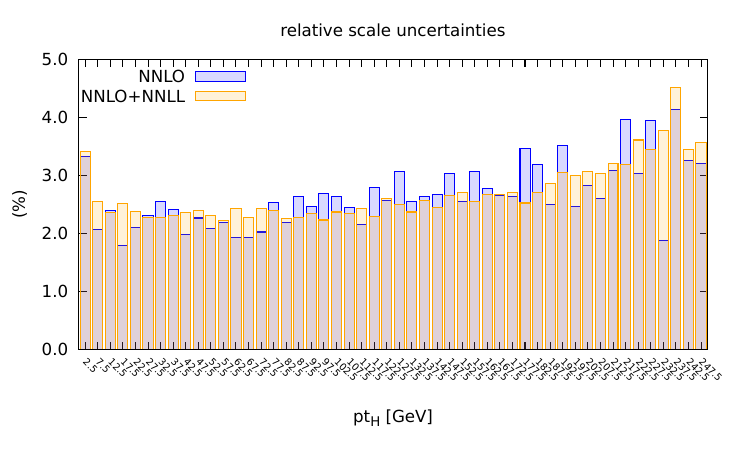}
\caption{Scale uncertainties for the histograms from Fig.~\ref{3:fig:ptH-NNLO}, normalised to their central values.}
\label{3:fig:ptH-scale-unc-rel}
\end{figure}

%%%%%%%%%%%%%%%%%%%%%%%%%%%%%%%%%%%%%%%%%%%%%%%%%%%%%%%%%%%%%%%%%%%%%%%%%%%%%%%%%%%%%%%%%%%%%%%%%%%%%%%%%%%%%%%%%%%%
%\begin{figure}
%  \centering
%  \includegraphics[page = 2, width = 0.9\linewidth]{images/plot_ptH_bands_new.pdf}
%  \caption{Comparison between the NNLO determination of the $\pt{H}$ spectrum in $HW$ associated production with a veto on the leading jet $\pt{JV}= 20$ GeV (blue) and its NNLO+NNLL matched counterpart with the resummation of $\pt{j}$. In both distributions, recoil effects are included according to Sec.~\ref{3:ssec:recoil-effect}. The difference amounts to a small correction of around 2\% across the bump of the distribution. The error bars represent scale variations of $\muF^2$, $\muR^2$ and $Q^2$ of a factor 2.}
%  \label{3:fig:ptH-recoilon}
%\end{figure} THIS PLOT IS COMPLETELY NONSENSICAL

\begin{figure}
\centering
\includegraphics[page = 9, width = 0.9\linewidth]{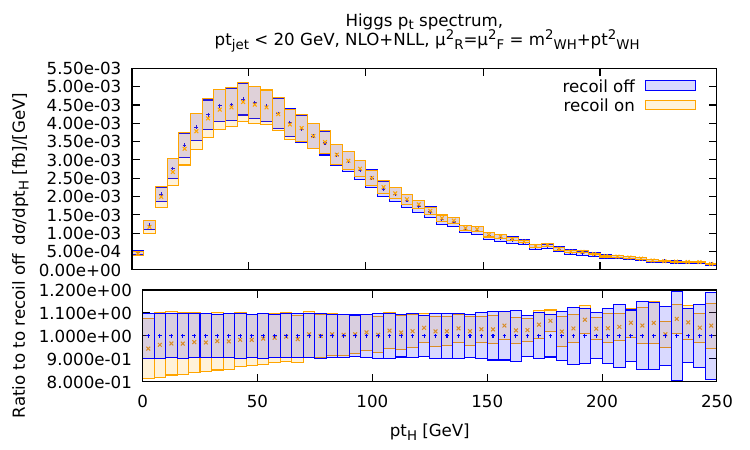}
\caption{Comparison of the NLO determination of the $\pt{H}$ spectrum in $HW$ associated production with $\qt$-subtraction with a veto on the leading jet $\pt{JV}= 20$~GeV. In the both histograms the NLL resummation of $\pt{j}$ is matched with the fixed order according to Eq.~\ref{3:eq:ptH_matched}. The blue (yellow) histogram do (not) include the $\calO \frac{\qt}{Q}$ correction discussed in Sec.~\ref{3:ssec:recoil-effect}. We see that this amount to a small correction of around 2\% across the peak of the distribution. The error bars represent scale variations of $\muF^2$, $\muR^2$ and $Q^2$ of a factor 2. }
\label{3:fig:ptH_recoil_NLO}
\end{figure}

\begin{figure}
\centering
\includegraphics[page = 3, width = 0.9\linewidth]{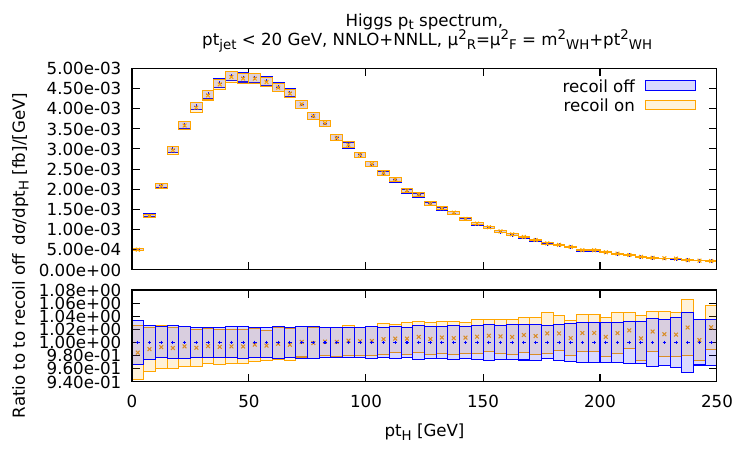}
\caption{Comparison of the NNLO determination of the $\pt{H}$ spectrum in $HW$ associated production with $\qt$-subtraction with a veto on the leading jet $\pt{JV}= 20$~GeV. In the both histograms the NNLL resummation of $\pt{j}$ is matched with the fixed order according to Eq.~\ref{3:eq:ptH_matched}. The blue (yellow) histogram do (not) include the linear power correction discussed in Sec.~\ref{3:ssec:recoil-effect}. We see that this amount to a small correction of around 2\% across the peak of the distribution. The error bars represent scale variations of $\muF^2$, $\muR^2$ and $Q^2$ of a factor 2. }
\label{3:fig:ptH_recoil_NNLO}
\end{figure}

%\begin{figure}
%\centering
%\includegraphics[page = 14, width = 0.9\linewidth]{images/plot_ptH_bands_new.pdf}
%\caption{Scale uncertainties for the histograms from Fig.~\ref{3:fig:ptH-NNLO}, normalised to their central values.}
%\label{3:fig:ptH-scale-unc-rel}
%\end{figure}

After completing the validation of our computational framework, we move on to consider a veto on the leading jet $\pt{}$ and the resummation of corresponding logarithms.
First, we enforce the jet veto condition by subtracting the region with  $\pt{j} > 20$ GeV from our $\qt$-subtraction determination, explicitly
\begin{equation}\label{3:eq:ptH-vetoed}
\frac{ \dd\sigma_{\HW}^{\rm N(NLO)}}{\dd \pt{H}}\( \pt{j} < 20 ~\text{GeV}\) =   \frac{ \dd\sigma_{\HW}^{\rm N(NLO)}}{\dd \pt{H}} - \frac{\dd\sigma_{\HW\text{+1jet}}^{\rm N(NLO)}}{\dd \pt{H}}\(\pt{j} > 20 ~\text{GeV}\) \, ,
\end{equation}
then the resulting distribution is used to match the Higgs spectrum with the N(NLL) resummation of jet veto logarithms. This is implemented in \RadISH~ using an analytical method from Ref.~\cite{Monni:2019yyr}.\footnote{ An implementation of $\pt{j}$-resummation analogous to the $\pt{}$ one in Sec.~\ref{3:ssec:RadISH} is also available. Since the two, differs for terms N3LL and beyond, they are equivalent for the purpose of this study and point the reader to Ref.~\cite{Kallweit:2020gva} for more details).}  
We use an additive scheme to perform the matching
\begin{align}
 \frac{ \dd\sigma_{\HW}^{\rm N(NLO)+N(NLL)}}{\dd \pt{H}}{\(\pt{j} < 20 ~\text{GeV}\)} & =  \frac{\dd\sigma_{\HW}^{\rm N(NLL)}}{\dd \pt{H}} {\(\pt{j} < 20 ~\text{GeV}\)} + \frac{\dd\sigma^{\rm N(NLO)}_{\HW\text{+jet}}}{\dd \pt{H}}{\(\pt{j} < 20 ~\text{GeV}\)} \nn
 & -\bigg[ \frac{\dd \sigma_{\HW}}{\dd \pt{H}}^{\rm N(NLL)} {\(\pt{j} < 20 ~\text{GeV}\)} \bigg]_{\rm N(NLO)}\, . \label{3:eq:ptH_matched} 
\end{align}
Unlike the previous tests, in this case we preserve the selection cut on the decay products of $W^+$, but use dynamical scales $ \muF^2=\muR^2=m_{HW}^2+\pt{HW}^2$ with no other cut on the invariant mass.
An estimate of the missing higher orders is evaluated by varying the scales $\muF^2$, $\muR^2$ by a factor 2 under the condition $\frac{1}{2} \leq \frac{\mu_R}{\mu_F} \leq 2$.
Additionally, the \RadISH\ resummation scale introduced by the modified logs in Eq.~\eqref{3:eq:modified-log} is set to  $Q = \frac{m_{HW}}{2}$, and variations are considered for fixed $\mu_R,\,\mu_F$.
The $\pt{j}$-resummation generates a suppression in the peak of the $\pt{H}$ spectrum, around 2\% relative to the NNLO fixed order, Fig.~\ref{3:fig:ptH-NNLO}. While this effect is small, it is in line with the size of radiative correction shown in Fig.~\ref{3:fig:NLO-to-NNLO}. Indeed, the plot shows that matching the NNLL jet veto resummation to the NLO distribution accounts for around half of the difference between NLO and NNLO. The perturbative convergence can be also probed by comparing the matched NLO+NLL and NNLO+NNLL distributions directly, finding acceptable agreement up to $\pt{H} \sim 125$ GeV in Fig.~\ref{3:fig:ptH-NNLO-3}.
Similarly, a comparison between the NLO+NNLL and the NNLO+NNLL histograms in Fig.~\ref{3:fig:ptH-NNLO-2} shows that the two are compatible within the quoted scale uncertainties.
It is worth noticing that the uncertainty induced by scale variations ranges between 2\% and 4\% for both the NNLO and NNLO+NNLL distributions, as shown in Fig.~\ref{3:fig:ptH-scale-unc-rel}, and 
this estimate is somewhat larger than the value found in other studies on HW associate production Ref.~\cite{Caola:2017xuq,Ferrera:2017zex,Gauld:2021ule}.
This is partly explained by the presence of a additional scale, $Q$, to consider in the uncertainties.
Another possible source of this effect may be identified in the two-steps combination procedure of Eqs.~\eqref{3:eq:master} and~\eqref{3:eq:ptH_matched}, and a way to diagnose this effect is to reproduce the same study using the full joint-resummation formalism of Ref.~\cite{Monni:2019yyr}. 

The effect of linear power corrections $\calO \(\frac{\qt}{m_{HW}}\)$ on the $\pt{H}$ spectrum is considered by applying the recoil prescription described in section~\ref{3:ssec:recoil-effect} in the computation of Eq.~\eqref{3:eq:master}. The results are shown in Figs.~\ref{3:fig:ptH_recoil_NLO} and \ref{3:fig:ptH_recoil_NNLO} for the NLO+NLL and NNLO+NNLL distributions respectively. The effect is roughly a downward correction of $1\%$ in the region $\pt{H} < 80$ GeV and an increase of few percent in the tail of the distribution, $\pt{H} > 150$ GeV.
The small size of these contribution is reasonably explained by the high scale of the process and the broad shape of the distribution already at fixed order.

        \chapter[PDF determination with a quantum statistical model]{Parton distribution function determination with a quantum statistical model}
	\label{ch4}
	Almost any theoretical prediction at hadron colliders requires knowledge of the internal structure of colliding hadrons (i.e. protons or heavy ions for the LHC). Through the use of the collinear factorisation framework we saw in chapter~\ref{ch1}, this information is encoded in the PDF. These depend on two variables, $Q^2$ and $x$. While the first scale dependence can be described perturbatively through the DGLAP evolution equation, the second one can not as it includes the non-perturbative side of QCD dynamics governing strong interaction at and  below its natural scale $Q\sim \Lambda_{\rm QCD}$.

\begin{figure}
  \centering
  \includegraphics[width=0.7\linewidth]{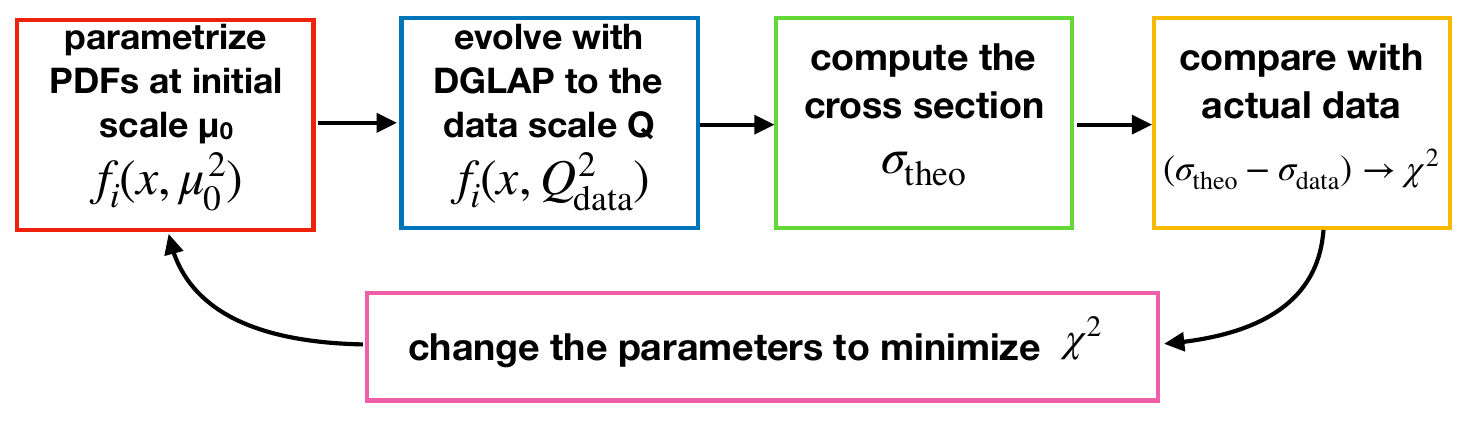}
  \caption{PDF fitting algorithm flowchart: the PDF is parametrised at an initial low-energy model ($\mu_{0}$) and evolved using DGLAP equations to match measured data scale ($Q_{\textnormal{data}}^2$). The partonic cross section is computed in perturbation theory and combined with the PDF to compare with experimental data. The PDF parameters are optimised by minimising a $\chi^2$ figure-of-merit to match with experimental input.}
  \label{5:fig:pdffitflow}
\end{figure}

Non-perturbative techniques, chiefly Lattice QCD, can be used, in principle, to compute PDFs\footnote{Properly speaking, the usual collinear PDFs cannot be probed on a lattice as they are light-cone object an thus they vanish in the Euclidean sector used to computed correlation functions on the lattice. Despite this, it is still possible to define more general lattice objects that capture some properties of the PDF in the correct limit~\cite{Gao:2022ytj}.}~\cite{Wilson:1974sk,Lin:2017snn,Rossi:2018stq,Constantinou:2020pek}. However the standard methodology for their determination is a fit to experimental data, schematised in figure \ref{5:fig:pdffitflow}. Briefly, PDFs are parametrised at an initial scale $\mu_0\sim \Lambda_{\rm QCD}$ with an almost arbitrary functional form depending on a set of parameters $\lbrace p\rbrace$ which is then evolved to the energy scale of the data to be fit using DGLAP evolution.
A theoretical prediction for the hadron-level process is then built by convolution of the PDFs with hard-scattering cross-sections from perturbative QCD.
Finally, this construction is compared with the corresponding experimental data and the best fit for $\lbrace p \rbrace$ are finally determined by minimising an appropriate figure of merit, usually a $\chi^2$ distribution.

The resulting parton distribution functions (PDFs) are influenced by various factors. First, the accuracy of the theory used to calculate partonic cross sections and DGLAP splitting functions is the primary determining factor of the extracted PDFs. This accuracy is not limited to the perturbative order and the inclusion of resummation effects in applicable kinematic regimes, but also depends on the scheme used, treatment of heavy quarks, choice of unphysical scales such as the renormalisation scale. Second, the dataset used to extract the PDFs is another significant factor, including data from DIS experiments and the HERA electron-proton collider~\cite{Abramowicz:2015mha}, as well as proton-(anti)proton machines, LHC and Tevatron colliders. Third, the choice of parametrisation also plays an essential role. An ideal parametrisation should describe the data without any bias, except when motivated by physical expectations, while also having a minimal number of parameters to prevent over-fitting. Finally, there is additional dependence on input parameters such as quark masses and strong coupling, fitting methodology including choice of $\chi^2$ definition, minimisation methods, and uncertainty estimation, and various technical aspects, which are less relevant for our discussion. Consequently, making accurate predictions in high-energy phenomenology requires accurate knowledge of PDFs, which in turn demands constant refinement of all the aforementioned ingredients.

This chapter is divided in two main parts. The first one~\ref{4:sec:pdfdet} is devoted to a review of the entire process of PDF determination and its current status. In the second section~\ref{4:sec:QSPDF} we introduce the Quantum Statistical PDF (QSPDF), a parametrisation for PDF based on a statistical model of the proton structure. We explain the main features of the parametrisation and perform a quality test by using it to fit the HERA dataset with NLO QCD predictions. We conclude the chapter by comparing the fit performance against those of the \HERAPDF\ parametrisation. 

\section{Review of PDF determination}\label{4:sec:pdfdet}

In this section we will briefly go over the current status for each of these inputs in PDF global analyses, borrowing the overview offered in Refs.~\cite{Gao:2017yyd,Rottoli:2018nma}. A more detailed discussion can be found in the literature~\cite{Roberts:1990ww,Forte:2010dt,DeRoeck:2011na,Perez:2012um,Forte:2013wc,Rojo:2019uip,NNPDF:2021njg,Khalek:2021ulf,PDF4LHCWorkingGroup:2022cjn,Amoroso:2022eow}.

\subsection{Experimental data}\label{4:ssec:expdata}
The selection of the dataset used in the PDF analysis is a crucial factor in determining parton densities. The choice is based on the constraining power of the data, which depends on a combination of PDF sensitivity, experimental precision, and the theoretical calculations for the examined processes. The observables considered in PDF fits are generally inclusive enough to ensure reliable leading-twist factorisation. Similarly, the scale of the data included is usually high enough to ensure that higher-twist contributions can be neglected.

%%%%%%%%%%%%%%%%%%%%%%%%%%%%%%%%%%%%%%%%%%%%%%%%%%%%%%%%%%%%%%%%%%%%%%%%%%%%%%%%
\begin{figure}
\centering
   \includegraphics[width=\linewidth]{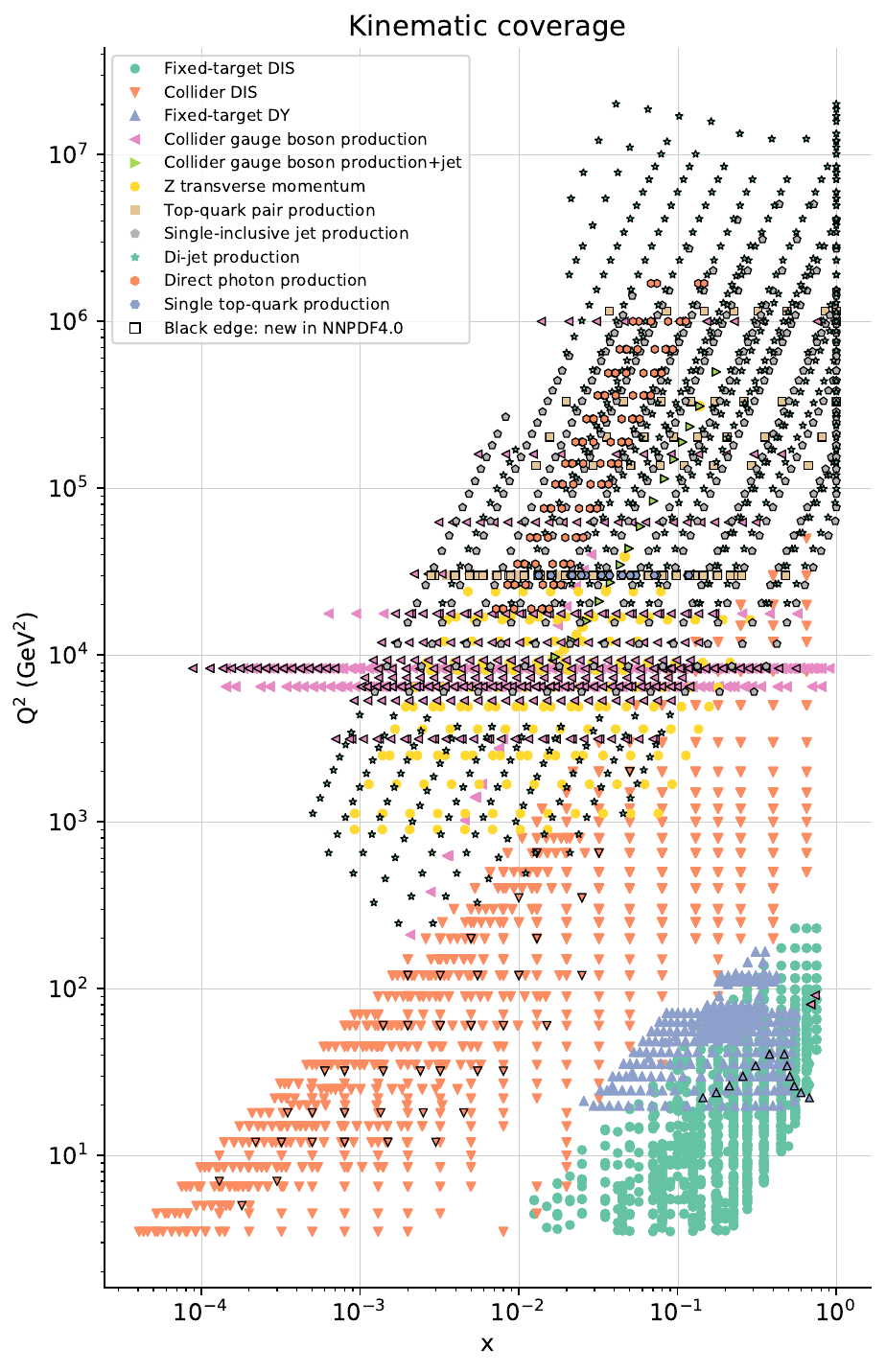}
   \caption{Kinematic coverage of a state-of-the art global fit (taken from Ref.~\cite{NNPDF:2021njg} in the $(x,Q^2)$ plane).}
 \label{4:fig:kinplot4}
\end{figure}
%%%%%%%%%%%%%%%%%%%%%%%%%%%%%%%%%%%%%%%%%%%%%%%%%%%%%%%%%%%%%%%%%%%%%%%%%%%%%%%%
Global PDF fits~\cite{Ball:2012cx,Jimenez-Delgado:2014zga,Harland-Lang:2014zoa,Dulat:2015mca,Accardi:2016qay,Alekhin:2017kpj,Ball:2017nwa,Hou:2019efy,Bailey:2020ooq,NNPDF:2017mvq,NNPDF:2021njg} use data from a variety of measurements, ranging from fixed-target DIS and Drell-Yan processes to jet and top-quark production at hadron colliders. The extensive data available is depicted in Figure~\ref{4:fig:kinplot4}, which displays the coverage of a modern global analysis in the $(x,Q^2)$ plane. Fixed target experiments~\cite{Aubert:1982tt,Arneodo:1996kd,Arneodo:1996qe,Whitlow:1991uw,Benvenuti:1989rh,Onengut:2005kv,Goncharov:2001qe,Moreno:1990sf,Webb:2003ps,Towell:2001nh,Abulencia:2007ez,Abazov:2007jy,Aaltonen:2010zza,Samoylov:2013xoa,Abazov:2013rja,D0:2014kma,Dove:2021ejl} cover the high-$x$ and low-$Q^2$ region, forming the foundation of the earliest PDF fits. Collider DIS experiments~\cite{Abramowicz:1900rp,ZEUS:2002nms,ZEUS:2006xvn,Aaron:2009af,ZEUS:2010vyw,H1:2014cbm,Abramowicz:2014zub,H1:2016goa,Abramowicz:2015mha,H1:2018flt} occupy the low-$x$ and low-to-medium-$Q^2$ region, while recent data from collider experiments at the LHC dominate the high-$Q^2$ region. In the following, we will review the key processes used in global PDF fits, emphasising especially collider DIS.

\paragraph{\itshape\mdseries Fixed-target and collider Deep Inelastic Scattering.}
Prior to the emergence of QCD as the renormalisable quantum field theory of the strong interaction, DIS experiment outcomes were interpreted under the assumptions the quark parton model alone. In this scenario, it was customary to combine the parton level theory with PDFs to form the so-called DIS structure functions. This formalism remains a widespread method to demonstrate the PDF sensitivity of this process~\cite{Patrignani:2016xqp,Gao:2017yyd}. Consider the case of Neutral Current (NC) DIS, the LO definition for the structure functions $F_2$ and $F_3$ are 
\begin{equation}
\frac{\dd^2 \sigma^{\rm NC,\ell^{\pm}}}{\dd x \dd Q^2} (x,y,Q^2)=\frac{2\pi \alpha^2}{ x Q^4}
Y_+ F_2^{NC}(x,Q^2) \mp Y_- x F_3^{NC}(x,Q^2)%-y^2 F_L^{NC}(x,Q^2)\rc \; ,
\label{4:eq:ncxsect}
\end{equation}
where we have defined 
\begin{equation}
 Y_{\pm}=1\pm (1-y)^2\; ,
\label{4:eq:ypmdef}
\end{equation}
with $x$ and $y$ being the momentum fraction and rapidity for the scattering. 
and the quark parton model expressions are given by
\begin{subequations}
\begin{align}
 \lb F_2^{\gamma},F_2^{\gamma Z}, F_2^Z\rb & = x\sum_{i=1}^{n_f}\lb e^2_i, 2e_ig_V^i, (g_V^{i})^2+(g_A^{i})^2\rb \( q_i+\bar{q}_i\) \, , \label{4:eq:partonmodel1} \\
 \lb F_3^{\gamma},F_3^{\gamma Z}, F_3^Z\rb & = x\sum_{i=1}^{n_f}\lb 0, 2e_ig_A^i, 2 g_V^i g_A^i\rb \( q_i-\bar{q}_i\) \, , \label{4:eq:partonmodel2}
\end{align}
\end{subequations}
the superscripts on the l.h.s. indicate the exchanged gauge boson (either a $\gamma^*$ or $Z$), as well as the contribution from the $\gamma Z$ interference term. $e_i$ is the electric charge of the quark of flavour $i$ and the weak couplings are given by $g_V^i=\pm \frac{1}{2}-2e_i\sin^2\theta^2_W$ and $g_A^i=\pm \frac{1}{2}$, where the $\pm$ corresponds to a $u$ or $d$ type quark. The sum runs over all the $n_f$ quarks that are active for the specific scale at which the scattering takes place. Crucially, the longitudinal structure function $F_L=F_2-2xF_1$ vanishes at this level due to the Callan-Gross relation.

Eqs.~\eqref{4:eq:partonmodel1} and~\eqref{4:eq:partonmodel2} show that the NC structure functions are limited in their ability to provide information on flavour separation, and cannot distinguish between quarks and antiquarks unless the $Z$-mediated contribution is significant, which occurs only at sufficiently high $Q^2$ scales. Historically this issue was circumvented by assuming isospin symmetry and including NC DIS on deuterium targets, which provides a LO measure of the triplet contribution $T^3 \equiv \( u+\bar{u} - d-\bar{d}\)$. This, leveraging the relation between proton and neutron valence partons, $u^{(p)}=d^{(n)}$ and $u^{(d)}=d^{(p)}$, allows to disentangle the quark and antiquark distributions.

Charged Current DIS allows to probe this difference more directly. Indeed, in the energy window between charm and top thresholds and assuming that CKM-suppressed transitions can be neglected, the corresponding CC DIS structure functions in the parton model are given by
\begin{subequations}
\begin{align}
F_2^{W^-}&=2x\( u+\bar{d}+\bar{s}+c \) \, , \nn
F_3^{W^-}&=2x\( u-\bar{d}-\bar{s}+c \) \, , \nn
F_2^{W^+}&=2x\( d+\bar{u}+\bar{c}+s \) \, , \nn
F_3^{W^+}&=2x\( d-\bar{u}-\bar{c}+s \) \, , 
\end{align}
\end{subequations}
where again the longitudinal structure function $F_L^{W^\pm}=0$ vanishes at this level.
Crucially, this time the $F_3^W$ structure function, which provides information on the difference between quark flavours, is not suppressed with respect to $F_2^W$.

For this reason, the inclusion of CC structure functions, both from HERA and from neutrino fixed--target experiments, is a staple of global fits in order to improve the discrimination between quarks and anti--quarks and to probe strangeness. 

Since the gluon does not couple to electroweak final states, the gluon distribution is probed in DIS experiments directly through the small contribution which enters at  $\mathcal O(\as)$, or indirectly via scaling violations encoded in DGLAP evolution. 
Scaling violations represent a small effect at medium and large-$x$, while they are a significantly more important effect  at small-$x$, in particular in the region covered by HERA. 
Therefore, scaling violations from the HERA $F_2^p$ data provide direct information on the small-$x$ gluon, while at medium to large $x$  a DIS-only fit does not strictly constrain gluon behaviour.
Additional information on the gluon PDF can be obtained from the longitudinal structure function $F_L$, which vanishes only at LO, and is non-zero at NLO.
Indeed, the Altarelli-Martinelli relation~\cite{Altarelli:1978tq}
\begin{equation}
F_L(x,Q^2)=\frac{\alpha_s(Q^2)}{\pi}\lb \frac{4}{3}\int_x^1 \frac{\dd y}{y}\( \frac{x}{y}\)^2F_2(y,Q^2)
+2\sum_i e_i^2 \int_x^1 \frac{\dd y}{y}\( \frac{x}{y}\)^2 \( 1-\frac{x}{y}\) g(y,Q^2)\rb \, ,
\end{equation}
establishes that $F_L$ measurements are directly sensitive to the gluon, and especially so at  small $x$, provided experimental uncertainties are competitive.
One more constraint to the gluon PDF is given by measurements of heavy-quark structure functions with the requirement that charm or bottom mesons are reconstructed in the final state, since the LO contribution (under the assumption of no intrinsic $c$,$b$ components in the proton) is $g\gamma \rightarrow Q \bar Q$.

Beside their role in PDF determination, Heavy Quark structure functions also play a large role in the determination of heavy quark mass and their effects in theoretical calculation~\cite{Bertone:2016ywq,Gao:2013wwa,Alekhin:2012vu}.
 
\paragraph{\itshape\mdseries Fixed-target and Collider Drell-Yan.}
The electroweak production of a lepton pair by annihilation of a quark with an antiquark, plays a primary role in PDF analysis due to its constraining power on light flavour decomposition, including strangeness.
The relevant partonic processes involved in inclusive EW boson production are
\begin{align}
	\qquad u \bar d, c \bar s \rightarrow W^+, \qquad d \bar u, s \bar c \rightarrow W^-, \qquad  q \bar q \rightarrow \gamma^*/Z \, ,
\end{align}
if we consider only LO QCD and neglect Cabibbo-suppressed channels.
Since each flavour channel carries a different weight, their combination offers a way to probe the flavour composition of the proton. 
Customarily, the rapidity and the invariant mass of the electroweak boson can be used to map the kinematics of this process into the momentum fraction $x_1$, $x_2$, which appear in the PDFs at LO.
They are related by the hyperbolic relation
\begin{equation}\label{4:eq:DYkin}
	x_{1,2} = \frac{M}{\sqrt{s}}e^{\pm Y} \quad \iff \quad \frac{M}{s}=x_1x_2, \: Y = \frac{1}{2}\log\(\frac{x_1}{x_2}\),
\end{equation}
where $M$ is the invariant mass of the virtual boson and $Y$ is its rapidity, which are experimentally accessible\footnote{This is not exactly correct in the CC case as the rapidity of the virtual boson cannot be reconstructed directly and must be reconstructed off the one of the lepton and other measurements in the event.}. 

Discarding for the moment the $s,c$ contributions and focusing instead on the ones from $u$ and $d$ in a hadron collider experiment, we observe that the ratio of $W^+$ and $W^-$ production differential in rapidity and their asymmetry are privileged observables to extract information on the flavour separation  in the charged current case,
\begin{subequations}
  \begin{align}
    R_{\pm} & = \frac{\dd \sigma (W^+)/\dd Y}{\dd \sigma (W^-)/\dd Y} \, ,\\
    A_{W} & = \frac{\dd \sigma (W^+)/\dd Y - \dd \sigma (W^-)/\dd Y}{\dd \sigma (W^+)/\dd Y+\dd \sigma (W^-)/\dd Y}.
  \end{align}
\end{subequations}
In the central rapidity region, where $x_1=x_2=x_0 \sim M/\sqrt{s}$ and one can approximate $\bar u = \bar d$, these two become
\begin{equation}\label{4:eq:RandA}
	R_{\pm} \simeq \frac{u(x_0)}{d(x_0)}, \qquad A_{W} \simeq \frac{u_V(x_0)-d_V(x_0)}{u_V(x_0)+d_V(x_0)},
\end{equation}
where the valence quark is defined as $q_V(x) = q(x) - \qb (x)$. We can see that, $R_{\pm}$ is sensitive to the ratio between $u$ and $d$, whilst $A_{W}$ is sensitive to the combinations of valence quarks.
Similarly the forward region $Y \gtrsim 2, \, x_1 \gg x_2,\, q(x_1)\sim q_V(x_1),\,\overline{u}(x_2)\sim \overline{d}(x_2)\,$, gives equivalent constraints
\begin{subequations}
  \begin{align}
    R_{\pm} & \sim \frac{u_V(x_1)}{d_V(x_1)}\;,\\
    A_W    & \sim  \frac{u_V(x_1)-d_V(x_1)}{u_V(x_1)+d_V(x_1)}\;.
  \end{align}
\end{subequations}

By studying neutral current Drell-Yan processes, it is possible to obtain comparable information about the distribution of $u$ and $d$ quarks in the proton. Additionally, as one moves away from the $Z$ peak and explores the low-invariant mass region, more constraints on PDFs can be placed at low and intermediate values of $x$. Here the virtual-photon exchange dominates
\begin{equation}
	\frac{\dd \sigma}{\dd Y} \simeq \sum_q e_q^2 (q (x_1) \bar q(x_2) + q(x_2) \bar q(x_1)),
\end{equation}
and as a consequence the $u \bar u $ channel is enhanced with respect to the $d \bar d$. 

The low-invariant mass region of neutral current Drell-Yan production is not only sensitive to the $u$ and $d$ content of the proton, but also to the gluon PDF, especially if final-state lepton cuts emphasise higher-order corrections. Additionally, fixed deuteron targets can be utilised in neutral-current Drell-Yan production to constrain the $\bar u/d$ ratio. In the valence region where $x_1 \gtrsim 0.1$, and $q(x_1) \sim q_V(x_1)$, isospin symmetry dictates that
\begin{equation}
\frac{\sigma^{pn}}{\sigma^{pp}}	\simeq \frac{\bar d (x_2)}{\bar u (x_2)}.
\end{equation}

\paragraph{\itshape\mdseries Inclusive jet production.}
In order to constrain the gluon PDF at large $x$, inclusive jet production measurements have been used at hadron colliders since the earliest measurements at Tevatron. These jet cross sections are reconstructed experimentally using a suitable jet algorithm that must be IRC safe to be compared with theoretical predictions. Although other algorithms are used, the anti-$\kt$ algorithm is the most common choice for data collected at the LHC. The PDF sensitivity of jet production depends on the kinematics and definition of the observable. PDF fits usually include double-differential single-inclusive jet cross-section data in $\pt{}$ and rapidity $Y$. In each event, all jets are considered and included in the same distribution. At low $\pt{}$, the gluon-induced contribution dominates in these measurements, but it is also significant at high $\pt{}$. Since DIS and DY measurements already provide constraints on the quark content of the proton, jet data are useful in handling the gluon PDF at medium and large $x$. Several double-differential single-inclusive dataset can be included in PDF fits, and a comprehensive list can be found in Ref.~\cite{Gao:2017yyd}.

\paragraph{\itshape\mdseries Transverse momentum of $Z$ boson.}
Very precise measurements of the inclusive transverse momentum of the $Z$ boson have recently been released by the ATLAS, CMS and LHCb collaborations based on the combined data of LHC Run I,II and the initial part of III. These are interesting in PDF analysis to probe the gluon in the medium-$x$ region, which is only partly constrained by collider DIS data and jets data. Indeed, the dominant contribution in the region of moderate and large transverse momentum for this process is gluon and quark scattering subprocesses
\begin{equation}
q\bar q\rightarrow Zg,\,\, gq\rightarrow Zq,\,\, g\bar q\rightarrow Z\bar q \; .
\end{equation}
In the leptonic channel where experimental measurements are the cleanest, the kinematics of the $Z$ boson, namely the transverse momentum $\pt{}$ and rapidity $Y$, can be reconstructed from the momenta of the lepton pair produced in the $Z$ decay. The momentum fractions of the initial--state partons are given by (at LO)
\begin{equation}
x_{1,2}=\frac{m_T}{\sqrt s}e^{\pm Y}+\frac{\pt{}}{\sqrt s}e^{\pm y_j}\;,\, 
%\,\,\,x_2=\frac{m_T}{\sqrt s}e^{-y_Z}+\frac{p_{T}}{\sqrt s}e^{-y_j}\; , 
\end{equation}
where $\sqrt s$ is again the centre-of-mass energy of the two incoming hadrons, $m_T=\sqrt{M_Z^2+\pt{}^2}$ is the transverse mass of the $Z$ boson and $y_j$ is the rapidity of the recoiling parton.
For inclusive production with respect to the hadronic recoil, that is integrated over $y_j$, these momentum fractions are therefore not uniquely determined, although for LO kinematics lower limits can be derived.
%%%%%%%%%%%%%%%%%%%%%%%%%%%%%%%%%%%%%%%%%%%%%%%%%%%%%%%%%%%%%%%%%%%%%
\begin{figure}[t]
  \begin{center}
  \includegraphics[width=\textwidth]{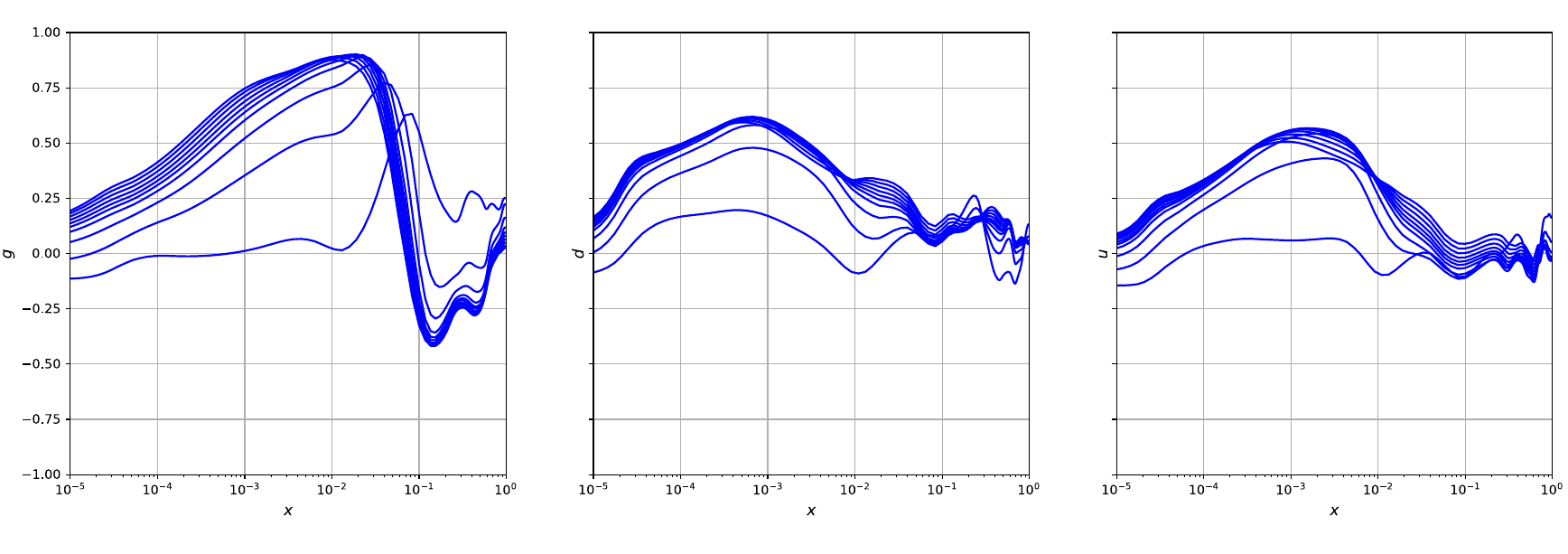}
  \caption{Correlations coefficients between the cross sections in various $\pt{}$ bins and the gluon, down and up quark PDFs as a function of $x$, from Ref.~\cite{Boughezal:2017nla}.
    The binning corresponds to the ATLAS 8 TeV measurement~\cite{Aad:2015auj} within the rapidity interval $0<|Y|<0.4$. }
  \label{4:fig:ptz4}
\end{center}
\end{figure}
%%%%%%%%%%%%%%%%%%%%%%%%%%%%%%%%%%%%%%%%%%%%%%%%%%%%%%%%%%%%%%%%%%%%%

Usually, at the LHC, the measurement of double-differential cross-sections in $\pt{}$ and $Y$ is done at the $Z$ peak. However, the off-shell region can also be studied, where contributions from virtual photons may be significant. The cross-sections at moderate and high transverse momentum are mainly due to the scattering of gluons and quarks and are closely related to the gluon PDF in the region relevant to Higgs boson production at the LHC. This correlation is shown in Figure \ref{4:fig:ptz4}, which displays the cross-sections in different $\pt{}$ bins, with $0<|Y|<0.4$, and their connections with the gluon, down and up quark PDFs at various $x$ values. The figure demonstrates that correlations with the gluon at $x\sim 10^{-2}$ are nearly as high as 0.9. The quark PDFs also show moderate correlations at $x\sim 10^{-3}$. The $Z$ $\pt{}$ distribution can potentially provide insights into the gluon for $x$ values between those covered by HERA structure functions and those covered by inclusive jets and $t\bar{t}$ production.

\paragraph{\itshape\mdseries Top quark production.}
The production of top quark pairs at the LHC is dominated by gluon-gluon fusion, which represents around 85\% of the total cross section.\footnote{Unlike the lower $\sqrt{s}$ region of the Tevatron, where top quark pair production is dominated instead by quark anti-quark annihilation.}
Therefore, top quark production data into the global PDF fit has the potential to constrain the gluon in the large--$x$ region, provided other sources of theoretical uncertainties such as missing higher orders and the values of the top mass $m_t$ can be kept under control.

\begin{figure}[t]
\begin{center}
  \includegraphics[width=\textwidth]{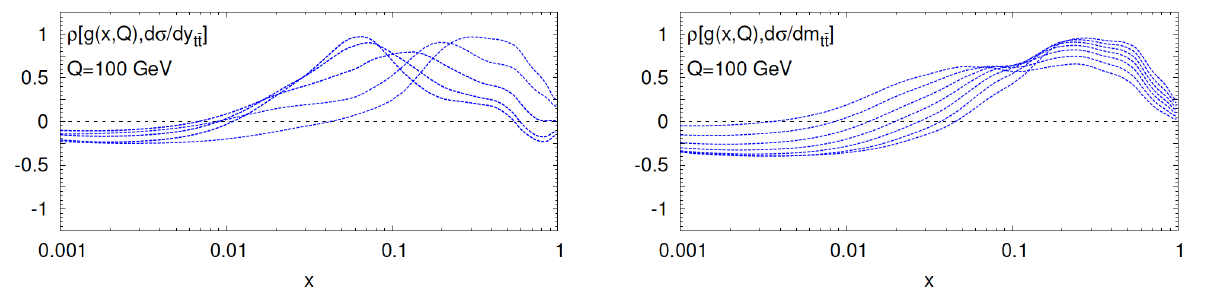}
  \caption{ The correlation coefficient between the gluon PDF at $Q=100$ GeV and the theory predictions for the absolute differential distributions in $y_{t\bar{t}}$ (left) and $m_{t\bar{t}}$ (right plot) at $\sqrt{s}=8$ TeV, as a function of $x$.   Each curve corresponds to a specific measurement bin. The higher the absolute value of the correlation coefficient, the bigger the sensitivity to the gluon for those specific values of $x$.~\cite{Gao:2017yyd}}
    \label{4:fig:topdiffcorr}
\end{center}
\end{figure}

We borrow Fig.~\ref{4:fig:topdiffcorr} from Ref.~\cite{Gao:2017yyd} to show the kinematical sensitivity of top quark pair production to the gluon. In the figure, the higher the absolute value of the correlation coefficient,  $\rho\[ g(x,Q), \dd \sigma\]$, the greater the sensitivity to the gluon for those specific values of $x$. The sensitivity to the gluon in this process is particularly strong for $x$ values up to approximately 0.6-0.7, which is beyond the range of other gluon-sensitive processes. Additionally, it should be noted that the use of differential distributions allows for a wider range of kinematic coverage beyond what is provided by the total inclusive cross sections.
%%%%%%%%%%%%%%%%%%%%%%%%%%%%%%%%%%%%%%%%%%%%%%%%%%%%%%%%%%%%%%%%%%%%%
%%%%%%%%%%%%%%%%%%%%%%%%%%%%%%%%%%%%%%%%%%%%%%%%%%%%%%%%%%%%%%%%%%%%%

\paragraph{\itshape\mdseries Heavy-flavour pair production.}
%Finally, the small-$x$ gluon can be further constrained by including in PDF fits charmed-meson production at hadron colliders, as the dominant channel for this process is $gg\rightarrow c\bar c$. LHCb data~\cite{Aaij:2013mga,Aaij:2015bpa,Aaij:2015bpa} provide information on the gluon PDF down to very small values of $x\sim10^{-6}$, a region which is also important for neutrino astrophysics~\cite{CooperSarkar:2011pa,Gauld:2015yia,Gauld:2015kvh,Garzelli:2016xmx,Gauld:2016kpd}.
\begin{figure}[t]
  \centering
  \includegraphics[width=0.75\textwidth]{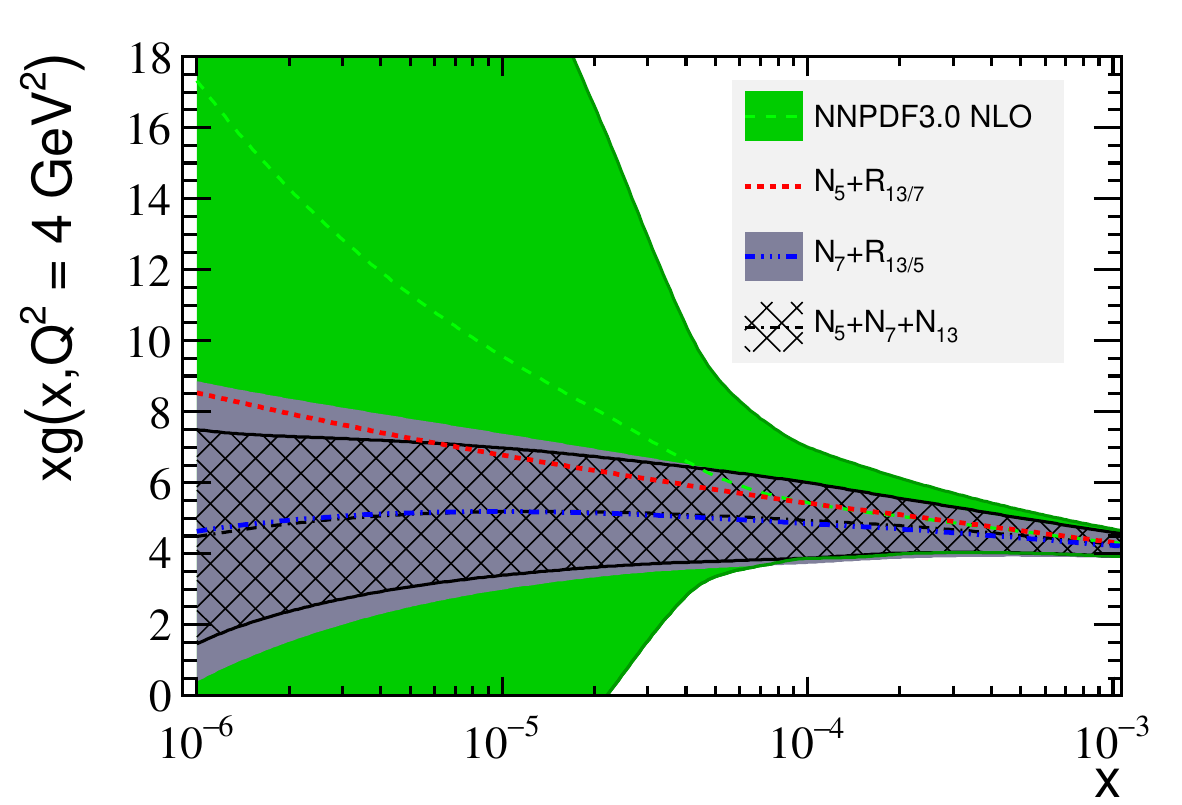}
  \caption{The NLO gluon in NNPDF3.0 and its uncertainty compared against the effect adding various combination of LHCb charm production data, at $Q^2=4$ GeV$^2$.Taken from Ref.~\cite{Gauld:2016kpd}}
  \label{4:fig:GluonOnCharm}
\end{figure}
The production of charmed mesons at hadron colliders is dominated by the $gg\to c\overline{c}$ subprocess, and therefore it provides a sensitive probe to the gluon PDF at small-$x$. In particular, the forward measurements from the LHCb experiment at 5, 7, and 13 TeV~\cite{Aaij:2013mga,Aaij:2016jht,Aaij:2015bpa} provide information on the gluon at values of $x$ as small as $x\simeq 10^{-6}$, well below the kinematic reach of the HERA structure function data, and thus in a region where PDF uncertainties are large due to the limited amount of experimental information available, Fig.~\ref{4:fig:GluonOnCharm}. Several studies of the impact of the LHCb charm measurements on the gluon PDF using different theoretical frameworks~\cite{Gauld:2016kpd,Cacciari:2015fta,Zenaiev:2015rfa,Zenaiev:2016kfl}, although to this day no PDF fit analysis includes small-$x$ resummed theory for hadron collider data, which is expected to be relevant in this kinematical regime~\cite{Bonvini:2022xio}

\paragraph{\itshape\mdseries Other processes.}

Beside these mainstays in global PDF analyses, contemporary fitting effort include several other processes that provide additional constraints on several combinations of PDF, which however we do not discuss further. A few example are:  prompt-photon production from the QCD Compton scattering $qg \rightarrow q \gamma$ as a constraint on the the large-$x$ gluon,
$W^\pm+c$ production to ascertain the strangeness and its asymmetry in the proton, Single-top production and central exclusive production of vector mesons as probes for the $b$ quark content and small-$x$ gluons respectively. Further discussion can be found in Refs.~\cite{Gao:2017yyd,NNPDF:2021njg}. 

\subsection{Theoretical accuracy and resummed fits}

Since parton densities are based on a combination of experimental data and perturbative QCD calculations, the accuracy of a PDF fit also depends on that of the reliability of theoretical predictions used in the analysis, that is the DGLAP splitting functions and the partonic cross-sections which are convoluted with the PDFs. 
PDFs are typically extracted using fixed-order perturbation theory.
For most of the processes we have reviewed in section~\ref{4:ssec:expdata} the coefficient functions have been computed to NNLO accuracy, and in some cases also to N$^3$LO~\cite{Gao:2017yyd}.
An inexhaustive list for NNLO includes, $W$ and $Z$ rapidity distributions in Drell-Yan ~\cite{Anastasiou:2003ds,Anastasiou:2003yy,Melnikov:2006di,Melnikov:2006kv}, single-inclusive jet and inclusive dijet production~\cite{Currie:2016bfm,Currie:2017eqf}.
$Z$ $p_T$ distributions have recently been computed at N3LO~\cite{Chen:2022cgv}.
Moreover, DIS is known up to N$^3$LO accuracy in the massless limit~\cite{Moch:2004xu,Vermaseren:2005qc} and to NNLO including mass effects~\cite{Laenen:1992zk,Laenen:1992xs}. The corresponding coefficient functions are available through a number of public codes like \texttt{APFEL}~\cite{Bertone:2013vaa} and \texttt{QCDNUM}~\cite{Botje:2010ay}. 

The state of the art theory input in global PDFs fits combines NNLO fixed-order prediction combined with NNLO accurate DGLAP evolution~\cite{Cridge:2021qfd,Hou:2019efy,NNPDF:2017mvq,NNPDF:2021njg,NNPDF:2021uiq,PDF4LHCWorkingGroup:2022cjn,Alekhin:2017kpj,ATLAS:2021vod}, and very recently the first analysis boasting consistent approximate N$3$LO theory appeared~\cite{McGowan:2022nag}. Beside the cutting edge, NLO and optimised LO PDFs are widely used in automated NLO Monte Carlo codes to produce NLO-accurate simulations for many different processes.

In most cases of processes analysed using PDF, fixed-order theory provides an accurate description of the data. However, there are instances, which were discussed in Chapters~\ref{ch2} and \ref{ch3}, where logarithmic-enhanced terms are present at all orders and disrupt the convergence of the perturbative series. In such cases, it is necessary to incorporate resummation of the enhanced terms to all orders in addition to the fixed-order description.
One such case is the application of small-$x$ resummation in the determination of PDFs, which affects the small-$x$ behaviour of both QCD coefficient functions and DGLAP splittings, and therefore influencing the extracted PDFs.
\begin{figure}[t]
  \centering
  \includegraphics[width=0.6\textwidth,page=2]{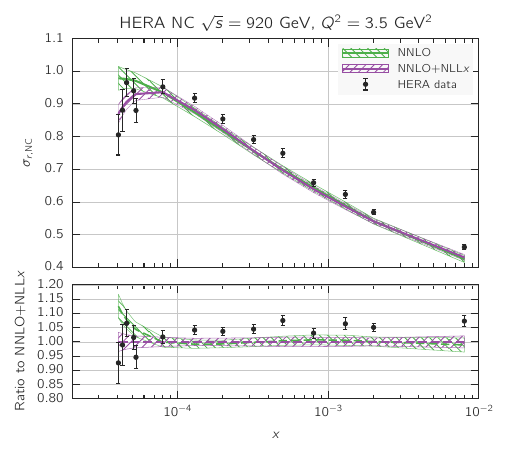}
  \caption{The $Q^2=3.5$~GeV$^2$ bin of the HERA neutral current data, together with the theoretical predictions  at fixed order (green) and with resummation (purple)~\cite{Ball:2017otu}.}
  \label{4:fig:HERA}
\end{figure}
A tension due to these logarithms was present in the small-$x$ region of DIS data from the HERA collider (which represents the main dataset used to determine PDFs).
The issue found a solution in a series of independent PDF fits including small-$x$ resummation~\cite{Ball:2017otu,Abdolmaleki:2018jln,Bonvini:2019wxf}. These results provide a confirmation of the importance of the resummation in the description of the small-$x$ dynamics, and led to important differences in the extracted PDFs.

As an example, figure~\ref{4:fig:HERA} shows one of the HERA data subset at low $Q^2$, which extends down to small $x$. Crucially, we see that the data present a characteristic turnover at $x\sim10^{-4}$, which is incompatible with a fixed-order theoretical prediction. Conversely, the resummed theory manages to reproduces the features of the data and achieve a significant reduction of the $\chi^2$ of the PDF fit.

These resummation effects have a complex interplay with other aspects of the PDF determination machinery.
Indeed, many of  the constraints on the PDFs at low $x$ come exclusively from the HERA dataset.
These are localised at low-$x$ and low $Q^2$, see figure~\ref{4:fig:kinplot4}, and are potentially sensitive to missing higher (logarithmic) orders, and perhaps also to non-perturbative corrections.
The best, and perhaps only, way to disentangle the effects of resummation from the other two in PDFs fits is to include  additional low-$x$ data at higher $Q^2$ in the fit, leveraging the low-mass production processes (i.e. Drell-Yan and heavy quark pair production) at LHC, which does cover the interesting region. While the inclusion of these data is already done extensively in most PDF analyses, the corresponding small-$x$ resummation theory is still only partially developed~\cite{Silvetti:2022hyc} and this remains the main hurdle for a PDF analysis of small-$x$ resummed PDFs beyond the HERA dataset.

\subsection{Fit quality and minimisation techniques}

The figure of merit used to measure the fit quality is typically the log-likelihood function $\chi^2$
\begin{equation}
  \label{4:eq:chi2form1}
  \chi^2 (\lbrace p_i\rbrace ) = \sum_{i,j}^{N} (D_i - T_i(\lbrace p_i \rbrace)) ({\rm cov}^{-1} )_{ij} (D_j - T_j(\lbrace p_i\rbrace )),
\end{equation}
where $D_i$ are the data points, $T_i$ are the theoretical predictions expressed in terms of the PDF parameters $\lbrace p_i \rbrace $, and the experimental covariance matrix ${\rm cov}$ and its inverse ${\rm cov}^{-1}$ are defined by 
\begin{align}
	({\rm cov})_{ij} &= (\sigma^{\rm uncorr}_{i})^2 \delta_{ij} + \sum_{\kappa=1}^{N_{\rm corr}} \sigma^{\rm corr}_{\kappa, i } \sigma^{\rm corr}_{\kappa, j },\\
	({\rm cov}^{-1})_{ij} &= \frac{\delta_{ij}}{(\sigma^{\rm uncorr}_{i})^2} - \sum_{\kappa,\lambda=1}^{N_{\rm corr}}  \frac{\sigma^{\rm corr}_{\kappa, i }}{(\sigma^{\rm uncorr}_{i})^2} A^{-1}_{\kappa \lambda}  \frac{\sigma^{\rm corr}_{\lambda, j }}{(\sigma^{\rm uncorr}_{j})^2}, & A_{\kappa \lambda} = \delta_{\kappa \lambda} + \sum_{i}^{N_{\rm dat }} \frac{\sigma^{\rm corr}_{\kappa, i } \sigma^{\rm corr}_{ \lambda,i } }{\sigma^{\rm uncorr}_{i}}.
\end{align}
Here $i=1, \ldots N  $ indicates the individual data points, affected by uncorrelated uncertainty $\sigma^{\rm uncorr}_{i}$ and $\kappa=1, \ldots {N_{\rm corr}}$ sources of correlated uncertainty $\sigma^{\rm corr}_{\kappa, i } $.

The $\chi^2$ Eq.~\eqref{4:eq:chi2form1} can be recast introducing nuisance parameters $r_\kappa$
\begin{subequations}
  \begin{align}
    & \chi^2 (\lbrace p_i\rbrace, \lbrace r_\kappa \rbrace) = \sum_{i=1}^{N}\( \frac{\hat D_i - T_i (\lbrace p_i\rbrace )}{(\sigma^{\rm uncorr}_{i})^2}\right)^2 + \sum_{\kappa=1}^{N_{\rm corr}} r_k^2 \label{4:eq:chi2form2} \, ,\\
    & \hat D_i  = D_i -  \sum_{\kappa=1}^{N_{\rm corr}} r_\kappa \sigma^{\rm corr}_{\kappa, i } \label{4:eq:sysshift} .
  \end{align}
\end{subequations}
Minimising eq~\eqref{4:eq:chi2form2} by assuming purely Gaussian errors with respect to the nuisance parameters one finds
\begin{align}
r_\kappa\bigg|_{\text{min}} \equiv \tilde r_\kappa = \sum_{i=1}^{N}\frac{D_i - T_i}{(\sigma^{\rm uncorr}_{i})^2} \sum_\lambda^{N_{\rm corr}} A_{\kappa \lambda}^{-1} \sigma^{\rm corr}_{\kappa, i }.
\end{align}

Although mathematically equivalent to equation~\eqref{4:eq:chi2form2} when $\{r_\kappa \}=\{\tilde r_\kappa \}$, equation~\eqref{4:eq:chi2form2} can be useful for examining the behaviour of the shifts at the minimum to ensure that they follow a normal distribution. Additionally, by shifting the data points according to equation~\eqref{4:eq:sysshift} and comparing them to theoretical predictions, one can better visualise the impact of correlated uncertainties~\cite{Pumplin:2002vw}. However, when dealing with multiplicative uncertainties, downward fluctuating data points get a smaller normalisation uncertainty~\cite{DAgostini:1993arp}.
Consequently, using  equation~\eqref{4:eq:chi2form1} directly tends to give biased results. Various solutions to this problem have been proposed in the literature, for example by redefining the covariance matrix~\cite{Ball:2009qv}. 

Once an appropriate figure of merit has been established, the next step is to locate the global minimum in the parameter space defined by the PDF parameters $\{ p_i\}$. Different groups use different methods to find the minimum depending on the number of free parameters involved. If the number of parameters is manageable, gradient-based methods are typically employed, one example being the variable metric one provided by the \texttt{MINUIT} package~\cite{James:1975dr}.

The neural-network based fits of the NNPDF collaboration, and especially its latest iteration NNPDF4.0~\cite{NNPDF:2021njg}, use a modular architecture to set the features of the fit and optimise its performance.
Foremost, fast fitting performance is achieved thanks to the use of customised stochastic gradient descent methods provided by the {\tt TensorFlow} library~\cite{tensorflow2015:whitepaper,Carrazza:2019mzf,Carrazza:2019agm,Cruz-Martinez:2020tte}. Moreover, all aspects of the neural network PDF parametrisation and optimisation (such as neural net architecture, learning rates or minimisation algorithm) are selected through a separate hyperparameter optimisation procedure~\cite{Carrazza:2019mzf}, to select the optimal methodology in the space of available models.

\subsection{Error propagation}
Reliable tools to estimate the uncertainty related to PDF is a key requirement of hadronic precision physics. We will now touch on two of the most common methods to evaluate PDF error: the Hessian method and the Monte Carlo method. Further discussion and additional techniques are available in the relevant literature~\cite{DeRoeck:2011na,Hartland:2014nha,Gao:2017yyd}.

\paragraph{\itshape\mdseries Hessian method.}
In the Hessian method the uncertainty for the fitted PDFs is obtained by  studying the perturbations of the $\chi^2$ around its minimum under variations of the fitting parameters $\lbrace p_i \rbrace$. The uncertainty on the physical observables is then determined geometrically by considering the values computed using the perturbed parameters.
Explicitly, consider $\tilde \chi^2$, the best-fit value for the figure-of-merit, the variations of the $\chi^2$ can be approximated as quadratic
\begin{equation}
  \Delta \chi^2 \equiv \chi^2 - \tilde \chi^2 = \sum_{i,j}^{N_{\rm par}} (p_i - \tilde{p}_i ) H_{ij} (p_j - \tilde{p}_j )\, .
\end{equation}
The uncertainty on an observable $\mathcal F$ that depends on PDFs can be calculated by linearly propagating the errors. This is done using the formula
\begin{subequations}
\begin{align}
  H_{ij} & \equiv \frac{1}{2} \frac{\partial^2 \chi^2}{\partial p_i \partial p_j } \Bigg|_{\{  p_i\} = \{ \tilde p_i\}}, \label{4:eq:hessianlinear} \\
  \sigma^2_{\mathcal F} & = T \left( \sum_{i,j}^{N_{\rm par}} \frac{\partial \mathcal F}{\partial p_i} (H)^{-1}_{ij}  \frac{\partial \mathcal F}{\partial p_i} \right) \, ,
  \end{align}        
\end{subequations}
where $T=\sqrt{\Delta \chi^2}$ is the tolerance factor used to match the range of variation of fit parameters to the confidence interval associated with the PDF uncertainties.

In practice evaluating Eq.~\eqref{4:eq:hessianlinear} presents some difficulties since the partial derivatives of the observables with respect to the fit parameters are generally unknown and a numerical evaluation can prove unstable depending on the underlying features of the $\chi^2$~\cite{DeRoeck:2011na}. Usually, this difficulty is circumvented simply by diagonalising the Hessian matrix~\cite{Pumplin:2000vx,Pumplin:2001ct}, so the error on an observable $\mathcal F$ is given by
\begin{equation}
	\sigma_{\mathcal F}^2 = \frac{1}{2} \left( \sum_{i}^{N_{\rm par}} \left[ \mathcal F(S^+_i) - \mathcal F(S^-_i)  \right] \right)^{2},
\end{equation}
where $S^\pm$ are PDF set constructed along the eigenvector directions, displaced by the desired $\Delta \chi^2=T^2$.

\paragraph{\itshape\mdseries The Monte Carlo method.} 

The Monte Carlo (MC) method is a complementary method for determination of the PDF uncertainty, which is based on a MC procedure in the space of the experimental data. The method is designed to construct a faithful representation of the uncertainties present in the initial data without any assumption on their nature.

The first step in the MC method is the construction of an ensemble of $N_{\rm rep}$ of artificial data replicas (dubbed {\it pseudodata} replicas) for every data point included in the fit, generated according to the probability distribution of the initial data. For a given experimental measurement of a generic observable $\mathcal F^{\rm exp}$, characterised by a total uncorrelated uncertainty $\sigma^{\rm uncorr}$ and $\kappa = 1, \ldots N_{\rm corr}$ correlated uncertainties $\sigma^{\rm corr}_\kappa$, the artificial MC replicas $\mathcal F^{{\rm art},k}$ are constructed as~\cite{Forte:2002fg}.
Each of these set is then fitted on its own, building $N_{\rm rep}$ equally probable PDF sets which reliably describe the probability density of PDF based on the original experimental errors. Then, the best estimate of the observable $\mathcal F$ are obtained as the average and the variance over the replica ensemble
\begin{align}
	\langle \mathcal F \rangle &= \frac{1}{N_{\rm rep}} \sum_{\ell=1}^{N_{\rm rep}} \mathcal F^\ell\, ,\\ 
	\sigma^2_{\mathcal F} &= \frac{1}{N_{\rm rep}-1} \sum_{\ell=1}^{N_{\rm rep}} (\mathcal F^\ell- \langle \mathcal F^{\ell} \rangle)^2\, ,
\end{align}
where $\mathcal F^{\ell}$ denotes the theoretical predictions of the observable $\mathcal F$ evaluated with the PDF replica $\ell$.

In case of fully-consistent dataset, both the MC method and the Hessian method with $\Delta \chi^2=1$ are known to produce compatible results~\cite{Dittmar:2009ii} and procedures to convert an Hessian set in a MC representation and vice versa are known in literature~\cite{Watt:2012tq,Carrazza:2015aoa}

\noindent \paragraph{\itshape\mdseries}
It is important to note that the term ``PDF uncertainty'' usually encompasses the experimental uncertainty associated with the data used to extract the PDF, as well as other potential methodological uncertainties. This leaves out theory uncertainty, for example in the form of missing higher-orders in fixed-order perturbation theory. In the past, the theory uncertainty was considered negligible compared to the experimental and methodological uncertainties. However, due to the abundance of highly precise data and the constant methodological improvements, the theory uncertainty has become comparable to the PDF uncertainty in a significant range of $x$ and $Q^2$, and it will eventually become the most significant source of uncertainty.

In recent years, the consideration of these so-called Missing Higher Order Uncertainties (MHOUs), and how to estimate them, has received great attention~\cite{Bendavid:2018nar,NNPDF:2019ubu,Harland-Lang:2018bxd,Gao:2017yyd,Ball:2021icz}. One method to solve this issue is using scale variations. This is well motivated by the renormalisation group invariance of physical observables and was already successfully implemented in an NLO PDF fit~\cite{NNPDF:2019ubu}. However, it has been argued~\cite{Harland-Lang:2018bxd,Bonvini:2020xeo,Ball:2021icz} that difficulties arise is the arbitrary nature in the scale variation, as well as the choice of central scale as well as the effect of various classes of logarithms (e.g.  small-$x$, mass threshold and leading large-$x$ contributions) present at higher orders.
An alternative method to the above is to parametrise the missing higher orders with a set of nuisance parameters, using the available (albeit incomplete) current knowledge~\cite{Tackmann:review,McGowan:2022nag}.

\subsection{PDF Parametrisation}
On top of the lack of a closed form for PDFs from QCD first principles, it is well known that the determination of a set of arbitrary functions from a finite-size ensemble of sampling points does not admit a unique solution. 
Therefore, at the start of the PDF fitting process, it is necessary to adopt an {\it ansatz} at the initial scale $\mu_0 \sim 1-2$ GeV to obtain a concrete solution. Although, every sufficiently smooth function would suffice and provide \textit{equivalent} fitting results, in practice any specific choice may introduce an artificial bias.
In general we can start from the expression
\begin{equation}
  \label{4:eq:PDFpar}
  x f_i(x,\mu_0^2) = x^{\alpha_i} (1-x)^{\beta_i}  F_i (x, \lbrace p_i \rbrace),
\end{equation}
where $F_i$ is a smooth function and the parameters $\alpha_i, \,  \beta_i$ and $\{p_i \}$ are determined by the fit. 
This choice is motivated by the theoretical, non-perturbative expectation that PDFs should behave as a power law at asymptotic values of $x$ (see e.g.~\cite{Ball:2016spl}).
In particular, the power behaviour at $x\rightarrow 0$ is predicted by Regge theory~\cite{Regge:1959mz}, whereas the behaviour at $x\rightarrow 0$ is constrained by quark counting rules~\cite{Brodsky:1973kr,Devenish:2004pb,Roberts:1990ww,Devenish:2004pb}.

The parametrisation Eq.~\eqref{4:eq:PDFpar} is adopted for all the PDFs entering the fit, these usually include the gluon, $u,d,s$ quarks and their anti-quark counterparts, thus using seven independent PDFs. The heavier flavours are generated perturbatively by DGLAP evolution. Recently, fits including an intrinsic charm component, under the assumption that $c(x,Q_0^2)=\bar c(x,Q_0^2)$, have become available in literature~\cite{Ball:2022qks}.

The form of $ F_i \(x,\{p_i\}\)$ must be determined from the fit, but the specific choice for the interpolation function $ F_i$ is not uniquely defined and several alternatives are common. The historically simplest one is to assume a polynomial or an exponential polynomial of $x$ or $\sqrt{x}$~\cite{H1:2015ubc}. 
%More complex choices involve the use  Chebyshev and Bernestein polynomials~\cite{Bailey:2020ooq,Hou:2019efy}\FS{is this actually true?}.
Adding logarithms to the palette of elementary functions in $ F_i$ has proven to be valuable to improve the fitting performance in the small-$x$ region~\cite{Bonvini:2019wxf}.

Finally, the NNPDF collaboration has adopted a substantially different approach by replacing $F_i$ with a collection of neural networks. This allows for great flexibility and lack of bias due to their redundancy~\cite{DelDebbio:2007ee,Bertone:2017bme,NNPDF:2021njg}. Recently~\cite{Carrazza:2021yrg}, the same framework has been used to remove the prefactor scaling appearing in Eq.~\eqref{4:eq:PDFpar}, arguing for better fitting performance and further reduction of parametrisation bias across the entire $x$ range.  

In conclusion, the number of free parameters to be determined in the fitting has important consequences on the minimisation strategies used by each collaboration. Moreover, large parameters sets are prone to overfitting issues or, in general, to a higher computational cost in the figure-of-merit minimisation.

\section{A quantum-statistics inspired model}\label{4:sec:QSPDF}
In this section, we will present a PDF set based on an alternative parametrisation built around the principle of minimising the number of parametres and using physical arguments to model $F_i$ at the initial scale. In a sense, this can be thought as \textit{maximum-bias} parametrisation.
\subsection{\textit{Ab initio} models of proton structure}
Generally speaking a complete calculation of parton distributions distributions emerging from a bound-state system in a relativistic field theory is a complicated~\cite{Drell:1971vx} and largely unsolved problem in QCD. Instead, we reinstate the naive parton model following the premises of Refs.~\cite{Cleymans:1986gy,Mac:1989ki,Bhalerao:1996fc,Sohaily:2017etw}. That is, considering DIS in the infinite momentum frame of reference, the target nucleon has a simple structure and the motion of its constituents is sufficiently slowed down by the time dilatation effect that the incident lepton scatters instantaneously and incoherently off the partons.
As far as the hard scattering interaction is concerned, the partons can be considered to be free particles in an ideal quantum gas characterised by an effective ``temperature'' $T$ and volume $V$.
Thus the main idea of the model is to build the parametrisation of the PDFs out of Bose-Einstein and Fermi-Dirac Distributions, with the addition of ad hoc scaling factors to ensure the proton sum rules keep holding.

The number of parton in this system will be given by
\begin{equation}
  N_f = \int f\(E\) \dd^3k \, , \label{4:eq:PartonNum}
\end{equation}
with $E$ being the energy of a given parton and $f$ the corresponding distribution
\begin{equation}
  f\(E\) = \frac{g_f V}{\(2\pi\)^3}\[\eu^{\frac{ \(E-\mu_f\)}{T}}\pm 1\]^{-1} \, , \label{4:eq:BEFDdistr}
\end{equation}
where the degeneracy coefficient $g_f$ is $6$ for quarks and $16$ for gluons. Similarly, quarks are assigned Fermi statistics with the $+$ sign and gluon Bose statistics with $-$. We rewrite Eq.~\eqref{4:eq:PartonNum} as
\begin{align}
  N_f & =  \int f\(E\) \delta \(E - \sqrt{\kt{}^2 + (k^3)^2 + m_f^2}\) \dd E \dd^2 \kt{} \dd k^{(3)} \, , \nn
  & = \int \tilde{f}\(x,\kt{}^2\)\dd x \dd^2 \kt{} \, ,
\end{align}
where $x=\frac{M}{2\spd{k}{n}}$ is ratio between $M$ the nucleon mass and the momentum fraction of a given parton along the +-branch of the light-cone.
In general, the distributions will still depend on the parton mass $m_f$ and transverse momentum $\kt{}$~\cite{Bourrely:2001du,Buccella:2014wpa}, but for the purpose of this discussion we set these quantities to zero. Thus in this  ``longitudinal approximation'' we redefine Eq.~\eqref{4:eq:BEFDdistr} as 
\begin{subequations}
  \begin{align}\label{4:eq:StatAnsatz1}
    f_q\(x\) & = \frac{6 M V}{2 \(2\pi\)^3} \[\eu^{\frac{x-\mu'_q}{\xb}} + 1\]^{-1}\vartheta \(x\) \, , \\
    f_g\(x\) & = \frac{16 M V}{2 \(2\pi\)^3} \[\eu^{\frac{x-\mu'_g}{\xb}} - 1\]^{-1}\vartheta \(x\) \, ,
  \end{align}
\end{subequations}
where $\mu'_f,\,\xb$ are the chemical potential and temperature parametres, rescaled by the mass of the nucleon. From these expressions one could proceed in statistical determination of an ``equilibrium'' configuration of the nucleon by setting
\begin{equation}\label{4:eq:Eq-potentials}
  \mu'_q = -\mu'_{\bar{q}},\quad \mu_g = 0
\end{equation}
and enforcing the suitable sum rules. In the case of the proton, these read 
\begin{subequations}\label{4:eq:SumRules}
  \begin{align}
    & \int_0^1 \dd x \[f_u(x)-f_{\ub} (x)\] = 2 \, ,\\
    & \int_0^1 \dd x \[f_d(x)-f_{\db} (x)\] = 1 \, ,\\
    & \int_0^1 \dd x \sum_{i} x f_i (x) = 1 \, .   
  \end{align}
\end{subequations}
Finally, the four statistical parameters $V,\xb,\mu'_u,\mu'_d$ can be determined by defining an appropriate entropy function for the model and maximising it~\cite{Mac:1989ki,Sohaily:2017etw}.

On the other hand, this rough modelling of the proton as an equilibrium system is clearly lacking as it does not account for QCD interactions and hard-wires the proton size to a fixed-volume, neglecting many finite-size corrections~\cite{Bhalerao:1996fc}.
To address this, we follow the deformation introduced in~\cite{Bourrely:1993wq} and write
\begin{equation}\label{4:eq:StatAnsatz2}
  f\(x\)  = \frac{A x^{b-1}}{\eu^{\frac{x-\mu'_f}{\xb}} \pm 1} \, ,
\end{equation}
where we replaced the overall factor with a flavour insensitive normalisation and introduced a power suppression $\(b>0\)$, to ensure that the valence distribution $f_q-f_{\bar{q}}$ is vanishing in the $x\rightarrow0$ limit, as motivated by Refs.~\cite{Buccella:2014wpa}. If instead of describing the unpolarised distributions we are interested in separating the distribution by helicity, we need to introduce more distributions and chemical potentials
\begin{subequations}
    \begin{align}
      & h \(x; b, \xb, X\) = \frac{x^b}{\exp\(\frac{x-X}{\xb}\)+1} \, , \nonumber \\
      & x f_{q^\uda} \(x\) = A X_q^\uda h\(x; b, \xb, X_q^\uda\)  \, , \label{4:eq:quark-distr} \\ 
      & x f_{\qb^\uda} \(x\) = \Ab {\frac{1}{X_q^\dua}} h\(x ;\bb, \xb,  -X_q^\dua\) \, , \label{4:eq:antiquark-distr} \\
      & \qquad \qquad \text{with} \:\: q \in \lbrace u,\, d \rbrace\, ,
    \end{align}
\end{subequations}
The $X_q^\uda$ are the chemical potentials for each combination of flavour $(q)$ and helicity $\(\uda\)$, the equilibrium condition Eq.~\eqref{4:eq:Eq-potentials} apply as $X_{\bar{q}}^\uparr = -X_q^\dwarr $.
Similarly the normalisation factors are modified in \eqref{4:eq:quark-distr} and \eqref{4:eq:antiquark-distr} to differentiate opposite helicities, and to reproduce the ordering in the polarised distributions
\begin{equation}
   u^{\uparr} > d^{\dwarr} \sim u^{\dwarr} > d^{\uparr}\,, 
\end{equation}
 implied at equilibrium by the valence and momentum sum rules~\eqref{4:eq:SumRules} and the defect in the Gottfried one~\cite{Bourrely:2001du,Buccella:2022tmb}.
Finally, the small-$x$ region is characterised by a rise of the gluon and sea-quark distributions governed by Regge theory~\cite{Forshaw:1997dc} which is absent by construction in the expressions modelled thus far. To amend this issue, the authors of Ref.~\cite{Bourrely:2001du}, introduce an additional universal term in the quark and antiquark distribution.
Explicitly
\begin{subequations}\label{4:eq:highen}
  \begin{align}
  & x f_{\(q,\bar{q}\)^\uda} \(x\) \rightarrow  x f_{\(q,\bar{q}\)^\uda} \(x\) + \frac{\tilde{A}x^{\btil}}{\eu^{x/\xb}+1}  \, ,\\
  & x f_g\(x\) = \frac{A_g x^{b_g}}{\exp\(x/\xb\)-1} .
  \end{align}
\end{subequations}
This addition dominates the small-$x$ behaviour where the valence term vanishes. Moreover, at high-energy the quark sea and the gluon distribution must grow together due to the mixing of gluon and singlet sector from DGLAP evolution~\ref{app:2:DGLAP}. Thus if we expand for small $x$
\begin{subequations}
  \begin{align}
  & x f_{\(q,\bar{q}\)^\uda} \(x\) \simeq  \frac{\Atil}{2}x^{\btil} +\calO \(x^{\btil+1}\)  \, ,\\
  & x f_g\(x\) \simeq  \frac{A_g \xbar }{2}x^{b_g-1} +\calO \(x^{b_g}\),
  \end{align}
\end{subequations}
we can constrain $b_g=\btil+1$.
All in all, the modifications we introduced in the distribution increase the number of free parametres from the original $4$ to $12$. From here one could follow the same approach of Ref.~\cite{Bhalerao:1996fc,Sohaily:2017etw} and try to constrain all of them with sum rules and state functions. Alternatively, one can use the same expression as an initial scale parametrisation for a PDF analysis, like it was attempted in Refs.~\cite{Bourrely:2001du,Bourrely:2005kw,Buccella:2019yij,Bellantuono:2022hqp}. We will showcase our own attempt of performing this fit against the HERA dataset using the \xfitter\ PDF infrastructure~\cite{Alekhin:2014irh,xfitter}.
\subsection{Our \QSPDF\ setup}
We will now discuss an application of this model as a parametrisation for PDF, which we will call \QSPDF\ along the lines of Ref.~\cite{Buccella:2019yij}. First  we start by summarising the expression from the model together as
\begin{subequations}
  \begin{align}
    & h \(x; b, \xb, X\) = \frac{x^b}{\exp\(\frac{x-X}{\xb}\)+1} \, , \nonumber \\
    & x f_{q^\uda} \(x,Q_0^2\) = A X_q^\uda h\(x; b, \xb, X_q^\uda\) + \tilde{A} h\(x;\tilde{b}, \xb, 0\)  \, , \\
    & x f_{\qb^\uda} \(x,Q_0^2\) = \Ab {\frac{1}{X_q^\dua}} h\(x ;\bb, \xb,  -X_q^\dua\) + \tilde{A} h\(x;\tilde{b},\xb,0\) \, ,\\
    & \qquad \qquad \text{with} \:\: q \in \lbrace u,\, d \rbrace\, , \nonumber\\
    & x f_g\(x, Q_0^2\) = \frac{A_g x^{1+\btil}}{\exp\(x/\xb\)-1} .
    \label{4:eq:params}
  \end{align}
\end{subequations}
where $Q_0$ is the initial parametrisation scale. Since we are interested only in fitting unpolarised DIS data from HERA, we combine the spin-dependent distributions
\begin{equation}
  f_{q}\(x,Q_0^2\) = f_{q^\uparr}\(x,Q_0^2\) + f_{q^\dwarr}\(x,Q_0^2\) \, ,
\end{equation}
and then rewrite the unpolarised valence and sea contributions ($q \in \left\lbrace u,d\right\rbrace$)
\begin{subequations}\label{4:eq:unpdistr}
    \begin{align}
      x f_{q_{v}} \(x, Q_0^2\) & = f_q\(x, Q_0^2\) - f_{\qb} \(x, Q_0^2\) \nonumber \\
      & =  A\[X_q^\uparr h\(x; b, \xb, X_q^\uparr\) + X_q^{\dwarr} h \(x; b,\xb, X_q^{\dwarr}\) \]  \nonumber \\
      & - \Ab \[ \frac{1}{X_q^\dwarr} h \(x;\bb, \xb, -X_q^{\dwarr}\) + \frac{1}{X_q^\uparr} h  \(x; \bb,\xb, -X_q^\uparr\) \] \, \\
      x f_{\qb} \(x, Q_0^2\) & = \Ab \[\frac{1}{X_q^\dwarr} h \(x;\bb,\xb, -X_q^{\dwarr}\) + \frac{1}{X_q^\uparr} h \(x; \bb,\xb, -X_q^\uparr\) \] \nonumber \\
      & + 2 \tilde{A} h \(x;\btil, \xb, 0\) \, , \\
      x f_g\(x,Q_0^2\) & = \frac{A_g x^{1+\btil}}{\exp\(x/\xb\)-1} \, .
    \end{align}
\end{subequations}
Given the limited power of the HERA dataset alone in resolving different parton flavours, we model the strange quark distribution as a fraction of the down one, 
\begin{equation}
  f_s \(x, Q_0^2\) = f_{\sbar} \(x; b, \xb, X,Q_0^2\) = \frac{\rmf_s}{1-\rmf_s} f_{\db} \(x; b, \xb, X,Q_0^2\)\, , 
\end{equation}
 with $\rmf_s=0.4$, which is the same choice of the \HERAPDF\ parametrisation.

 In total, there are $12$ parametres: $\left\lbrace \xb,A_g,A,\Ab,\Atil,X_u^\uda,X_d^\uda,b,\bb,\btil \right\rbrace$. However, we choose to constrain $\bb =b$. This has no intrinsic physical motivation but is consistent with previous tests of the parametrisation~\cite{Bourrely:2001du}.
\footnote{More precisely the older literature used $\bb=2b$, however after repeated testing we concluded that the additional factor had little to no effect in the results of the fit. While it would be interesting to release this constraint and allow more diversity between the quark and antiquark distributions, we found that different values of $\bb$ do not alter the PDF shapes significantly.}
Valence and momentum sum rules from Eqs.~\eqref{4:eq:SumRules} allow to fix the normalisation parameters $\left\lbrace A_g,A,\Ab \right\rbrace $. Unlike other polynomial parametrisations, $A,\Ab$ are not multiplicative factors in the valence distributions. So, to fit we have to solve the system
\begin{align}
& \begin{pmatrix} 2 \\ 1 \end{pmatrix}
  = \begin{pmatrix}
    K_u & \Kb_u\\
    K_d & \Kb_d
  \end{pmatrix}
  \begin{pmatrix} A \\ \Ab \end{pmatrix}
  %= \mathcal{K} \begin{pmatrix} A\\ \Ab \end{pmatrix}
  \; ,  \\
& K_q = \int_0^1 \dd x \[C^\uparr_q f\(x;b-1, X^\uparr_q \) + C^\dwarr_q f\(x;b-1, X^\dwarr_q \)\] \; ,\\ 
& \Kb_q = -\int_0^1 \dd x \[\Cb^\uparr_q f\(x;\bb-1, -X^\dwarr_q \) + \Cb^\dwarr_q f\(x;\bb-1, X^\uparr_q \)\] \; ,
\end{align}
where the integrals $K_q,\, \Kb_q$ can be evaluated numerically at every step of the fitting process. This leaves only $8$ free parameters to fit: $\left\lbrace \xb,\Atil,X_u^\uda,X_d^\uda,b,\btil \right\rbrace $.
\begin{figure}
  \centering
  \includegraphics[width=0.7\linewidth]{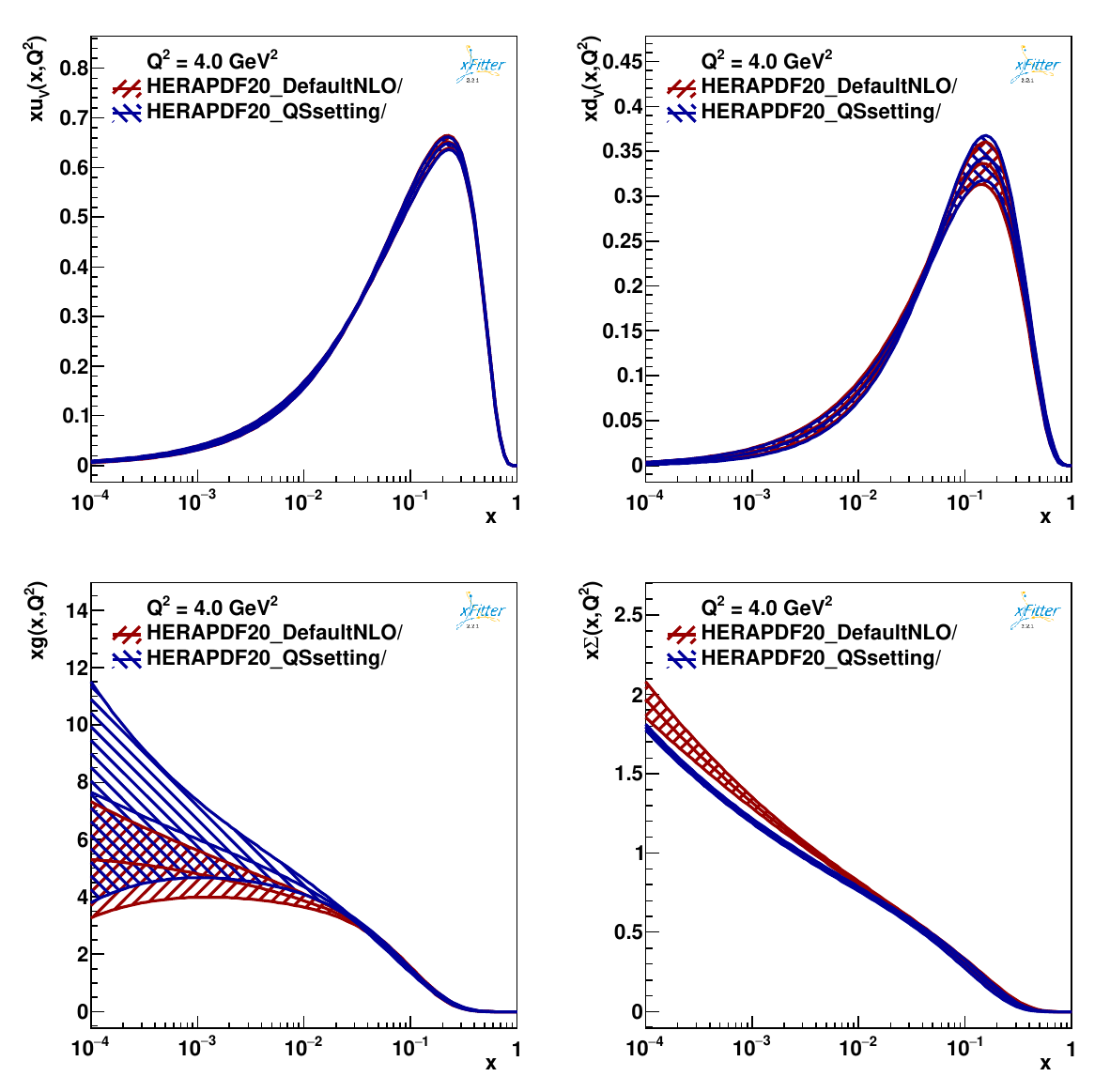}
  \caption{Benchmark fit between default \HERAPDF\ NLO configuration red in \xfitter\ and our settings blue.}
  \label{4:fig:BenchFitPlot}
\end{figure}
\begin{table}
  \centering
  \caption{$\chi^2$ breakdown of the default \HERAPDF\ NLO fit (left column) in \xfitter\ and our settings (right column). We emphasise the improved description of NCep 920 with the latter.}
  \includegraphics[width=0.8\linewidth]{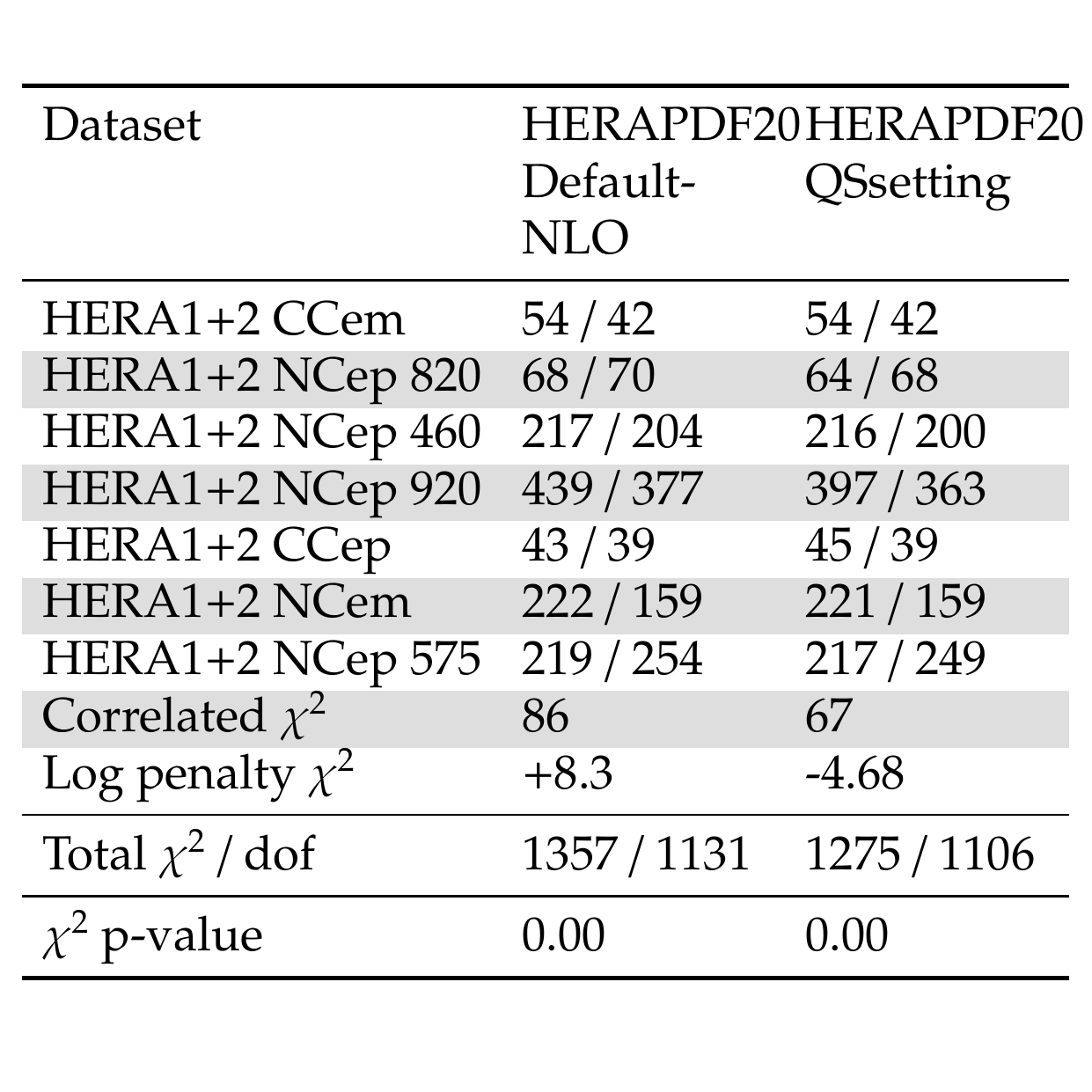}
  \label{4:tab:BenchFitTab}
\end{table}

Given that the model behind the formulation of the parametrisation does not account for QCD corrections, we perform our PDF analysis using NLO theory from \texttt{APFEL}~\cite{Bertone:2013vaa}. This choice is suitable for a proof-of-concept test and will be relaxed to study in greater detail the parametrisation in future work. We set the initial scale $Q_0^2=4 \; \text{GeV}$, as suggested in Ref.~\cite{Bourrely:2001du}. Consequently, the heavy flavours are accounted for using the FONLL-B scheme and with the charm mass raised to $m_{\text{ch}} = 1.46$ GeV and threshold $ t_{\text{ch}} = 2.0148 $ GeV. Moreover, this choice of $Q_0^2$ implies cutting out of the fitting data the $Q^2=3.5$ GeV bin, to avoid backward scale evolution. To assess the impact of this configuration, we perform a benchmark fit of the \HERAPDF\ parametrisation and compare it against the output of the NLO analysis in Ref.~\cite{H1:2015ubc}. The resulting valence, singlet and gluon distributions are plotted in Fig.~\ref{4:fig:BenchFitPlot}, and are mostly in agreement.
The uncertainty of the gluon PDF at small-$x$ the is enlarged with respect to \HERAPDF\ due to the use of the Hessian representation for the uncertainty bands. Table~\ref{4:tab:BenchFitTab} contains the $\chi^2$ contributions for both fits, showing overall comparable quality except for the Neutral Current $920\, \text{GeV}$ bin, which is better fitted with our modified configuration.
\subsection{Results}
\begin{figure}
  \centering
  \includegraphics[width=0.7\linewidth]{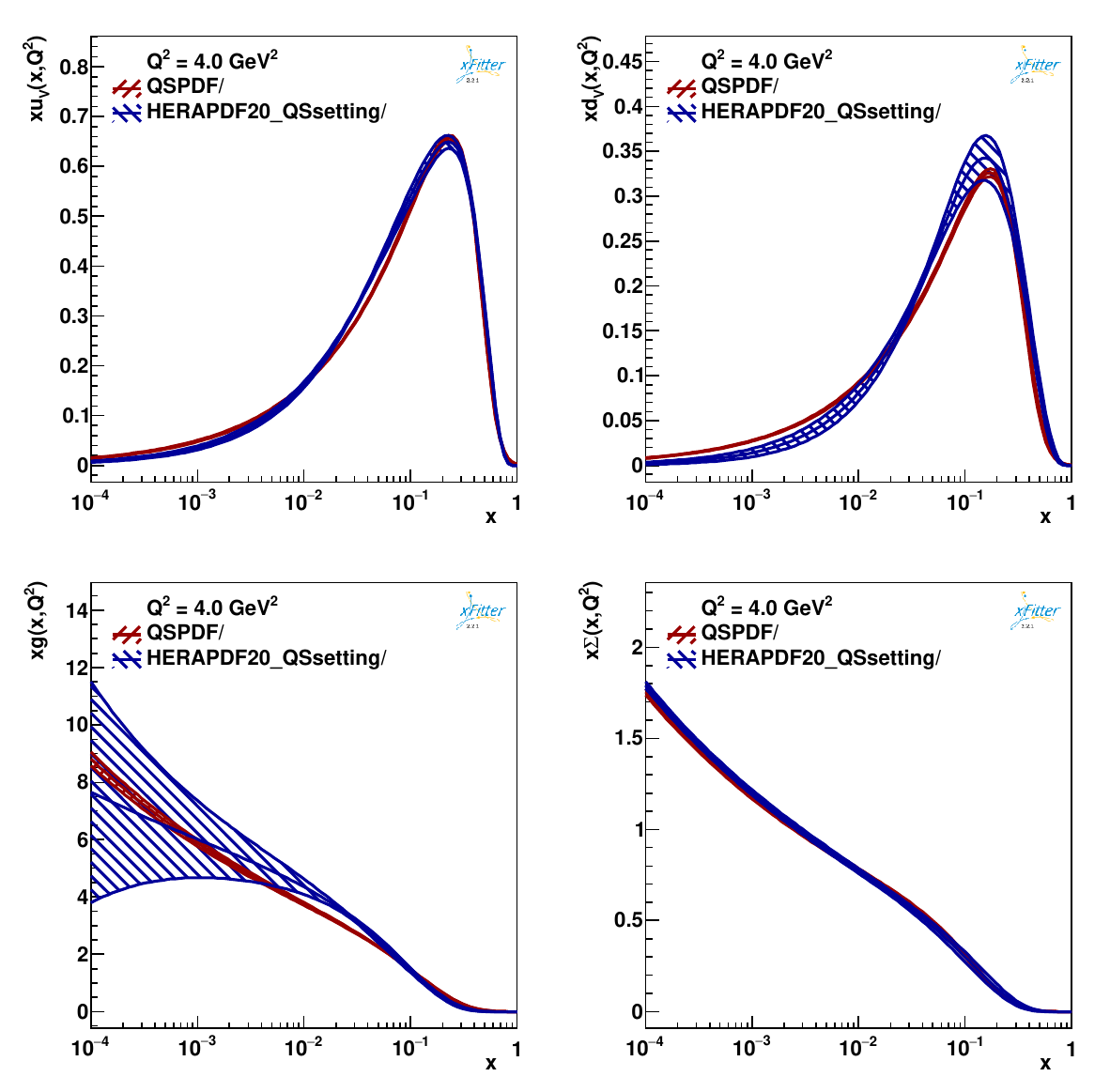}
  \caption{In blue the benchmark fit with the \HERAPDF\  and in red the fit of \QSPDF}
  \label{4:fig:QSPDFFitFig}
\end{figure}
\begin{table}
  \centering
  \caption{$\chi^2$ breakdown for the benchmark fit with the \HERAPDF\ (left)  and  \QSPDF\ (right)}
  \includegraphics[width=0.8\linewidth]{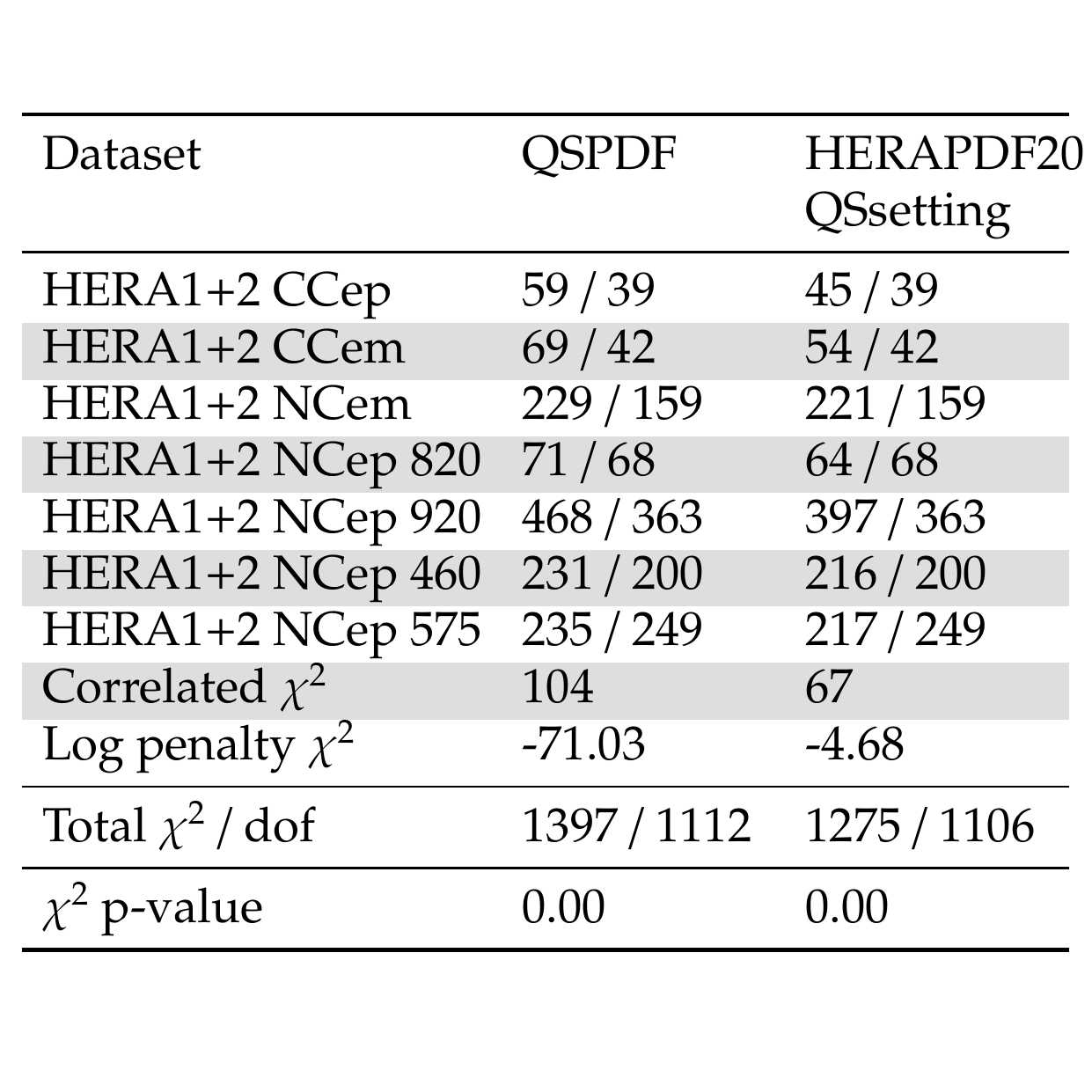}
  \label{4:tab:QSPDFFitTab}
\end{table}
After clearing this validation stage, we proceeded to fit the \QSPDF\ against the HERA DIS dataset. The resulting distributions are plotted together with the ones from the benchmark fit in Fig.~\ref{4:fig:QSPDFFitFig}. We observe that the \QSPDF\ parametrisation produces smaller error bands.\footnote{At the present time only experimental error is accounted for, no study for variations of the fitting parametres (masses, cuts etc...) is included}
\begin{table}
  \centering
  \caption{Summary of the $\chi^2$ for the \QSPDF\ and benchmark fit}
  \label{4:tab:QSPDFchi2}
  \begin{tabular}{|c|c|c|}
    \hline
    &  \QSPDF & \HERAPDF \\
    \hline
    \hline
    \# param. & $8$ & $14$ \\
    $\chi^2$/D.O.F. & $1.26$ & $1.15$ \\
    \hline
  \end{tabular}
\end{table}
We summarise the $\chi^2$ performance of the fit in Tabs~\ref{4:tab:QSPDFFitTab} and \ref{4:tab:QSPDFchi2}. The main observation is that the reduced $\chi^2$ of the \QSPDF\ parametrisation is worse than the one of \HERAPDF. On the other hand, the former has a lower free parametres count than the latter, hinting at the possibility that the more constrained expressions of Eqs.~\eqref{4:eq:unpdistr} are able to model the HERA dataset just as well as a more general polynomial.
\begin{figure}[t]
  \centering
  \begin{subfigure}{0.49\linewidth}
    \centering
    \includegraphics[ width = \linewidth, trim = 0 10cm 10cm 0, clip]{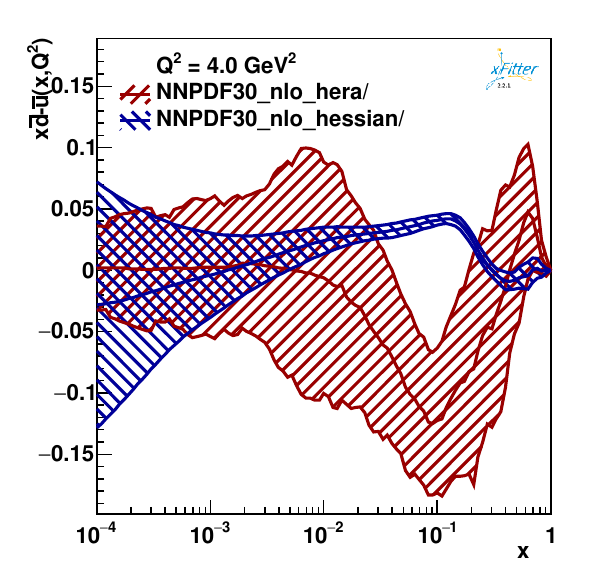}
    \caption{Comparison of \texttt{NNNPDF}$3.0$ fits with the HERA dataset only and the default dataset, (NLO theory)}
    \label{4:fig:D-UPlotsNNPDF30}
  \end{subfigure}
  \begin{subfigure}{0.49\linewidth}
    \centering
    \includegraphics[ width = \linewidth, trim = 0 10cm 10cm 0, clip]{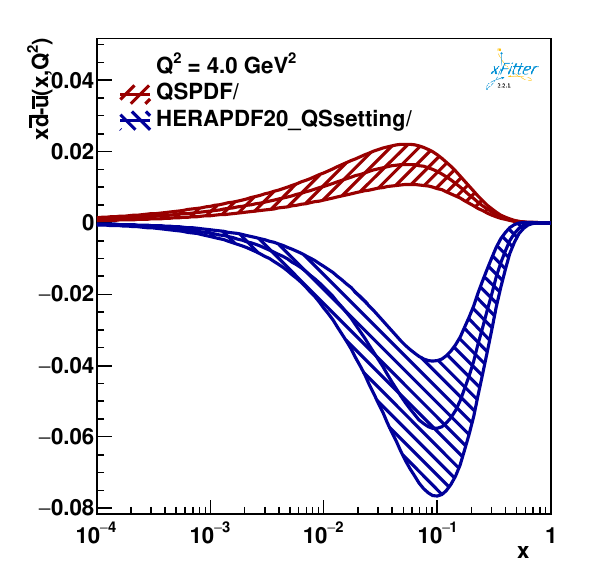}
    \caption{ In blue the benchmark fit with the \HERAPDF\ parametrisation and in red the fit of \QSPDF}
    \label{4:fig:D-UPlotsQSPDF}
  \end{subfigure}
  \caption{}
  \label{4:fig:D-UPlots}
\end{figure}
In figure~\ref{4:fig:D-UPlotsQSPDF} we show the $\db -\ub $ distribution, highlighting that, unlike \HERAPDF, the \QSPDF\ fit manages to reproduce, at least qualitatively, the correct flavour separation distribution using only from HERA data. This is likely a result of the stiffness of the parametrisation and the very small set of parameters shared by the PDFs. More flexible parametrisations usually fail to capture this feature without the input of data from $pp$ colliders. As an example, in figure~\ref{4:fig:D-UPlotsNNPDF30},  we show the same flavour separation in the NLO sets from \text{NNPDF}$3.0$~\cite{NNPDF:2014otw,Carrazza:2015aoa}. We see that the HERA only fit in red does not capture the same shape of the full dataset, in blue, which contains additional input from LHC and other colliders\footnote{We showcased this example with \text{NNPDF}$3.0$ for convenience, but this pattern is reproduced by many parametrisations when hadron-hadron collider data is introduced in the analysis}.
%No information on polarised PDF (unlike previous work)
\begin{table}
  \centering
  \caption{Best fit parametres for the \QSPDF, compared against the ones from the 2002 iteration of the model. The normalisation computed with the sum rules are marked in blue.}
  \label{4:tab:QSPDFParam}
  \begin{tabular}{|c|c|c|}
    \toprule
    Parameter    & QSPDF  &   Ref.~\cite{Bourrely:2001du}\\ 
    \midrule\midrule
    $A$          &  $\textcolor{blue}{ 3.04 }$ & 1.75 \\ 
    $\Ab$        &  $\textcolor{blue}{ 0.12 }$ & 1.91 \\ 
    $A_g$       &  $\textcolor{blue}{ 33.52 }$  & 14.28 \\ 
    $\Atil$      &  $0.133 \pm 0.004$           & 0.083 \\
    $X_d^\uparr$   & $0.14 \pm 0.02$              & 0.23  \\
    $X_d^\dwarr$   & $0.284 \pm 0.007$            & 0.302 \\
    $X_u^\uparr$   & $0.419 \pm 0.007$            & 0.461 \\
    $X_u^\dwarr$  & $0.21 \pm 0.02$               & 0.298 \\
    $b = \bb$   & $0.52 \pm 0.01$               & 0.41  \\
    $\btil = b_g-1$ & $-0.173 \pm 0.003$        & -0.253 \\
    $\xb$       & $0.092 \pm 0.001$             &  0.099 \\
    \bottomrule
  \end{tabular}
\end{table}

Finally, in table~\ref{4:tab:QSPDFParam} we collect the best fit values from our fit and compare them with the values from a similar test carried out in Ref.~\cite{Bourrely:2001du}.
We recover a general agreement between the two. Moreover we observe that in both cases the ``chemical potentials'' reproduce the ordering
\begin{equation}
  X_u^\uparr > X_d^\dwarr \sim X_u^\dwarr > X_d^\uparr\, ,
\end{equation}
matching the theoretical expectation of a flavour asymmetry in the sea quark distributions, with  $\bar{d}$ bigger than $\bar{u}$, as a consequence of the Pauli principle~\cite{Field:1976ve}.

%In conclusion, this PDF analysis is mid af.
In conclusion, our NLO analysis of Parton Distribution Functions with the \QSPDF\ parametrisation has led us to some interesting findings. We have shown that a custom parametrisation, inspired by physical arguments, can successfully be used to fit the proton content to an acceptable degree. Moreover, it can reproduce, at least partially, some flavour separation patterns in DIS data that are typically missed by more a flexible parametrisation.
While our custom parametrisation did not achieve the same level of overall quality-of-fit as the benchmark parametrisation, our results suggest that it is a promising avenue for further exploration, perhaps starting by including higher order theory inputs and more data from other experiments. 
Overall, our study highlights the opportunity of considering physical motivations when developing a parametrisation for Parton Distribution Functions, and we hope that our findings will inspire further investigation into the development of more physically motivated models.

	\chapter{Conclusions and outlook} %\addcontentsline{toc}{chapter}{Conclusions and outlook}
	At the conclusion of this thesis, we recap the main results of the work presented.
In chapter ~\ref{ch2}, we presented a generalisation of the \hell\ resummation formalism for triple differential distributions based on \cite{Silvetti:2022yed,Silvetti:2022hyc}. We showcased a phenomenological application to heavy quark pair production at the LHC and presented predictions in both the kinematical variables of a single final-state quark or the entire outbound pair. These results are of great interest as they form a basis for repeating the same resummation study to other LHC processes, chiefly Drell-Yan. For future work, this new resummation framework allows to include resummed theory predictions in PDF determination software. This would allow to leverage a wealth of LHC data to perform the analysis of \cite{Ball:2017otu,xFitterDevelopersTeam:2018hym} with more data to constrain the small-$x$ gluon.

In chapter~\ref{ch2a} we proposed a possible  strategy to achieve NLL$x$ resummation in coefficient functions. So far, complications in NLO off-shell computation for our case study of Higgs-induced DIS prevented us to come to a definitive answer. Hopefully, some of the issue highlighted in the chapter can be resolved with an improved implementation of the computation and leveraging a similar computation from Ref.~\cite{Celiberto:2022fgx}.

A study of the Higgs $\pt{}$ spectrum in $HW^+$ associate production under a jet veto was presented in chapter~\ref{ch3}, using the \RadISH\ framework and $\qt$-subtraction to build a NNLO fixed order prediction.
We complemented the latter to account for resummation of NNLL logarithms of the jet veto, as well as linear power corrections in $\pt{HW}$. We found that both these effects account for correction around $1\sim2\%$ between the resummed computation and its fixed-order counterpart. Furthermore, a natural follow-up would be to repeat the same combination of \NNLL resummation and $\qt{}$ subtraction using fixed-order NNLO predictions instead of NLO ones. This would result in the first determination of the Higgs $\pt{}$ for this process at N$3$LO. 

Finally in chapter~\ref{ch4}, we introduced a simple Parton Distribution Function parametrisation (\QSPDF) based on quantum-statistical arguments on the proton structure. We validated this proposal with a fit against the HERA dataset using NLO QCD theory. We find that, \textit{ceteris paribus}, the fit quality is slightly worse than the one of the general HERAPDF$2.0$ parametrisation we picked as a benchmark. Instead, the highly constrained form of \QSPDF\ manages to produce the correct shape in the flavour separation of $f_{\ub}(x)-f_{\db}(x)$ using only HERA data, which more flexible and unbiased PDF parametrisations do not capture without extra input from LHC data. This initial success motivates further test of the \QSPDF\ parametrisation, starting with a fit using NNLO theory.
An interesting opportunity to investigate is to perform a fit of \QSPDF\ using resummed NNLO+NLL$x$ theory. Indeed, since the physical argument at the core of the parametrisation holds true especially in the high-$x$ region, the most accurate treatment possible of low-$x$ data may allow to put its performance to a strict test.

        \begin{appendices}
        \chapter{Useful formulae}
\section{Kinematics of NLO radiative corrections}\label{app:NLO-corr}
In this appendix we expand the computation in section \ref{1:ssec:Radiative-corrections}.
We start by writing the amplitude for the real emission in figure~\ref{1:fig:RadR}
\begin{equation}
  \Acal_{\textnormal{R}}  =  \Mcal (p-k) \frac{\slashed{p} - \slashed{k} }{(p-k)^2}  \( g_s t^a \gamma^\mu\) \epsilon_{\mu}^a (k) u (p)  \label{app1:eq:correctionsR} \, . 
\end{equation}
The kinematics of the extra emission can be written with a Sudakov parametrisation
\begin{subequations} \label{app1:eq:kindef}
  \begin{align}
	k &= \(1-z\) p_1 + \kt{} + z_2 p_2 \, , \label{app1:eq:kindef1} \\
	p_{1,2} &= \frac{\sqrt{S}}{2} \(1,\vec{0}, \pm 1\) \, , \label{app1:eq:kindef2} \\
	\kt{} &= \(0, \vec{k}, 0\) \, .
  \end{align}
\end{subequations}
So, we can recast the gluon phase space as
\begin{align}
  \frac{\dd^3 k }{2 E_k \(2 \pi\)^3} & = \frac{\dd \kvec}{2 (2 \pi)^3} \frac{\dd z }{1-z} \, ,\\
  z_2 & = \frac{| \vec{k} |^2}{2 (1-z) \spd{p_1}{p_2 } } \, , \\
  (p-k)^2 & = \frac{\vec{k}^2}{1-z} \, .
\end{align}
To highlight the singularity structure in \eqref{app1:eq:correctionsR}, we rewrite it using Eqs.~\eqref{app1:eq:kindef} (excluding $\as$ and the colour algebra) and  keep only the singular part for $\kt{}^2 \rightarrow 0$. This gives
\begin{align}
  \Acal_{\text{R}}  & = \Mcal (p- k) \frac{z \slashed{k} - \slashed{\kt{}}}{\vec{k}^2} \gamma^\mu  u(p) \epsilon_\mu (k) \nn
  & = \Mcal{} (p- k) \frac{- 2 z \kt{}^\mu - (1-z) \slashed{k}_t \gamma^\mu}{\vec{k}^2}  u(p) \epsilon_\mu (k) \, , \label{app1:eq:correctionR1}
\end{align}
where we have taken advantage of the gauge and on shell conditions $k^\mu \epsilon_{\mu} (k) = 0$ and $ \slashed{p} u (p) = 0$.
We see that the collinear singularity in~\eqref{app1:eq:correctionR1},  $\frac{1}{\left|{\vec{k}}\right|}$, is weaker than what first appeared in Eq.~\eqref{app1:eq:correctionsR}, $\frac{1}{\left|{\vec{k}}\right|}$.
After taking the modulus squared and the sum over the final state polarisations, we apply the phase space integral and the coupling plus colour factor ($4 \pi \as C_F $) to get the cross section
% and the initial state flux $ 2 P$
\begin{equation}
  \sigh_{1R}  = \frac{\as C_F}{2 \pi } \int_0^\infty \int_0^1 \frac{\dd \vec{k}^2 }{\vec{k}^2} \dd z \sigh_0 (z p) \frac{ 1+ z^2 }{1- z}  \, .
\end{equation}
This last expression contains a double singularity in the soft and collinear limit. Naturally, this should not be the case in a renormalisable field theory due to the Kinoshita-Lee-Nauenberg theorem\cite{Schwartz:2013pla}. At this simple level, we can even enforce the cancellation without explicitly computing the virtual part $\sigh_{1V}$. As the quark current is conserved the total combination of the quark emitting a gluon with momentum fraction $1-z$ plus the one of having no emission at all must be 1
\begin{equation}
  \int_0^1 \dd z \[ \frac{C_F \as}{2\pi } \frac{1+x^2}{1-x} + \delta\(1-z\) (1+A)\] = 1 \, ,
\end{equation}
where the $A\delta\(1-z\)$ term represents the double divergent term from the virtual correction and allows us to symbolically write in collinear limit
\begin{align}
\sigh_{1V} = - \frac{\as C_F}{2 \pi } \sigh_0 (p) \int_0^\infty \int_0^1 \frac{\dd \vec{k}^2 }{\vec{k}^2} \dd z \delta (1-z)\frac{ 1+ z^2 }{1- z}  \, , 
\end{align}
and combine them into
\begin{equation}
\sigh_1 = \frac{\as C_F}{2 \pi } \int \(\frac{1+z^2}{1-z}\)_+ \sigh_0 (zp)   \ln(\frac{Q^2}{\LambdaQ^2}) \dd z \, . \label{app1:eq:sigh1}
\end{equation}
Eq.~\ref{app1:eq:sigh1}  is now finite in the soft limit (but not in the collinear one!) and we introduced the $+$-distribution $\[f\(z\)\]_+$, which we define next, for compactness.

\section{The plus-distribution}\label{app1:ssec:Ppresc}
Consider $f(z;z_0)$ is an arbitrary function of $z\in\(0,1\)$ that diverges at $z_0$ like $\(z-z_0\)^{-a}$ with $a<2$ or less.
We define the $+$-distribution as
\begin{equation}
      \int_0^1 \dd z \[ f(z;z_0)\]_+ g (z) = \int_0^1 \dd z f (z;z_0) \[g (z) - g (z_0) \]\, ,
\end{equation}
so that for any function $g(z)$ regular in $z_0$. The subtraction of the $f(z;z_0)g(z_0)$ in the integral ensures it remains finite.

If $g(z)$ is regular, we also have the following identities then
\begin{align}
\[g(z)\,f(z;z_0)\]_+ &= g(z)\,\[f(z;z_0)\]_+ - \delta(z-z_0) \int_0^1 dy \; g(y)\,\[f(y;z_0)\]_+ \label{eq:plus_useful1}, \\
g(z)\,\[f(z;z_0)\]_+ &= g(z_0)\,\[f(z;z_0)\]_+ + \[g(z)-g(z_0)\]_+ f(z;z_0) \;.\label{eq:plus_useful2}
\end{align}
In the last term of second line, the plus distribution is redundant as $g(z)-g(z_0)$ regularises $f(z;z_0)$.

\section{Evolution basis and DGLAP kernels}\label{app:2:DGLAP}
These objects, as we have seen in our simple example, are computable in perturbation theory as
\begin{equation}
         P_{ij}(x, \as) = \sum_{n=1}^\infty \(\frac{\alpha_s}{2 \pi }\)^n P_{ij}^{\(n-1\)}(x) \, ,
\end{equation}
with the lowest order splittings given by ref. \cite{Altarelli:1977zs}
\begin{subequations} \label{eq:Pij0}
	\begin{align}
                P_{qq}^{(0)}(z) & = C_F \(\frac{1+z^2}{1-z}\)_+ \, , \\
	        P_{qg}^{(0)}(z) & = T_F \[z^2 + \(1-z^2\)\] \, , \\
	        P_{gq}^{(0)}(z) & = C_F \frac{1 + \(1-z\)^2}{z} \, , \\
	        P_{gg}^{(0)}(z) & = 2 C_A \[ \frac{z}{\(1-z\)_+} + \frac{1-z}{z} + z\(1-z\) + \(\frac{11}{12} - \frac{n_f}{2 C_A}\) \delta (1-z) \] \, .
	\end{align}
\end{subequations}
The next-to-leading order (2-loop) splitting functions have been computed long ago~\cite{Curci:1980uw,Furmanski:1980cm}, and the 3-loop ones have been known for a while~\cite{Moch:2004pa,Vogt:2004mw}  some impressive progress towards the computation of next-to-next-to-next-to-leading order (4-loop) splitting functions has been made~\cite{Davies:2016jie,Moch:2017uml}, and even some partial results at 5 loops have been reported~\cite{Herzog:2018kwj}.

We stress that the beyond leading order the splitting functions of different quark flavors will not have the same expression. For this reason the parton distributions and their respective splitting functions are usually rewritten in terms of singlet $\(\Sigma\)$ and non-singlet $\(V_i, T_3, T_8, T_{15}, T_{24}, T_{35}\)$.
\begin{subequations}
	\begin{align}
		\Sigma (x, \muF^2) & = \sum_{ i = \left\lbrace u,d,s,c,b,t\right\rbrace} f_{i}^+ (x, \muF^2) \, , \\
		V_i (x, \muF^2) & = f_{q_i}^- (x, \muF^2) \, , \\
		T_3 (x, \muF^2) & =  f_{u}^+ (x, \muF^2) - f_{d}^+ (x, \muF^2) \, , \\
		T_8 (x, \muF^2) & = f_{u}^+ (x, \muF^2)  + f_{d}^+ (x, \muF^2) - 2 f_{s}^+ (x, \muF^2) \, , \\
		T_{15} (x, \muF^2) & = f_{u}^+ (x, \muF^2) + f_{d}^+ (x, \muF^2) + f_{s}^+ (x, \muF^2) - 3f_{c}^+ (x, \muF^2) \, , \\
		T_{24} (x, \muF^2) & = f_{u}^+ (x, \muF^2) + f_{d}^+ (x, \muF^2) + f_{s}^+ (x, \muF^2) + f_{c}^+ (x, \muF^2) - 4f_{b}^+ (x, \muF^2) \, , \\
		T_{35} (x, \muF^2) & = f_{u}^+ (x, \muF^2) + f_{d}^+ (x, \muF^2) + f_{s}^+ (x, \muF^2) \nonumber \\
		& \:  + f_{c}^+ (x, \muF^2) + f_{b}^+ (x, \muF^2) - 5f_{t}^+ (x, \muF^2) \, , 
	\end{align}
\end{subequations}
where for any flavor $f_{i}^\pm \(x, \muF^2\) = f_{q_i} \(x, \muF^2\) \pm f_{\overline{q}_i} \(x, \muF^2\)$. All these objects except $\Sigma$ evolve independently of one another with its own equation like \eqref{1:eq:DGLAP}. The singlet DGLAP evolution instead is coupled to the gluon distribution, as all quark flavors interact with gluons in the same fashion,
\begin{equation}
  \muF^2 \frac{\dd}{\dd \muF^2} \begin{pmatrix} \Sigma \(x, \muF^2\) \\ g \(x, \muF^2\) \end{pmatrix} =  \int_{x}^{1} \frac{\dd z}{z}
  \begin{pmatrix} P_{qq}\( \frac{x}{z}, \as (\muF^2) \) & 2 n_f P_{qg}\(\frac{x}{z}, \as (\muF^2)\) \\
      P_{gq}\(\frac{x}{z}, \as (\muF^2)\) & P_{gg}\(\frac{x}{z}, \as (\muF^2)\) \end{pmatrix}
  \begin{pmatrix} \Sigma \(z, \muF^2\) \\ g \(z, \muF^2\)  \end{pmatrix}  \, .
\end{equation}

\section{Mellin transforms}\label{app2:Mellintransf}
In this appendix we give the definition of the Mellin Transform used troughout the thesis.
We begin from the definition
\begin{equation}
f(N)\equiv \int_0^1 \dd x x^{N-1} f(x)\,, 
\end{equation}
which is effectively a Laplace transform after a change of variable $x=\eu^{-t}$.
Its inverse is given by
\begin{equation}
  f(x)=\int_{C-i\infty}^{C+i\infty} \frac{\dd N}{2\pi i} x^{-(N+1)}f(N) ,
\end{equation}
as long as the coefficient $C$ is such that the integration path lies beyond the rightmost pole of $f(N)$.

This transform maps logarithms of $x$ into poles of $N$
\begin{equation*}
    \int_0^1 \dd x x^{N-1} \frac{\log^{k-1}(x)}{x} = \frac{(-1)^{k-1}(k-1)!}{(N-1)^{k}} \, .
\end{equation*}
Additionally it has the property of factorising integral of the form
\begin{equation}
  h(x) = \int_x^1\frac{\dd}{z} f\(\frac{x}{z}\)g(z) \: \rightarrow \: h(N) = f(N)g(N).
\end{equation}
%\begin{align}\label{app2:eq:MellFourCL}
%\frac{\dd C_{ij}}{\dd Q^2\dd y\dd\qt^2}\(N,Q^2,b,\qt^2,\as, \frac{Q^2}{\muf^2}\)
%  &= \int_0^1\dd x\,x^N \int_{-\infty}^\infty \dd y\,\eu^{iby}\frac{\dd C_{ij}}{\dd Q^2\dd y\dd\qt^2} \(x,Q^2,y,\qt^2,\as, \frac{Q^2}{\muf^2}\) \nonumber\\
%\end{align}

%The last equality we have used the definition Eq.~\eqref{eq:lumi}and changed variable from $\xl,\yl$ to $\hat x_{1,2}=\sqrt{\xl}\myexp{\pm\yl}$ and used explicitly the $\theta$ function to obtain the product of two Mellin transforms
%\begin{equation}
%f_i(N,\muf^2) = \int_0^1\dd \hat x_{1,2}\,\hat x_{1,2}^N f_i(\hat x_{1,2},\muf^2).
%\end{equation}

\chapter{Heavy-quarks and small-$x$}\label{app:HQ}
This appendix contains more details on the computation for the triple-differential resummation of coefficient functions and its application to heavy-quark pair production.
\section{All partonic channels}
\label{sec:channels}

In the resummed expression Eq.~\eqref{2:eq:resCggzy} the key ingredient is the evolution function in $x$ space.
This object is obtained in \hell\ as the inverse Mellin transform of  Eq.~\eqref{2:eq:UABFht} and is a distribution. Indeed, the zeroth order term in the $\as$-power expansion of $U(N,Q^2\xi,\muF^2)$  is just $1$, whose inverse Mellin is $\delta(1-x)$.
For convenience, we choose to manipulate this piece separately
\begin{equation}\label{eq:Ureg}
U(N, Q^2\xi,\muF^2) = 1+U_{\rm reg}(N, Q^2\xi,\muF^2) \:\Leftrightarrow\: 
U(x, Q^2\xi,\muF^2) = \delta(1-x)+U_{\rm reg}(x, Q^2\xi,\muF^2),
\end{equation}
where $U_{\rm reg}$ contain the rest of the inverse Mellin transform and is just a function.

There are some subtleties when taking the $\xi$-derivative in Eq.~\eqref{2:eq:Foff}. First, a factor $\theta(\xi)$ must be introduced in the definition, $U(N, Q^2\xi,\muF^2) = \theta(\xi)\[1+U_{\rm reg}(N, Q^2\xi,\muF^2)\]$. In principle this makes no difference since $Q^2\xi=\kvec^2$ is positive defined.
Then, the derivative yields
\begin{align}\label{eq:U'regN}
  U'(N, Q^2\xi,\muF^2)
  &=\delta(\xi) + \delta(\xi) U_{\rm reg}(N, 0,\muF^2) + \theta(\xi)U'_{\rm reg}(N, Q^2\xi,\muF^2) \nonumber\\
  &=\delta(\xi) - \delta(\xi) \int_0^{\frac{\muF^2}{Q^2}} \dd\xi'\,U'_{\rm reg}(N, Q^2\xi',\muF^2) + \theta(\xi)U'_{\rm reg}(N, Q^2\xi,\muF^2) \nonumber\\
  &=\delta(\xi) + \[U'_{\rm reg}(N, Q^2\xi,\muF^2)\]_+,
\end{align}
where in the second step we have used the fact that $U_{\rm reg}(N, \muF^2,\muF^2)=0$.
The lone $\delta(\xi)$ term can be interpreted as an undisturbed gluon, that remains on-shell and does not emit extra radiation.
This indeed corresponds to the term subtracted in the quark contribution to the unintegrated PDF, Eq.~\eqref{2:eq:Foff}.

The introduction of the plus distribution is actually redundant as $U_{\rm reg}(N, 0,\muF^2)$ appearing in the first line of Eq.~\eqref{eq:U'regN} is finite.
More precisely, $U(N, 0,\muF^2)$ vanishes by construction in Eq.~\eqref{2:eq:UABFht}, then we have $U_{\rm reg}(N, 0,\muF^2) =-1$, corresponding in $x$-space to
\begin{equation}\label{eq:Uregxi0}
U_{\rm reg}(x, 0,\muF^2) = -\delta(1-x).
\end{equation}
When this occurs, the first two terms in the first line of Eq.~\eqref{eq:U'regN} would cancel, simplifying the expression to $U'(N, Q^2\xi,\muF^2)=U'_{\rm reg}(N, Q^2\xi,\muF^2)$ which is what we would have obtained if we hadn't introduced the $\theta$ function in the first place.
This, however, negates the physical distinction between the no-emission contribution $\delta(\xi)$ and the other contribution $U'_{\rm reg}(N, Q^2\xi,\muF^2)$.
This is clearly undesirable, and may hint at a problem in the construction of the evolution function.

One crucial observation about this issue is that the $\xi\to0$ limit of $U'_{\rm reg}$, Eq.~\eqref{eq:Uregxi0},is localised at large $x$.
But the evolution function at large $x$ is constructed to resum large logarithmic contributions
at small $x$ and is not necessarily accurate away from there.
As a matter of fact, the function $U_{\rm reg}(x,Q^2\xi,\muF^2)$ (and thus its $\xi$-derivative) are multiplied by a damping factor function of the form $(1-x)^a$ at large $x$ (we use $a=2$ in the code). With this modification the evolution function satisfies the condition
\begin{equation}\label{eq:Uregx1}
U_{\rm reg}(x=1, Q^2\xi,\muF^2) = 0,
\end{equation}
for any value of $\xi$, including $\xi=0$.
In this way, we obtain $U_{\rm reg}(x, 0,\muF^2) =0$ and thus $U_{\rm reg}(N, 0,\muF^2) =0$,
implying that the second term in the first line of Eq.~\eqref{eq:U'regN} vanishes, thus giving
\begin{align}\label{eq:U'reg}
  & U'(N, Q^2\xi,\muF^2) = \delta(\xi)+U'_{\rm reg}(N, Q^2\xi,\muF^2) \nn
  & \quad \Leftrightarrow\: U'(x, Q^2\xi,\muF^2) = \delta(\xi)\delta(1-x)+U'_{\rm reg}(x, Q^2\xi,\muF^2).
\end{align}
In other words the plus distribution is rendered ineffective by the addition of the damping.
Indeed, this relation was verified by making sure that the integral of $U'_{\rm reg}(x, Q^2\xi,\muF^2)$
from zero to $\muF^2/Q^2$ is vanishing for all values of $x$.

Coming back to the coefficient function, the unintegrated PDF Eq.~\eqref{2:eq:Foff} can be rewritten using Eq.~\eqref{eq:U'reg} as
\begin{equation}\label{eq:Foff2}
{\cal F}_g(N,\xi) = \[U'_{\rm reg}\(N,Q^2\xi,\muF^2\)+\delta(\xi)\] f_g(N,\muF^2) + \frac{C_F}{C_A} U'_{\rm reg}\(N,Q^2\xi,\muF^2\) f_q(N,\muF^2).
\end{equation}
In this last equation, the $\delta(\xi)$ contribution represents again the (on-shell) gluon that does not emit extra radiation,
 and thus produces no logarithms. It coincides with the fixed-order contribution, and it also reproduces the on-shell result.
The other term, $U'_{\rm reg}$, is the term containing at least one emission, and thus at least one small-$x$ log.

Starting from Eq.~\eqref{eq:Foff2} and proceeding as in section~\ref{2:ssec:diff-hell}, we can write the coefficient function including the quark contributions as well.
We obtain the following expressions
\begin{subequations}\label{eq:resC}
\begin{align}
  \frac{\dd C_{gg}}{\dd Q^2\dd y\dd\qt^2}\(x,Q^2,y,\qt^2,\as, \frac{Q^2}{\muF^2}\)
  &= \int_0^\infty \dd\xi_1 \int_0^\infty \dd\xi_2 \int_x^1\frac{\dd z}z \int_{-\frac12\log\frac zx}^{\frac12\log\frac zx} \dd\bar\eta\,\nonumber\\
  &\times \frac{\dd{\cal C}}{\dd Q^2\dd \eta\dd\qt^2}(z,\xi_1,\xi_2,Q^2,y-\bar\eta,\qt^2,\as) \nonumber\\
  &\times \[U'_{\rm reg}\(\sqrt{\frac xz}\eu^{\bar\eta},Q^2\xi_1,\muF^2\)+\delta(\xi_1)\delta\(1-\sqrt{\frac xz}\eu^{\bar\eta}\)\]\nonumber\\
  &\times \[U'_{\rm reg}\(\sqrt{\frac xz}\eu^{-\bar\eta},Q^2\xi_2,\muF^2\) +\delta(\xi_2)\delta\(1-\sqrt{\frac xz}\eu^{-\bar\eta}\)\], \\
  \frac{\dd C_{qg}}{\dd Q^2\dd y\dd\qt^2}\(x,Q^2,y,\qt^2,\as, \frac{Q^2}{\muF^2}\)
  &= \frac{C_F}{C_A}\int_0^\infty \dd\xi_1 \int_0^\infty \dd\xi_2 \int_x^1\frac{\dd z}z \int_{-\frac12\log\frac zx}^{\frac12\log\frac zx} \dd\bar\eta\,\nonumber\\
  &\times \frac{\dd{\cal C}}{\dd Q^2\dd \eta\dd\qt^2}(z,\xi_1,\xi_2,Q^2,y-\bar\eta,\qt^2,\as) \nonumber\\
  &\times U'_{\rm reg}\(\sqrt{\frac xz}\eu^{\bar\eta},Q^2\xi_1,\muF^2\)\nonumber\\
  &\times \[U'_{\rm reg}\(\sqrt{\frac xz}\eu^{-\bar\eta},Q^2\xi_2,\muF^2\) +\delta(\xi_2)\delta\(1-\sqrt{\frac xz}\eu^{-\bar\eta}\)\], \\
  \frac{\dd C_{gq}}{\dd Q^2\dd y\dd\qt^2}\(x,Q^2,y,\qt^2,\as, \frac{Q^2}{\muF^2}\)
  &= \frac{C_F}{C_A}\int_0^\infty \dd\xi_1 \int_0^\infty \dd\xi_2 \int_x^1\frac{\dd z}z \int_{-\frac12\log\frac zx}^{\frac12\log\frac zx} \dd\bar\eta\,\nonumber\\
  &\times \frac{\dd{\cal C}}{\dd Q^2\dd \eta\dd\qt^2}(z,\xi_1,\xi_2,Q^2,y-\bar\eta,\qt^2,\as) \nonumber\\
  &\times \[U'_{\rm reg}\(\sqrt{\frac xz}\eu^{\bar\eta},Q^2\xi_1,\muF^2\)+\delta(\xi_1)\delta\(1-\sqrt{\frac xz}\eu^{\bar\eta}\)\]\nonumber\\
  &\times U'_{\rm reg}\(\sqrt{\frac xz}\eu^{-\bar\eta},Q^2\xi_2,\muF^2\), \\
  \frac{\dd C_{qq}}{\dd Q^2\dd y\dd\qt^2}\(x,Q^2,y,\qt^2,\as, \frac{Q^2}{\muF^2}\)
  &= \(\frac{C_F}{C_A}\)^2\int_0^\infty \dd\xi_1 \int_0^\infty \dd\xi_2 \int_x^1\frac{\dd z}z \int_{-\frac12\log\frac zx}^{\frac12\log\frac zx} \dd\bar\eta\,\nonumber\\
  &\times \frac{\dd{\cal C}}{\dd Q^2\dd \eta\dd\qt^2}(z,\xi_1,\xi_2,Q^2,y-\bar\eta,\qt^2,\as) \nonumber\\
  &\times U'_{\rm reg}\(\sqrt{\frac xz}\eu^{\bar\eta},Q^2\xi_1,\muF^2\)\, U'_{\rm reg}\(\sqrt{\frac xz}\eu^{-\bar\eta},Q^2\xi_2,\muF^2\).
\end{align}
\end{subequations}
These results can be written in a more compact form as
\begin{subequations}\label{eq:resC_reg_aux}
\begin{align}
\frac{\dd C_{gg}}{\dd Q^2\dd y\dd\qt^2}
&= \frac{\dd C_{\rm reg}}{\dd Q^2\dd y\dd\qt^2} + \frac{\dd C_{\rm aux\,+}}{\dd Q^2\dd y\dd\qt^2} + \frac{\dd C_{\rm aux\,-}}{\dd Q^2\dd y\dd\qt^2} 
  +\frac{\dd {\cal C}}{\dd Q^2\dd \eta\dd\qt^2}(x,0,0,Q^2,y,\qt^2,\as), \\
\frac{\dd C_{qg}}{\dd Q^2\dd y\dd\qt^2}
&= \frac{C_F}{C_A} \[ \frac{\dd C_{\rm reg}}{\dd Q^2\dd y\dd\qt^2} + \frac{\dd C_{\rm aux\,+}}{\dd Q^2\dd y\dd\qt^2}\], \\
\frac{\dd C_{gq}}{\dd Q^2\dd y\dd\qt^2}
&= \frac{C_F}{C_A} \[ \frac{\dd C_{\rm reg}}{\dd Q^2\dd y\dd\qt^2} + \frac{\dd C_{\rm aux\,-}}{\dd Q^2\dd y\dd\qt^2}\], \\
\frac{\dd C_{qq}}{\dd Q^2\dd y\dd\qt^2}
&= \(\frac{C_F}{C_A}\)^2 \frac{\dd C_{\rm reg}}{\dd Q^2\dd y\dd\qt^2}
\end{align}
\end{subequations}
having defined
\begin{align}\label{eq:resCreg}
  \frac{\dd C_{\rm reg}}{\dd Q^2\dd y\dd\qt^2}\(x,Q^2,y,\qt^2,\as, \frac{Q^2}{\muF^2}\)
  &= \int_0^\infty \dd\xi_1 \int_0^\infty \dd\xi_2 \int_x^1\frac{\dd z}z \int_{-\frac12\log\frac zx}^{\frac12\log\frac zx} \dd\bar\eta\,\nonumber\\
  &\times \frac{\dd{\cal C}}{\dd Q^2\dd \eta\dd\qt^2}(z,\xi_1,\xi_2,Q^2,y-\bar\eta,\qt^2,\as) \nonumber\\
  &\times U'_{\rm reg}\(\sqrt{\frac xz}\eu^{\bar\eta},Q^2\xi_1,\muF^2\)\, U'_{\rm reg}\(\sqrt{\frac xz}\eu^{-\bar\eta},Q^2\xi_2,\muF^2\)
\end{align}
and
\begin{align}\label{eq:resCaux}
  \frac{\dd C_{\rm aux\,\pm}}{\dd Q^2\dd y\dd\qt^2}\(x,Q^2,y,\qt^2,\as, \frac{Q^2}{\muF^2}\)
  &= \int_0^\infty \dd\xi \int_x^1\frac{\dd z}z 
  \frac{\dd{\cal C}}{\dd Q^2\dd \eta\dd\qt^2}\(z,\xi,0,Q^2,y\pm\frac12\log\frac zx,\qt^2,\as\) \nonumber\\
  &\times U'_{\rm reg}\(\frac xz,Q^2\xi,\muF^2\),
\end{align}
in the last equation we leveraged the symmetry $\xi_1\leftrightarrow\xi_2$ of the off-shell coefficient.
So the resummed expressions for all channels are written in terms of a ``regular'' resummed coefficient and two simpler ``auxiliary'' functions,\footnote{The choice of the ``auxiliary'' label follows the nomenclature of Ref.~\cite{Bonvini:2018iwt},  extended to differential distributions.}
each defined in terms of integrals over ordinary functions (and thus easy to implement numerically).
The dependence of the $gg$ coefficient function on the on-shell limit of the off-shell coefficient is present; however, in practical applications, this contribution is subtracted when matching the resummed result to a fixed-order computation, making it unnecessary.

It is worth noting that the auxiliary functions are derived by considering one of the incoming gluons to be on-shell. As a result, these functions represent a contribution where resummation, achieved through $\kt{}$-factorisation, acts on a single initial state parton, while the other parton follows the standard collinear factorisation. This bears resemblance to the hybrid factorisation discussed in previous works (Refs.~\cite{Deak:2009xt, Deak:2011ga, Celiberto:2020tmb, Celiberto:2022rfj, Celiberto:2022dyf, vanHameren:2022mtk}) used to describe forward production. We believe that our auxiliary contribution captures similar resummed contributions obtained from the hybrid factorisation. However, there might be some differences due to the distinct approaches to resummation, which we plan to investigate in future research.

\section{Matching to fixed order}
\label{sec:matching}

The resummed expression given by equation Eq.~\eqref{eq:resC} captures only the small-$x$ logarithms. However, when utilising this result for phenomenological purposes, it must be matched with a fixed-order calculation. To do so, we must expand the resummed expression in powers of $\as$ up to a certain order, subtract this expansion, and substitute it with the precise fixed-order result at the corresponding order.
 
Expanding the $\as$ from the integrand of Eq.~\eqref{eq:resC} requires some care. First, the off-shell coefficient function is needed only at the lowest non-trivial order to achieve LL accuracy, so the entire expansion actually comes from the evolutor. Let us consider the expansion of $U_{\rm reg}$ for starters. It is straightforward to see that the expansion contains powers of $\log\xi$ from Eqs.~\eqref{2:eq:UABFht} and \eqref{2:eq:UABF}.
\begin{align}\label{eq:Uregexp}
U_{\rm reg}(N,Q^2\xi,\muF^2) &= \as(\muF^2)\gamma_0(N)\log\frac{Q^2\xi}{\muF^2} \nonumber\\
&+ \as^2(\muF^2)\[\gamma_1(N)\log\frac{Q^2\xi}{\muF^2} + \frac12\gamma_0(N)\(\gamma_0(N)-\beta_0\)\log^2\frac{Q^2\xi}{\muF^2}\] + \calO\(\as^3\),
\end{align}
leveraging the expansion $\gamma(N,\as) = \as\gamma_0+\as^2\gamma_1+\calO\(\as^3\)$ for the resummed anomalous dimension from~\cite{Bonvini:2018iwt,Bonvini:2018xvt}.
Logarithms in the form $\log^k\xi/\xi$ emerge after taking the $\xi$-derivative and clearly divergent when $\xi\to0$. Thus, we need a regularisation procedure.

To do so, we return to Eq.~\eqref{eq:U'regN} and recall that the evolutor derivative also involves a plus distribution of $U'_{\rm reg}$. 
At the resummed level the plus distribution is ineffective because to all orders $U'_{\rm reg}(N,0,\muF^2)=0$, which does not hold in true order by order in $\as$, and so keeping it into account becomes mandatory.

With a slight abuse in Eq.~\eqref{eq:Uregexp}, we can still write the first couple of orders of the expansion of $U'_{\rm reg}$,
\begin{align}\label{eq:U'regexp}
U'_{\rm reg}(N,Q^2\xi,\muF^2) &= \as(\muF^2)\gamma_0(N)\(\frac1{\xi}\)_+ \nonumber\\
&+ \as^2(\muF^2)\[\gamma_1(N) \(\frac1{\xi}\)_+ + \gamma_0(N)\(\gamma_0(N)-\beta_0\) \(\frac{\log\frac{Q^2\xi}{\muF^2}}{\xi}\)_+\] + \calO\(\as^3\),
\end{align}
or, in $x$ space,
\begin{align}\label{eq:U'regexpx}
U'_{\rm reg}(x,Q^2\xi,\muF^2) &= \as(\muF^2)P_0(x)\(\frac1{\xi}\)_+ \nonumber\\
&+ \as^2(\muF^2)\[P_1(x) \(\frac1{\xi}\)_+ + \(P_{00}(x)-\beta_0P_0(x)\) \(\frac{\log\frac{Q^2\xi}{\muF^2}}{\xi}\)_+\] + \calO\(\as^3\),
\end{align}
with $P_{00}(x)$ being the Mellin convolution of two $P_0$'s and using the expansion $P(x,\as) = \as P_0(x) +\as^2P_1(x)+\calO\(\as^3\)$ (the inverse Mellin transform of the resummed anomalous dimension $\gamma(N,\as)$).

Substituting into Eq.~\eqref{eq:resCreg} and Eq.~\eqref{eq:resCaux} we get the expansion of the resummed result up to relative $\calO(\as^2)$,
\begin{align}
  \frac{\dd C_{\rm reg}}{\dd Q^2\dd y\dd\qt^2}\(x,Q^2,y,\qt^2,\as, \frac{Q^2}{\muF^2}\)
  &= \int_0^\infty \dd\xi_1 \int_0^\infty \dd\xi_2 \int_x^1\frac{\dd z}z \int_{-\frac12\log\frac zx}^{\frac12\log\frac zx} \dd\bar\eta\,\nonumber\\
  &\times \frac{\dd{\cal C}}{\dd Q^2\dd \eta\dd\qt^2}(z,\xi_1,\xi_2,Q^2,y-\bar\eta,\qt^2,\as) \nonumber\\
  &\times \[\as^2(\muF^2) \(\frac1{\xi_1}\)_+ \(\frac1{\xi_2}\)_+ P_0\(\sqrt{\frac xz}\eu^{\bar\eta}\)\, P_0\(\sqrt{\frac xz}\eu^{-\bar\eta}\) \] \nn
  &+\calO\(\as^3\),\label{eq:resCregFO} \\
  \frac{\dd C_{\rm aux\,\pm}}{\dd Q^2\dd y\dd\qt^2}\(x,Q^2,y,\qt^2,\as, \frac{Q^2}{\muF^2}\)
  &= \int_0^\infty \dd\xi \int_x^1\frac{\dd z}z 
  \frac{\dd{\cal C}}{\dd Q^2\dd \eta\dd\qt^2}\(z,\xi,0,Q^2,y\pm\frac12\log\frac zx,\qt^2,\as\) \nonumber\\
  &\times \bigg\{\as(\muF^2)P_0\(\frac xz\)\(\frac1{\xi}\)_+ + \as^2(\muF^2)\bigg[P_1\(\frac xz\) \(\frac1{\xi}\)_+  \nn
  &+ \(P_{00}\(\frac xz\)-\beta_0P_0\(\frac xz\)\) \(\frac{\log\frac{Q^2\xi}{\muF^2}}{\xi}\)_+\bigg] + \calO\(\as^3\)\bigg\},\label{eq:resCauxFO}
\end{align}
which can be assembled into the expansions of the coefficient function in Eq.~\eqref{eq:resC_reg_aux}.
\section{The off-shell coefficient function}
\label{app:XS}

This Appendix gives a detailed overview of the off-shell coefficient function for heavy-quark pair production.
Consider the LO partonic subprocess
\begin{align}
g^*\(k_1\)+g^*\(k_2\) \rightarrow Q\(p_3\)+\bar Q\( p_4\).
\end{align}
where $Q$ and $\bar Q$ are the two heavy quarks of mass $m$.
We parametrise the momenta as
\begin{subequations}
\begin{align}
k_1 &= x_1 p_1 + \vec{k}_1 , \\
k_2 &= x_2 p_2 + \vec{k}_2 , \\
p_3 &= z_1 x_1 p_1 + z_2x_2 p_2 + \vec{p} , \\
p_4 &= (1-z_1)x_1 p_1 + (1-z_2) x_2 p_2 + \vec{k}_1 + \vec{k}_2 -\vec{p},
\end{align}
\end{subequations}
where, in the collider centre-of-mass frame, the protons momenta are $p_{1,2}= \frac{\sqrt{S}}{2}\(1,0,0,\pm1\)$.

This setup  already accounts for momentum conservation, and implicitly picks the reference frame of the laboratory.
There are 7 initial-state parameters ($s,x_1,x_2,\vec{k}_1,\vec{k}_2$) and 4 final-state parameters ($z_1,z_2,\vec{p}$).
While there are two on-shell conditions,
\begin{subequations}\label{eq:onshellppbar1}
\begin{align}
m^2 = p_3^2 &= z_1z_2x_1x_2S-|\vec{p}|^2 ,\\
m^2 = p_4^2 &= (1-z_1)(1-z_2)x_1x_2S-|\vec{k}_1+\vec{k}_2-\vec{p}|^2,
\end{align}
\end{subequations}
setting the squared momenta $p_3^2$ and $ p_4^2$ to the \emph{same} mass $m^2$, leaves three independent parameters rather than two.

The partonic off-shell coefficient function is computed in the ``partonic'' reference frame,
as if the two incoming gluons were on shell, or simply $\vec{k}_1=\vec{k}_2=0$.
Thus,  the partonic frame is related to the collider one by a longitudinal boost
\begin{equation}
\bar\eta= \frac12\log\frac{x_1}{x_2}.
\end{equation}
In this frame, the partonic coefficient can only depend on $x_1$, $x_2$ and $s$ through the product $x_1x_2s$.
Moreover we can freely integrate over one azimuthal angle, as long as we assume unpolarised protons.
Thus the coefficient can only depend on 4 out of the 7 initial-state parameters, which we pick as
\begin{subequations}\label{eq:initialstatevar}
\begin{align}
z &\equiv \frac{Q^2}{x_1x_2s} ,\\
\xi_1 &\equiv \frac{|\vec{k}_1|^2}{Q^2} ,\\
\xi_2 &\equiv \frac{|\vec{k}_2|^2}{Q^2} ,\\
\varphi &\equiv \text{angle between $\vec{k}_1$ and $\vec{k}_2$}.
\end{align}
\end{subequations}
Here, $Q^2$ is a stand in for ``the hard scale'', of the final state. We set $Q^2=q^2$, where $q$ is the momentum of the distribution we want to produce.
For example,  to study the kinematics of the heavy-quark pair, we set $q=p+\bar p$ and $Q^2$ is the squared invariant mass of the pair, while for the single heavy quark then $q=p$ and $Q^2=m^2$.
Observe that that $x_1x_2s$ does not coincide with $\hat s=(k_1+k_2)^2=x_1x_2s-|\vec{k}_1+\vec{k}_2|^2$, as long as the gluon have a transverse component.
Instead, the full partonic centre-of-mass energy $\hat s$ with the off-shell gluons is
\begin{equation}\label{eq:shatnewvar}
\hat s = Q^2\[\frac1z-\xi_1 -\xi_2-2\sqrt{\xi_1\xi_2}\cos\varphi\]
\end{equation}
 which reduces to the usual expression $\hat s = Q^2/z$ when the gluons are on shell.

\section{Kinematics for the single quark}
\label{app:singlekin}
Here we consider the differential distribution in the kinematics of one of the final-state heavy quarks.
We pick the heavy quark of momentum $p$, but the process is symmetric so the procedures applies equally to the antiquark $\bar{p}$
\begin{subequations}\label{eq:finalstatevars}
\begin{align}
Q^2 &\equiv p^2 = z_1z_2x_1x_2s-|\pvec|^2 = m^2 \\
\eta &\equiv \frac12\log\frac{p^0+p^3}{p^0-p^3}-\bar\eta= \frac12\log\frac{z_1}{z_2} \\
\pthat^2 &\equiv \frac{\pvec^2}{Q^2} = \frac{\pvec^2}{m^2} \\
\vartheta &= \text{angle between $\pvec$ and $\vec{k}_1+\vec{k}_2$}.
\end{align}
\end{subequations}
Because $Q^2=m^2$ is fixed, the most differential distribution\footnote{Note that from now on we are omitting the argument $\as$ from the off-shell distribution as we are interested in the lowest order result only.} we are interested in is ($\pt{}^2=\pthat^2Q^2$)
\begin{equation}\label{eq:triplediffxs}
\frac{\dd{\cal C}}{\dd \eta\, \dd\pt{}^2}(z,\xi_1,\xi_2,m^2, \eta,\pthat^2),
\end{equation}
which is integrated over $\vartheta$ and averaged over $\varphi$.

Let us consider the phase space. The two-body phase space is given by
\begin{align}
  \dd\phi_2(k_1+k_2;p,\bar p)
  &= \theta(\hat s-4m^2)\frac{\dd^4p}{\left(2\pi\right)^3}\frac{\dd^4\bar p}{\left(2\pi\right)^3}
  \delta\left(p^2-m^2\right) \nn
  &\: \delta\left(\bar p^2-m^2\right)\left(2\pi\right)^4\delta^{(4)}\left(k_1+k_2-p-\bar p\right) \theta(p^0) \theta(\bar p^0) \nonumber \\
  &=\theta(\hat s-4m^2)\frac{\dd^4p}{4\pi^2}\delta\(p^2-m^2\)\delta\left((k_1+k_2-p)^2-m^2\right)
    \theta(p^0) \theta(k_1^0+k_2^0-p^0),
\end{align}
with $\hat s = (k_1+k_2)^2$.
We need to express this phase space in terms of the new variables.
The variable $\hat s$ is given in Eq.~\eqref{eq:shatnewvar},
the integration element can be written as
\begin{equation}
\dd^4p = \frac{Q^2}4 \,\dd Q^2\,\dd \eta\,\dd\pthat^2\,\dd\vartheta,
\end{equation}
and the antiquark momentum squared is
\begin{align}\label{eq:pp2}
\bar p^2 &= (k_1+k_2-p)^2 \nonumber\\
&= (1-z_1)(1-z_2)x_1x_2s-|\vec{k}_1+\vec{k}_2-\pvec|^2 \nonumber\\
&= Q^2\bigg[1+\frac1z-\sqrt{\frac{1+\pthat^2}{z}}(\eu^{\eta}+\eu^{-\eta})-\xi_1-\xi_2-2\sqrt{\xi_1\xi_2}\cos\varphi
\nonumber\\ &\qquad\qquad
+2\sqrt{\(\xi_1+\xi_2+2\sqrt{\xi_1\xi_2}\cos\varphi\)\pthat^2}\cos\vartheta\bigg],
\end{align}
where we have used the inverse relations
\begin{equation}\label{eq:z12pty}
z_1 = \sqrt{z(1+\pthat^2)}\eu^{\eta},\qquad
z_2 = \sqrt{z(1+\pthat^2)}\eu^{-\eta}.
\end{equation}
The conditions imposed by the two theta functions in the energies translate easily into conditions on $z_1$ and $z_2$
that depend on $x_1$ and $x_2$, namely $z_1x_1+z_2x_2\geq0$ and $(1-z_1)x_1+(1-z_2)x_2\geq0$.
From the on-shell conditions Eq.~\eqref{eq:onshellppbar1} we also know that $z_1z_2x_1x_2\geq0$ and $(1-z_1)(1-z_2)x_1x_2\geq0$.
Because $x_1$ and $x_2$ are positive, it follows that $z_1$ and $z_2$ satisfy the conditions $0\leq z_{1,2}\leq1$,
that translate into
\begin{equation}\label{eq:z12condition}
z(1+\pthat^2)\leq \eu^{-2|\eta|}.
\end{equation}
After the trivial integration over $Q^2$, the phase space can thus be recast as
\begin{align}
\dd\phi_2
  &= \theta\(\frac1z-\xi_1-\xi_2-2\sqrt{\xi_1\xi_2}\cos\varphi-4\)\, \theta\(\frac1z-(1+\pthat^2)\eu^{2|\eta|}\)\,
    \frac1{16\pi^2}\,\dd \eta\,\dd\pthat^2\,\dd\vartheta\\
    &\times\delta\bigg(\frac1z-\sqrt{\frac{1+\pthat^2}{z}}(\eu^{\eta}+\eu^{-\eta}) \nn
    &\: -\xi_1-\xi_2-2\sqrt{\xi_1\xi_2}\cos\varphi+2\sqrt{\(\xi_1+\xi_2+2\sqrt{\xi_1\xi_2}\cos\varphi\)\pthat^2}\cos\vartheta\bigg). \nonumber
\end{align}
To simplify the notation, we introduce the function
\begin{equation}
\xi(\xi_1,\xi_2,\varphi) = \xi_1+\xi_2+2\sqrt{\xi_1\xi_2}\cos\varphi = |\vec{k}_1+\vec{k}_2|^2,
\end{equation}
and simply write $\xi$ without arguments for short.
Putting everything together we have
\begin{align}
\frac{Q^2\dd{\cal C}}{\dd \eta\, \dd\pt{}^2}(z,\xi_1,\xi_2,m^2, \eta,\pthat^2)
&= \sigma_0\frac{1}{2} \int_0^{2\pi}\frac{\dd\varphi}{2\pi} \int \frac{\dd\phi_2}{\dd \eta\, \dd\pthat^2}
\;|{\cal M}|^2 \\
&=\frac{\sigma_0}{32\pi^2} \int_0^{2\pi}\frac{\dd\varphi}{2\pi}\, \theta\(\frac{1}{z}-\xi-4\)\, \theta\(\frac{1}{z}-(1+\pthat^2)\eu^{2|\eta|}\) \int_0^{2\pi}\dd\vartheta
\;|{\cal M}|^2 \nonumber\\
&\quad \times\delta\(\frac{1}{z}-\sqrt{\frac{1+\pthat^2}{z}}(\eu^{\eta}+\eu^{-\eta})-\xi+2\sqrt{\xi\pthat^2}\cos\vartheta \),\nonumber
\end{align}
where in the first line $1/2$ is the flux factor, $\sigma_0=16\pi^2\as^2/Q^2$ and the $1/2\pi$ comes from the average over $\varphi$.
The matrix element squared $|{\cal M}|^2$ is given in Appendix~\ref{app:ME}.

It is most convenient to use the $\delta$ function to integrate over $\vartheta$,
as all other variables appear at least quadratically.
The fact that $|\cos\vartheta|\leq1$ produces the constraint
\begin{align}
\left|\frac1z-\sqrt{\frac{1+\pthat^2}{z}}(\eu^{\eta}+\eu^{-\eta})-\xi\right|
&\leq 2\sqrt{\xi\pthat^2}.
\end{align}
We then get
\begin{align}\label{eq:dCoffFinSQ}
\frac{Q^2\dd{\cal C}}{\dd \eta\, \dd\pt{}^2}(z,\xi_1,\xi_2,m^2, \eta,\pthat^2)
  &=\frac{\sigma_0}{32\pi^2} \int_0^{2\pi}\frac{\dd\varphi}{2\pi}\,
    \theta\(\frac{1}{z}-\xi-4\)
  \, \theta\(\frac{1}{z}-(1+\pthat^2)\eu^{2|\eta|}\)
\nonumber\\
&\qquad \times\theta\(2\sqrt{\xi\pthat^2}-\left|\frac1{z}-\sqrt{\frac{1+\pthat^2}{z}(\eu^{\eta}+\eu^{-\eta})-\xi}\right|\)
\nonumber\\
&\qquad \times\frac{|{\cal M}|^2_{\vartheta=\bar\vartheta}+|{\cal M}|^2_{\vartheta=2\pi-\bar\vartheta}}{\sqrt{4\xi\pthat^2-\(\frac{1}{z}-\sqrt{\frac{1+\pthat^2}{z}}(\eu^{\eta}+\eu^{-\eta})-\xi\)^2}} 
,\\
\bar\vartheta &= \cos^{-1}\frac{\xi-\frac{1}{z}+\sqrt{\frac{1+\pthat^2}{z}}(\eu^{\eta}+\eu^{-\eta})}{2\sqrt{\xi\pthat^2}},\qquad 0\leq\vartheta\leq\pi.
\end{align}
The theta functions in Eq.~\eqref{eq:dCoffFinSQ} may prove troublesome from a numerical point of view.
Indeed, if used as ``if'' conditions that set the integrand to zero when the theta functions are zero,
the numerical integration may become inaccurate.
It is much more convenient to translate them into integration limits of some variable.
To do so, we define
\begin{equation}
X = \frac1{\sqrt{z}} \geq 1
\end{equation}
so that the constraint imposed by the three theta functions become
\begin{subequations}
\begin{align}
  &X\geq\sqrt{4+\xi},\\
  &X\geq\sqrt{1+\pthat^2}\eu^{|\eta|},\label{eq:condition21}\\
  &-2\sqrt{\xi}\pthat\leq -X^2+2BX+\xi\leq 2\sqrt{\xi}\pthat, \qquad B \equiv \sqrt{1+\pthat^2}\cosh \eta \geq 1.
\end{align}
\end{subequations}
Focussing on $\xi$, we may write
\begin{subequations}
\begin{align}
  &\xi\leq X^2-4, \label{eq:condition11}\\
  &\xi+2\pthat\sqrt{\xi} + 2BX-X^2\geq0, \label{eq:condition12}\\
  &\xi-2\pthat\sqrt{\xi} + 2BX-X^2\leq0. \label{eq:condition13}
\end{align}
\end{subequations}
The functions $\xi\pm 2\pthat\sqrt{\xi} + 2BX-X^2$ represent two parabolae in $\sqrt{\xi}$ with centres (minima) in $\sqrt{\xi}=\mp\pthat$,
at which they both equal $-\pthat^2+2BX-X^2$. If this value is positive, there is no solution to the system,
so we have the condition
\begin{equation}\label{eq:condition22}
\pthat^2-2BX+X^2 \geq0
\end{equation}
that represents an equation for the other variables to be taken into account later, together with Eq.~\eqref{eq:condition21}.
Under this condition, the solution of the inequalities Eq.~\eqref{eq:condition12}, \eqref{eq:condition13} is the region
between the two right solution of the second inequality and the largest between the right solution of the first and the left solution of the second,
which are identical but have opposite sign. Thus we get
\begin{equation}\label{eq:conditionxifinal}
\boxed{\left|\pthat-\sqrt{\pthat^2-2BX+X^2}\right| \leq \sqrt{\xi} \leq \pthat+\sqrt{\pthat^2-2BX+X^2}.}
\end{equation}
The other condition Eq.~\eqref{eq:condition11} is always
automatically satisfied. Indeed, we can prove that
\begin{equation}
\sqrt{X^2-4} \geq \pthat+\sqrt{\pthat^2-2BX+X^2}
\end{equation}
for all meaningful values of $X$ (namely values for which the square roots are real).
Indeed this condition can be manipulated to
\begin{equation}
(B^2-\pthat^2)X^2-4BX+4(1+\pthat^2)\geq0,
\end{equation}
which is always satisfied because the minimum of the quadratic function, located at $X=2B/(B^2-\pthat^2)$, is always non-negative.
Indeed the minimum is proportional to $B^2-1-\pthat^2$ which is non-negative because $B^2\geq 1+\pthat^2$.
Therefore, Eq.~\eqref{eq:conditionxifinal} is the complete condition on $\xi$.

We now focus on the other variables, that must satisfy the inequalities
Eq.~\eqref{eq:condition21} and \eqref{eq:condition22}.
Let us focus on Eq.~\eqref{eq:condition22}, solving it for $X$. The parabola $X^2-2BX+\pthat^2$ has a minimum in $X=B$
where it equals $\pthat^2-B^2$. This is always negative, as by construction $B^2\geq1+\pthat^2>\pthat^2$. Therefore,
there are two separate solutions, $X\geq B+\sqrt{B^2-\pthat^2}$ and $X\leq B-\sqrt{B^2-\pthat^2}$.
However, since we always have
\begin{equation}
\sqrt{1+\pthat^2}\eu^{|\eta|} \geq B,
\end{equation}
the second solution is not compatible with Eq.~\eqref{eq:condition21}, and it is therefore forbidden.
We are thus left with the condition
\begin{align}\label{eq:condition22new}
  \boxed{X\geq B+\sqrt{B^2-\pthat^2}} \geq B+1\geq2,
\end{align}
together with Eq.~\eqref{eq:condition21}.
We can show that Eq.~\eqref{eq:condition21} is always compatible with Eq.~\eqref{eq:condition22new}.
Indeed the inequality
\begin{equation}
B+\sqrt{B^2-\pthat^2} \geq \sqrt{1+\pthat^2}\eu^{|\eta|}
\end{equation}
holds because we can manipulate it into
\begin{equation}
\sqrt{B^2-\pthat^2} \geq \sqrt{1+\pthat^2}\(\eu^{|\eta|}-\cosh\eta\) = \sqrt{1+\pthat^2}\sinh|{\eta}|
\end{equation}
and then, squaring both sides (which are both positive) and rearranging,
\begin{equation}
  (1+\pthat^2)\(\cosh^2|{\eta}|-\sinh^2|{\eta}|\)-\pthat^2\geq0 \qquad\Rightarrow\qquad 1+\pthat^2-\pthat^2\geq0
\end{equation}
which is clearly true. In conclusion, $X$ satisfies only the inequality Eq.~\eqref{eq:condition22new}
which automatically encodes all the others.

It is useful to mention also the conditions on the kinematic limits of the on-shell resummed coefficient,
as well as on the integration variables defining the resummed result.
From Eq.~\eqref{eq:resCreg}, recalling that the first argument of the evolution function
is a momentum fraction and is thus smaller than 1, we obtain the condition $x/z\leq \eu^{-2|\bar\eta|}$.
Similarly, from Eq.~\eqref{eq:z12condition} we also have $z(1+\pthat^2)\leq \eu^{-2|\eta|}$.
From the product of the two inequalities, we obtain the condition
\begin{equation}
A^2\equiv x(1+\pthat^2) \leq \eu^{-2|\eta|-2|\bar\eta|} \leq \eu^{-2|\eta+\bar\eta|} = \eu^{-2|y|}.
\end{equation}
The condition $A\leq \eu^{-|y|}$, with $A\equiv\sqrt{x(1+\pthat^2)}$, represents a constraint on the arguments
of the on-shell coefficient function. However, this is not the most stringent one.
Indeed, looking at the first inequality, we can derive the integration range of $\bar\eta$, which is given by
\begin{equation}
\frac{A\eu^y-x\pthat^2}{1-A\eu^{-y}} \leq \eu^{2\bar\eta} \leq \frac {1-A\eu^y}{A\eu^{-y}-x\pthat^2}.
\end{equation}
For this range to be non-trivial, the upper limit must be larger than the lower limit, leading to
the condition
\begin{equation}
\eu^{|y|}\leq\frac{1+x\pthat^2}{2A}+\sqrt{\frac{(1+x\pthat^2)^2}{4A^2}-1},
\end{equation}
which is smaller than 1/A in the region where the square root exists,
given by the condition $\pthat^2\leq\frac{1-2\sqrt x}{x}$ or, equivalently,
\begin{equation}
x\leq\(\frac{\sqrt{1+\pthat^2}-1}{\pthat^2}\)^2\leq\frac14.
\end{equation}

To conclude, we recall that the matrix element squared that we will present in appendix~\ref{app:ME}
must be expressed in terms of the variables defined here.
To achieve this, we need to express $z_1,z_2$ in terms of $\pthat, \eta$ through Eq.~\eqref{eq:z12pty},
and to write the product $\vec{k}_2\cdot\pvec$ appearing in Eqs.~\eqref{eq:T} and \eqref{eq:U} as
\begin{equation}
\frac{\vec{k}_2\cdot\pvec}{Q^2} = \sqrt{\xi_2\pthat^2}\cos(\vartheta+\varphi'),
\end{equation}
where $\varphi'$ is the angle of $\vec{k}_1+\vec{k}_2$ with respect to $\vec{k}_2$,
which can be computed from the Cartesian representation (aligning the $x$ axis along $\vec{k}_2$)
\begin{align}
\qvec = \dvec{|\qvec|\cos\varphi'}{|\qvec|\sin\varphi'}
= \dvec{|\vec{k}_2|+|\vec{k}_1|\cos\varphi}{|\vec{k}_1|\sin\varphi},
\end{align}
leading to
\begin{align}
\sin\varphi' &= \frac{\sqrt{\xi_1}\sin\varphi}{\sqrt{\qthat^2}} &
\cos\varphi' &= \frac{\sqrt{\xi_2}+\sqrt{\xi_1}\cos\varphi}{\sqrt{\qthat^2}},
\end{align}
which gives the result
\begin{equation}\label{eq:varphi'}
\varphi' =
\begin{cases}
  \cos^{-1}\(\frac{\sqrt{\xi_2}+\sqrt{\xi_1}\cos\varphi}{\sqrt{\qthat^2}}\) & \text{if }\sin\varphi\geq0 \\
  2\pi-\cos^{-1}\(\frac{\sqrt{\xi_2}+\sqrt{\xi_1}\cos\varphi}{\sqrt{\qthat^2}}\) & \text{if }\sin\varphi<0.
\end{cases}
\end{equation}

\section{Kinematics for the pair}
\label{app:pairkin}

We now consider the heavy-quark pair as a fictitious intermediate state, with momentum
\begin{align}
q &\equiv p+\bar p \nonumber\\
  &\equiv \alpha_1x_1 P_1 + \alpha_2 x_2 P_2 + \qvec &&\text{(generic parametrisation)}\nonumber\\
  &= k_1+k_2 &&\text{(momentum conservation)}\nonumber\\
  &= x_1 P_1 + x_2 P_2 + \vec{k}_1+ \vec{k}_2,
\end{align}
where by momentum conservation $\alpha_1=\alpha_2=1$ and $\qvec = \vec{k}_1+\vec{k}_2$.
For this intermediate state, we introduce the variables\footnote
{Note that we are using the same names $Q^2$ and $\eta$ that we used for the single quark kinematics, now referring to another momentum.}
\begin{subequations}\label{eqP:finalstatevars}
  \begin{align}
    Q^2 &\equiv q^2 = \alpha_1\alpha_2x_1x_2s-|\qvec|^2 = x_1x_2s-|\vec{k}_1+\vec{k}_2|^2 = (k_1+k_2)^2 \equiv \hat s, \\
    \eta &\equiv \frac12\log\frac{q^0+q^3}{q^0-q^3}-\bar\eta =\frac12\log\frac{\alpha_1}{\alpha_2}=0
        \qquad\text{(rapidity of $q$ in the partonic frame)}, \\
    \qthat^2 &\equiv \frac{\qvec^2}{Q^2}= \frac{|\vec{k}_1+\vec{k}_2|^2}{Q^2}, \\
    \thetapair &\equiv \text{angle between $\qvec$ and $\vec{k}_1+\vec{k}_2$} = 0.
  \end{align}
\end{subequations}
Our goal is to compute the parton-level off-shell coefficient function ($\qt^2=\qthat^2Q^2$)
\begin{equation}\label{eqP:triplediffxs}
\frac{\dd{\cal C}}{\dd Q^2\,\dd \eta\, \dd\qt^2}(z,\xi_1,\xi_2,Q^2,\eta,\qthat^2),
\end{equation}
which is integrated over $\thetapair$ and averaged over $\varphi$.

Let us consider the phase space.
The two-body phase space of the two final-state heavy quarks can be factorised
into the phase space of the pair and its ``decay'' as
\begin{align}\label{eqP:PS}
\dd\phi_2(k_1+k_2;p,\bar p) &= \theta(\hat s-4m^2)\int_{4m^2}^{\hat s}\frac{\dd q^2}{2\pi}\, \dd\phi_1(k_1+k_2;q) \, \dd\phi_2(q;p,\bar p),
\end{align}
where
\begin{align}
\dd\phi_1(k_1+k_2;q)&= \frac{\dd^4q}{\(2\pi\)^3}\delta\(q^2-\hat s\)\(2\pi\)^4\delta^{(4)}\(k_1+k_2-q\) \nonumber\\
&=2\pi\, \dd^4q\, \delta\(q^2-\hat s\) \delta^{(4)}\(k_1+k_2-q\),
\end{align}
with $\hat s=(k_1+k_2)^2$ the invariant mass of the pair, and
\begin{align}\label{eq:twobody}
\dd\phi_2(q;p,\bar p)&= \frac{\dd^4p}{\(2\pi\)^3}\frac{\dd^4\bar p}{\(2\pi\)^3}\delta\(p^2-m^2\)\delta\(\bar p^2-m^2\)\(2\pi\)^4\delta^{(4)}\(q-p-\bar p\) \theta(p^0) \theta(\bar p^0).
\end{align}
The full phase space Eq.~\eqref{eqP:PS} can be simplified using the delta function
of the one-body phase space to perform the $q^2$ integral, giving
\begin{align}
  \dd\phi_2(k_1+k_2;p,\bar p)
  &= \theta(\hat s-4m^2)\, \dd^4q \,\delta^{(4)}\(k_1+k_2-q\)\, \dd\phi_2(q;p,\bar p)\nonumber\\
  &= \theta(\hat s-4m^2)\, \dd Q^2\, \dd \eta\, \dd\qthat^2\, \dd\thetapair\, \dd\phi_2(q;p,\bar p)\nonumber\\
  &\qquad\times\delta\(Q^2-\hat s\)\, \delta(\eta)\, \delta\(\qthat^2-\xi_1-\xi_2-2\sqrt{\xi_1\xi_2}\cos\varphi\)\, \delta(\thetapair) \nonumber\\
  &= \theta(Q^2-4m^2)\, \dd Q^2\, \dd \eta\, \dd\qthat^2\, \dd\thetapair\, \dd\phi_2(q;p,\bar p)\nonumber\\
  &\qquad\times\frac1{Q^2}\delta\(1+\qthat^2-\frac1z\)\, \delta(\eta)\, \delta\(\qthat^2-\xi_1-\xi_2-2\sqrt{\xi_1\xi_2}\cos\varphi\)\, \delta(\thetapair),
\end{align}
where we have rewritten $\dd^4q$, the delta function and $\hat s$ in terms of the new variables.

The two-body phase space can be used to integrate the matrix element and remove the ``internal'' degrees of freedom of the pair,
while the one-body phase space can be used to obtain the desired differential observable.
Thus, we immediately find the relation
\begin{align}\label{eqP:triplediffxs2}
  \frac{Q^4\,\dd{\cal C}}{\dd Q^2\,\dd \eta\, \dd\qt^2\, \dd\varphi}(z,\xi_1,\xi_2,Q^2,\eta,\qthat^2,\varphi)
  &= \frac{\dd{\cal C}}{\dd\varphi}(z,\xi_1,\xi_2,Q^2,\varphi)\nonumber\\
  &\times\delta\(1+\qthat^2-\frac1z\)\, \delta(\eta)\, \delta\(\qthat^2-\xi_1-\xi_2-2\sqrt{\xi_1\xi_2}\cos\varphi\),
\end{align}
where we had to include also the explicit dependence on $\varphi$ as it appears in the delta function.
This result expresses the fully differential distribution in terms of the distribution
differential only in the angle $\varphi$ between $\vec{k}_1$ and $\vec{k}_2$.
The key object that we need is thus
\begin{equation}\label{eq:dCoffdphi}
\frac{\dd{\cal C}}{\dd\varphi}(z,\xi_1,\xi_2,Q^2,\varphi) = \theta(Q^2-4m^2)
\frac12\frac1{2\pi} \sigma_0 \int \dd\thetapair\,\delta(\thetapair)\int \dd\phi_2(q;p,\bar p)\;|{\cal M}|^2,
\end{equation}
where $\sigma_0=16\pi^2\as^2/Q^2$, the factor $1/2$ is the flux factor and the $1/2\pi$ comes from the $\varphi$ average.
The matrix element squared $|{\cal M}|^2$ is given in Appendix~\ref{app:ME}.
Note that because of the delta functions in Eq.~\eqref{eqP:triplediffxs2}
not all the variables of Eq.~\eqref{eq:dCoffdphi} are independent. In particular one can
write $1/z=1+\xi_1+\xi_2+2\sqrt{\xi_1\xi_2}\cos\varphi$ and use it to fix one of them in terms of the others.

We now focus on the computation of $\dd{\cal C}/\dd\varphi$. We observe that
the two-body phase space Eq.~\eqref{eq:twobody} contains two delta functions corresponding to the mass shell condition of the heavy quarks.
We write them in terms of the new variables, and get
\begin{subequations}\label{eq:onshellppbar}
\begin{align}
0=p^2-m^2 &= z_1z_2\frac{Q^2}{z}-|\pvec|^2-m^2, \\
0=\bar p^2-m^2 &= (1-z_1)(1-z_2)\frac{Q^2}{z}-|\qvec-\pvec|^2-m^2\nonumber\\
%&= (1-z_1)(1-z_2)\frac{Q^2}{z}-|\qvec|^2-|\pvec|^2+2\qvec\cdot\pvec-m^2\nonumber\\
&= (1-z_1-z_2)\frac{Q^2}{z}-|\qvec|^2+2\qvec\cdot\pvec,
\end{align}
\end{subequations}
where in the last step we have used the first on-shell condition.
The second condition contains a scalar product, and thus an angle, which is not ideal as this appears in the argument of the delta function.
In order to get rid of the scalar product, we use the first condition to fix $z_2$, through the equation
\begin{equation}
z_2 = z\frac{|\pvec|^2+m^2}{z_1Q^2},
\end{equation}
so that the second condition becomes
\begin{align}
0=\bar p^2-m^2
&= (1-z_1)\frac{Q^2}{z}-\frac{|\pvec|^2+m^2}{z_1}-|\qvec|^2+2\qvec\cdot\pvec.
\end{align}
We can now get rid of the scalar product by introducing a new vector $\Dvec$ defined by
\begin{equation}
\pvec = z_1\qvec+\Dvec,
\end{equation}
so that
\begin{align}
0=\bar p^2-m^2
&= (1-z_1)\frac{Q^2}{z}-\frac{|z_1\qvec+\Dvec|^2+m^2}{z_1}-|\qvec|^2+2z_1|\qvec|^2+2\qvec\cdot\Dvec\nonumber\\
&= (1-z_1)\(\frac{Q^2}{z}-|\qvec|^2\)-\frac{|\Dvec|^2+m^2}{z_1}\nonumber\\
&= (1-z_1)Q^2-\frac{|\Dvec|^2+m^2}{z_1},
\end{align}
that only depends on squared vectors (in the last step we have used $1+\qthat^2=\frac1z$).
This can be now used to fix
\begin{equation}
|\Dvec|^2 = z_1 (1-z_1)Q^2 -m^2.
\end{equation}
The two-body phase space can thus be rewritten as
\begin{align}\label{eq:dphi2}
  \dd\phi_2(q;p,\bar p)
  &= \frac{\dd^4p}{\(2\pi\)^3}\frac{\dd^4\bar p}{\(2\pi\)^3}\delta\(p^2-m^2\)\delta\(\bar p^2-m^2\)\(2\pi\)^4\delta^{(4)}\(q-p-\bar p\) \theta(p^0) \theta(\bar p^0) \nonumber \\
  &=\frac{\dd^4p}{4\pi^2}\delta\(p^2-m^2\) \delta\((q-p)^2-m^2\) \theta(p^0) \theta(q^0- p^0)\nonumber\\
  &=\frac{Q^2}{8\pi^2z}\delta\(z_1z_2\frac{Q^2}{z}-|\pvec|^2-m^2\) \delta\((1-z_1-z_2)\frac{Q^2}{z}-|\qvec|^2+2\qvec\cdot\pvec\) \dd z_1\, \dd z_2\, \dd^2\pvec  \nonumber\\
  &\quad\times\theta(z_1)\theta(z_2)\theta(1-z_1)\theta(1-z_2)  \nonumber\\
&=\frac1{8\pi^2} \delta\((1-z_1)Q^2-\frac{|\Dvec|^2+m^2}{z_1}\) \theta(z_1)\theta(1-z_1)\frac{\dd z_1}{z_1}\, \dd^2\Dvec  \nonumber\\
&=\frac1{16\pi^2} \theta\(z_1 (1-z_1)Q^2 -m^2\) \dd z_1\, \dd\omega \nonumber\\
&=\frac1{16\pi^2} \theta\(\sqrt{\frac14-\frac{m^2}{Q^2}}-|{\frac12-z_1}|\) \dd z_1\, \dd\omega \nonumber\\
&=\frac1{16\pi^2} \sqrt{\frac14-\frac{m^2}{Q^2}} \sin\beta\, \dd\beta\, \dd\omega,
\end{align}
where $\omega$ is the azimuthal angle of $\Dvec$ with respect to $\vec{k}_1+\vec{k}_2$.
Note that the condition $Q^2>4m^2$, needed to satisfy the theta function, is always verified in Eq.~\eqref{eq:dCoffdphi}.
If we wish to compute the $z_1$ integral numerically, it is convenient to change variable as
\begin{equation}\label{eq:z1beta}
z_1 = \frac12-\sqrt{\frac14-\frac{m^2}{Q^2}}\cos\beta,
\qquad \beta\in[0,\pi],
\end{equation}
which we used to obtain the last line of Eq.~\eqref{eq:dphi2}. Interestingly, in terms of these variables $|\Dvec|^2$ becomes
\begin{equation}
|\Dvec|^2 = \frac{Q^2 -4m^2}4\sin^2\beta \qquad\Rightarrow\qquad
|\Dvec| = \frac12\sqrt{Q^2 -4m^2}\sin\beta,
\end{equation}
where we do not need to include an absolute value, as in the allowed range $\sin\beta$ is always positive.

The form of the phase space Eq.~\eqref{eq:dphi2} is very convenient from a numerical point of view.
To be able to perform all integrations, we also need to express the matrix element squared appearing
in Eq.~\eqref{eq:dCoffdphi} in terms of the variables $\beta$ (or $z_1$) and $\omega$.
We start by rewriting
\begin{align}
z_2 &= z\frac{|z_1\qvec+\Dvec|^2+m^2}{z_1Q^2} \nonumber\\
&= z\frac{z_1^2|\qvec|^2+|\Dvec|^2+2z_1\qvec\cdot\Dvec+m^2}{z_1Q^2} \nonumber\\
&= z\[1-z_1(1-\qthat^2)+2\sqrt{\qthat^2}\sqrt{z_1 (1-z_1) -m^2/Q^2}\cos(\omega-\thetapair)\] \nonumber\\
&= z\[1-(1-\qthat^2)\(\frac12-\sqrt{\frac14-\frac{m^2}{Q^2}}\cos\beta\)+\sqrt{\qthat^2}\sqrt{1-4m^2/Q^2}\sin\beta\cos(\omega-\thetapair)\],
\end{align}
where $\thetapair=0$ for our choice of variables, Eq.~\eqref{eqP:finalstatevars}.
Finally, we will see in appendix~\ref{app:ME} that the matrix element depends on the scalar product $\vec{k}_2\cdot\pvec$
through the variables $T$ Eq.~\eqref{eq:T} and $U$ Eq.~\eqref{eq:U}.
We can write
\begin{align}
  \frac{\vec{k}_2\cdot\pvec}{Q^2}
  &=  \frac{\vec{k}_2\cdot(z_1\qvec+\Dvec)}{Q^2} \nonumber\\
  &=  \frac{z_1(\vec{k}_2^2+ \vec{k}_2\cdot\vec{k}_1)+\vec{k}_2\cdot\Dvec}{Q^2} \nonumber\\
  &=  z_1\xi_2+ z_1\sqrt{\xi_1\xi_2}\cos\varphi+\sqrt{\xi_2}\sqrt{\frac14-\frac{m^2}{Q^2}}\sin\beta\cos\omega',
\end{align}
where $\omega'$ is the angle between $\Dvec$ and $\vec{k}_2$. It is given by $\omega'=\omega+\varphi'$,
where $\varphi'$ is the angle of $\qvec=\vec{k}_1+\vec{k}_2$ with respect to $\vec{k}_2$, given in Eq.~\eqref{eq:varphi'}.
For the on-shell limit $\xi_2\to0$ it is also useful to write
\begin{align}
\pthat^2 &\equiv \frac{\pvec^2}{Q^2} = z_1^2\xi_1+\frac{|\Dvec|^2}{Q^2} + 2z_1 \sqrt{\xi_1}\frac{|\Dvec|}{Q}\cos\omega
\end{align}
in terms of the new phase-space variables.
In the fully on-shell limit the result simplifies further
\begin{align}
\pthat^2 &= \frac{|\Dvec|^2}{Q^2} = z_1(1-z_1)-\frac{m^2}{Q^2}.
\end{align}

\section{Matrix element}
\label{app:ME}

In this Appendix we report the matrix element squared for heavy quark pair production
from two off-shell gluons. This has been computed in Refs.~\cite{Catani:1990eg,Ball:2001pq}.
Here, we rewrite that result in terms of the variables that we have defined above.

The matrix element is separated into an Abelian and a non-Abelian parts as
\begin{equation}
  |{\cal M}|^2 = \frac1{2C_A}|{\cal M}|^2_{\rm ab} + \frac1{4C_F}|{\cal M}|^2_{\rm nab}
\end{equation}
with
\begin{equation}\label{eq:Mab}
|{\cal M}|^2_{\rm ab} = \frac1{z^2}\[\frac1{TU} - \frac1{\xi_1\xi_2}\(1+\frac{z_2(1-z_1)}{zT}+\frac{z_1(1-z_2)}{zU}\)^2\]
\end{equation}
and
\begin{align}\label{eq:Mnab}
  |{\cal M}|^2_{\rm nab}
  &= \frac1{z^2}\bigg[-\frac1{TU}+ \frac{2z}S +\frac{(T-U)(z_1-z_2)}{STU} \nonumber\\
  &\qquad\qquad+ \frac2{\xi_1\xi_2}
  \(\frac12+\frac{z_2(1-z_1)}{zT}-\frac\Delta S\)
  \(\frac12+\frac{z_1(1-z_2)}{zU}+\frac\Delta S\)
\bigg],
\end{align}
where
\begin{equation}
\Delta = \frac{z_1(1-z_2)}z - \frac{z_2(1-z_1)}z + \xi_1z_2-\xi_2z_1+\frac{z_2-z_1}{2z}+\frac{\pvec\cdot(\vec{k}_2-\vec{k}_1)}{Q^2}
\end{equation}
and
\begin{subequations}
\begin{align}
  S=\frac{\hat s}{Q^2} &= \frac{(k_1+k_2)^2}{Q^2} = \frac1z-\xi_1-\xi_2-2\sqrt{\xi_1\xi_2}\cos\varphi, \label{eq:S}\\
  T= \frac{t-m^2}{Q^2} &= \frac{(p-k_1)^2-m^2}{Q^2} = \frac{2\vec{k}_1\cdot\pvec}{Q^2} - \xi_1-\frac{z_2}z\\
                       &= \frac{(\bar p-k_2)^2-m^2}{Q^2} = -\frac{2\vec{k}_2\cdot\pvec}{Q^2} + \xi_2+2\sqrt{\xi_1\xi_2}\cos\varphi-\frac{1-z_1}z,\label{eq:T}\\
  U= \frac{u-m^2}{Q^2} &= \frac{(p-k_2)^2-m^2}{Q^2} = \frac{2\vec{k}_2\cdot\pvec}{Q^2} - \xi_2-\frac{z_1}z\label{eq:U}\\
                       &= \frac{(\bar p-k_1)^2-m^2}{Q^2} = -\frac{2\vec{k}_1\cdot\pvec}{Q^2} + \xi_1+2\sqrt{\xi_1\xi_2}\cos\varphi-\frac{1-z_2}z.
\end{align}
\end{subequations}
Note that in the case of the pair kinematics, where $q=p+\bar p$, we have $\hat s=Q^2$ and thus $S=1$.
We can use the expressions of $T$ and $U$ to rewrite
\begin{equation}
\frac{\pvec\cdot(\vec{k}_2-\vec{k}_1)}{Q^2} = \frac12\[U-T+\xi_2-\xi_1+\frac{z_1-z_2}z\],
\end{equation}
so that $\Delta$ simplifies to
\begin{equation}\label{eq:Deltadef}
\Delta = \frac{z_1-z_2}z + \xi_1z_2-\xi_2z_1+\frac{U+\xi_2-T-\xi_1}2.
\end{equation}
We also recall the relation
\begin{equation}
\hat s+t+u = 2m^2-|\vec{k}_1|^2-|\vec{k}_2|^2,
\end{equation}
namely
\begin{equation}
S+T+U+\xi_1+\xi_2 = 0.
\end{equation}
Thus, one can always express one of these variables in terms of the other four.

Note that the matrix element squared is symmetric under the simultaneous exchange
\begin{equation}
\vec{k}_1\leftrightarrow\vec{k}_2
,\qquad
z_1\leftrightarrow z_2.
\end{equation}
This implies that, in the pair kinematics, after integrating over the two $z_1$ and $z_2$ variables (which appear symmetrically in the phase space)
the off-shell coefficient is symmetric under the exchange of the two gluon virtualities.
Similarly, in the single-quark kinematics, the off-shell coefficient is symmetric under the exchange of the two gluon virtualities
and a sign change in the rapidity $\eta$.

\chapter{Computing the NLO off-shell coefficient function for HDIS}
\label{app:HDIS-NLO}

\section{Real corrections}
\begin{figure}
\centering
\begin{subfigure}{0.4\linewidth}
\centering
\includegraphics[page = 5]{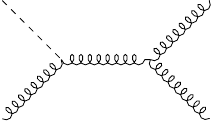}
\caption{}
\label{2:fig:NLOdias}
\end{subfigure}
\begin{subfigure}{0.4\linewidth}
\centering
\includegraphics[page = 3]{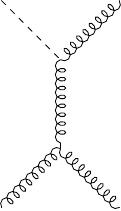}
\caption{}
\label{2:fig:NLOdiat}
\end{subfigure}\\
\begin{subfigure}{0.4\linewidth}
\centering
\includegraphics[page = 4]{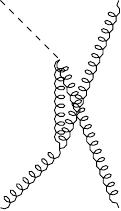}
\caption{}
\label{2:fig:NLOdiau}
\end{subfigure}
\caption{Diagrams entering the real correction of $g^* H \rightarrow g$ at NLO }
\label{2:fig:NLOdia}
\end{figure}
Several different contribution enter the real corrections in $A_1$, the contact interaction in Fig. ~\ref{2:fig:NLOcdia} plus the three diagrams of Fig. ~\ref{2:fig:NLOdia}. The amplitude of the former is reported in Eq. ~\ref{2:eq:EFF3vertex} for the first three are written as
\begin{subequations}\label{2:eq:realsNLOamplitudes} 
\begin{align}
A^{\mu \nu \rho}_{abc} & = V^{\mu \rho \theta}_{a,b,d}\(k, -p_3, p_3-k\) \frac{\iu \delta_{de} d_{\theta \eta}\(p_3-k, n\) }{t} M^{\eta \nu}_{ec}\(k-p_3, p_4\)\, , \label{2:eq:tch}\\
B^{\mu \nu \rho}_{abc} & = V^{\mu \nu \theta}_{a,c,d}\(k, -p_4, p_3-q\) \frac{\iu \delta_{de} d_{\theta \eta}\(q-p_3, n\) }{u} M^{\eta \rho}_{eb}\(p_3-q, p_3\)\, , \label{2:eq:uch}\\
C^{\mu \nu \rho}_{abc} & = M^{\mu \theta}_{ad}\(k, k+q\) \frac{\iu \delta_{de} d_{\theta \eta} \(k+q, n\)}{s} V^{\eta \nu \rho}_{ebc}\(k+q, -p_4, -p_3\)\, , \label{2:eq:sch}\\
%	D^{\mu\nu\rho}_{abc} & = \frac{C}{g_s} V^{\mu \rho \nu }_{abc}\(k, -p_3, -p_4\) \, , 
\end{align}
\end{subequations}
where $V$ is the usual trilinear gluon vertex and we explicitly write the polarisation tensor $d$ in the LCG gauge
\begin{align}
&V^{\mu \nu \rho}\(k_1, k_2, k_3\)  = g_s f_{abc}\[g^{\mu \nu} \(k_1^\rho - k_2 ^\rho\) + g^{\nu \rho}\(k_2^{\mu}- k_3^{\mu}\) + g^{\rho \mu}\(k_3^\nu - k_1^\nu\)\] \, , \label{2:eq:QCD3vertex} \\
&d_{\mu \nu} \(k, n\)  = -g_{\mu\nu} + \frac{n_\mu k_\nu + k_\mu n_\nu}{\spd{k}{n}} \, \label{2:eq:dLCG}. %- n^2 \frac{k^\mu k^\nu}{\spd{k}{n}^2} 
\end{align}
As mentioned in Sec. \ref{2:sec:strategy}, we are interested in computing only a subset of squared amplitudes which do not factorised across $t,u$ propagator. In practice, this means discarding the modulus squared contributions from Eqs. ~\ref{2:eq:tch} and ~\ref{2:eq:uch}. Then we break down the remaining contributions to $A_1^{\mu\nu}$ from the real emissions in the following way
\begin{align}
A_{1\text{real}}^{\mu\nu} & = A_{AB\text{int}}^{\mu\nu} + A_{CAB\text{int}}^{\mu\nu} + A_{CE}^{\mu\nu} \nn
& = \int\dd\Phi_2\(k+q,p_3,p_4\) \lb \tilA_{AB\text{int}}^{\mu\nu} + \tilA_{CAB\text{int}}^{\mu\nu} + \tilA_{CE}^{\mu\nu} \rb \, , \label{2:eq:A1realdec}
\end{align}
\begin{subequations}
\begin{align}
& \tilA_{AB\text{int}}^{\mu\nu}  = \frac{\delta_{mn}}{2 C_F C_A}d_{\alpha\gamma}\(p_4,n\)d_{\beta\delta}\(p_3,n\)\[ A^{\mu \alpha \beta}_{mab}B^{*\nu \gamma \delta}_{ncd} + B^{\mu \alpha \beta}_{mab}A^{*\nu \gamma \delta}_{ncd}\] \label{2:eq:ABint}\\
& \tilA_{CAB\text{int}}^{\mu\nu} = \frac{\delta_{mn}}{2 C_F C_A}d_{\alpha\gamma}\(p_4,n\)d_{\beta\delta}\(p_3,n\)\nn
&\quad \[\(C^{\mu \alpha \beta}_{mab}+E^{\mu \alpha \beta}_{mab}\)\(A^{*\nu \gamma \delta}_{ncd}+B^{*\nu \gamma \delta}_{ncd}\)+ \(A^{\mu \alpha \beta}_{mab} + B^{\mu \alpha \beta}_{mab}\)\(C^{*\nu\gamma\delta}_{ncd}+E^{*\nu\gamma\delta}_{ncd}\)\] \label{2:eq:CABint}\\
& \tilA_{CE}^{\mu\nu}  = \frac{\delta_{mn}}{2 C_F C_A}d_{\alpha\gamma}\(p_4,n\)d_{\beta\delta}\(p_3,n\)\[\(C^{\mu \alpha \beta}_{mab} + E^{\mu \alpha \beta}_{mab}\)\(C^{*\nu\gamma\delta}_{ncd}+E^{*\nu\gamma\delta}_{ncd}\)\]  \, , \label{2:eq:CEsq}
\end{align}
\end{subequations}
so that the second line corresponds to the interference between $t$ and $u$ channels, the ones on the third to the interference between the $s$ channel and the $t,u$ and finally the ones in the last one to the squared amplitude of the sum of $s$ channel amplitude and effective contact vertex.
The reason of this separation is rooted in the choice of regulators for the infrared divergence that will appear after the integration over the two-body phase space, $\dd\Phi_2$, as it will become apparent shortly.

\subsection{Kinematics and phase space}
We repeat the momenta parametrisation for consistency
\begin{subequations}
\begin{align}
& k = z p_1 +  \kt \, , \\
& q = p_2 - x p_1 \, , \\
& p_3 = z_1 (z-x) p_1 + z_2  p_2 + p_t \, , \\
& p_4 = (1-z_1) (z-x) p_1 + (1-z_2) p_2 + p_t \, , \\
& n = p_2 \, ,
\end{align}
\end{subequations}
where again $p_{1,2}= \frac{\sqrt{S}}{2}\(1,0,0,\pm 1 \)$. The phase space constraints imply

\begin{equation}\label{2:eq:PSconstraints}
  0 \leq z_{1,2} \leq 1 \, ; \:
  \rho < z \leq 1 \, ; \:
  0 \leq \xi \leq \frac{z-x}{x} \, ; \:
  0 \leq \cos^{-1} \(\frac{-\spd{\kt}{p_t} }{ \sqrt{\kt {}^2 p_t^2} } \) \leq 2\pi \, . \nonumber
\end{equation}
We can write the Mandelstam variables in the propagators as
\begin{subequations}
\begin{align}
& t = \(k-p_3\)^2 = \(q-p_4\)^2 \nn
& \quad = Q^2\[ \(1-z_2\)-1 -\frac{z-x}{x}\(1-z_1\) \] \, , \label{2:eq:mand-t} \\
& u = \(k-p_4\)^2 = \(q-p_3\)^2 \nn
& \quad = Q^2\[ z_2-1 -\frac{z-x}{x}z_1 \] \, , \label{2:eq:mand-u} \\
& s = \(k+q\)^2 = \(p_3+p_4\)^2 \nn
& \quad = Q^2\[ \frac{z-x}{x} - \xi \] \, , 
\end{align}
\end{subequations}
As explained in section~\ref{2:ssec:NLO-strat}, for each term entering Eq.~\ref{2:eq:A1realdec} we carry out the following operations:
\begin{itemize}
\item[1.] we recast the tensor to eliminate any vector-like dependence on $p_3$,
\item[2.] we perform the phase space integration of Eq. \ref{2:eq:PSspacepar}, using a suitable combination of off-shellness, dimensional regularisation and PV-value prescription to isolate the singularities emerging from the physical infrared QCD structure and non-covariant choice of gauge.
\end{itemize}

Here, we repeat the complete set of projectors used to obtain the decomposition of $A_1^{\mu\nu}$:
\begin{subequations}
\begin{align}
& \int \dd \Phi_2\(k+q; p_3,p_4\) A_{\text{I}}^{\mu\nu}= C_{\text{I},i} T^{\mu\nu}_i, \nn
& C_{\text{I},i} = P_{i,\alpha\beta}\int  \tilA_{\text{I}}^{\alpha\beta} \dd\Phi_2\(k+q;p_3,p_4\), \nn
  \nn& T_1^{\mu\nu} = g^{\mu\nu}, \: T_2^{\mu\nu} = p_1^\mu p_1^\nu,\: T_3^{\mu\nu} = p_2^\mu p_2^\nu,\: T_4^{\mu\nu}=\kt^\mu \kt^\nu,\nn
& T_5^{\mu\nu}= p_1^\mu p_2^\nu + p_1^\nu p_2^\mu, \: T_5^{\mu\nu} = \kt^\mu p_2^\nu + \kt^\nu p_2^\mu,\: T_7^{\mu\nu} = p_1^\mu \kt^\nu + p_1^\nu \kt^\mu, \label{2:eq:tensors} \\
\nn& P_1^{\mu\nu} = \frac{1}{D-3}\[g^{\mu\nu} - \frac{\kt^\mu \kt^\nu}{\kt^2}- \frac{p_1^\mu p_2^\nu+p_1^\nu p_2^\mu}{\spd{p_1}{p_2}}\],\: P_2^{\mu\nu} = \frac{p_2^\mu p_2^\nu}{\spd{p_1}{p_2}^2}\,\: P_3^{\mu\nu} = \frac{p_1^\mu p_1^\nu}{\spd{p_1}{p_2}^2}, \nn
& P_4^{\mu\nu} = \frac{1}{(D-3) \kt^2}\[(D-2)\frac{kt^\mu\kt^\nu}{\kt^2} - g^{\mu\nu} + \frac{p_1^\mu p_2^\nu+ p_1^\nu p_2^\mu}{\spd{p_1}{p_2}}\], \nn
& P_5^{\mu\nu} = \frac{1}{(D-3)\spd{p_1}{p_2}}\[(D-1)\frac{p_1^\mu p_2^\nu + p_1^\nu p_2^\mu}{2\spd{p_1}{p_2}} + \frac{\kt^\mu\kt^\nu}{\kt^2} - g^{\mu\nu}\], \nn
& P_6^{\mu\nu} = \frac{\kt^\mu p_1^\nu + \kt^\nu p_1^\mu}{2 \kt^2 \spd{p_1}{p_2} }, \: P_7^{\mu\nu} = \frac{\kt^\mu p_2^\nu  + \kt^\nu p_2^\mu}{2 \kt^2 \spd{p_1}{p_2}}, \label{2:eq:proj}
\end{align}
\end{subequations}
with the label I standing in for any of the three amplitude contributions Eqs. \eqref{2:eq:ABint}, \eqref{2:eq:CABint} and \eqref{2:eq:CEsq}.
As already mentioned in section~\ref{2:ssec:NLO-strat}, this basis of tensors is more general then what we actually require, as the contraction in Eq.~\eqref{2:eq:tensorfact2} ensures that $T_{3,5,6}^{\mu\nu}$ are vanishing.

The phase space integral in dimensional regularisation instead is rewritten as
\begin{align}
\dd \Phi_2\(k+q;p_3,p_4\) &= \frac{1}{2}\dd^Dp_3 \dd^Dp_4 \(2\pi\)^{2-D}\delta^D\(k+q-p_3-p_4\)\delta\(p_3^2\)\delta\(p_4^2\)\vartheta\(p_3^0\) \vartheta\(p_4^0\) \nn
& = \dd z_1 \dd z_2 \frac{Q^2\(z-x\)}{4x}\dd^{D-2}\pb\(2\pi\)^{2-D}\delta\[Q^2\(z_1z_2\frac{z-x}{x}  -\frac{\pb^2}{Q^2}\)\] \nn
& \quad \times \delta\[Q^2\((1-z_1)(1-z_2)\frac{z-x}{x}  -|\kb-\pb|^2\)\] \nn
& \quad \times \vartheta\(z_1\(z-x\)+z_2 \)\vartheta\((1-z_1)(z-x)+(1-z_2)\) \label{2:eq:PSspacepar}\, .
\end{align}
Following Refs. ~\cite{Catani:1990eg,Ball:2001pq,Bonvini:2017ogt} the most convenient variable choice to integrate the delta functions is to use translation invariance of $\pb$ to have
\begin{equation}
\pb' = \pb - z_1 \kb \, ,
\end{equation}
which then allows to fix
\begin{subequations}
\begin{align}
&\xi_p = \frac{\pb^{\prime 2}}{Q^2} = z_1 (1-z_1)\[\frac{z-x}{x}  - \xi\] = z_1(1-z_1)\frac{s}{Q^2}\, , \\
& z_2 = \frac{x}{z_1(z-x)(1+\alpha)}\[\xi_p+z_1^2\xi+2z_1\sqrt{\xi \xi_p} \cos(\theta)\]\, .
\end{align}
\end{subequations}
After these manipulations and integrating over all angular degrees of freedom except $\theta$ the phase space looks like
\begin{equation}
\dd \Phi_2\(k+q;p_3,p_4\) = \dd z_1 \dd \theta  \frac{\(Q^2\)^{\frac{D-4}{2}}}{8} \frac{\pi^\frac{D-5}{2}}{\Gamma\(\frac{D-3}{2}\)} \[z_1(1-z_1)\frac{s}{Q^2}\]^{\frac{D-4}{2}}   \label{2:eq:last-int}\, ,
\end{equation}
where $D=4-2\epsilon$ as usual in dimensional regularisation.

After illustrating all this setup we can go in detail on how to solve the remaining two integrals. The exact steps involved depend on which block from Eq.~\eqref{2:eq:A1realdec} we are considering.
\subsection{$s$ channel and contact diagram squared}
Eq.~\eqref{2:eq:CEsq} is the simplest case as the $s$-channel propagator has no explicit dependence on the angle $\theta$. Then, the corresponding integral is essentially polynomial in $\cos(\theta)$. After this step, the dependence on the final integration variable $z_1$ has the form
\begin{equation}
C_{\text{CE},i}= K_n\( z,\alpha,\xi \) \int_0^1 \dd z_1 z_1^{n-1-\epsilon}\(1-z_1\)^{-1-\epsilon} \, ,
\end{equation}
the singularities at $z_1 = 0,1$ are induced by the polarisation tensors for the outbound gluons in the LCG gauge. The use of dimensional regularisation would be enough to regularise them, but we introduce the Principal-value prescription to match the procedure used in the virtual contributions
\begin{equation}
I_n\(\epsilon,\delta\)=\int_0^1 \dd z_1 \frac{z_1^{n-1-\epsilon}\(1-z_1\)^{1-\epsilon}}{\(z_1^2 + \delta^2\)\(\(1-z_1\)^2+\delta^2\)} \, .
\end{equation}
The integral yields
\begin{align}
& I_n\(\epsilon,\delta\)=\int_0^1 \dd z_1 \frac{z_1^{n-1-\epsilon}\(1-z_1\)^{1-\epsilon}}{\(z_1^2 + \delta^2\)\(\(1-z_1\)^2+\delta^2\)} \nn
& \quad = \frac{\iu}{2 \delta^2} \Gamma \(2-\epsilon\) \Gamma \(2+n-\epsilon\)\bigg[\frac{1}{\iu+2\delta} {}_2\tilF_1\(1,2+n-\epsilon,4+n-2\epsilon,\frac{-\iu}{\delta}\) \nn
&\qquad + \frac{1}{\iu-2\delta} {}_2\tilF_1\(1,2+n-\epsilon,4+n-2\epsilon,\frac{\iu}{\delta}\) \nn
&\qquad + \frac{\delta}{ \delta (3\iu - 2\delta) + 1}{}_2\tilF_1\(1,2+n-\epsilon, 4+n-2\epsilon,\frac{-\iu}{\delta-\iu}\) \nn
& \qquad + \frac{\delta}{\delta(3\iu + 2 \delta) -1} {}_2\tilF_1\(1,2+n-\epsilon, 4+n-2\epsilon,\frac{\iu}{\delta + \iu }\)\bigg].
\end{align}

\subsection{Interference between $t$ and $(s+\text{contact})$ }
The subgroup of diagrams in Eqs. \eqref{2:eq:CABint} still includes one $s$-channel propagator, thus are sensitive to the same collinear divergence we saw in the previous computation. This time however, an additional complications is introduced by higher complexity of the angular integration. First notice that we can use the decomposition
\begin{equation}
\tilA_{\text{CABint}}^{\mu\nu} = \frac{1}{s}\[\frac{\tilN_{\text{CAint}}^{\mu\nu}}{t} + \frac{\tilA_{\text{CBint}}^{\mu\nu}}{u} \],  
\end{equation}
where we made explicit the dependence on the Mandelstam of the two interference. Furthermore, after projecting with Eqs. \eqref{2:eq:tensors},\eqref{2:eq:proj} we can get the scalar coefficients
\begin{align}
  C_{\text{CABint}}^{i} & = \int \dd \Phi_2\(k+q; p_3,p_4\) \frac{1}{s}\[ \frac{\tilC^i_{\text{CA}}\(k,q,p_3,p_2\)}{t} +  \frac{\tilC^i_{\text{CB}}\(k,q,p_3,p_2\)}{u} \] \nn
  & = \int \dd \Phi_2\(k+q; p_3,p_4\) \frac{2}{s} \frac{\tilC^i_{\text{CA}}\(k,q,p_3,p_2\)}{t}\, ,
\end{align}
taking advantage of the exchange symmetry between the final state gluons. Since every $\tilC$ depends only on scalar combinations of their arguments and the dependence on the angle $\theta$ is introduced only through the combinations $\spd{\kt}{p_t}$ and $\spd{p_1}{p_3}$, we can trade the for expressions of the Mandelstam and other parameters. Then we arrive to a finite power series, for example
\begin{equation}
\frac{\tilC^i_{\text{CA},\text{CB}}\(k,q,p_3,p_2\)}{t} = \tilK^i_n t^{n-1} \, ,
\end{equation}
where, crucially, each $\tilK^i_n$ does not depend on $\theta$. Now, both Mandelstam $t,u$ are binomial expressions in $\cos\(\theta\)$, for all $n \geq 1 $, its integration is, again, simply polynomial. We can gather one more time all contributions by powers of $z_1$ and $1-z_1$ this time getting the expression
\begin{equation}
\int^1_0 \dd z_1 \int_0^{2\pi} \dd \theta \tilK^i_n t^{n-1}\(1-\delta_{0n}\) = \tilK^{\prime,i}_{n,k}\(1-\delta_{0n}\)\int_0^1 \dd z_1 \frac{z_1^{k+1-\epsilon}\(1-z_1\)^{1-\epsilon}}{\(z_1^2+\delta^2\)\((1-z_1)^2+\delta^2\)} \frac{1}{z-z_0} \, ,
\end{equation}
where $z_0 = \frac{z}{z-x}$ corresponds to the denominator of the polarisation tensor for the $t$-channel propagator of momentum $k-p_3$. We observe that $z_0 \in \(1,\infty\)$, so no additional poles are present in this type of integrals.
We can consider more generally the integrals of the form
\begin{align}\label{2:eq:PVIntegrals}
I_{kj} = \int_0^1 \dd x x^{k-1-\epsilon}(1-x)^{j-1-\epsilon}f(x)\,, \text{with} f(x)\neq 0 \forall x \in \(0,1\)\, .
\end{align}

On the other hand the case $n=0$ yields the expression
\begin{equation}\label{2:eq:nastythetaintegral}
\tilK^i_0 \int_0^{2\pi}  t^{-1} \dd \theta = \frac{-2\pi x \(z-x\) \tilK^i_0 }{\sqrt{\(z_1 - a\)^2 +b^2}} \, ,
\end{equation}
with $a = \frac{2 \xi  x ^3-x  z^2+z^3-3 \xi  x ^2 z}{(z-x ) \left(z^2-4 \xi  x ^2\right)},\, b = 2 \sqrt{-\frac{\xi  x ^4 \left(\xi (\xi +1) x ^2-\xi  x  z\right)}{(z-x )^2 \left(z^2-4 \xi  x ^2\right)^2}}$ .
If we focus first on the $\frac{1}{\epsilon}$ contribution corresponding to the Born kinematics we can impose the additional constraint $z=x\(1+\xi\)$ and Eq.~\eqref{2:eq:nastythetaintegral} reduces to the simpler
\begin{equation}
\int_0^1 \dd z_1 \tilK^i_0 \int_0^{2\pi}  t^{-1}=  \tilK^{\prime i}_{0,kj} \int_0^1 \dd z_1 z_1^{k-1-\epsilon}\(1-z_1\)^{j-1-\epsilon} \frac{1}{\(z_1-z_0\)\(z_1-z_t\)} \, ,
\end{equation}
where $z_t = \frac{-\xi}{1-\xi}$ and $z_0\xrightarrow[z=x(1+\xi)]{} \frac{1+\xi}{\xi}$.
With some manipulations, it is possible to obtain closed form for these integrals in both configurations and we elucidate the necessary assembly in Sec.~\ref{app:bad-integrals}.
\subsection{$t$-$u$ channel interference}
The final block entering the $2$GI decomposition is Eq.~\eqref{2:eq:ABint}. This case can be represented using the same pieces from the previous section. After applying the projectors~\eqref{2:eq:proj} we can get the scalar coefficients
\begin{align}
  C_{\text{ABint}}^{i} & = \int \dd \Phi_2\(k+q; p_3,p_4\) \frac{1}{tu} \frac{\tilC^i_{\text{CA}}\(k,q,p_3,p_2\)}{\spd{(k-p_3)}{n}\spd{(k-p_4)}{n}}  \nn
  & = \int \dd \Phi_2\(k+q; p_3,p_4\) \frac{\tilC^i_{\text{AB}}\(k,q,p_3,p_4\)}{k^2+q^2-s}\[\frac1t + \frac1u\]\frac{-1}{2z_0-1}\( \frac{1}{z_1-z_0} - \frac{1}{z_1-(1-z_0)}\) \nn
  & = \frac{-1}{(k^2+q^2-s)(2z_0-1)}\nn
  &\times\int \dd \Phi_2\(k+q; p_3,p_4\) \frac{2 \tilC^i_{\text{AB}}\(k,q,p_3,p_4\)}{t} \[ \frac{1}{z_1-z_0} - \frac{1}{z_1-(1-z_0)}\]\, ,
\end{align}
which can be evaluated with the same technique of the previous section up to replacing the numerators and switching $z_0\rightarrow1-z_0$ in the second term.

\subsection{Elementary integrals representation}\label{app:bad-integrals}
We need to establish the analytical solution of this integral
\begin{align}
  & \int_0^1 \dd z \frac{z^{k+1-\epsilon}(1-z)^{1-\epsilon}}{(z^2+\delta^2)((1-z)^2+\delta^2)}f(z) = \NN
  = \frac{1}{1+4\delta^2}\[ \int_0^1 \dd z z^{k+1-\epsilon}(1-z)^{1-\epsilon}\( \frac{1-2z}{z^2+\delta^2} + \frac{1-2(1-z)}{(1-z)^2+\delta^2} \) \]f(z) \NN
  = \frac{1}{1+4\delta^2}\bigg[ I_0(\delta)\( \delta_{0k} f(0) + f(1) \) + \delta_{0k}\int_0^1 \dd z \(\frac{1}{z}\)_+ (1-z)(1-2z)f(z)   \NN
    + (1-\delta_{0k})\int_0^1 \dd z z^{k-1} (1-z)(1-2z)f(z) + \int_0^1 \dd z \(\frac{1}{1-z}\)_+ z^{k+1}(1-2(1-z))f(z) \bigg] \NN
  + \frac{-\epsilon}{1+4\delta^2}\bigg[ \delta_{k0} I_1(\delta) f(0) + \delta_{0k} \int_0^1 \dd z \(\frac{\log (z)}{z}\)_+ (1-z)(1-2z)f(z) \NN
    + \delta_{0k} \int_0^1 \dd z \(\frac{1}{z}\)_+ (1-z)\log(1-z)(1-2z)f(z) \NN
    + (1-\delta_{0k}) \int_0^1 \dd z z^{k-1} \log(z) (1-z)(1-2z)f(z) \NN
    + (1-\delta_{0k}) \int_0^1 \dd z  z^k (1-z) \log(1-z) (1-2z) f(z) \NN
    + \int_0^1 \dd z \(\frac{\log(1-z)}{1-z}\)_+ z^{k+1} (1-2(1-z))f(z) \NN
    + \int_0^1 \dd z\(\frac{1}{1-z}\)_+ z^{k+1} \log(z) (1-2(1-z)) f(z) \bigg],
\end{align}
with the following definitions
\begin{align}
  I_0(\delta)  & = \int_0^1 \dd z \frac{z}{z^2+\delta^2} = - \log (\delta) + \calO (\delta)\,, \nn
  I_1 (\delta) & = \int_0^1 \dd z \frac{z \log (z)}{z^2 + \delta^2} = \frac{-1}{24}\(12 \log^2(\delta) + \pi^2\) + \calO (\delta) \, . \nonumber
\end{align}
\subsubsection{``Born'' configuration}
In the Born configuration we can rely on the following simplification
\begin{equation}
  f(z) = \frac{1}{(z-z_0)(z-z_t)} = \frac{1}{z_0-z_t}\[\frac{1}{z-z_0} - \frac{1}{z-z_t}\] \, ,
\end{equation}
with $z_0 = \frac{1+\xi}{\xi}$ and $ z_t = \frac{\xi}{\xi-1} $ ($0 < \xi $) so both lie outside the integral domain. So the integrals one by one evaluate to
\begin{align}
  & \int_0^1 \dd z \(\frac{1}{z}\)_+ (1-z)(1-2z)f(z) = \frac{1}{z_0-z_t}\[\frac{B_A(z_0)}{z_0} - \frac{B_A(z_t)}{z_t}\] \NN
  B_A(z') = \int_0^1 \dd z \frac{z' (1-z)(1-2z)+(z-z')}{z(z-z')} = 2 z' + (z'-1)(2z'-1)\log\(\frac{z'-1}{z'}\),
\end{align}
\begin{align}
  & \int_0^1 \dd z z^{k-1} (1-z)(1-2z)f(z) = \frac{B_B^k(z_0)-B_B^k(z_t)}{z_0-z_t} \NN
  B_B^k(z') = \int_0^1 \dd z z^{k-1} \frac{(1-z)(1-2z)}{z-z'} \NN
  \quad = -z^{\prime (k-1)}\[ B_{\frac{1}{z'}}\(k,0\)-3 z'B_{\frac{1}{z'}}\(k+1,0\)+2 z^{\prime 2}B_{\frac{1}{z'}}\(k+2,0\) \] \, ,
\end{align}
with $B_x(a,b)= \int_0^x \dd s s^{a-1}(1-s)^{b-1}$ being the incomplete Euler beta function,
\begin{align}
 & \int_0^1 \dd z \(\frac{1}{1-z}\)_+ z^{k+1}(1-2(1-z))f(z) =\frac{1}{z_0-z_t}\[ \frac{B_C^k(z_0)}{1-z_0}-\frac{B_C^k (z_t)}{1-z_t} \] \NN
  B_C^k(z') = \int_0^1 \dd z \frac{z^{k+1}(1-2(1-z))(1-z')-(z-z')}{(1-z)(z-z')} \NN
  \quad = \frac{-1}{z^{\prime 2}(k+2)(k+3)(k+4)}\bigg[ -(k^2 + 6 k +8 ) \hypF \( 1,k+3; k+4;\frac{1}{z'} \) \NN
    \quad + 2 (k^2 + 5k +6 )\hypF \( 1,k+4;k+5;\frac{1}{z'} \) \NN
    \quad + \gE k^2 (k+4) z^{\prime 2} + (k+4) (k^2+5k+6) z^{\prime 2} \polyg (k+2) \NN
    \quad + 5 \gE k (k+4) z^{\prime 2} + 2 k (k+4) z^{\prime 2} + 6 (k+4) z^{\prime 2} + (k+1)(k+4)z' \bigg],
\end{align}
\begin{align}
  & \int_0^1 \dd z \(\frac{\log(z)}{z}\)_+ (1-z)(1-2z)f(z) = \frac{1}{z_0-z_t}\[\frac{B_D(z_0)}{z_0}-\frac{B_D(z_t)}{z_t}\] \NN
  B_D (z') = \int_0^1 \dd z \frac{\log(z)}{z(z-z')}(z'(1-z)(1-2z)+(z-z')) \NN
  \quad = -2 z' + (1-z')(1-2z') \dilog \(\frac{1}{z'}\),
\end{align}
\begin{align}
  & \int_0^1 \dd z \(\frac{1}{z}\)_+ (1-z)\log(1-z) (1-2z) f(z) = \frac{B_E(z_0)-B_E(z_t)}{z_0-z_t} \NN
  B_E(z') = \int_0^1 \dd z \frac{(1-z)\log (1-z) (1-2 z)}{z(z-z')} \NN
  \quad = \frac{1}{6 z'}\bigg[ \(-12 +\pi^2 (3-2z')\)z' +12 \iu \pi (1-z')(2z'-1) \arcoth (1-2z') \NN
    \quad -6 (1-z')(2z'-1)\log(z'-1)\(\log(z'-1)-\log(z')\) \NN
    \quad -6 (1- z')(2z'-1)\dilog \(\frac{z'}{z'-1}\)\bigg],
\end{align}
\begin{align}
  & \int_0^1 \dd z z^{k-1} \log(z) (1-z)(1-2z) f(z)  = \frac{B_F^k(z_0)-B_F^k(z_t)}{z_0-z_t} \NN
  B_F^k(z') = \int_0^1 \dd z \frac{z^{k-1} \log(z)(1-z)(1-2z)}{z-z'} \NN
  \quad = \frac{1}{z'}\[\Phi\(\frac{1}{z'},2,k\)-3\Phi\(\frac{1}{z'},2,k+1\)+2\Phi(\frac{1}{z'},2,k+2)\],
\end{align}
where $\Phi\(z,s,a\) = \sum_{k=0}^\infty \frac{z^k}{(k+a)^s}$ is the Lerch transcendent series.
\begin{align}
  &\int_0^1\dd z z^k (1-z)\log(1-z)(1-2z)f(z)= \frac{1}{z_0-z_t}\[B_G^k(z_0)-B_G^k(z_t)\] \NN
  B_G^k(z') = \int_0^1 \dd z \frac{z^k(1-z)\log(1-z)(1-2z)}{z-z'}, %\NN  \quad = \text{add later}
\end{align}
\begin{align}
  &\int_0^1\dd z \(\frac{\log(1-z)}{1-z}\)_+ z^{k+1}(1-2(1-z)) f(z)= \frac{1}{z_0-z_t}\[B_H^k(z_0)-B_H^k(z_t)\] \NN
  B_H^k(z') = \int_0^1 \dd z \frac{\log(1-z)}{1-z} \frac{(1-z')z^{k+1} (1-2(1-z)) - (z-z')}{(z-z')(1-z')}, %\NN  \quad = \text{add later}
\end{align}
\begin{align}
  & \int_0^1 \dd z \(\frac{1}{1-z}\)_+ z^{k+1}\log(z)(1-2(1-z))f(z) = \frac{1}{z_0-z_t}\[B_I^k(z_0)-B_I^k(z_t)\] \NN
  B_I^k(z') = \int_0^1 \dd z \frac{z^{k+1}\log(z) (1-2(1-z))}{(1-z)(z-z')} \NN
  \quad = \frac{1}{z' (z'-1)}\[(1-2z')\Phi\(\frac{1}{z'}, 2 ,k+2\)+z'\polygp (k+2)\],
\end{align}
\subsubsection{General configuration}
In absence of the Born kinematics constraints the integrand function reads
\begin{equation}
  f(z) = \frac{1}{(z-z_0) \sqrt{(z-a)^2+b^2} } \, ,
\end{equation}
which makes the evaluation of the integrals substantially more complicated. Luckily, a suitable change of variables allows to rationalise f(z).
\begin{equation}
  z(t) = \frac{2[t(a^2+b^2)-a]}{t^2(a^2+b^2)-1} = \frac{2(t-t_M)}{(t-t_{+1})(t-t_{-1})},
\end{equation}
with $t \in (t_m,t_M)$ and $t_m= 1-\sqrt{1-\frac{2a-1}{a^2+b^2}} ,\: t_M = \frac{a}{a^2+b^2}$ and $t_{\pm1} = \frac{\pm 1}{\sqrt{a^2 + b^2}}$.
Under this change of variable we can rewrite several pieces of the various integrals as
\begin{subequations}
\begin{align}
  & 1-z = \frac{(t-t_m)(t-t_2)}{(t-t_{+1})(t-t_{-1})}, \\
  & J_t = \left| \frac{\dd z}{\dd t} \right|  = \frac{(t-t_{+3})(t-t_{-3})}{(t-t_{+1})^2(t-t_{-1})^2}, \\
  & f(z) = (f\cdot z)(t) = \frac{1}{z_0\sqrt{a^2+b^2}} \frac{(t-t_{+1})^2(t-t_{-1})^2}{(t-t_{+3})(t-t_{-3})(t-t_{+4})(t-t_{-4})}, \\
    &  t_2 = 1+ \sqrt{1-\frac{2a-1}{a^2+b^2}}, \\
    & t_{\pm 3} = \frac{1}{a \pm \iu b}, \\
  & t_{\pm 4} = \frac{1}{z_0}\[ 1 \pm \frac{z_0}{\left|z_0 \right|}\sqrt{1-z_0\frac{2a-z_0}{a^2+b^2}} \], \\
  & t_{\pm 5} = 2 \pm \sqrt{4 - \frac{4a-1}{a^2+b^2}},
\end{align}
\end{subequations}
then it is more convenient to rescale all this quantity so that the integral domain is $(0,1)$. Explicitly
\begin{equation}
  t(s) = \(t_M-t_m\)s + t_m,\quad s_h = \frac{t_h-t_m}{t_M-t_m}, \quad J_s = \left|\frac{\dd s}{\dd t} \right| = t_M-t_m \,.
\end{equation}
Finally we can write the decomposition
\begin{subequations}
  \begin{align}
    & \int_0^1 \dd z \(\frac{1}{z}\)_+ (1-z)(1-2z)f(z)  \NN
    = \frac{1}{z_0\sqrt{a^2 +b^2}}\int_0^1\dd s\(\frac{(s-s_{+3})(s-s_{-3})}{s-1}\)_+\frac{s(s-s_2)(s-s_{+5})(s-s_{-5})}{(s-s_{+3})(s-s_{-3})(s-s_{+4})(s-s_{-4})}, \nonumber \\
    & \int_0^1 \dd z z^{k+1}(1-z)(1-2(1-z))f(z) \NN
    = \(\frac{2}{J_s}\)^k\frac{\partial_{s_{+1}}^k \partial_{s_{-1}}^k \[G_1^k\(s_{+4}\)-G_1^k\(s_{-4}\)\]}{z_0\sqrt{a^2+b^2}(s_{+4}-s_{-4})(k!)^2} \NN
    G_1^k(s') = \int_0^1 \dd s \frac{s(s-1)^{k-1}(s-s_2)(s-s_{+5})(s-s_{-5})}{(s-s_{+1})(s-s_{-1})(s-s')}, \nonumber \\
    & \int_0^1 \dd z \(\frac{1}{1-z}\)_+ z^{k+1}[1-2(1-z)]f(z) \NN
    = \frac{-1}{z_0\sqrt{a^2+b^2}}\(\frac{2}{J_s}\)^{k+2}\bigg[\frac{\partial_{s_{+1}}^k \partial_{s_{-1}}^k}{k!^2} \sum_{h\in\pm1,2\pm4}\(\prod_{j\neq h}\frac{1}{s_h-s_j}\) G_2^k(s_h) \NN
      - \frac{(-1)^{k+1}s_{+5}s_{-5}}{(s_{+1}s_{-1})^ks_{+4}s_{-4}s_{+3}s_{-3}}\sum_{h \in \pm 1, 2}\(\prod_{j \neq h} \frac{1}{s_h-s_j}\)G_3(s_h)  \bigg], \nonumber \\
    & G_2^k\(s_h\) = \int_0^1 \dd s \(\frac{1}{s}\)_+ \frac{(s-1)^{k+1}(s-s_{+5})(s-s_{-5})}{s-s_h} \NN
    = \begin{cases}
      \frac{-1}{2} + s_h -s_{+5}-s_{-5} & \\
       +\frac{1}{s_h}\[(s_h-1)(s_h-s_{-5})(s_h-s_{+5})(\log(1-s_h)-\log(-s_h))\] & \hspace{-5em}k = 0\\
      \frac{1}{6}\[2+6s_h^2+9s_{-5}+9s_{+5}-3s_h(3+2s_{-5}+2s_{+5})+6s_{-5}s_{+5}\] & \\
       +\frac{\(1-s_h\)^2}{s_h}(s-s_{+5})(s-s_{-5})(\log(1-s_h)-\log(-s_h)) & \hspace{-5em}k = 1\\
      \frac{1}{s_h}\bigg[ \frac{1}{12} \bigg(12 s_h^4 -6 s_h^3 (5 + 2s_{-5}+2s_{+5})+2 & \\
        s_h^2(11+15s_{+5}+3s_{-5}(5+2s_{+5})) -s_h (3+22s_{+5}+s_{-5}(22+30s_{+5})) & \\
        +s_{-5}(11+15 s_{+5})\bigg) -(s_h-1)^3(s-s_{+5})(s-s_{-5}) & \\
        \times(\log(1-s_h)-\log(-s_h))\bigg] & \hspace{-5em}k = 2\\
      \frac{-1}{s_h}\bigg[ \frac{1}{60}\bigg( -60 s_h^5 +30 s_h^4(7 + 2s_{-5}+2s_{+5})-10 s_h^3(26 + 21 s_{+5} + 3 s_{-5} (7+2s_{+5})) & \\
        + 5 s_h^2(25 +52s_{+5} +s_{-5} (52 +42 s_{+5})) -s_h (12+125 s_{+5} 5s_{-5}(25 +52 s_{+5})) & \\
        - 25 s_{+5}\bigg) - (-1+s_h)^4 (s_h -s_{+5})(s_h-s_{-5})(\log(1-s_h)-\log(-s_h)) \bigg] & \hspace{-5em}k =3
    \end{cases}\\
    & G_3\(s_h\) = \int_0^1\dd s \(\frac{1}{s}\)_+ \frac{(s-s_{+3})(s-s_{-3})}{s-s_h} \NN
    = 1 + \frac{(s_h-s_{-3})(s_h-s_{+3})}{s_h}\[\log(1-s_h)-\log(-s_h)\].
  \end{align}
\end{subequations}
\section{Virtual corrections}
\begin{figure}[h]
  \centering
  \begin{minipage}[c]{0.3 \textwidth}
    \centering
    \includegraphics[width=0.7 \linewidth]{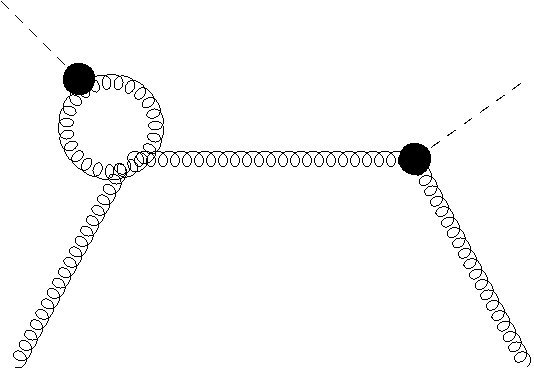}
    %\caption{First diagram}
    %\label{fig:doubleprop}
  \end{minipage}
  \hspace{0.01 \textwidth}
  \begin{minipage}[c]{0.3 \textwidth}
    \centering
    \includegraphics[width=0.88 \linewidth]{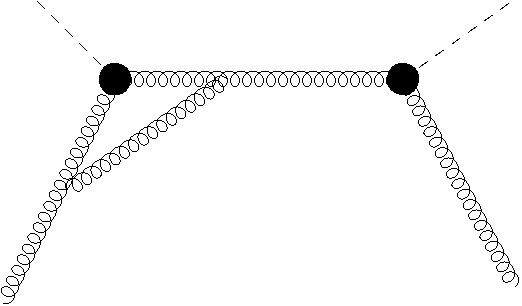}
    %\caption{Second diagram}
    %\label{fig:triangle}
  \end{minipage}
  \hspace{0.01 \textwidth}
  \begin{minipage}[c]{0.3 \textwidth}
    \centering
    \includegraphics[width=1 \linewidth]{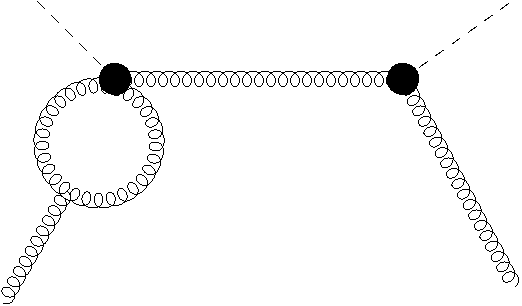}
    %\caption{Third diagram}
    %\label{fig:k1bubble}
  \end{minipage}
  \caption{Diagrams contributing to one-loop virtual correction to Higgs DIS.}
  \label{fig:diagramsoneloop}
\end{figure}

Now we focus on the virtual contribution to the $2$GI part of $A_1^{\mu\nu}$ in the light-cone gauge. The diagrams contributing to this amplitude are shown in Fig.(\ref{fig:diagramsoneloop}) and in the following sections we will briefly present how we computed each contribution.

\subsection{Diagrams}\label{2:ssec:Vdiagrams}
The contribution from the first diagram in Fig.(\ref{fig:diagramsoneloop}) is given by
\begin{align}  \label{eq:dubleprop}
  L_0^{\mu \nu} &=\frac{1}{2}\[2\pi \delta\((k+q)^2\)\] \frac{\delta_{mn}}{2 C_A C_F }\intv M^{\alpha\beta}_{ab}\(v,q+v\)\frac{d_{\alpha\alpha'}\(v,n\)\delta_{ac}d_{\beta\beta'}\(v+q,n\)\delta_{bd}}{v^2\(v+q\)^2} \nn
  &\times W^{\mu\alpha'\beta'\rho}_{mcdr}\(k,-v ,q+v ,-(k+q)\)d_{\rho\rho'}\(k+q,n\)\delta_{rs}M^{*,\nu \rho'}_{nr}\(k,k+q\)\, ,
\end{align}
where
\begin{align}
  W^{\alpha\beta\gamma\delta}_{abcd}\(k_1,k_2,k_3,k_4\) &= - \iu g_s^2 \bigg[ f_{xac}f_{xbd}\(g^{\alpha\beta}g^{\gamma\delta}-g^{\alpha\delta}g^{\beta\gamma}\) +  f_{xad}f_{xbc}\(g^{\alpha\beta}g^{\gamma\delta}-g^{\alpha\gamma}g^{\beta\delta}\) \nn
    &+  f_{xab}f_{xcd}\(g^{\alpha\gamma}g^{\beta\delta}-g^{\alpha\delta}g^{\beta\gamma}\) \bigg]\, ,
\end{align}
is the usual QCD four gluon vertex, the first square bracket is the remaining factor from the the Born phase-space and $\frac{1}{2}$ is overall combinatorial factor.
After rewriting the tensors in the base of \eqref{2:eq:tensors} and evaluating the loop integrals we are left with
\begin{equation}
  L_0^{\mu \nu} =  \frac{\pi C^2 Q^2}{4} \frac{\as}{4 \pi} C_A \[-\frac{1}{2}\(\xi-1\)Q^4 g^{\mu \nu} + \frac{x Q^2}{2} \(p_1^\mu \kt^\nu + p_1^\nu \kt^\mu\) + \frac{Q^2}{\xi} \kt^\mu \kt^\nu\]\,. 
\end{equation}
    Remarkably, this result is quite simple and free from infrared and ultraviolet divergences.
    %\FS[I am not sure the overall factor is the correct one, double check with Anna?]
    The contribution from the second diagram in Fig.(\ref{fig:diagramsoneloop}) is the more complicated of the three. The overall expression can be written as
    \begin{align}    \label{eq:triangint}
        L_1^{\mu\nu}(k,q,n)&= \[2\pi \delta\((k+q)^2\)\] \frac{\delta_{mn}}{2 C_A C_F } \intv V^{\mu\alpha\beta}\(k,v-k,-v\)f_{mab} \nn
        &\times M^{\alpha'\gamma'}_{cd}\(k-v,k+q-v\)V^{\gamma\beta'\epsilon}\(k+q-v,v,-k-q\)f_{fgh} \nn
        &\times \frac{d_{\alpha \alpha'}\(k-v,n\)\delta_{ac}d_{\beta\beta'}\(v,n\)\delta_{df}d_{\gamma\gamma'}\(k+q-v,n\)\delta_{bg}}{\(k-v\)^2v^2\(k+q-v\)^2} \nn
        &\times d_{\epsilon\epsilon'}\(k+q,n\)\delta_{hi} M^{*,\nu\epsilon'}_{ni}\(k,k+q\)\,.
    \end{align}
    Again we decompose the integral into tensors and compute the integrals. In this case the result is quite more complicated, it can be written as
    \begin{equation}
      L_1^{\mu\nu} =  \left[T_{k_1k_1} k_1^\mu k_1^\rho +T_{k_2k_2} k_2^\mu k_2^\rho +\frac{1}{2}T_{k_1k_2}( k_1^\mu k_2^\rho+k_2^\mu k_1^\rho) +T_{g} g^{\mu\rho} +\ldots \right].
    \end{equation}
    Finally, the contribution from the last diagram in Fig.(\ref{fig:diagramsoneloop}) is written as
    \begin{align}     \label{eq:bubbleint}
        L_2^{\gamma\nu}(k,q,n) &= \[2\pi \delta\((k+q)^2\)\] \frac{\delta_{mn}}{2 C_A C_F } \intv V^{\mu\alpha\beta}\(k,v-k,-v)\)f_{mab} \nn
        & \times \frac{d_{\alpha\alpha'}\(k-v\)\delta_{ac}d_{\beta\beta'}\(v\)\delta_{bd}}{\(k-v\)^2v^2}E^{\alpha'\beta'\rho}\(k-v,v,-(k+q)\)f_{cde} \nn
        & \times d_{\rho\rho'}\(k+q,n\)\delta_{ef}M^{*,\nu\rho'}_{nf}\(k,k+q\)\,,
    \end{align}
    the tree level vertex, $T_{\gamma \delta}(k_1,k_2)$, and the sum over polarisation of the on-shell gluon, $d_{\nu\delta}(k_2,n)$, are the same as the ones defined in the previous sections, while in this case the numerator is 
    \begin{equation}
      \begin{aligned}
        N^{\mu \nu\alpha\sigma\beta\gamma\rho\delta}(k_1,k_2,n) = \mathrm{i}g^2 c^2\, C_A \delta_{a,b} &\left[g^{\mu \rho } (k+k_1)^\sigma +g^{\mu \sigma } (k-2 k_1)^\rho +g^{\rho \sigma } (k_1-2 k)^\mu  \right]\\
        &\left[ g^{\gamma \delta } (k_2-2 k)^\nu +g^{\gamma\nu } (k-2 k_2)^\delta +g^{\delta \nu } (k+k_2)^\gamma \right]\\
        &\left[ (k_1 -k)^\beta (k_2-k)^\alpha -g^{\alpha \beta } (k_1-k)\cdot(k_2-k) \right].
      \end{aligned}
    \end{equation}
    
    \subsection{Scalar integrals for loops}\label{2:ssec:ScalarLoopsLCG}
    The integrals in Eqs. \eqref{eq:dubleprop},\eqref{eq:triangint} and \eqref{eq:bubbleint} can be reduced to scalar integrals by means of Passarino-Veltman decomposition and then projected on the tensor basis chosen in the previous section. Then, in general we need to evaluate integrals in the form
    \begin{equation*}
      \int \frac{d^d v}{(2 \pi)^d}\frac{N(v,l_1,\ldots,l_m, n, q_1\ldots q_j)}{v^2(v-l_1)^2(\ldots) (v-l_m)^2 (v \cdot n - q_1 \cdot n )^{\alpha_1}(\ldots)(v \cdot n - q_j \cdot n )^{\alpha_j}},
    \end{equation*}
    where there are  $m + 1$ covariant denominators and $j$  non-covariant ones. The numerator depends on all external momenta through $l_m,q_j$. Since we are interested mostly in the loops with two or three propagators, we introduce the following notation:
    \begin{equation}
      \begin{aligned}
	&B^{\mu_1\ldots \mu_i}_{\beta, i} = \int \frac{d^d v }{(2 \pi)^d}\frac{v^{\mu_1} \ldots v^{\mu_i}}{v^2 (v-l)^2(v_+ - q_{1\,+})^{\alpha_1}\ldots (v_+ -  q_{j\,+})^{\alpha_j}},\\
	&T^{\mu_1\ldots \mu_i}_{\beta, i} = \int \frac{d^d v }{(2 \pi)^d}\frac{v^{\mu_1} \ldots v^{\mu_i}}{v^2 (v-l)^2 (v-p)^2(v_+ - q_{1\,+})^{\alpha_1}\ldots (v_+ -  q_{j\,+})^{\alpha_j}},
      \end{aligned}
	 \label{2:eq:integrals}
    \end{equation}
    where $\beta = \sum_{m=1}^j \alpha_m$ and $p_+ = p \cdot n$ is a shorthand for the contraction of a vector with the light cone gauge axis.
    Now, as an example we consider explicitly
          \begin{equation}
	    B_{10} = \int \frac{d^d v}{(2 \pi)^d} \frac{1}{v^2 (v-l)^2 (k \cdot n)},
	  \end{equation}
	  where we assume in general $l^2 \neq 0$. After Feynman parametrisation, the integral becomes
	 \begin{equation}
	   B_{1,0} = \int_{0}^{1}dx \int\frac{d^d k}{(2 \pi)^d}\frac{1}{k\cdot n}\frac{1}{\left[k^2 -2P\cdot n + M^2\right]},
	 \end{equation}
	 where $P^\mu = x\, l^\mu$ and $M^2 = x l^2$.
         
         Here we can apply the general identity from Ref.~\cite{Pritchard:1978ts}
	 \begin{equation}\label{2:eq:PriSti-GenInt}
	   \int \frac{\dd^D v}{(2 \pi)^d}\frac{1}{(v^2 -2 P \cdot v + M^2)^\alpha}\frac{1}{(v\cdot n)^\beta} = \frac{i(-1)^{\alpha}}{(4\pi)^{\frac{d}{2}}}\frac{\Gamma\left(\alpha - \frac{d}{2}\right)}{\Gamma(\alpha)}\frac{1}{(P^2-M^2)^{\alpha -\frac{d}{2}}(P\cdot n)^\beta},
	 \end{equation} 
	 where $\alpha$ and $\beta$ are natural numbers. Then we obtain
	 \begin{equation}
	   B_{1,0} = \frac{i}{16\pi^2}\left(\frac{4 \pi}{-l^2}\right)^\epsilon\frac{\Gamma(1+\epsilon)}{\epsilon} \int_{0}^{1}\dd x \frac{1}{x l_+}x^{-\epsilon}(1-x)^{-\epsilon},
	 \end{equation}
	 where we set $D=4-2\epsilon$. Note that if $l^2=0$, this integral vanishes. By combining more Feynman parametrisations with Eq.~\eqref{2:eq:PriSti-GenInt}, it's straightforward to show a generalisation of the integral for any number of non-covariant denominators to any natural power.
         \begin{align}
           B_{\beta,0} & = \intv \frac{1}{v^2\(v-l\)^2}\frac{1}{\(v_+-q_{1+}\)^{\alpha_1}\(v_+ - q_{2+}\)^{\alpha_2}\ldots\(v_+-q_{3+}\)^{\alpha_n}} \nn
           & = \frac{\iu}{\(4 \pi\)^2}\(\frac{4 \pi}{-l^2}\)^\epsilon \frac{\Gamma\(1+\epsilon\)}{\epsilon}\int_0^1\dd z \frac{z^{-\epsilon}(1-z)^{-\epsilon}}{\(zl_+-q_{1+}\)^{\alpha_1}\(zl_+ - q_{2+}\)^{\alpha_2}\ldots\(zl_+-q_{3+}\)^{\alpha_n}}, \label{2:eq:B-0}
         \end{align}
	with $\beta = \sum_{i=1}^n \alpha_i$. The second integral in Eq.~\eqref{2:eq:integrals} can be evaluated with the same, albeit longer, manipulations, yielding
        \begin{align}	     \label{2:eq:Tmassive}
          T_{\beta,0} & = \frac{i}{16\pi^2}\left(\frac{4 \pi}{-2 l\cdot p }\right)^\epsilon\frac{1}{-2 l\cdot p}\frac{\Gamma(1+\epsilon)}{\epsilon}\frac{b^{-1-\epsilon}}{x} \nn
          &\times \int_{0}^{x}d z f(l_{+} \,z) z^{-\epsilon}(1-z)^{-1-\epsilon}{}_2F_1\left(1-\epsilon,1;1+\epsilon; \frac{(1-b\,x)}{b\,x}\frac{z}{1-z} \right) \nn
          &+\frac{i}{16\pi^2}\left(\frac{4 \pi}{2 l\cdot p }\right)^\epsilon\frac{1}{2 l\cdot p}\frac{\Gamma(1+\epsilon)}{\epsilon}\,b^{-\epsilon}(b\,x-1)^{\epsilon}\nn
          &\times \int_{x}^{1}d y f(l_{+} y)(1-y)^{-\epsilon}(b\,x-y)^{-1-\epsilon}{}_2F_1\left(1+\epsilon,-\epsilon;1-\epsilon; \frac{(1-y)b\,x}{b\,x-y} \right)\nn
          &+\frac{i}{16\pi^2}\left(\frac{4 \pi}{-2 l\cdot p }\right)^\epsilon\frac{1}{-2 l\cdot p}\frac{\Gamma(1+\epsilon)}{\epsilon}\,\frac{(b\,x-1)^{\epsilon}(1-x)^\epsilon}{ (b-1)^{\epsilon} }\nn
          &\times\int_{x}^{1}d y f(l_{+} y)(1-y)^{-\epsilon}(b\,x-y)^{-1-\epsilon}{}_2F_1\left(1+\epsilon,-\epsilon;1-\epsilon; \frac{(b-1)\,x}{(1-x)}\frac{(1-y)}{(b\,x-y)} \right),
	\end{align}
	where $f(k_+) = \frac{1}{(k_+-q_{1+})^{\alpha_1}\ldots(k_+ - q_{n+})^{\alpha_n}}$, $b=\frac{l^2}{2\spd{l}{p}}$, and $\beta = \sum_{i=1}^n \alpha_i$. Given Eqs.~\eqref{2:eq:B-0} and \eqref{2:eq:Tmassive}, the expressions for the general integrals~\eqref{2:eq:integrals} can be obtained by differentiation.
         \subsection{Ultraviolet effective counterterm}
         Naturally the loop diagrams will include a UV divergent part. This will need to be subtracted by a suitable renormalisation of the effective vertex, which we perform with an ad-hoc counterterm.
         \begin{align}
           &CT^{\mu\nu}(k,k_2=k+q,n)  =  -\frac{\alpha_s C\, C_A }{4 \pi \epsilon}\bigg[CT_{11}(k_1,k_2,n) k_1^\mu k_1^\nu + CT_{22}(k_1,k_2,n) k_2^\mu k_2^\nu  \nn
             & \qquad +  CT_g(k_1,k_2,n) g^{\mu \nu}+CT_{12}(k_1,k_2,n) k_1^\mu k_2^\nu + CT_{21}(k_1,k_2,n) k_2^\mu k_1^\nu +\ldots \bigg],
           \label{eq:CTtotal}
         \end{align}
         where the ellipses stand for the tensor structures proportional to the gauge vector $n$. Indeed, these terms, will always give a vanishing contribution when contracted with the lower part of Eqs.~\eqref{2:eq:tensorfact}.
         Schematically, the coefficients of the other parts can be extracted by computing each of the diagrams in Figs.~\ref{fig:diagramsoneloop} in dimensional regularisation, $D=4-2\epsilon$, while keeping every external leg off-shell. The off-shellness and PV-regulator $\delta$ will keep at bay the IR and gauge-induced singularities, so that the $\frac1\epsilon$ poles can be traced back to the ultraviolet contribution.         
         We report the explicit expression for the coefficients in Eq.(\ref{eq:CTtotal}), as provided by the author of Ref.~\cite{Rinaudo:2022nar}:
         \begin{subequations}
           \begin{align}
             &CT_{1 1} =-4 x \ln (\delta )- 2x +\frac{x^3 \ln (x)\left[(k_1^2)^2 \left(x^2-2\right)-4 k_1^2 k_1\cdot k_2 (x-2)-4 (k_1\cdot k_2)^2\right]}{2  (x (k_1^2 x-2 k_1\cdot k_2)+k_2^2)^2}\nn
             &\quad+\frac{x k_2^2 \ln (x)\left[ x \left(k_1^2 \left(x^2+3 x-6\right)+2 k_1\cdot k_2 \left(2 x^2-7 x+7\right)\right)+k_2^2 (3 x-4)\right]}{2  (x (k_1^2 x-2 k_1\cdot k_2)+k_2^2)^2}\nn
             &\quad+\frac{x k_1^2\ln (1-x)\left[ 4 k_1\cdot k_2 x^3-x^4k_1^2+ k_2^2 \left(3 x^3-9 x^2+6 x-2\right)\right]}{2 (x (k_1^2 x-2 k_1\cdot k_2)+k_2^2)^2} \nn
             &\quad+\frac{x \ln (1-x)\left[-4x^2 (k_1\cdot k_2)^2 -2 k_1\cdot k_2\, k_2^2 x(x-3) +(k_2^2)^2 (x-2)\right]}{2 (x (k_1^2 x-2 k_1\cdot k_2)+k_2^2)^2} \nn
             &\quad+\frac{k_2^2 (x-1) x \ln (x-1) \left[x (k_1^2 (3 x-2)-4 k_1\cdot k_2 x+2 k_1\cdot k_2)+k_2^2 (2 x-1)\right]}{2 (x (k_1^2 x-2 k_1\cdot k_2)+k_2^2)^2}; \label{eq:ctk1k1}\\
             &CT_{2 2} = -\frac{4 \ln (\delta )}{x}-\frac{2}{x} +\frac{x k_1^2 \ln (x)\left[k_1^2  \left(5 x^2-x-2\right)+2 k_1\cdot k_2 (5-9 x) \right]}{2 (x (k_1^2 x-2 k_1\cdot k_2)+k_2^2)^2}\nn
             &\quad+\frac{\ln (x)\left[k_1^2 k_2^2 \left(2 x^3-2 x^2+11 x-7\right)+2 x (k_2^2-2 k_1\cdot k_2)^2\right]}{2 (x (k_1^2 x-2 k_1\cdot k_2)+k_2^2)^2}\nn
             &\quad+\frac{x^2\ln (x-1)\left[ (k_1^2)^2 x (2 x-1)+k_1^2 k_1\cdot k_2 (2-6 x)+4 (k_1\cdot k_2)^2\right]}{2 x (x (k_1^2 x-2 k_1\cdot k_2)+k_2^2)^2}\nn
             &\quad+\frac{k_2^2\ln (x-1)\left[ x \left(k_1^2 \left(2 x^3-6 x^2+9 x-3\right)-4 k_1\cdot k_2\right)+(k_2^2)^2\right]}{2 x (x (k_1^2 x-2 k_1\cdot k_2)+k_2^2)^2}\nn
             &\quad+\frac{k_1^2 (x-1) \ln (1-x) (k_1^2 x (x-2)-2 k_1\cdot k_2 (x-2)+k_2^2 (2 x-3))}{2 (x (k_1^2 x-2 k_1\cdot k_2)+k_2^2)^2};\label{eq:ctk2k2} \\
             & CT_{12} =4 \ln (\delta )+\frac{2 x^2+x+2}{x} \nn
             &\quad+\frac{k_1^2 x^3\ln (1-x)\left[ k_1^2 x \left(2 x^2-3 x+2\right)-2  k_1\cdot k_2 \left(5 x^2-9 x+6\right)\right]}{2 x^2 (x (k_1^2 x-2 k_1\cdot k_2)+k_2^2)^2}\nn
             &\quad+\frac{x^2\ln (1-x)\left[4 (k_1\cdot k_2)^2 \left(3 x^2-6 x+4\right)+ k_1^2 k_2^2  \left(-3 x^3+12 x^2-15 x+8\right)\right]}{2 x^2 (x (k_1^2 x-2 k_1\cdot k_2)+k_2^2)^2}\nn
             &\quad+\frac{k_2^2\ln (1-x)\left[2 k_1\cdot k_2 x\left(x^3-7 x^2+12 x-8\right)-k_2^2 \left(x^3-4 x^2+6 x-4\right)\right]}{2 x^2 (x (k_1^2 x-2 k_1\cdot k_2)+k_2^2)^2}\nn
             &\quad+\frac{x^2k_1^2\ln (x-1)\left[ k_1^2 x \left(4 x^3-6 x^2+4 x-1\right)-2  k_1\cdot k_2 \left(8 x^3-12 x^2+7 x-1\right)\right]}{2 (x (k_1^2 x-2 k_1\cot k_2)+k_2^2)^2} \nn
             &\quad+\frac{x \ln (x-1)\left[4 (k_1\cdot k_2)^2x \left(4 x^2-6 x+3\right)+ k_1^2k_2^2 \left(8 x^3-15 x^2+12 x-3\right)\right]}{2 (x (k_1^2 x-2 k_1\cdot k_2)+k_2^2)^2}\nn
             &\quad+\frac{k_2^2\ln (x-1)\left[ -2 k_1\cdot k_2 x\left(6 x^2-9 x+5\right)+k_2^2 \left(2 x^2-3 x+2\right)\right]}{2 (x (k_1^2 x-2 k_1\cdot k_2)+k_2^2)^2}\nn
             &\quad+\frac{x^2 k_1^2 \ln (x)\left[k_1^2  \left(-4 x^4+6 x^3-6 x^2+x+2\right)+2 k_1\cdot k_2  \left(8 x^3-12 x^2+11 x-5\right)\right]}{2 (x (k_1^2 x-2 k_1\cdot k_2)+k_2^2)^2} \nn
             &\quad+\frac{x \ln (x)\left[k_1^2 k_2^2 \left(-8 x^3+11 x^2-12 x+7\right)-4 (k_1\cdot k_2)^2 x \left(4 x^2-6 x+3\right)\right]}{2 (x (k_1^2 x-2 k_1\cdot k_2)+k_2^2)^2}\nn
             &\quad+\frac{x k_2^2 \ln (x)\left[2 k_1\cdot k_2  \left(6 x^2-5 x+1\right)+k_2^2 (3-4 x)\right]}{2 (x (k_1^2 x-2 k_1\cdot k_2)+k_2^2)^2}; \label{eq:ctk1k2}\\
             &CT_{21} =4 \ln(\delta )+\frac{k_1^2 k_2^2 (x-1)^3 x \ln (x-1)}{(x (k_1^2 x-2 k_1\cdot k_2)+k_2^2)^2}-\frac{k_1^2 k_2^2 (x-1)^3 \ln (1-x)}{(x (k_1^2 x-2 k_1\cdot k_2)+k_2^2)^2}\nn
             &\quad -\frac{(x-1) \ln (x)}{(x (k_1^2 x-2 k_1\cdot k_2)+k_2^2)^2} \bigg(k_2^2 x \left(k_1^2 \left(x^2+3\right)-4 k_1\cdot k_2\right)\nn
             &\qquad+k_1^2 x^2 (k_1^2 x+k_1^2-4 k_1\cdot k_2)+(k_2^2)^2 (x+1)\bigg); \label{eq:ctk2k1}\\
             CT_{g} &=-4 k_1\cdot k_2 \ln (\delta )\nn
             &\quad+\frac{k_2^2 \ln(x-1) \left(k_1^2 x \left(2 x^3-3 x^2+4 x-1\right)-2 k_1\cdot k_2 x \left(x^2+1\right)+k_2^2 \left(x^2+1\right)\right)}{2 x (x (k_1^2 x-2 k_1\cdot k_2)+k_2^2)}\nn
             &\quad +\frac{k_1^2\ln (x)\left[k_1^2 \left(x-x^3\right)+4 x k_1\cdot k_2 (x-1) + k_2^2 \left(-2 x^3+x^2-4 x+3\right)\right]}{2 (x (k_1^2 x-2 k_1\cdot k_2)+k_2^2)} \nn
             &\quad+\frac{2 k_2^2\ln (x)\left[ k_1\cdot k_2  \left(x^2+2 x-1\right)- x k_2^2 \right]}{2 (x (k_1^2 x-2 k_1\cdot k_2)+k_2^2)}. \label{eq:g}
           \end{align}
         \end{subequations}
         Unlike the case of computing the counterterm in a covariant gauge, our setup of gauge-fixing and regulators makes it so that Eq.~\eqref{2:eq:tensorfact} depends on all the momenta of the process.
         \subsection{Virtual contribution to the coefficient function}
         in this last section we report the singular part of the virtual contribution from section~\ref{2:ssec:Vdiagrams}, computed using the tools in Sec.~\ref{2:ssec:ScalarLoopsLCG} and after the subtraction of the counterterm from the previous section.
         \begin{equation}
           A_{1V}^{\mu\nu} = \frac{C_A}{\epsilon}\(\frac{\pi C^2 Q^2}{4}\)\[C_{0V} g^{\mu\nu} + C_{1V} \frac{p_{1}^{\mu}p_{1}^{\nu}}{Q^2} + C_{2V} \frac{\(p_{1}^\mu \kt^\nu + p_{1}^\nu \kt^\mu\)}{\sqrt{\xi}Q^2} + C_{3V} \frac{\kt^\mu \kt^\nu}{\xi Q^2}\] \, ,
         \end{equation}

         \begin{subequations}
           \begin{align}
             & C_{0V} = \frac{\left(\xi ^3+\xi ^2-2\right)   \log   \left(\frac{\xi }{\xi +1}\right)}{(\xi +1)^2 },\\
             & C_{1V} = -\frac{2    \xi  x ^2   \log \left(\frac{\xi }{\xi   +1}\right)}{(\xi +1) },\\
             & C_{2V} = \frac{-\sqrt{\xi } (3 \xi   +5) x    \log \left(\frac{\xi }{\xi +1}\right)}{\sqrt{2} (\xi +1)},\\
             & C_{3V} = -\frac{ \xi  \left(2 \xi ^2+5 \xi +5\right)     \log \left(\frac{\xi }{\xi +1}\right)}{(\xi +1)^2}.
           \end{align}
         \end{subequations}
	
        \end{appendices}

	\backmatter

        \phantomsection
 	\addcontentsline{toc}{chapter}{Bibliography}
	\bibliographystyle{sapthesis}
	\bibliography{Thesis_references_complete}

\end{document}